\documentclass[twoside,12pt]{article}
\usepackage{ulem}
\usepackage{graphicx}
\usepackage{type1cm}
\usepackage{eso-pic}
\usepackage{epsfig}
\usepackage{color}
\usepackage{cite}
\usepackage{amsmath}
\usepackage{wrapfig}
\usepackage{amssymb}
\usepackage{multirow}
\usepackage{braket}
\usepackage{slashed}
\usepackage{url}
\usepackage[utf8]{inputenc}
\usepackage{booktabs}
\usepackage{dsfont}

\newcommand{\et}{\mbox{$\eta$}}
\newcommand{\etp}{\mbox{$\eta'$}}
\newcommand{\etap}{\mbox{$\eta'$}}
\newcommand{\po}{\mbox{$\pi^0$}}
\newcommand{\pio}{\mbox{$\pi^0$}}
\newcommand{\pim}{\mbox{$\pi^-$}}
\newcommand{\pip}{\mbox{$\pi^+$}}
\newcommand{\M}{\mathcal{M}}
\newcommand{\Order}[1]{\mathcal{O}#1}
\newcommand{\Lagr}{\mathcal{L}}
\newcommand{\BR}{{\cal B}}
\newcommand{\T}{\,\text{T}}
\newcommand{\m}{\,\text{m}}
\newcommand{\cm}{\,\text{cm}}
\newcommand{\mm}{\,\text{mm}}

\newcommand{\MeV}{\,\text{MeV}}
\newcommand{\GeV}{\,\text{GeV}}

\newcommand{\nb}{\,\text{nb}}
\newcommand{\pb}{\,\text{pb}}
\newcommand{\fb}{\,\text{fb}}
\newcommand{\ab}{\,\text{ab}}
\newcommand{\BW}{\text{BW}}

\newcommand{\diff}{\text{d}}
\renewcommand{\Re}{\text{Re}\,}
\renewcommand{\Im}{\text{Im}\,}

\DeclareMathOperator{\arccot}{arccot}

\newcommand{\CT}{$c$--$\tau$}
\newcommand{\FV}{F_\pi^V}
\newcommand{\alphaem}{\alpha}

\begin{document}
\title{ \vspace{1cm} What can we learn about light-meson interactions
at electron--positron colliders?}
\author{Shuang-shi Fang$^{1,2}$, Bastian Kubis$^3$, Andrzej Kup\'s\'c$^{4,5}$ 
\\
\small $^1$Institute of High Energy Physics, Chinese Academy of Sciences, \\
\small 100049 Beijing, People's Republic of China\\
\small $^2$University of  Chinese Academy of Sciences, 
100049 Beijing, People's Republic of China\\
\small $^3$Helmholtz-Institut f\"ur Strahlen- und Kernphysik (Theorie) and
Bethe Center for Theoretical Physics,\\ \small Universit\"at Bonn, 53115 Bonn, Germany\\
\small $^4$Uppsala University, Box 516, 75120 Uppsala, Sweden\\
\small $^5$National Centre for Nuclear Research, 02-093 Warsaw, Poland
}
\maketitle

\begin{abstract}
Precision studies at electron--positron colliders with center-of-mass energies in the charm--tau region and below have strongly contributed to our understanding of light-meson interactions at low energies. We focus on the processes involving two or three light mesons with invariant masses below nucleon--antinucleon threshold. A prominent role is given to the interactions of the nine lightest pseudoscalar mesons (pions, kaons, $\eta$, and $\eta'$) and the two narrow neutral isoscalar vector mesons $\omega$ and $\phi$.  Experimental methods used to produce the mesons are reviewed as well as theory tools to extract properties of the meson--meson interactions. Examples of recent results from the DA$\Phi$NE, BEPCII, and VEPP-2000 colliders are presented. In the outlook we briefly discuss prospects for further studies at future super-charm--tau factories. 
\end{abstract}

\tableofcontents
\newpage

\section{Introduction}
\subsection{Light mesons and their interactions}\label{sec:intro}

In principle, the spectrum and interactions of all hadrons can be deduced from Quantum Chromodynamics (QCD), the  fundamental theory of the strong interactions. In general, up to this date this is an impossible task, we are often surprised by the discoveries of unexpected hadronic states~\cite{Klempt:2007cp,Chen:2016qju,Esposito:2016noz,Lebed:2016hpi,Ali:2017jda,Olsen:2017bmm,Karliner:2017qhf,Guo:2017jvc}
and do not understand their interactions. In the long term, the most promising approach to connect experimental data and the fundamental short-distance processes is lattice QCD. The vast majority of lattice QCD calculations to date have focused on the properties of states that are stable with respect to the strong forces, and these calculations have matured to the level where they can be considered realistic: they reproduce the masses of known hadronic states or calculate form factors and pseudoscalar meson decay constants~\cite{Aoki:2019cca}. A much more difficult task for lattice QCD is the determination of meson--meson scattering and three-body decays of resonances~\cite{Briceno:2017max,Hansen:2019nir}.

Many present applications require a more precise description of the low-energy strong interactions than currently achievable from lattice QCD. One example is the unprecedented high-statistics data on hadronic systems from the presently running or planned facilities like LHCb~\cite{Bediaga:2018lhg}, Belle-II~\cite{Kou:2018nap}, JLab~\cite{Dudek:2012vr}, and PANDA~\cite{Lutz:2009ff}. Precision information on the interactions of the lightest mesons is crucial input for searches of physics beyond the Standard Model. There are two prominent examples of such applications: the interpretation of the muon $g-2$ measurement~\cite{Bennett:2006fi}, where the calculation of hadronic contributions within the Standard Model dominates the uncertainty~\cite{Jegerlehner:2009ry,Aoyama:2020ynm};
and studies of $CP$ symmetry violations in $s$- and $c$-quark hadrons. 
Pion--pion final-state interactions have a large impact on the determination of $\epsilon'/\epsilon$ from kaon decays~\cite{Pallante:1999qf,Cirigliano:2011ny,Buras:2016fys,Gisbert:2017vvj}, and final-state interactions predominantly between pions and/or kaons similarly affect $D$- and $B$-meson multihadonic decays~\cite{Cheng:2004ru,Bediaga:2009tr,Aaij:2019jaq}. 
Detailed knowledge of interactions of the known mesons is also necessary for the interpretation of the recently discovered exotic hadronic states. In particular, the most important question is if the states can be explained in terms of the long-range physics, such as final-state interaction of the known hadrons, molecular states, or threshold effects, or are the unambiguous signature of  multiquark states or glueballs.

\begin{figure}[t]
\centering
\includegraphics[width=0.35\textwidth]{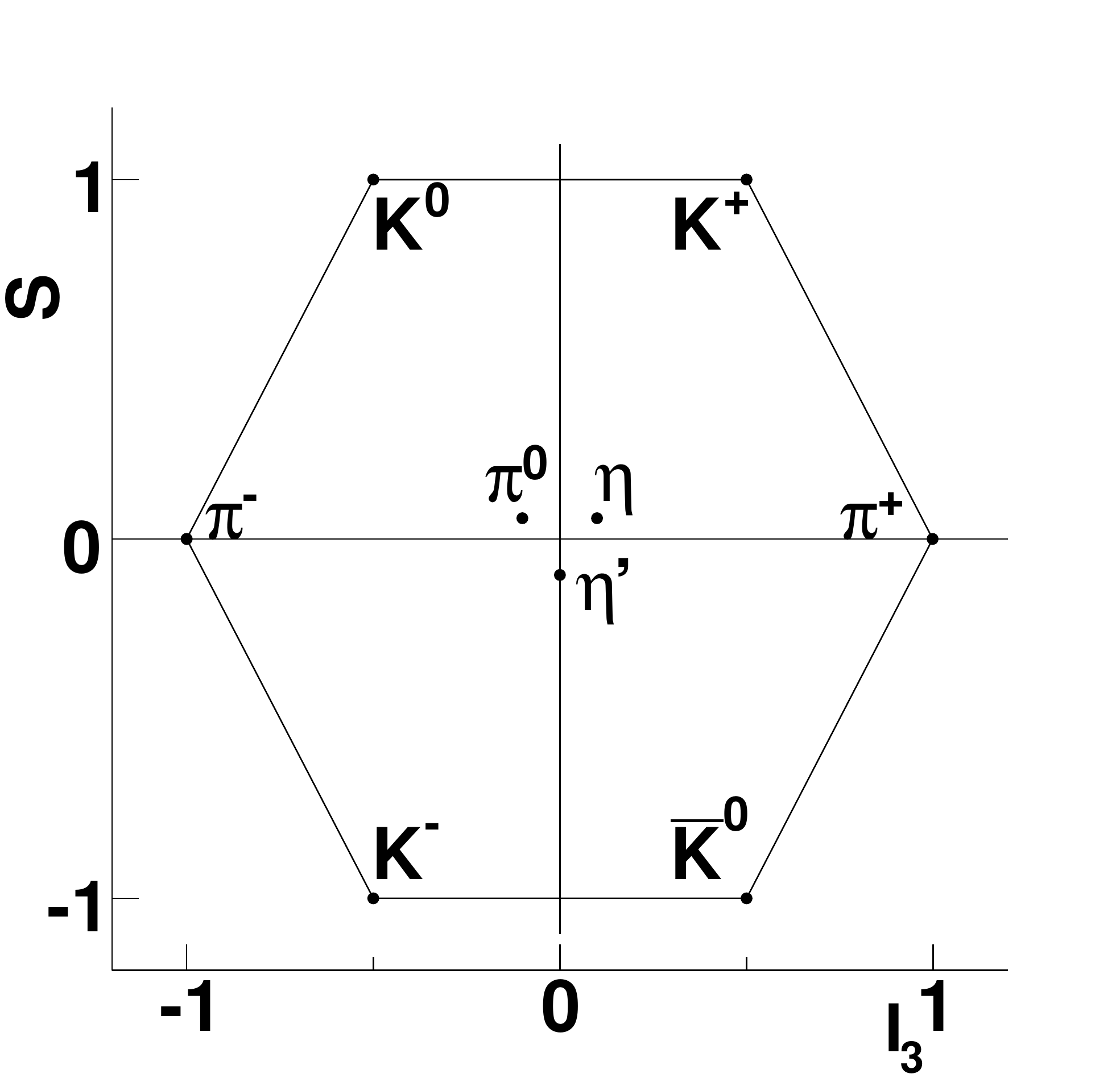}  
\includegraphics[width=0.35\textwidth]{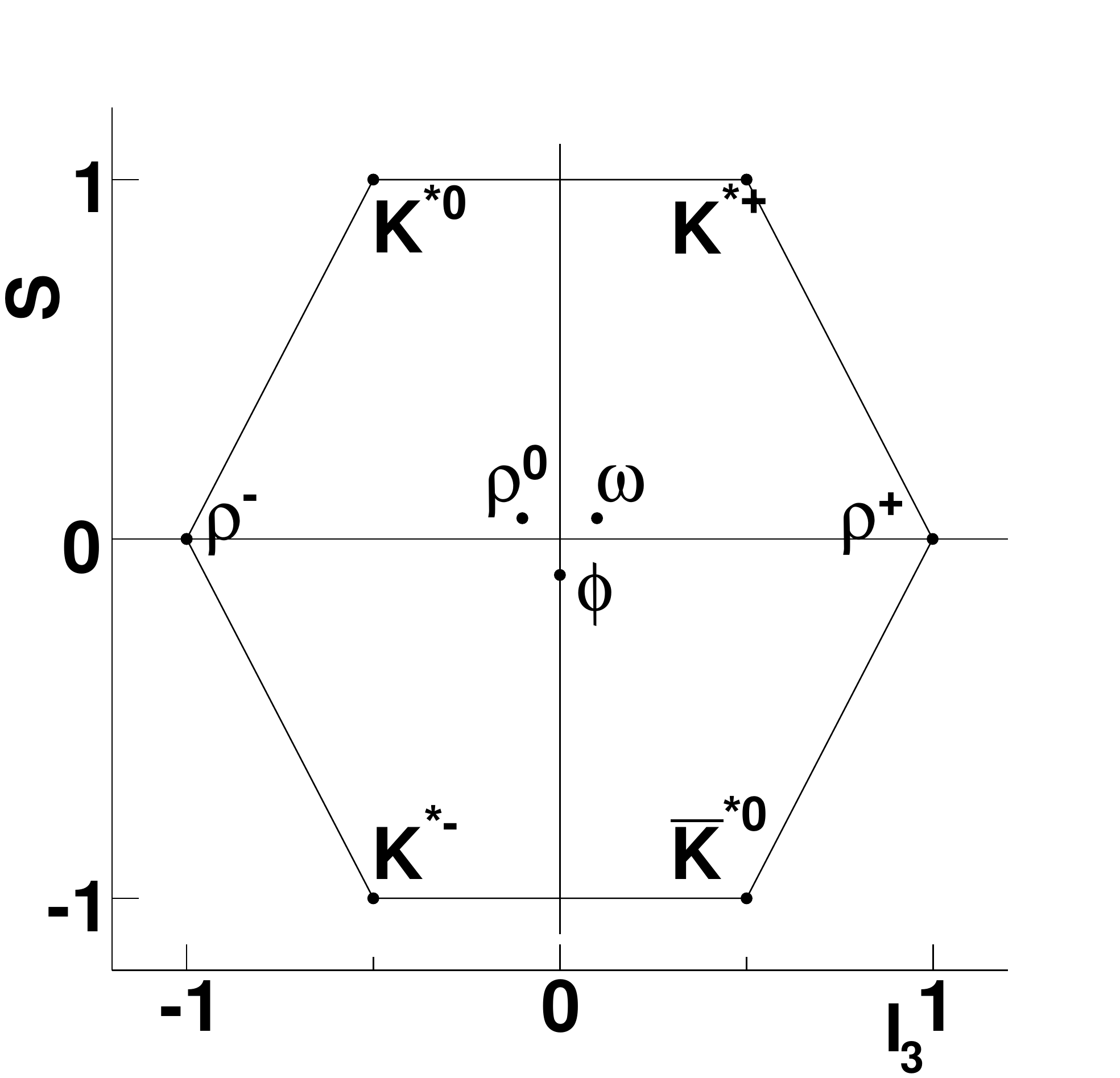}  
  \caption[Lowest-mass meson nonets]{Lowest-mass meson nonets in the quark model: (a) pseudoscalar mesons; (b) vector mesons. \label{fig:mesons}}
\end{figure}

\begin{figure}[t]
    \centering

    \includegraphics[width=0.22\linewidth]{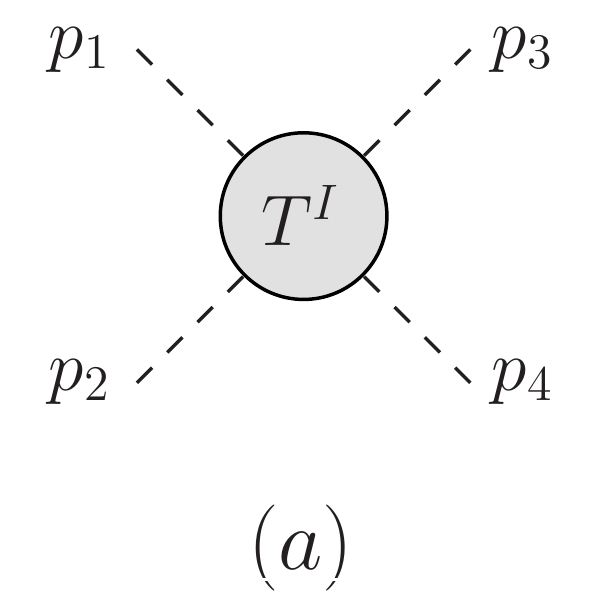} \hspace*{1cm}
    \includegraphics[width=0.25\linewidth]{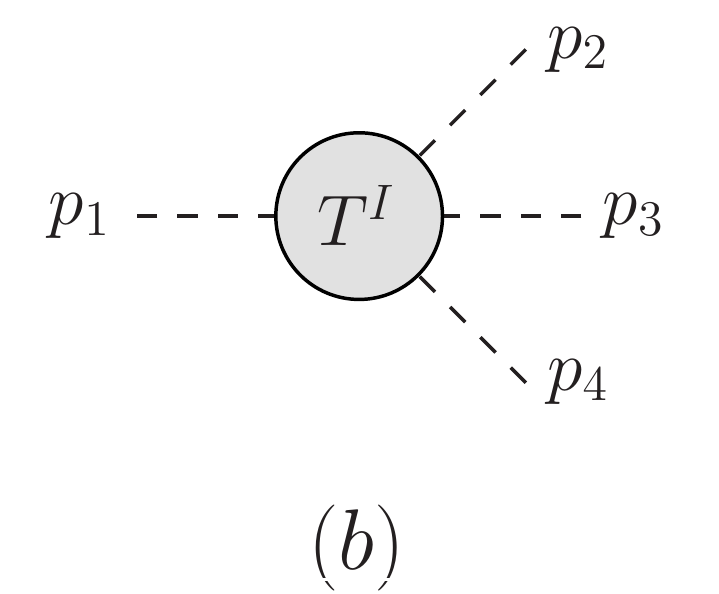}
    
    \caption[Generic meson--meson processes]{(a) A generic meson--meson scattering amplitude (of isospin $I$); (b) a three-body meson decay, linked to (a) by crossing symmetry. 
    }
    \label{fig:MMAmplitude}
\end{figure}

The required precision for the above applications can instead be reached by approaches based on effective field theories (EFT) using approximate low-energy symmetries of QCD, and general properties of field theories such as analyticity and unitarity (the mathematical consequences of causality and probability conservation) allow us to describe and interrelate results from various precision experiments. The main objects of the present review are the lowest-mass mesons, see Fig.~\ref{fig:mesons}, and their interaction amplitudes, see Fig.~\ref{fig:MMAmplitude}, the basic building blocks needed to study interactions in hadronic systems. One question is how to extract these amplitudes from data, and the other how to apply the extracted amplitudes to study multibody hadronic systems. 

\begin{figure}
    \centering
\includegraphics[width=0.98\textwidth]{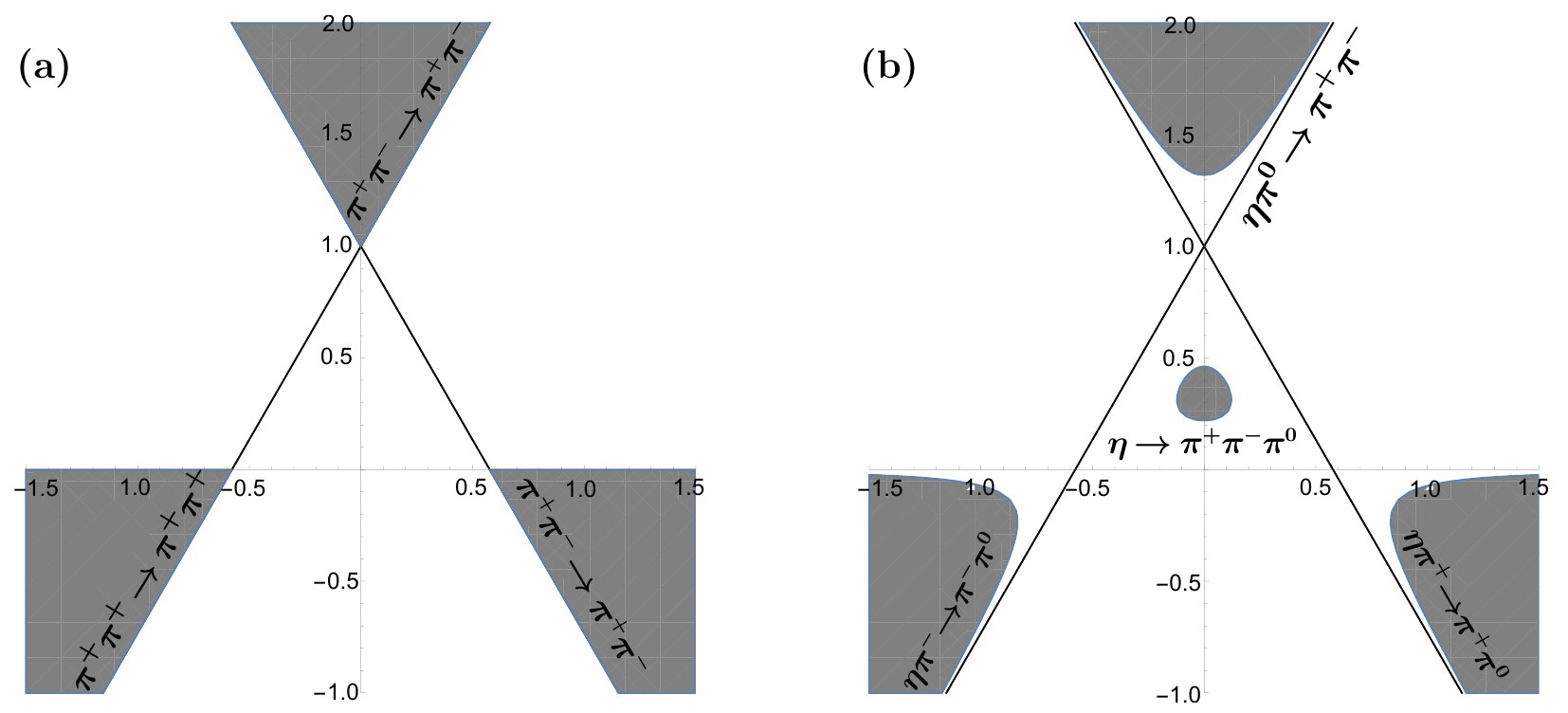}
    \caption[Mandelstam plot for meson--meson scattering]{Mandelstam planes indicating physical regions as shaded areas for  (a) $\pi^+\pi^-\to\pi^+\pi^-$  and (b)
   $\eta\pi^0\to\pi^+\pi^-$  and the corresponding crossed channels. 
   The units are chosen such that the vertical axes correspond to $s/(3s_0)$ and the horizontal ones to $(t-u)/(6s_0)$; see Eq.~\eqref{eq:s0}. The borders of the regular triangle denote the lines $s=0$, $t=0$, and $u=0$.
   The physical regions for the $s$-, $t$-, and $u$-channels are located at the top, right, and left, respectively, with the decay region for $\eta\to\pi^+\pi^-\pi^0$ at the center of the Mandelstam plane. }
    \label{fig:Mandelstam}
\end{figure}

We start with the first question. QCD at very low energies can be expressed in terms of the pseudoscalar meson octet fields, see Fig.~\ref{fig:mesons}(a): the effective degrees of freedom connected to approximate, spontaneously broken chiral symmetry. Therefore the goal is to determine interactions of the pseudoscalar meson octet like $\pi\pi$, $K\pi$, and $K \bar K$. The two-body meson--meson scattering $1+2\to 3+4$ shown schematically in Fig.~\ref{fig:MMAmplitude}(a) cannot be studied directly in experiments, since it is not possible to have mesonic targets and the availability of mesonic beams is limited to kaons and charged pions. 
Therefore the interaction will in general be a sub-process that should be extracted from more complicated reactions. Ideally it will be well isolated or embedded in a well-understood process (in particular, electromagnetic or weak). In addition it should have a cross section large enough to provide a sufficient number of events for precision studies. Finally, it should provide access to a reasonable range of kinematic variables in the meson--meson scattering sub-process. These variables can be expressed in terms of the following Mandelstam invariants defined as 
\begin{align}
    s=(p_1+p_2)^2 \,, \qquad 
    t=(p_1-p_3)^2 \,, \qquad
    u=(p_1-p_4)^2 \,,\label{eqn:Mandelstam}
\end{align}
where $p_i$ are the four-momenta as indicated in Fig.~\ref{fig:MMAmplitude}(a). The sum of the three variables is equal to the sum of squares of the masses of the initial and final particles:
\begin{equation}
    s+t+u=m_1^2+m_2^2+m_3^2+m_4^2 \equiv 3s_0 \,.\label{eq:s0}
\end{equation}
The kinematic regions are illustrated in Fig.~\ref{fig:Mandelstam}. The colored regions represent experimentally accessible processes related by crossing symmetry: $1+2\to 3+4$ ($s$-channel), $1+\bar3\to \bar 2+4$ ($t$-channel),
and  $1+\bar4\to \bar 2+3$ ($u$-channel). In addition, if $m_1>m_2+m_3+m_4$ the decay process $1\to \bar 2+3+4$ shown in Fig.~\ref{fig:Mandelstam}(b) is also possible. The partial-wave decomposition of the amplitude for spinless particles in $1+2\to 3+4$ scattering of total isospin $I$ reads 
\begin{equation}
    T^I(s,\cos\theta)=16\pi S \sum_{\ell=0}^\infty(2\ell+1)t_\ell^I(s)P_\ell(\cos\theta) \,,
\end{equation}
where $\theta$ is the scattering angle between particles 3 and 1 in the $1+2$ center-of-mass (c.m.)\ system directly related to the variable $t$, $P_\ell$ are the Legendre polynomials, and the symmetry factor is given by $S=2$ for identical particles and $S=1$ otherwise. The partial-wave amplitude $t_\ell^I$ corresponding to angular momentum $\ell$ can be expressed in terms of phase $\delta_\ell^I$ and inelasticity $\eta_\ell^I$ according to 
\begin{equation}
    t_\ell^I=\frac{\eta_\ell^I\exp(2i\delta_\ell^I)-1}{2i}\,.\label{eqn:tIj}
\end{equation}

\begin{figure}
    \centering
\includegraphics[width=0.32\linewidth]{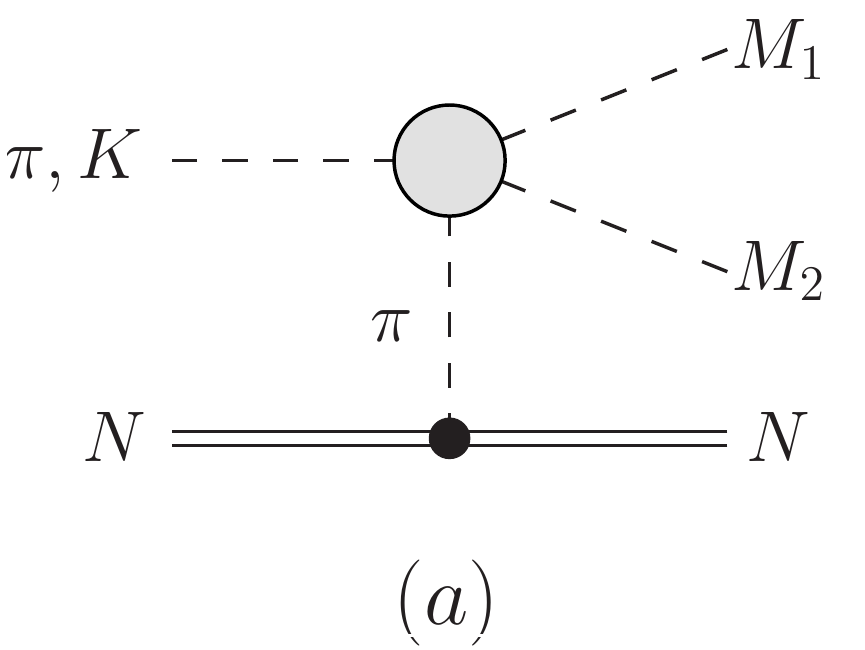}
\hspace*{2cm}
\includegraphics[width=0.4\linewidth]{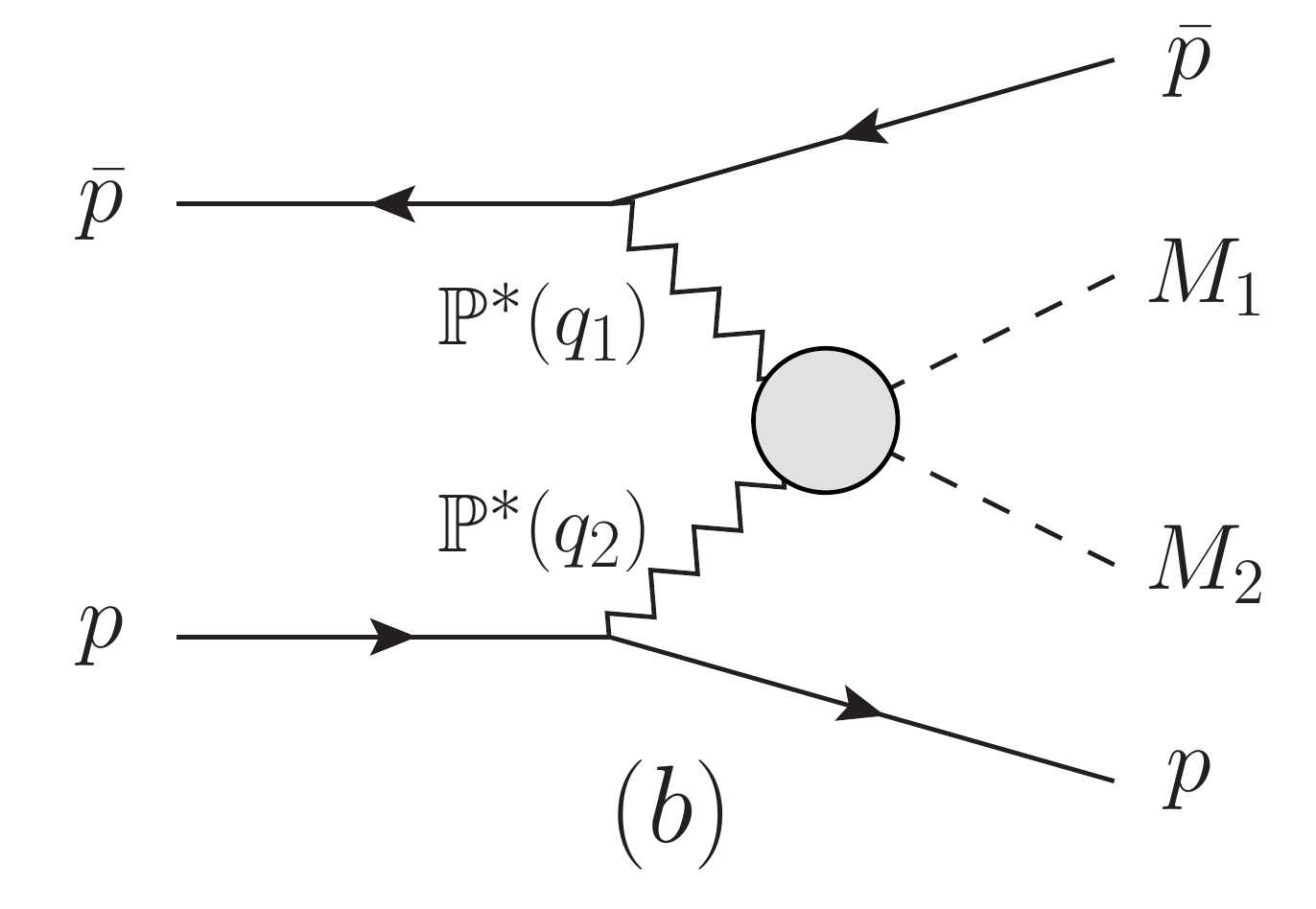}

\includegraphics[width=0.4\textwidth]{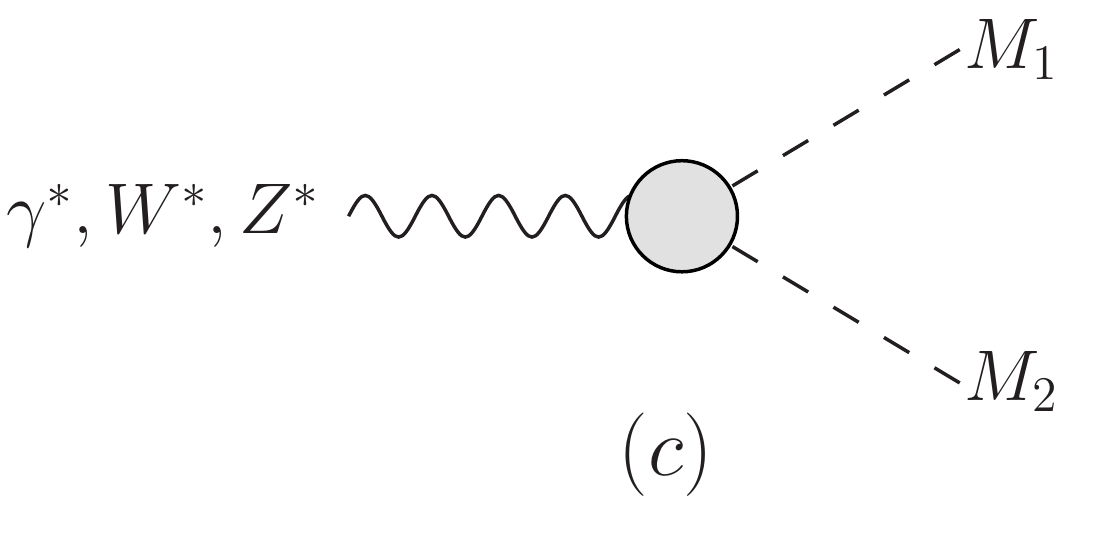} 

\caption[Meson production]{Examples how two-meson isolated systems can be produced for studies of their interactions: (a) one-pion-exchange mechanism to study $\pi\pi$ and $K\pi$ scattering amplitudes; (b)  double-Pomeron production;
(c) production of hadronic systems via an elementary gauge boson. 
}
    \label{fig:onepionex}
\end{figure}

Traditional experimental approaches to study meson scattering with pions use reactions like $\pi^\pm N\to \pi\pi N$,  $\pi^\pm N\to K\bar K N$, or $K N\to K\pi N$. 
For low four-momentum transfer squared $t$, one-pion-exchange is assumed to be the main contribution as shown in Fig.~\ref{fig:onepionex}(a), and meson--meson scattering is supposed to be kinematically well separated. Many such experiments were performed in the past, see Refs.~\cite{Protopopescu:1973sh,Estabrooks:1974vu,Grayer:1974cr,Hyams:1973zf}, but they only
allow to study $\pi\pi$ and $K\pi$ scattering and are often plagued by ambiguities or large systematic effects. Another example of kinematic separation at high-energy hadron accelerators is central production; e.g., the reaction $pp\to p(MM)p$ at large c.m.\ energies where the regions with low four-momentum transfer to the final protons are selected~\cite{Akesson:1985rn,Barberis:1999ap,Barberis:1999an,Bellazzini:1999sj}. The two mesons produced are well separated from the protons in rapidity, and their production is well described within the Regge model~\cite{Albrow:2010yb}. The reaction can be regarded as $(p\bar p)(p\bar p)\to\pi\pi$, where the $(p\bar p)$ exchanges have vacuum quantum numbers carried by the ``pomeron'', see Fig.~\ref{fig:onepionex}(b). Such a reaction is in many ways similar to the $\gamma\gamma\to\pi\pi$ process.
Another well-known process is the semileptonic kaon decay $K^\pm\to\pi^+\pi^-e^\pm\nu$, the so-called $K_{e4}$ decay mode, where a recent example comes from the NA48/2 experiment~\cite{Batley:2007zz}. The processes involve only two hadrons in the final state. The formalism for studies of such reactions was given by Cabibbo and Maksymowicz in 1965~\cite{Cabibbo:1965zzb}.

The aim of this review is to show a broad spectrum of precision light-hadron experiments carried out at the modern high-intensity electron--positron colliders. 
Here, the elementary electroweak production mechanisms shown in Fig.~\ref{fig:onepionex}(c) can provide very clean conditions to study the strong interactions in few-meson systems. The initial approach was to study the energy dependences of the single-photon annihilation processes into $\pi^+\pi^-$, $K^+K^-$, and $K_SK_L$ pairs, which is closely related to $P$-wave two-meson scattering amplitudes. However, a variety of options are available to
determine  two-meson scattering amplitudes in a direct and clean way, using reactions where only two hadrons are involved. These reactions are in the first place higher-order electromagnetic processes involving initial- or final-state photon radiation. Examples are radiative decays of neutral vector mesons directly produced in electron--positron annihilation, such as $\phi\to\pi^0\pi^0\gamma$~\cite{Aloisio:2002bt,Ambrosino:2006hb} or $J/\psi \to \pi^0 \pi^0 \gamma$~\cite{Ablikim:2015umt}. 
These radiative decays give access to meson systems with scalar quantum numbers, whose investigation was motivated by the search for a scalar glueball candidate in the $1 \GeV$ range,
and the quest for an understanding of the nature of the lowest-mass scalar mesons. The decay channels of such potentially exotic systems are essentially restricted to the $\pi\pi$ and $K\bar K$ final states.  Radiative decays 
of quarkonia were advertised as an ideal place to search for bound states created in gluon--gluon fusion~\cite{Koller:1978qg,Becker:1986zt,Augustin:1987fa,Bai:1996wm,Bai:1998tx}.
 However, as it was pointed out by Au, Morgan and Pennington~\cite{Au:1986vs}
any production process is very tightly related to scattering reactions, $\pi\pi\to \pi\pi$ and  $\pi\pi\to K\bar K$, by unitarity. Therefore it cannot reveal effects that could not be seen in the scattering reactions, but  it can act as a filter to suppress backgrounds and enhance the signal.  
Other examples are two-photon formation processes such as  $\gamma\gamma\to \pi^+ \pi^-$ as studied at Belle~\cite{Mori:2006jj}. Electron--positron colliders are also sources of $\tau$ lepton pairs, which allow studies of hadronic weak decays like $\tau^-\to \pi^-\pi^0\nu_\tau$ or $\tau^-\to K^-\pi^0\nu_\tau$. 

Another option is amplitude analyses of multi-meson decays. Such hadronic decays have large branching fractions and often provide ``infinite'' statistics, in the sense that the statistical uncertainties are negligible. In recent years there has been a lot of progress in the development of precision analysis tools for processes involving three and more hadrons in the final state. There is a trend to base the tools on sound conjectures such as the analytic properties of the transition matrix. Analyses of such reactions normally use two-meson interaction amplitudes as their building blocks. However, in case the amplitudes are not known sufficiently well, it is possible to validate or even extract information on the two-body meson processes. We will discuss three-meson decays in which the decaying particles are abundantly produced and easily tagged at electron--positron colliders. Some examples include multi-hadron decays:  weak decays of $\tau$ leptons or $K$ and $D$ mesons as well as strong and radiative decays of mesons.

The scope of this review is limited to the interactions in systems of two and three stable mesons with invariant mass below $\sim2 \GeV$, i.e., below baryon--antibaryon pair production threshold. This choice is motivated by the applicability of the methods used for the analysis and extraction of meson interaction amplitudes (dispersive as well as based on chiral symmetry).
\begin{figure}
\centering
\includegraphics[width=0.45\linewidth]{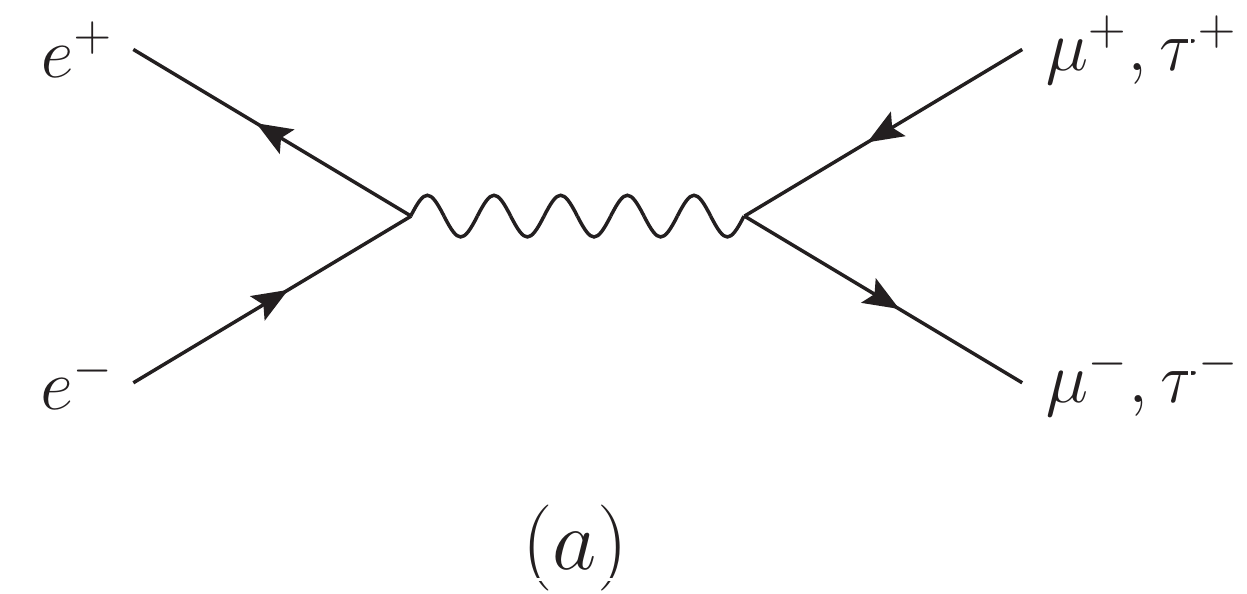}
\hspace*{1cm}
\includegraphics[width=0.45\linewidth]{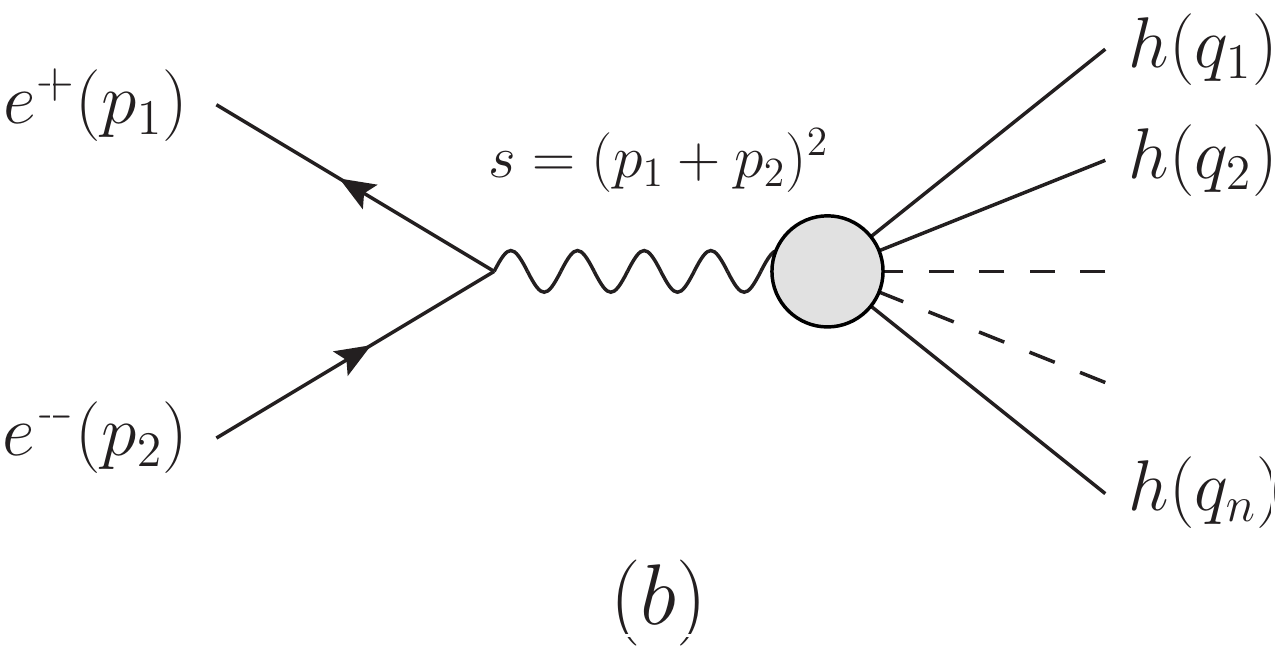}

\caption[Single-photon production]{Lowest-order diagrams representing particle production in single-photon 
    electron--positron annihilation: a) lepton--antilepton pairs $\mu^+\mu^-$ or $\tau^+\tau^-$; b)
    production of a multi-hadron system.
    \label{fig:mech}}
\end{figure}
For colliders with c.m.\ energies up to the charm--tau region, the main annihilation process into a lepton--antilepton or quark--antiquark pair proceeds via an intermediate photon. The leading order of such annihilation processes $e^+e^-\to\mu^+\mu^-$ or $\tau^+\tau^-$ is shown in Fig.~\ref{fig:mech}(a). The contribution from $Z$ or $Z$--$\gamma$ interference in the propagator at these energies is negligible. However, with c.m.\ energies close to narrow vector resonances, one should consider hadronic vacuum polarization contributions in the propagator, as we will discuss in Sec.~\ref{sec:HVP}.

\begin{figure}
    \centering
\includegraphics[width=0.98\textwidth]{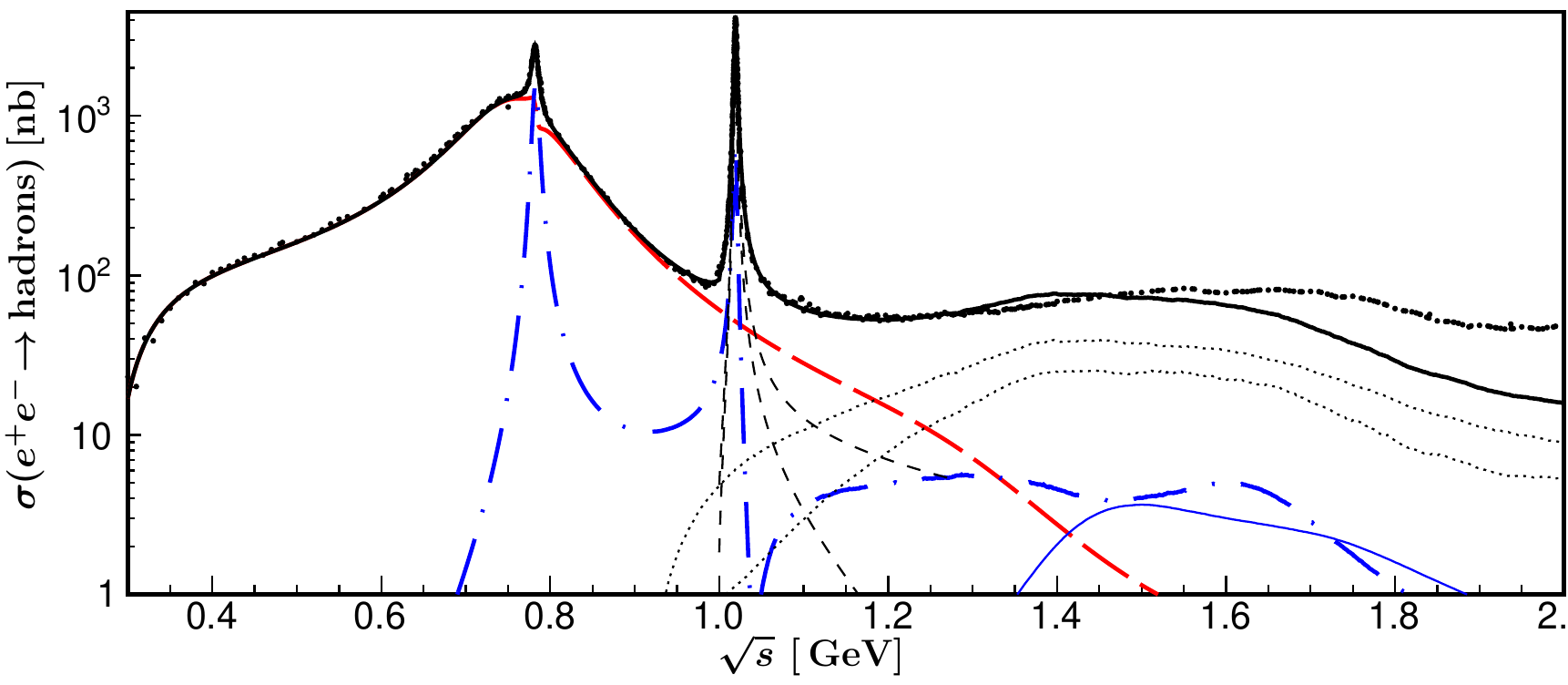}
    \caption[ Cross sections for electron--positron annihilation to hadrons ]{Cross sections for electron--positron annihilation to hadrons. Data points are from the Particle Data Group compilation~\cite{PDG}. Lines represent parameterizations of the individual contributions from the main final states from Phokara~\cite{Rodrigo:2001kf}. Included final states are: $\pi^+\pi^-$ (red long-dashed line), $\pi^+\pi^-\pi^0$ (blue dashed--dotted line), $K^+K^-$ and $K_SK_L$ (black dashed lines), $\pi^+\pi^-\pi^+\pi^-$ and $\pi^+\pi^-\pi^0\pi^0$ (black dotted lines), and $\eta\pi^+\pi^-$ (blue solid line). The sum, shown as the black solid line, approximately saturates the total cross section up to $1.2 \GeV$.
    }
    \label{fig:csexcl}
\end{figure}

The main mechanism for the production of hadronic systems at electron--positron colliders is shown in Fig.~\ref{fig:mech}(b). The electron--positron pair annihilates via a single virtual photon, and a hadronic system is produced. The system has therefore the same $J^{PC}=1^{--}$ quantum numbers as the photon. The lightest hadronic system that can be created in such an annihilation is $\pi^+\pi^-$, with the c.m.\ electron--positron energies above $\sqrt{s}=2 m_{\pi}$. However, the lightest hadron, the $\pi^0$ meson, can be created in a higher-order radiative process $e^+e^-\to\pi^0\gamma$ already at $\sqrt{s}>m_{\pi^0}$, but the cross section is very low.
For c.m.\ energies at which hadrons can be produced, the mass of the electron is negligible and the virtual photon has only helicities $\pm1$ (no 0 projection).  The cross section of $e^+e^-\to \text{hadrons}$ is dominated by $e^+e^-\to\pi^+\pi^-$ and $e^+e^-\to\pi^+\pi^-\pi^0$ for the isospin states $I=1$ and $I=0$, respectively, at low energies. The contribution of the two channels to the total cross section is shown in Fig.~\ref{fig:csexcl}.
In addition, the final states $K^+K^-/K_SK_L$, $\eta\pi^+\pi^-$, and $\pi^+\pi^-\pi^+\pi^-/\pi^+\pi^-\pi^0\pi^0$ are included. The sum of these approximately saturates the total cross section up to $1.2 \GeV$. The result is not exact since for the exclusive data, we used an arbitrary parameterization included in the Phokara program~\cite{Rodrigo:2001kf}.

When the concept of the electron--positron collider was first proposed (see Sec.~\ref{sec:history}), the main physics motivation was to study the time-like pion vector form factor directly accessible in the process $e^+e^-\to\pi^+\pi^-$. Precise studies  of this process are still most relevant today in the context of the hadronic contribution to the muon anomalous magnetic moment~\cite{Aoyama:2020ynm}. 
The energy dependence of the total cross section for electron--positron annihilation into hadrons directly reveals the crucial role of the neutral vector mesons with quantum numbers $J^{PC}=1^{--}$, identical to those of the photon. The production yield of hadronic systems is greatly enhanced at  c.m.\ energies tuned to one of such resonances. Decays of the mesons represent one of the main approaches to study hadronic systems at low-energy electron--positron colliders discussed in the present review.

In more detail we will discuss processes where two and three pseudoscalar mesons are produced both in the continuum and at certain (narrow) resonances. A special case is when the produced vector meson decays in sequence, where the first step is a radiative two-body process involving narrow hadronic state like $\eta$, $\eta'$, or $\chi_{cJ}$. Another important class is decays of $K$ and $D$/$D_s$ mesons, which are produced in single-photon processes as entangled pairs. Again the cross section is enhanced when the beams are at resonances such as the $\phi$ for $K\bar K$ or $\psi(3770)$ for $D\bar D$ production. Finally, the light-meson systems can be
produced in  decays of $\tau$ leptons, which are in turn produced via $e^+e^-\to\tau^+\tau^-$.

\subsection{Historical note}\label{sec:history}

The concept of electron--positron colliders was triggered by technical developments in electron accelerators, the idea of colliding beams~\cite{Wideroe:1953aa,ONeill:1956iga,Kerst:1997rs}, and by discoveries of the first vector meson resonances in the beginning of the 1960s. The culmination of this first stage of development was the discovery of the $J/\psi$ particle during the ``November revolution'' in 1974: at BNL~\cite{Aubert:1974js} (November 12th) and at SLAC~\cite{Augustin:1974xw} (November 13th). A week after, the discovery was confirmed at Laboratori Nazionali di Frascati (LNF) Frascati~\cite{Bacci:1974za} (November 18th). Two of the observations were made using electron--positron colliders, demonstrating the potential of the new machines. 

We try to give a brief account of the crucial events before the November revolution~\cite{Bauer:1977iq,Greco:2005aa,Baier:2006ye,Bonolis:2011wa,Bernardini:2015wja,Pancheri:2017kdu,Shiltsev:2019rfl}. The advantages of the center-of-mass collision scheme  were well known in the mid-1950s~\cite{Kerst:1997rs}, and by the end of the decade the idea of a particle collider was technically mature to be considered seriously.  Gerard O'Neill had proposed an electron--electron storage ring collider at SLAC~\cite{ONeill:1956iga} and the plans were presented at CERN in June 1959. The physics motivation was high-precision experiments and tests of the predictions of Quantum Electrodynamics (QED), in particular tests of the space-like photon propagator~\cite{Brown:1960}. This collider project was presented during a seminar at LNF in October 1959 by Pief Panofsky. One of the listeners was Bruno Touschek, who during the discussion reportedly asked the question~\cite{Bernardini:2015wja}: ``Why not try electrons against positrons?''   Touschek's vision was ``that electron--positron annihilations---reactions proceeding through a state of well-defined quantum numbers---would be the pathway to new physics''~\cite{Pancheri:2017kdu}. For the collider concept the particle and antiparticle beams with opposite charges and the equal masses are particularly simple and appealing since they can be stored in one single ring, while the electron--electron collider requires two separate rings.  This is considered to be the starting point of the chain of events leading to the first prototype electron--electron storage ring (AdA---Anello di Accumulazione) and further projects at LNF.

The question was what purpose such a machine could be used for. The first ideas were in fact related to tests of the electron's space-like propagator. To understand the emergence of the concept of the unique role of electron--positron colliders due to the ``time-like one-photon channel dominating, to first order of Quantum Electrodynamics, the production of final states''~\cite{Bernardini:1997aa}, let us turn to the status of the theory of strong interactions at that time. 

From the available data on pion production in pion-induced reactions above energies of $1 \GeV$ it had become clear that a small momentum transfer to the nucleon is preferred.  This was seen in the nucleon angular distributions, which are sharply peaked in the backward direction. These results suggested the importance of large-impact-parameter collisions in such processes. The simplest mechanism that could contribute is the beam pion colliding with a virtual (nearly real) pion emitted by the target nucleon. The quantitative aspects of such collisions were discussed by Chew and Low~\cite{Chew:1958wd}, as well as Salzman and Salzman~\cite{Salzman:1960zz}, as means of extracting information on the  $\pi\pi$ interactions from $\pi N\to\pi\pi N$ data~\cite{Goebel:1958zz}.

On the other hand, experimental studies of elastic electron scattering off protons and deuterons, carried out by Hofstadter et al.~\cite{Hofstadter:1956qs}, had raised the question of photon interactions with nucleons. It seemed logical to assume that also here the long-range domain would be dominated by the interaction with systems of the lightest hadrons---pions. The role of the photon--two-pion vertex for the explanation of the nucleons' anomalous magnetic moments was investigated by G.~F.~Chew et al.~\cite{Chew:1958zjr} in 1958, using dispersion relations developed for the analyses of the pion--nucleon scattering. 
In a follow-up paper by P.~Federbush et al.~\cite{Federbush:1958zz}, the dispersive analysis was extended and the concept of the pion form factor $\FV(s)$ was introduced to represent the structure of the photon--two-meson vertex. An important part of the photon--nucleon interactions can be understood in this way via an isovector $\pi^+\pi^-$ pair, linking the isovector nucleon form factors to the pion form factor and $\pi\pi\to N\bar N$ $P$-wave amplitudes.  This relation has been exploited repeatedly, with ever more accurate experimental input, in the following decades~\cite{Frazer:1960zzb,Hohler:1976ax,Mergell:1995bf,Belushkin:2005ds,Hoferichter:2016duk} (see also Ref.~\cite{Pacetti:2015iqa} for a broader review on nucleon form factors). 
The properties of the pion form factor $\FV(s)$ can in turn be related to $\pi^+\pi^-$ elastic rescattering using dispersion relations. With 
\begin{equation}
    \FV(s)=1+\frac{s}{\pi}\int_{4m_\pi^2}^\infty \diff x\frac{{\Im} \FV(x)}{x(x-s-i\epsilon)}\,, \label{eq:DR-FpiV}
\end{equation}
setting $\FV(s)=\exp\left\{i\varphi(s)\right\}|\FV(s)|$ and assuming the phase $\varphi(s)$ to be known, the expression for the form factor is
\begin{equation}
    \FV(s)=P(s)\exp\left\{\frac{s}{\pi}\int_{4m_\pi^2}^\infty \diff x\frac{\varphi(x)}
    {x(x-s-i\epsilon)}\right\} \,, \label{eq:Omnes}
\end{equation}
where $P(s)$ is an arbitrary polynomial. In case of elastic $\pi\pi$ rescattering, $\varphi(s)=\delta_1^1(s)$, the pion--pion $P$-wave phase shift of isospin $I=1$. The paper points out that there was no experimental or theoretical  
information on the pion--pion scattering phase shifts. The phases $\delta_1^1$ are real 
for $s<16m_\pi^2$, i.e., below threshold for the production of two additional pions,
since production of any odd number of pions is forbidden by $G$ parity~\cite{Lee:1956sw}.
In February 1960, Cabibbo and Gatto proposed to use the reaction $e^+e^-\to\pi^+\pi^-$ to 
study the pion form factor in the time-like region~\cite{Cabibbo:1960zza}. 
Meanwhile the dispersive analyses were further extended and an isovector $\pi\pi$ resonance was proposed by Frazer and Fulco~\cite{Frazer:1959gy,Frazer:1960zzb}. 
Such a structure should be directly observed as a peak in electron--positron collisions. We will continue a detailed  discussion of the properties of $\FV(s)$ in Sec.~\ref{sec:eePP}.

On the other hand, to explain the neutron charge structure emerging from the Hofstadter experiments, a strong coupling of the photon to an isoscalar hadronic current was required. The lightest such system consists of three pions $\pi^+\pi^-\pi^0$ in the $J^{PC}=1^{--}$ state. Here, each pair of the pions is in a relative $P$-wave. The dispersive treatment of such three-body systems is much more complicated, but already in 1957 Nambu~\cite{Nambu:1957vw} speculated about the existence of an isoscalar three-pion bound state or a strongly decaying resonance (if the mass was sufficiently high). Such a state would be narrow and thus simplify the analysis of the isoscalar part of the electromagnetic current. The idea of both isovector and isoscalar $J^{PC}=1^{--}$ mesons was then extended on $SU(3)$ flavor symmetry grounds to include an additional neutral isoscalar vector meson by Sakurai in 1960~\cite{Sakurai:1960ju}, leading to the concept of vector meson dominance (VMD). These three neutral vector mesons were discovered soon after. In 1961 a broad $\rho^0$ isovector vector resonance was observed by Erwin et al.~\cite{Erwin:1961ny}, and an isoscalar $\omega$ meson by Maglic et al.~\cite{Maglic:1961nz}.  In the following year in July 1962, the third  meson $\phi(1020)$ was observed by Bertanza et al.~\cite{Bertanza:1962zz} and confirmed in 1963 by two experiments~\cite{Schlein:1963zz, Connolly:1963pb}. In parallel to these discoveries a comprehensive physics program for electron--positron colliding beam experiments was formulated by Cabibbo and Gatto in the so-called ``Bible''~\cite{Cabibbo:1961sz}.
 
The technological milestones towards electron--positron colliders can be summarized as follows.  A proof-of-concept storage ring AdA at the LNF was proposed in  February 1960~\cite{Bernardini:1960osh}, with beam energies up to $250 \MeV$ and a diameter of $1.3\m$. The first electrons and positrons were accumulated in AdA in February 1961~\cite{Bernardini:1962yyy,Bernardini:2004rp}. Tests continued in 1963 at the Orsay Laboratory where the ring was moved and resulted, e.g., in the discovery of the so-called Touschek effect~\cite{Bernardini:1997sc,Bernardini:1964lqa}. Soon after the AdA, research electron--positron colliders were proposed, both in the USA and in Europe, with higher energies and luminosities.
At LNF the construction of ADONE with $1.5 \GeV$ energy per beam was proposed as early as January 1961, ACO with $550 \MeV$ beam energies at Orsay, and SPEAR~\cite{Richter:1970gf} in the USA.
Parallel independent efforts took place in the Soviet Union, where Gersch Budker and collaborators had been active in electron--electron colliders since the mid 1950s, and later in electron--positron collisions. A historic account of these events in Novosibirsk is given in Refs.~\cite{Baier:2006ye,Levichev:2018cvd}. The first Novosibirsk $e^+e^-$ collider VEPP-II with beam energies of $700 \MeV$ began to produce 
physics data in 1966; shortly after ACO came into operation, and the first data at ADONE was taken in 1969.
These physics topics and detectors of the early period of electron--positron accelerators are well described in a report from 1976~\cite{PerezYJorba:1977rh}. In some sense our review can be treated as a modern revision of the physics topics discussed there. 
\subsection{Meson resonances}\label{sec:mesons}
\begin{figure}
    \centering
    \includegraphics[width=0.7\textwidth]{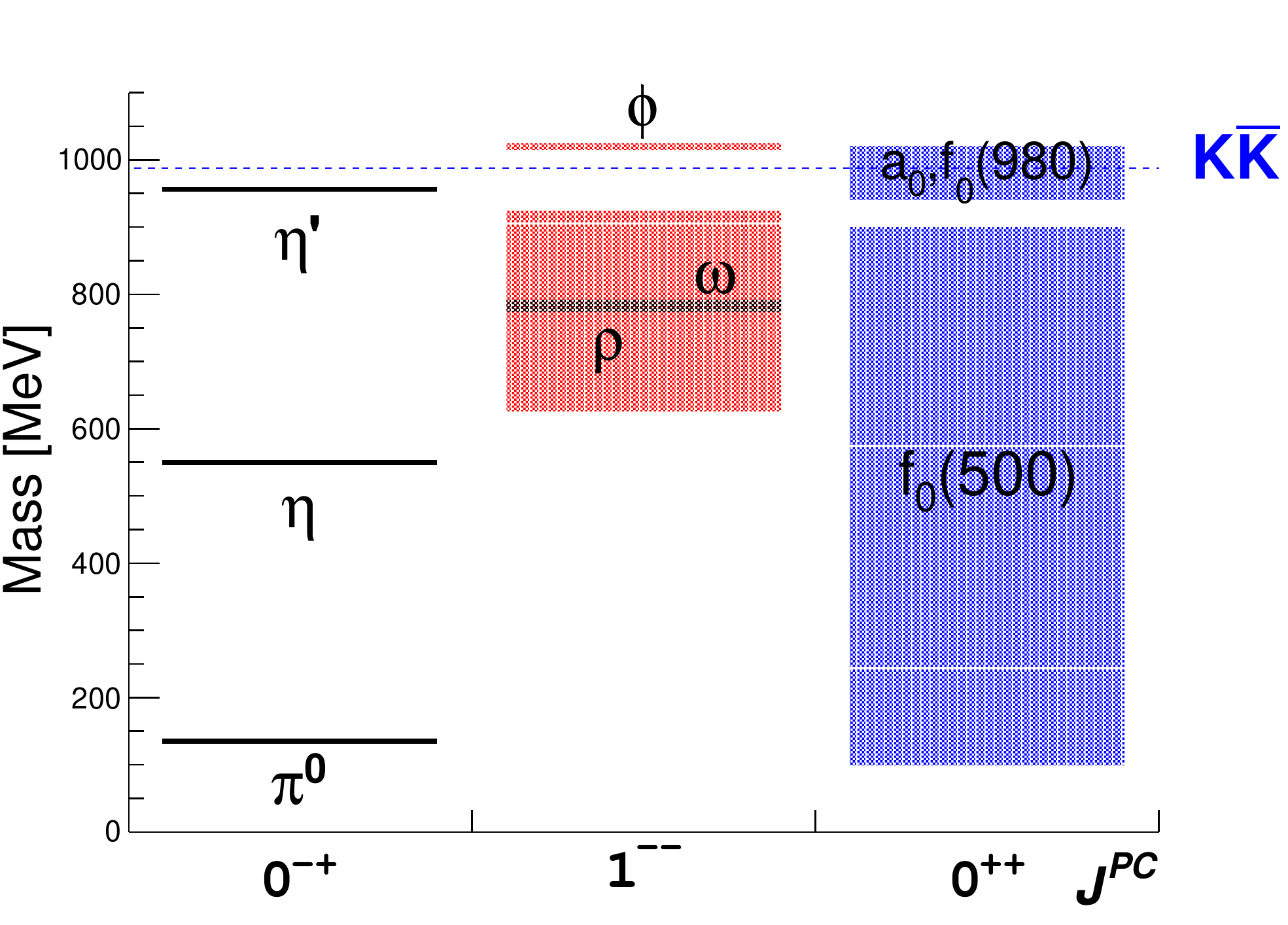}
\caption[Flavor-neutral mesons below $1\GeV$]{Spectrum of the flavor-neutral mesons with masses below $1\GeV$. The masses and widths of the lightest states are represented schematically. In addition, the $K\bar{K}$ threshold for open strangeness is indicated explicitly.}
    \label{fig:ChPTspectrum}
\end{figure}
The review focuses on interactions of stable  mesons, i.e., mesons that do not decay strongly or whose strong decays are suppressed. These are in the first place pseudoscalar mesons from the ground state octet and the $\eta'$ meson. They play a special role in the description of the low-energy domain of QCD. Among the few meson resonances with masses up to the $\phi(1020)$, see Fig.~\ref{fig:ChPTspectrum}, there are only two narrow states the neutral vector mesons $\omega$ and $\phi$. We try to treat the remaining light vector mesons $\rho(770)$ and $K^*(892)$, as well as the scalar states $f_0(500)$ (formerly known as $\sigma$), $f_0(980)$, and $a_0(980)$ as resonances in the systems of the pseudoscalar mesons, as they are much broader, and the narrow-width approximation much less reliable. 
Interactions between pairs of pseudoscalar mesons are a main source of information about the spectrum and structure of scalar mesons. From the perspective of the present review, the low-mass scalar meson spectrum is of interest as a probe of the interactions of the nonet of pseudoscalar mesons, since it allows us to test chiral perturbation theory and dispersive methods.

\begin{figure}[t]
    \centering
 \includegraphics[width=0.9\textwidth]{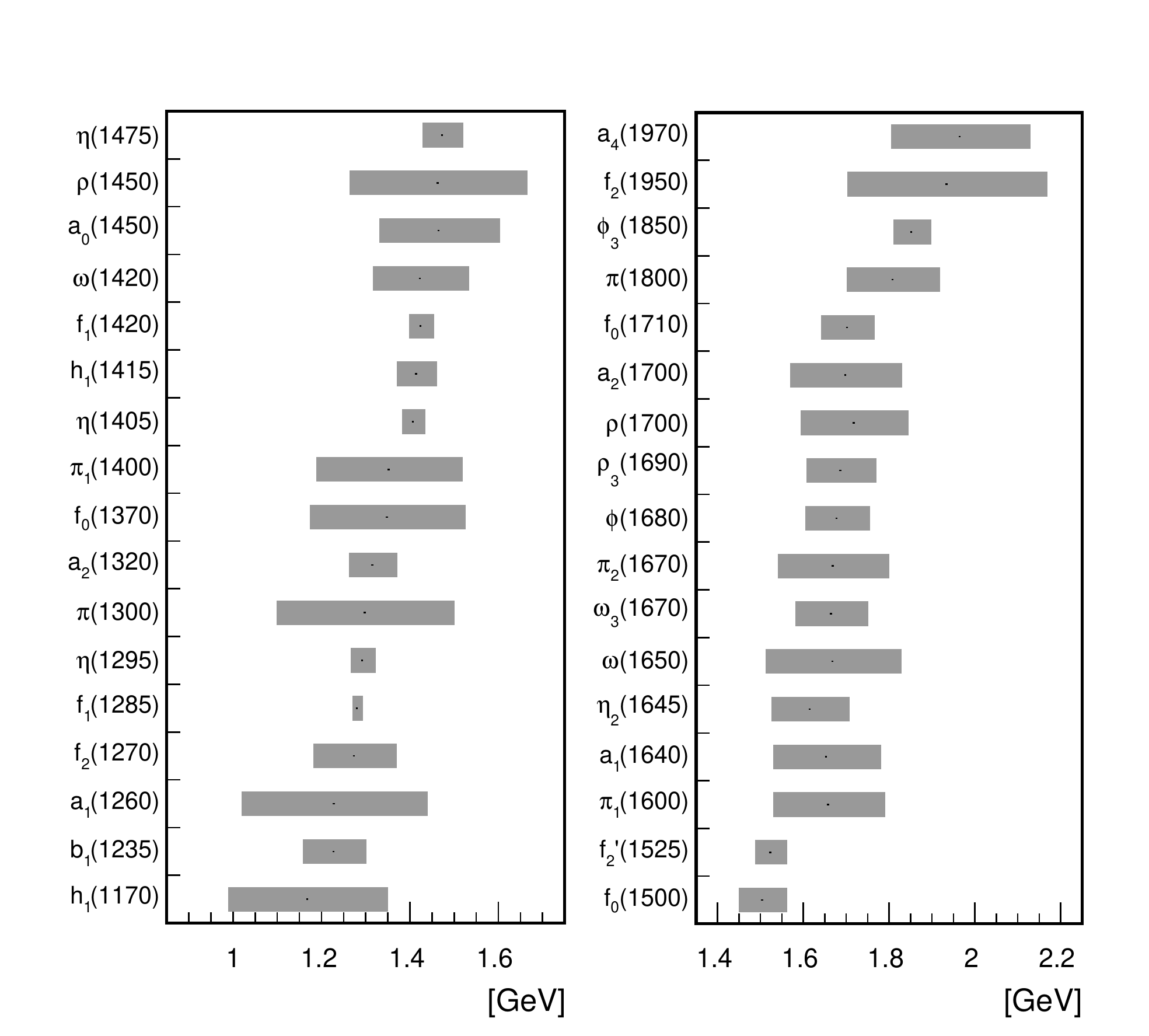}
    \caption[Flavor-neutral meson resonances with masses $1$--$2\GeV$]{Spectrum of the well-established flavor-neutral meson resonances with masses in the range  $1$--$2\GeV$. The
    mass of a resonance is represented by the center of the shaded box, the decay width by its horizontal size.}
    \label{fig:Mes12spectrum}
\end{figure}

Above $1\GeV$ there are a large number of meson resonances. The spectrum of the flavor neutral states with masses below $2\GeV$ is shown in Fig.~\ref{fig:Mes12spectrum},\footnote{All values for masses and  widths of resonances as well as branching fractions in the review are taken from the Particle Data Group~\cite{PDG} unless otherwise stated.} and the naming convention introduced by the Particle Data Group (PDG) given in Table~\ref{tab:pdgMesons}.
\begin{table}[t]
\caption[Naming convention for meson states from PDG]{\label{tab:pdgMesons}Naming convention for meson states  according to Particle Data Group~\cite{PDG}. }
\begin{center}
\renewcommand{\arraystretch}{1.3}
\begin{tabular}{lcccccrr}\toprule
${J}$ & ${PC}$ & $I=1$ && $I=0$  \\ \midrule
$0,2,\ldots$ & ${-+}$ & $\pi$ &  & $\eta,\eta'$  \\
$1,3,\ldots$ & ${+-}$ & $b$ &  & $h,h'$  \\
${0,1,2,\ldots}$ & ${++}$ & $a$ & & $f,f'$\\
${1,2,3,\ldots}$ & ${--}$ & $\rho$      & & $\omega,\phi$  \\
\bottomrule 
\end{tabular}
\renewcommand{\arraystretch}{1.3}
\end{center}
\end{table}
In this energy range, the lightest purely gluonic bound state, a {\it glueball}, is expected with scalar quantum numbers~\cite{Morningstar:1999rf}.  The observation of such a state would be the evidence that gluon self-interactions can generate a massive meson. Unfortunately, glueballs may mix with conventional quark bound states, making the interpretation of candidate resonances challenging. Despite the availability of a large amount of data on $\pi\pi$ and $K\bar K$ scattering in the low-mass region, the existence and characteristics of isoscalar scalar ($I^{G}J^{PC}=0^{+}0^{++}$) and tensor ($0^{+}2^{++}$) states remain controversial~\cite{Klempt:2007cp}. This lack of understanding is due in part to the presence of broad, overlapping resonances, which are poorly described by most analytical methods. The PDG reports eight $0^{+}0^{++}$ mesons, which have widths between $100$ and $450\MeV$. Several of these states, including the $f_{0}(1370)$, are characterized only by ranges of values for their masses and widths.

\begin{figure}[t]
    \centering
    \includegraphics[width=0.6\textwidth]{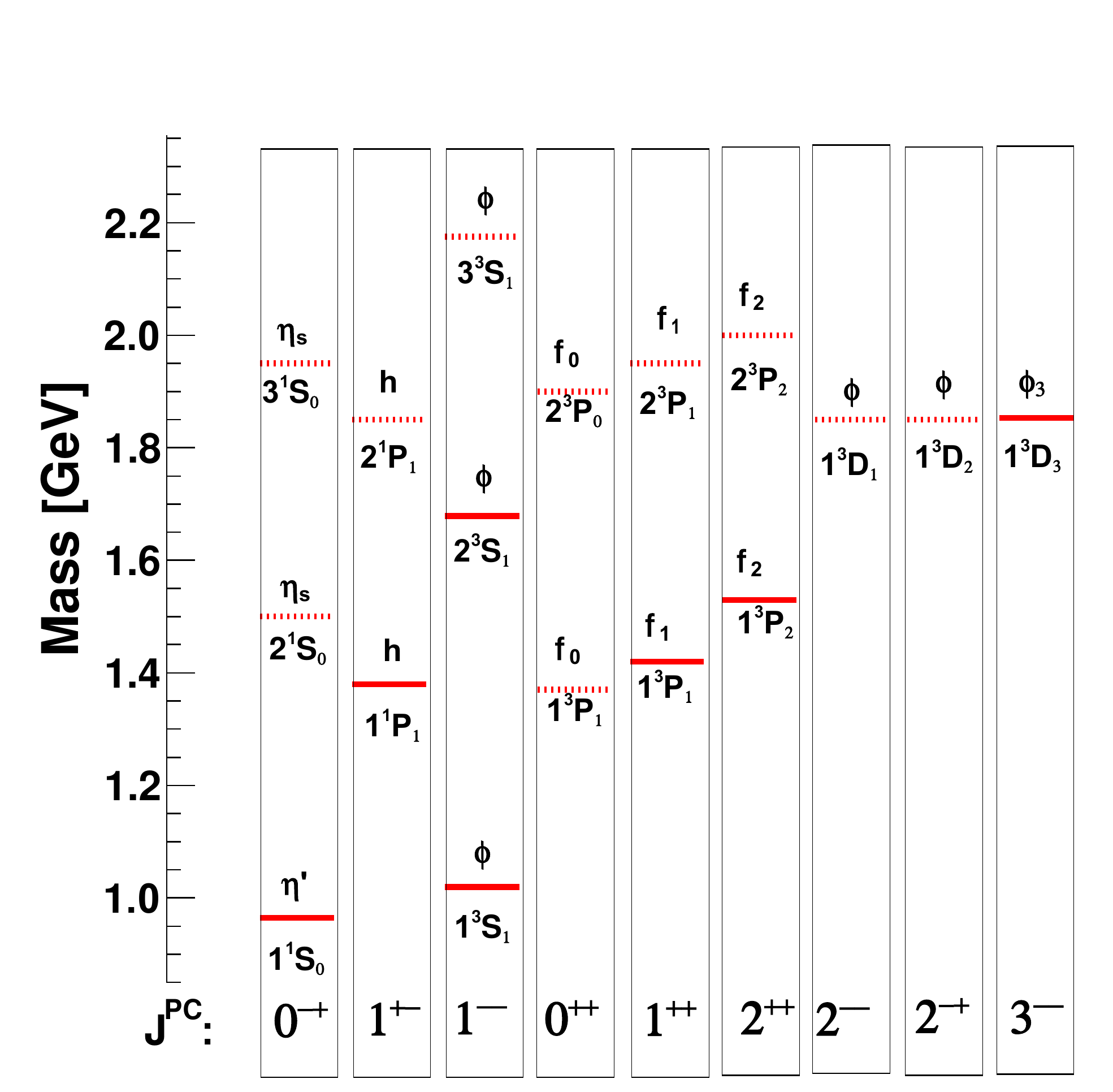}
    \caption[Strangeonium spectrum]{Spectrum of $s\bar s$ states with spin $J<4$, predictions from Ref.~\cite{Barnes:2002mu}. Experimentally confirmed states are indicated by solid lines. 
    }
    \label{fig:ssbar}
\end{figure}

For the energy region discussed in this review, the $s\bar{s}$ mesons, strangeonia, are also of relevance. In Fig.~\ref{fig:ssbar} the expected spectrum  of the states with spin $J<4$ predicted by the $^{3}P_{0}$ model~\cite{Godfrey:1985xj,Barnes:2002mu} is shown.  Only seven states (indicated with the solid lines) have been
assigned to the observed mesons and many members of the spectrum are still missing.
The established states are $\eta'(954)$ [$1^1S_0$], $\phi$ [$1^{3}S_{1}$],
 $h_1(1380)$ [$1^1P_1$], $f_1(1420)$ [$1^3P_1$],  $f^\prime_2(1525)$ [$1^3P_2$],
$\phi(1680)$ [$2^3S_1$], $\phi_3(1850)$ [$1^3D_3$]. Another possible strangeonium is the $\phi(2170)$, which can be either a $3^3S_1$ or a $2^3D_1$ state. It was first observed in the initial-state radiation (ISR) process $e^+e^-\rightarrow \gamma \phi f_0(980)$~\cite{Aubert:2006bu}
with a mass $m=2175(18)\MeV$ and width $\Gamma=58(26)\MeV$.
The poor experimental situation is due to the small production rates in low-energy hadronic processes, the large widths of the strangeonia since all states except for the $\eta'$ are above $K\bar K$ threshold,
and significant mixing with $u\bar u$ and $d\bar d$ states.  High-statistics electron--positron machines in the charm region are especially well suited to study these states~\cite{Liu:2015zqa}, since there is a significant fraction of $s\bar s$ states produced by the electromagnetic current (compare the discussion in Sec.~\ref{sec:production}).
\subsection{Meson--meson scattering}\label{sec:meson-meson}

Isospin and crossing symmetry allow us to relate amplitudes of different 
scattering processes of the $\pi$, $K$, and $\eta$ mesons, such that all possible amplitudes can be expressed in terms of only eight independent ones.
Pion--pion scattering in the isospin limit is strongly constrained by crossing symmetry, see Figs.~\ref{fig:MMAmplitude}(a) and \ref{fig:Mandelstam}(a).  It can be expressed in terms of a single function $A(s,t,u)$ according to
\begin{equation}
    M(\pi^a \pi^b \to \pi^c \pi^d ) = \delta^{ab}\delta^{cd} A(s,t,u) 
+  \delta^{ac}\delta^{bd} A(t,u,s) 
+  \delta^{ad}\delta^{bc} A(u,s,t) \,. \label{eq:defMpipi}
\end{equation}
The amplitudes of definite isospin are given in terms of this by
\begin{align}
T^{I=0} &= 3A(s,t,u)+A(t,u,s)+A(u,s,t) \,,\nonumber\\
T^{I=1} &= A(t,u,s)-A(u,s,t) \,, \nonumber\\
T^{I=2} &= A(t,u,s)+A(u,s,t) \,. \label{eq:Ipipi}
\end{align}
Kaon--pion scattering is obviously less symmetric, but similarly given in terms of a single function: crossing relates $K\pi\to K\pi$ to $K\bar K\to \pi\pi$.  The possible isospin amplitudes $I=1/2, 3/2$ and $I=0,1$, respectively, can be 
expressed by one $I=3/2$ amplitude corresponding to the $K^+\pi^+\to K^+\pi^+$ process~\cite{Petersen:1971ai,GomezNicola:2001as}. 
$\bar KK$ scattering can be of total isospin $I=0,1$, which can be obtained as linear combinations of elastic $K^+K^-$ as well as charge-exchange $K^+K^-\to K^0\bar K^0$ scattering~\cite{GomezNicola:2001as}.  
The remaining processes involve just one isospin amplitude each:
crossing symmetry relates $K\pi\to K\eta$ ($I=1/2$) to $\bar KK\to\pi\eta$ ($I=1$), $K\eta\to K\eta$ ($I=1/2$) to $\bar KK\to\eta\eta$ ($I=0$), and
$\pi\eta\to \pi\eta$ ($I=1$) to $\pi\pi\to\eta\eta$ ($I=0$); the final process is $\eta\eta$ scattering. 

Traditionally the experimental information about pseudoscalar meson scattering comes from $\pi N$ and $K N$ meson production experiments, where peripheral events are assumed to be dominated by the $t$-channel single-meson exchange. In particular phase shifts for $\pi\pi$ scattering for low partial 
waves were studied in detail in the past. 
\begin{figure}
    \centering
     \includegraphics[height=0.9\textheight]{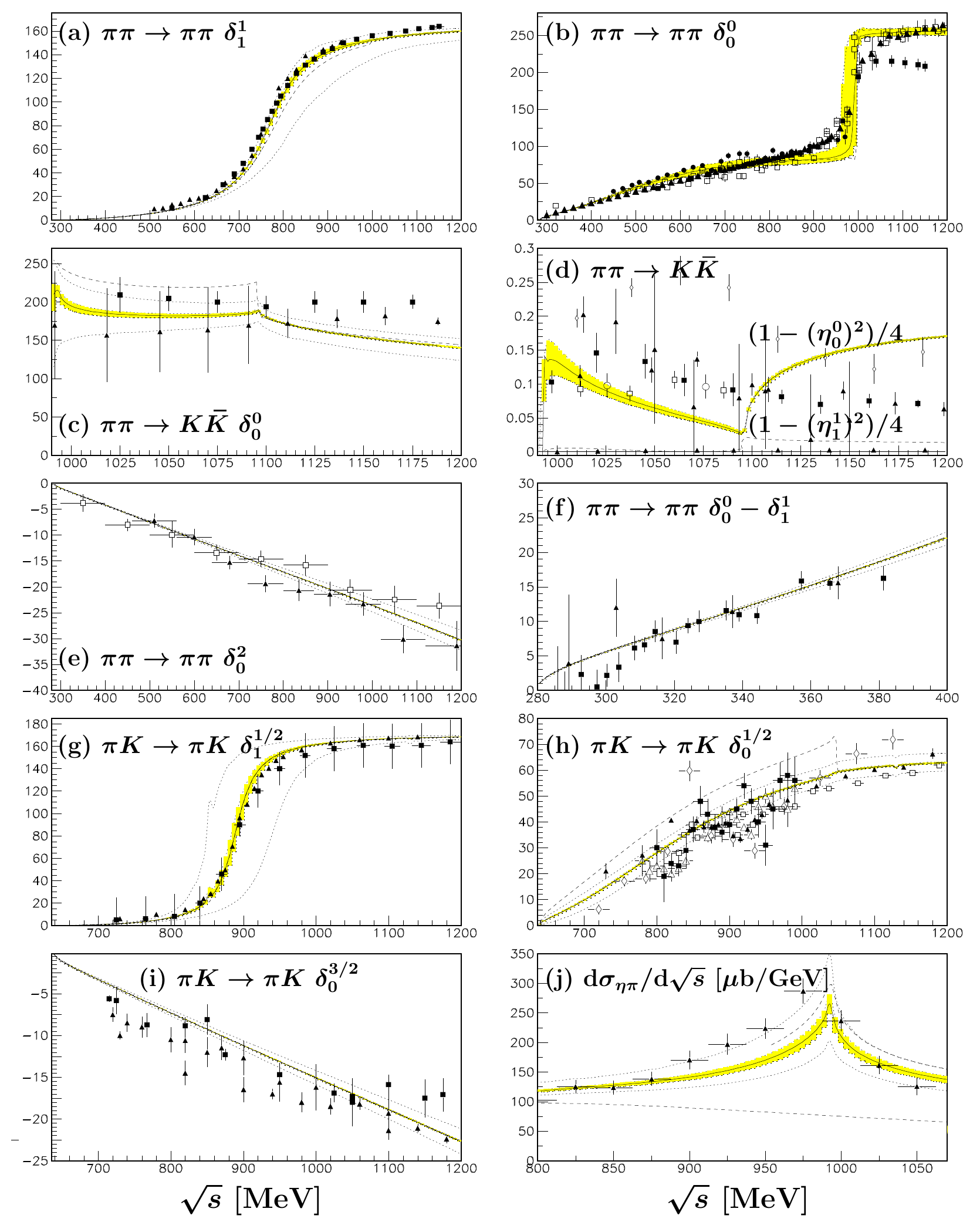}
    \caption[Meson--meson phase shifts]{Compilation of experimental data on pseudoscalar meson--meson scattering amplitudes. See description in the text. Figure adapted from~\cite{GomezNicola:2001as}. 
    }
    \label{fig:PPphase}
\end{figure}
A compilation of meson--meson scattering amplitudes $t_\ell^I(s)$ (related to phase shifts $\delta_\ell^I(s)$ and inelasticity parameters $\eta_\ell^I(s)$ via Eq.~\eqref{eqn:tIj}) for $\sqrt{s}<1.2\GeV$ and $\ell=0,1$ from Ref.~\cite{GomezNicola:2001as} is shown in Fig.~\ref{fig:PPphase}. 
The theoretical curves shown for comparison are calculated using the inverse-amplitude method, which we will briefly introduce in Sec.~\ref{sec:IAM}.
Here is an extensive caption of this figure:
\begin{itemize}
\item Amplitude $t_1^1(s)$, states $\pi\pi$ and $K\bar K$. In 
Fig.~\ref{fig:PPphase}(a) data on the $\pi\pi$ scattering phase shift
from Refs.~\cite{Protopopescu:1973sh} (squares) and~\cite{Estabrooks:1974vu} (triangles) is shown.
This channel is completely dominated by the $\pi\pi$ state and the $\rho$ resonance therein, and there is almost no inelasticity due to
$K\bar{K}$ production below $1200\MeV$; in fact, the inelasticity is rather dominated by $4\pi$, especially clustered as $\pi\omega$ above its threshold~\cite{Niecknig:2012sj}. The  $(1-|\eta_1^1|^2)/4$ points from Ref.~\cite{Martin:1979gm} are shown in the lowest part  of Fig.~\ref{fig:PPphase}(d).

\item Amplitude $t_0^0(s)$, states $\pi\pi$, $K\bar K$, and $\eta\eta$. 
By general consent, the coupling of the $\eta\eta$ state is believed to be rather weak, such that this amplitude is often described as a two-channel system $\pi\pi\leftrightarrow K\bar K$.
In this case,
there are three observables with several sets of data, which, as
can be seen in Figs.~\ref{fig:PPphase}(b), (c), and (d), are somewhat incompatible. For the $\pi\pi$ scattering
phase shift $\delta_0^0$, see Fig.~\ref{fig:PPphase}(b), the experimental data
is from Refs.~\cite{Hyams:1973zf} (open squares),
\cite{Protopopescu:1973sh} (solid squares), \cite{Estabrooks:1974vu} (solid triangles), and
\cite{Froggatt:1977hu} (solid circles). Concerning the $\pi\pi\to
K\bar{K}$ phase, the data in Fig.~\ref{fig:PPphase}(c) corresponds to
Refs.~\cite{Martin:1979gm} (solid triangles) and~\cite{Cohen:1980cq} (solid squares); both are reasonably compatible.  Figure~\ref{fig:PPphase}(d) shows data for $(1-|\eta_0^0|^2)/4$, which contains the information on the modulus of the $\pi\pi\to K\bar{K}$ amplitude.  At higher energies, further inelastic channels become relevant, in particular $4\pi$.

\item Amplitude $t_0^2(s)$, state $\pi\pi$. In Fig.~\ref{fig:PPphase}(e)
the $\delta^2_0$ phase shifts from Refs.~\cite{Hoogland:1977kt} (open squares) and~\cite{Losty:1973et} (solid triangles) are displayed.

\item Amplitude $t_1^{1/2}(s)$, states $K\pi$ and $K\eta$.
The $K\pi$ $P$-wave resonating in the $K^*(892)$ couples very weakly 
to $K\eta$, which is usually ignored; the inelasticity at higher energies is rather due to $K\pi\pi$ intermediate states (see, e.g., Ref.~\cite{Moussallam:2007qc}).
Fig.~\ref{fig:PPphase}(g) shows data on $\delta_1^{1/2}(s)$ from Refs.~\cite{Mercer:1971kn} (solid squares) and~\cite{Estabrooks:1977xe} (solid triangles).  --- The amplitude $t_1^{3/2}(s)$ ($K\pi$) is very weak and usually disregarded.

\item Amplitude $t_0^{1/2}(s)$, states $K\pi$ and $K\eta$.
Again, $K\eta$ is phenomenologically in fact not the most relevant inelasticity for the $K\pi$ $S$-wave, which is rather $K\eta'$ (at a much higher threshold)~\cite{Jamin:2001zq}.
The data in Fig.~\ref{fig:PPphase}(h) comes from Refs.~\cite{Mercer:1971kn} (solid squares), \cite{Bingham:1972vy} (open triangles), \cite{Baker:1974kr}
(open diamonds),
\cite{Estabrooks:1977xe} (solid triangles), and~\cite{Aston:1987ir} (open squares).

\item Amplitude $t_0^{3/2}(s)$, state $K\pi$.
The data sets from~\cite{Estabrooks:1977xe} (solid squares) and~\cite{Linglin:1973ci} 
 (solid triangles) are plotted in Fig.~\ref{fig:PPphase}(i).

\item Amplitude $t_0^1(s)$, states $\pi\eta$ and $K\bar K$.
Fig.~\ref{fig:PPphase}(j) shows the $\pi\eta$ effective-mass distribution  from the $pp\to p(\eta\pi^+\pi^-)p$ reaction~\cite{Armstrong:1991rg}.
\end{itemize}
The extrapolations cause most of the deviations between the analyses, where a single data set can lead to contradictory results due to different assumptions and approximations. Some of the $K\pi\to K\pi$ production experiments are also limited by statistics.

Beyond the attempted ``direct'' access to meson scattering data, theoretical considerations provide very strong constraints on at least some of the meson scattering channels discussed above, and extremely accurate representations of the leading partial waves at low-to-moderate energies are available in particular for pion--pion~\cite{Ananthanarayan:2000ht,Colangelo:2001df,GarciaMartin:2011cn,Caprini:2011ky,Pelaez:2019eqa}
and kaon--pion~\cite{Buettiker:2003pp,Pelaez:2018qny,Pelaez:2020gnd} scattering.
We will briefly sketch the principles behind those analyses in Sec.~\ref{sec:Roy} on Roy equations.

In the course of this review, we will not so much discuss scattering reactions as such, but rather production reactions, in particular via electroweak currents, whose final-state interactions allow to infer detailed information on scattering phase shifts.  A prime example for this are the decays $K\to\pi\pi l\nu_l$ ($K_{l4}$), which---according to the Pais--Treiman method~\cite{Pais:1968zza}---allow to extract the $\pi\pi$ scattering phase shift difference $\delta_0^0-\delta_1^1$ in a model-independent way.  Corresponding data is shown in Fig.~\ref{fig:PPphase}(f) from Ref.~\cite{Rosselet:1976pu} (solid triangles) and, of higher precision, from Refs.~\cite{Truoel:2000ev,Pislak:2001bf} (solid squares); yet more recent data from the NA48/2 collaboration~\cite{Batley:2010zza} is not included.  As these data are necessarily at very low energies ($\sqrt{s}<m_K$), $K_{l4}$ decays fix the low-energy properties of the $\pi\pi$ scattering amplitude as predicted from chiral perturbation theory with high precision~\cite{Colangelo:2001sp}, which in turn, e.g., determines the position of the $f_0(500)$ resonance very accurately~\cite{Caprini:2005zr,GarciaMartin:2011jx,Pelaez:2015qba}.

There are other examples of $\pi\pi$ scattering information extracted from production, most obviously the $P$-wave as constrained from data on the pion vector form factor~\cite{DeTroconiz:2001rip,deTroconiz:2004yzs,Colangelo:2018mtw,Colangelo:2020lcg}.  Also, $B$-decay data has been used to extract the (strange) scalar pion form factors at energies up to $2\GeV$, thereby also constraining the underlying $S$-matrix for isoscalar $S$-wave scattering~\cite{Ropertz:2018stk}.

For other meson pairs, production data is essentially the only access we have to their scattering phase shifts; this is, e.g., the case for $\pi\eta$ scattering, see Fig.~\ref{fig:PPphase}(j).  Modern parameterizations of the $S$-wave based on chiral constraints~\cite{Albaladejo:2015aca} have most recently been adjusted~\cite{Lu:2020qeo} to $\gamma\gamma\to\pi^0\eta$ data from Belle~\cite{Uehara:2009cf}
(compare also Ref.~\cite{Danilkin:2017lyn} for a similar effort); a suggested similar extraction from $B$-decays has not been put into practice yet~\cite{Albaladejo:2016mad}. 

There are first results for meson--meson scattering from the lattice, where the status until 2017 is reviewed in Ref.~\cite{Briceno:2017max}.  Among more recent results we wish to emphasize studies of isovector pion--pion scattering and determinations of the $\rho$ meson properties~\cite{Werner:2019hxc,Erben:2019nmx,Fischer:2020fvl}; see also Ref.~\cite{Niehus:2020gmf} for the link to chiral perturbation theory. This process is treated as a benchmark and recent developments are motivated by the goal to constrain hadronic contributions to $g-2$ of the muon~\cite{Aoyama:2020ynm}. 

Traditionally, amplitude analyses at higher energies rely on modeling of the $s$-dependence of the meson--meson interaction as a coherent sum of resonant  Breit--Wigner amplitudes $\BW_k(s)$, described in Sec.~\ref{sec:eePP}:
\begin{equation}
  t_\ell(s)=\sum_{k=1}^Nc_k \BW_k(s) \,,  
\end{equation}
where $c_k$  represent complex coupling parameters  of the various contributions. Such modeling  provides a practical way to parameterize the data accurately by including additional resonances until adequate agreement is achieved. However, this approach violates the relation between magnitude and phase of the amplitude based on unitarity and analyticity. This is particularly problematic for broad, overlapping resonances. Therefore, the extracted parameters lack direct correspondence to the underlying meson--meson amplitude and it is difficult to relate different experiments.
In the $K$-matrix formalism~\cite{Aitchison:1972ay}, contributions from broad overlapping resonances 
can be combined preserving unitarity. Here the scattering amplitude is expressed as 
\begin{equation}
    t_\ell(s)=K(s)\left[1-i\sigma(s) K(s)\right]^{-1} \,,
\end{equation}
where $K(s)$ is a real function and $\sigma(s)$ the phase space factor including the branch cut for the corresponding channel.
The $K$ functions describing individual resonances can be combined as a sum, while unitarity will still be preserved. Scattering of two of the final-state particles to different coupled channels is described in this method by considering $K$-matrices representing the interactions between the channels. 
Such a $K$-matrix method allows one, e.g., to relate and constrain data from coupled channels like $\pi\pi\to\pi\pi$, $\pi\pi\to K\bar K$, and $K\bar K\to K\bar K$ channels in $I=0$, $J=0$~\cite{Au:1986vs}. 
In Sec.~\ref{sec:VtoPPg}, we will show some results for  $J/\psi$ radiative decays into $\pi\pi$, $K_SK_S$, and $\eta\eta$ coupled channels analyzed in such a way. An extended coupled-channel analysis using the $K$-matrix formalism has been 
performed using data from pion production, $p\bar{p}$ and $n\bar{p}$ annihilation, and $\pi\pi$ 
scattering~\cite{Anisovich:2002ij}, resulting in $K$-matrix amplitudes of the transitions
into channels $\pi\pi$, $K\bar K$ , $\eta\eta$, $\eta\eta'$, and $4\pi$.

A special case of such a coupled-channel analysis is the $f_0(980)$ decay into 
$\pi\pi$ and $K\bar K$ final states. The resonance is located close to the  $K\bar K$ threshold and the shape can be parameterized by the Flatt{\'e} formula~\cite{Flatte:1976xu}:
\begin {equation}
f(s) = \frac {1}{m_{f_0}^2 - s - i\left[g_1\sigma _{\pi}(s) + g_2\sigma _{K}(s)\right]} \,,\label{eq:Flatte}
\end {equation}
where $\sigma_P (s)= 2q_P/\sqrt {s}$ is the phase space factor for each decay channel, with $q_P$ the c.m.\ momentum in the resonance rest frame.

Once a set of strongly interacting particles is created in a production process, they will interact in the final state in the way dictated by the scattering phases.
In general, to relate multiparticle final-state interactions to the two-body interactions, one has to solve complicated equations for which not all required input is available. If there are only two hadrons in the final state or the interaction for a certain pair of the final hadrons is much stronger, unitarity relations for two-particle system can be used. They were first derived by Watson~\cite{Watson:1952ji}. The amplitudes for such two-body processes can be represented as a product of a universal and a reaction-specific part. The universal part $\Omega (s)$ is given by the so-called Omn\`es function~\cite{Omnes:1958hv} describing final-state interactions between the mesons generated by the scattering phase.

\subsection{Theory of low-energy strong interactions}\label{sec:theoryintro}
\subsubsection{Chiral perturbation theory}\label{sec:ChPT}
QCD, the theory of the strong interactions, is formulated in terms of fundamental quark and gluon fields,
with local  $SU(3)_c$ gauge symmetry related to color charge.  The main parts of its Lagrangian are given by
\begin{equation}
\mathcal{L}_{\rm QCD} = -\frac{1}{4} G_{\mu \nu}^a G^{a \mu \nu} + \sum_{q} \bar{q} \left(i  \slashed{D} - m_q \right) q  \, , \label{eq:L_QCD}
\end{equation}
where the sum extends over all quark flavors $q$. $D_\mu = \partial_\mu + i g_s \lambda^a G_\mu^a/2$ is the covariant derivative, $G_\mu^a$ stands for the gluon field with strong coupling $g_s$, 
$G_{\mu \nu}^a = \partial_\mu G_\nu^a- \partial_\nu G_\mu^a - g_s f^{abc}G_\mu^b G_\nu^c$ is the gluon field strength tensor, 
and $\lambda^a$ are the Gell-Mann matrices.  We omit gauge-fixing terms etc.\ for simplicity.

A crucial feature of
this nonabelian theory, as compared to QED, is the self-coupling of the gauge bosons, the gluons, which carry color charge themselves.  As a consequence, the behavior of the strong coupling constant under the renormalization group, i.e., its change with the energy scale, is characteristically different: QCD shows
asymptotic freedom, which means that the coupling strength becomes small, hence the theory as such perturbative, for large-momentum-transfer processes.  This feature allows
to calculate QCD processes at high energies using the well-known perturbative methods in the coupling constant $\alpha_s = g_s^2/(4\pi)$.
At the other end, however, the running coupling grows at small energies, which is presumably related to the phenomenon of confinement
of the quark and gluon degrees of freedom inside color-neutral hadronic states.  In this regime, for low-momentum-transfer reactions, a large coupling in particular prohibits predicting static properties of hadrons and low-energy interactions using a perturbation expansion in $\alpha_s$. There is no direct (analytic) link between QCD's fundamental degrees of freedom and the observed spectrum of hadrons. 

\begin{sloppypar}
The key concept to understand this energy regime is chiral symmetry.  
To illustrate this, we decompose the quark fields into their chiral components
\begin{equation}
q = \frac{1}{2}(1-\gamma_5)q + \frac{1}{2}(1+\gamma_5)q
\equiv P_{L} q + P_{R} q \equiv q_{L} + q_{R} \,, 
\end{equation}
whereby we can rewrite the QCD Lagrangian Eq.~\eqref{eq:L_QCD} according to
\begin{align}
 \Lagr_{\rm QCD} = \Lagr_{\rm QCD}^0 - \Lagr_{\rm QCD}^m + \ldots  \,, \quad 
\Lagr_{\rm QCD}^0 &= -\frac{1}{4} G_{\mu\nu}^a G^{\mu\nu,a} + 
i \bar q_L \slashed{D} q_L + i \bar q_R \slashed{D} q_R \,, \nonumber\\
\Lagr_{\rm QCD}^m &= \bar q_L \M q_R + \bar q_R \M^\dagger q_L \,,
\end{align}
where $q$ now denotes a vector collecting the light quark flavors $q^T = (u,d,s)$,
and $\M = {\rm diag}(m_u,\,m_d,\,m_s)$ is the quark mass matrix.
The ellipsis denotes the heavier quark flavors.
We note that, besides the obvious symmetries like
Lorentz invariance, $SU(3)_c$ gauge invariance, 
and the discrete symmetries $P$, $C$, $T$, $\Lagr_{\rm QCD}$ displays
a {chiral} symmetry in the limit of vanishing quark masses (which is hence
called ``chiral limit''):
$\Lagr_{\rm QCD}^0$ is invariant under chiral $U(3)_L \times U(3)_R$ flavor transformations, 
\begin{equation}
    (q_L, q_R) \longmapsto (L q_L, R q_R) \,, \quad L,R \in U(3)_{L,R} \,.
\end{equation}
As the masses of the three light quarks are small on the typical 
hadronic scale,
\begin{equation}
m_{u,d,s} \ll 1\GeV \approx \Lambda_{\text{hadr}} \,,
\end{equation}
there is hope that the real world is not too far from the chiral limit, such that
one may invoke a perturbative expansion in the quark masses.  
\end{sloppypar}

If we rewrite the symmetry group according to
\begin{equation}
U(3)_L \times U(3)_R = 
SU(3)_L \times SU(3)_R \times U(1)_V \times U(1)_A \,,
\end{equation}
where we have introduced vector $V=L+R$ and axial vector $A=L-R$ transformations,
and consider the Noether currents associated with this symmetry group,
it turns out that the different parts of it are realized in very different ways in nature. 
The $U(1)_V$ current $V_\mu^0 = \bar q \gamma_\mu q$,
the quark number or baryon number current, is a conserved current in the Standard Model; on the other hand, 
the $U(1)_A$ current $A_\mu^0 = \bar q \gamma_\mu \gamma_5 q$ is
broken by quantum effects, the  $U(1)_A$ anomaly, and 
is not a conserved current of the quantum theory.
As far as the chiral symmetry group $SU(3)_L \times SU(3)_R$ 
and its conserved currents 
\begin{equation}
V_\mu^a = \bar q \gamma_\mu \dfrac{\lambda^a}{2} q \,,  \qquad
A_\mu^a = \bar q \gamma_\mu \gamma_5 \dfrac{\lambda^a}{2} q \,,  \qquad
\partial^\mu V_\mu^a  = \partial^\mu A_\mu^a = 0 \,, \qquad
   a=1,\ldots, 8 \,,
\end{equation}
are concerned, they are certainly broken {explicitly} by the quark masses, but this is expected to be a small effect.

In nature, chiral symmetry is realized in the Goldstone mode, i.e., the symmetry is spontaneously broken: we find (approximate) $SU(3)_V$ multiplets, but no parity doubling is observed. The symmetry breaking pattern is
\begin{equation}
SU(3)_L \times SU(3)_R \stackrel{{\rm SSB}}{\longrightarrow} 
SU(3)_V \,.
\end{equation}
Accordingly, the axial charges commute with the Hamiltonian, but do not leave the ground state invariant.
As a consequence, massless excitations, so-called ``Goldstone bosons'' appear,
which are noninteracting for vanishing energy.
In the case at hand, the eight Goldstone bosons should be pseudoscalars, which
the lightest hadrons in the spectrum indeed are, namely 
$\pi^\pm$, $\pi^0$, $K^\pm$, $K^0$, $\bar K^0$, and $\eta$. 

The task is therefore to construct a low-energy theory for these Goldstone bosons, and
the first step to this end is to find a Lagrangian that fulfills all QCD symmetries.  
This can be constructed in terms of the matrix field~\cite{Coleman:1969sm,Callan:1969sn}
\begin{equation}
U=\exp \bigg(\frac{i\phi}{F_0}\bigg) \,, \quad
\phi = \sqrt{2} \left( \begin{array}{ccc} 
\frac{\pi^0}{\sqrt{2}} + \frac{\eta}{\sqrt{6}} & \pi^+& K^+ \\
\pi^- & -\frac{\pi^0}{\sqrt{2}} + \frac{\eta}{\sqrt{6}} & K^0 \\
K^- & \bar K^0 & -\frac{2\eta}{\sqrt{6}} \end{array} \right) \,,\label{eq:defU}
\end{equation}
that transforms according to $U \mapsto L U R^\dagger$ under chiral transformations, $L\in SU(3)_L$, 
$R\in SU(3)_R$,
and $F_0$ is a dimensionful constant that can be identified with the pion
decay constant (in the $SU(3)$ chiral limit), 
$F_0 \approx F_\pi = 92.2\MeV$.\footnote{We have neglected the isospin-breaking effect of $\pi^0$-$\eta$ mixing in Eq.~\eqref{eq:defU}.}
As we want to construct a low-energy effective theory,
the guiding principle is to use the power of momenta or derivatives to order 
the importance of various possible terms.  ``Low energies'' here refer to 
a scale well below $1\GeV$, i.e., an energy region where the Goldstone bosons
are the only relevant degrees of freedom.
Lorentz invariance dictates that Lagrangian terms can only come in even powers 
of derivatives, hence $\Lagr$ is of the form
\begin{equation}
    \Lagr = \Lagr^{(0)} + \Lagr^{(2)} + \Lagr^{(4)} + \ldots \,.
\end{equation}
However, as $U$ is unitary, $U U^\dagger = \mathds{1}$, 
$\Lagr^{(0)}$ can only be a constant.  
Therefore, in accordance with the Goldstone theorem, the leading term in the Lagrangian
is $\Lagr^{(2)}$, which already involves derivatives.  It can be shown to consist of 
one single term,
\begin{equation}
 \Lagr^{(2)} = \frac{F_0^2}{4} \langle \partial_\mu U \partial^\mu U^\dagger \rangle \,, \label{eq:L2}
\end{equation} 
where $\langle\ldots\rangle$ denotes the trace in flavor space.
Expanding $U$ in powers of $\phi$, 
we find the canonical kinetic terms
\begin{equation} 
\Lagr^{(2)} = \partial_\mu \pi^+ \partial^\mu \pi^- + \partial_\mu K^+ \partial^\mu K^- + \ldots  \,,
\end{equation}
and indeed no mass terms: in the chiral limit, the Goldstone bosons are massless.  

\begin{sloppypar}
To render the theory more realistic, we need to reintroduce the quark
mass matrix $\M=\text{diag}(m_u,m_d,m_s)$ as a perturbation.
At leading order, i.e., to linear order in the quark masses and
without any further derivatives, one finds exactly one term in the chiral Lagrangian,
such that $\Lagr^{(2)}$ is then of the complete form
\begin{equation}
\Lagr^{(2)} = \frac{F_0^2}{4} \langle \partial_\mu U \partial^\mu U^\dagger 
+2B_0 \bigl( \M U^\dagger + \M^\dagger U \bigr) \rangle \,. \label{eq:L2mass}
\end{equation}
$B_0$ is an additional parameter that is related to the quark condensate
in the chiral limit, $B_0 = - \langle 0|\bar uu|0\rangle/F_0^2$.
Expanding once more in powers of $\phi$, we can read off the mass terms and find, e.g., 
\begin{equation}
m_{\pi^\pm}^2=B_0(m_u+m_d) \,, \quad
m_{K^\pm}^2  = B_0(m_u+m_s) \,, \quad
m_{K^0}^2   = B_0(m_d+m_s) \,. \label{eq:GMOR}
\end{equation}
This demonstrates why, in a common power counting scheme that assigns the generic order $p$ to both energies/momenta and Goldstone boson masses, quark masses are counted as $\Order(p^2)$.
\end{sloppypar}

\begin{figure}
\centering
\includegraphics[width=0.5\textwidth]{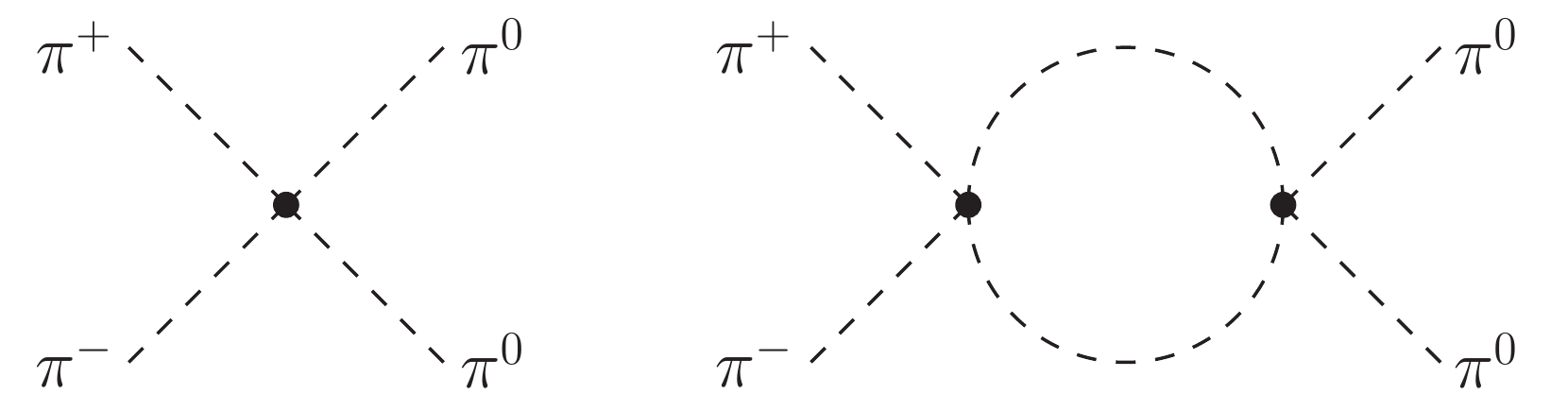}
\caption[Pion--pion interactions]{\label{fig:pipi} Pion--pion scattering in the current algebra approximation, 
i.e., in the lowest order of ChPT (left), as well as an example of a one-loop or next-to-leading-order (NLO) contribution  (right).}
\end{figure}

Due to the nonlinear field representation in $U$, the Lagrangian $\Lagr^{(2)}$ contains much more information than the masses, but also interaction terms. 
In particular, we can calculate the invariant amplitude $A(s,t,u)$ of pion--pion scattering, see Eq.~\eqref{eq:defMpipi}, to be
\begin{equation}
 A(s,t,u) = \frac{s-m_\pi^2}{F^2_{0}} \,, \label{eq:pipiCA}
\end{equation}
a parameter-free prediction~\cite{Weinberg:1966kf}.
If we furthermore define the $S$-wave scattering lengths of definite isospin, proportional to the scattering amplitudes of Eq.~\eqref{eq:Ipipi} at threshold, 
$ a_0^I = T^I (s=4 m_\pi^2,t=u=0)/32\pi $,
we find
\begin{equation}
    a_0^0 = \frac{7 m_\pi^2}{32\pi F_\pi^2} = 0.16 \,, \qquad
a_0^2 = - \frac{m_\pi^2}{16\pi F_\pi^2} = -0.045 \,. \label{eq:a0Op2}
\end{equation}
Diagrammatically, this is represented by the tree-level diagram of Fig.~\ref{fig:pipi}(left).

The Lagrangian~\eqref{eq:L2mass} as well as the relations~\eqref{eq:GMOR}, \eqref{eq:pipiCA}, and \eqref{eq:a0Op2} 
are only the leading terms in a systematic approach called chiral perturbation theory (ChPT)~\cite{Weinberg:1978kz,Gasser:1983yg,Gasser:1984gg}, for which there are many comprehensive reviews available~\cite{Bijnens:1994qh,Ecker:1994gg,Pich:1995bw,Manohar:1996cq,Scherer:2002tk,Gasser:2003cg,Bernard:2006gx,Scherer:2012xha}.
Equation~\eqref{eq:pipiCA} immediately renders the necessity of higher-order corrections obvious: the scattering amplitude is real, while the optical theorem (applied to partial waves  $t_\ell^I$) requires imaginary parts of the form 
\begin{equation}
{\rm Im} \,t_\ell^I = \sqrt{1-\frac{4 m_\pi^2}{s}} \left| t_\ell^I \right|^2 \,. \label{eq:PWunitarity} 
\end{equation}
It is well known from the Born series in quantum-mechanical scattering that the optical theorem links terms of different orders in perturbation theory to each other: if (the real part of) $t_\ell^I$
is of $\Order(p^2)$, Eq.~\eqref{eq:PWunitarity} shows that ${\rm Im} \,t_\ell^I$ will start at $\Order(p^4)$, as a loop correction.  
Furthermore, all higher-order terms in the effective Lagrangian are allowed: they ought to be smaller at low energies, but for precision predictions and estimates of theoretical uncertainties, these corrections should be evaluated.
In addition, the loop corrections entail divergences in the real parts of amplitudes, which have to be removed by appropriate counterterms.  The theory is nonrenormalizable in the conventional sense: it contains interaction terms of negative mass dimension, and hence cannot be rendered finite to all orders by a finite number of counterterms: new operators have to be introduced at each order in perturbation theory.  ChPT is however renormalizable, and therefore predictive, order by order in the low-energy expansion.
The number of independent terms and corresponding free coupling constants
increases rapidly at higher orders, with the $SU(3)$ version of it containing 2 ($F_0$, $B_0$), 10~\cite{Gasser:1984gg}, and 90~\cite{Bijnens:1999sh} terms at orders $p^2$, $p^4$, and $p^6$, respectively.

As a perturbative expansion needs to proceed in powers of some dimensionless ratio $p/\Lambda_\chi$, let us briefly discuss the size of the typical breakdown scale $\Lambda_\chi$.  From a renormalization-group-type of analysis, it can be shown that loop corrections are generically expected to generate a scale $\Lambda_\chi = 4\pi F_0$.  On the other hand, we have to remember that we
have constructed an effective theory for Goldstone bosons, 
which are the only dynamical degrees of freedom.
The effective theory must fail once the energy reaches the resonance region, hence for momenta of the order of $\Lambda_\chi \approx m_{\rm res}$.  Both of these arguments hence point towards a breakdown scale of the order of $1\GeV$.

The appearance of many unknown free Lagrangian parameters, usually called low-energy constants (LECs), at higher orders does not preclude the theory from making predictions.  This is due to the fact that the same operators still contribute to various observables at the same time, and thus provide nontrivial relations between them.
As an illustration for this principle, we briefly discuss the next-to-leading-order corrections to the pion--pion $S$-wave scattering length of isospin 0, $a_0^0$ (which is actually even known at next-to-next-to-leading order~\cite{Bijnens:1995yn,Bijnens:1997vq}).  An example for a loop diagram contributing therein is given in Fig.~\ref{fig:pipi}(right). 
The $\Order(p^4)$ corrections are of the form~\cite{Gasser:1983yg}
\begin{equation}
a_0^0 = \frac{7 m_\pi^2}{32\pi F_\pi^2} \Bigl\{ 1+\epsilon + \Order(m_\pi^4) \Bigr\} \,, \qquad
\epsilon = \frac{5 m_\pi^2}{84\pi^2 F_\pi^2} \left( {\bar\ell_1} + 2{\bar\ell_2} 
+ \frac{3}{8}{\bar\ell_3}
+ \frac{21}{10}{\bar\ell_4} + \frac{21}{8} \right) \,. \label{eq:a00p4}
\end{equation}
Here, $\bar\ell_i$ refers to the renormalized, i.e., finite and scale-independent, LECs~\cite{Gasser:1983yg}.
There are two different types of LECs in the expression~\eqref{eq:a00p4}.
$\bar \ell_1$ and $\bar \ell_2$ come with structures containing four derivatives, i.e., they survive in the chiral limit
and can be determined from the momentum dependence of the $\pi\pi$ scattering amplitude,
namely from $D$-waves. $\bar \ell_3$ and $\bar \ell_4$ however are 
symmetry breaking terms that specify the quark mass dependence, therefore they
cannot be determined from $\pi\pi$ scattering alone.
The LEC $\bar\ell_3$ is particularly interesting as it determines the extent to which the pion mass is given by the leading term in the quark mass expansion, see Eq.~\eqref{eq:GMOR}:
\begin{equation}
m_{\pi}^2 = m^2 - \frac{m^4}{32\pi^2F_\pi^2}\bar\ell_3 + \Order(m^6) \,,
\end{equation}
where we have denoted the $\Order(p^2)$ relation by $m^2=2B\hat m \equiv B(m_u+m_d)$ (and $B$ is the $SU(2)$ analogue of $B_0$).  
The constant $\bar \ell_4$, on the other hand, is related to the scalar form factor of the pion $\Gamma(s)$,
\begin{equation}
   \langle \pi^a(p) \pi^b(p') \,|\, \hat m(\bar u u+\bar d d)\,|\,0 \rangle
= \delta^{ab} {\Gamma(s)} \,, \quad s=(p+p')^2 \,.
\end{equation} 
At tree level, one has  ${\Gamma(s)} = 2B\hat m = m_\pi^2 + \Order(p^4) $
in accordance with the Feynman--Hellmann theorem~\cite{Hellmann:1937,Feynman:1939zza},
$
{\Gamma(0)} = \langle \pi | \hat m\, \bar qq |\pi\rangle = \hat m \,{\partial m_\pi^2}/{\partial\hat m} 
\,.
$
At next-to-leading order, one  defines the scalar radius $\langle r^2\rangle_\pi^S$ according to
\begin{equation} 
{\Gamma(s)} = \Gamma(0) \left\{ 1+\frac{s}{6}{\langle r^2\rangle_\pi^S} + \Order(s^2) \right\}  
\,, \qquad
{\langle r^2\rangle_\pi^S} = \frac{3}{8\pi^2 F_\pi^2} \left( {\bar\ell_4} - \frac{13}{12} \right) 
+ \Order(m_\pi^2) \,,
\end{equation}
therefore the scalar radius is directly linked to $\bar \ell_4$.
Although the scalar form factor is not directly experimentally accessible,
one can analyze $\Gamma(s)$ in dispersion theory~\cite{Donoghue:1990xh,Moussallam:1999aq,DescotesGenon:2000ct,Hoferichter:2012wf} and extract $\langle r^2\rangle_\pi^S$ that way.
These dispersively reconstructed scalar form factors are important for the phenomenology of $S$-wave pion--pion final-state interactions~\cite{Daub:2015xja,Ropertz:2018stk}, as well as for the description of hadronic matrix elements important in beyond-the-Standard-Model physics searches~\cite{Daub:2012mu,Celis:2013xja,Monin:2018lee,Winkler:2018qyg}. 

Traditionally, the relation laid out above have been utilized to determine the size of the quark condensate (via its relation to $\bar\ell_3$) through a determination of $a_0^0$~\cite{Colangelo:2001sp}.
Both $S$-wave scattering lengths have been predicted to remarkable precision
by combining their ChPT expansion with Roy equations, see Sec.~\ref{sec:Roy}, where they appear as subtraction constants~\cite{Colangelo:2000jc}:
\begin{equation}
    a_0^0 = 0.220(5) \,, \qquad a_0^2 = - 0.0444(10) \,. \label{eq:Roy-pipiscatt}
\end{equation}
Experimentally, these predictions have been tested most accurately by the NA48/2 collaboration, combining analyses of $K_{e4}$ decays~\cite{Batley:2010zza} and the cusp effect in $K^+\to\pi^+\pi^0\pi^0$~\cite{Batley:2000zz}, arriving at~\cite{Batley:2010zza}
\begin{equation}
    a_0^0 = 0.2210(47)_{\rm stat}(40)_{\rm syst} \,, \qquad a_0^2 = - 0.0429(44)_{\rm stat}(28)_{\rm syst} \,,
\end{equation}
hence demonstrating remarkable agreement with Eq.~\eqref{eq:Roy-pipiscatt}.
To arrive at this result, subtle radiative corrections need to be applied in the experimental analyses~\cite{Bissegger:2008ff,Colangelo:2008sm}.
We note that $a_0^2$ is very close to the leading-order ChPT prediction in Eq.~\eqref{eq:a0Op2}, while $a_0^0$ is enhanced quite significantly: this sizable correction can be understood in terms of a chiral logarithm with a large coefficient contained already in the one-loop correction, see Eq.~\eqref{eq:a00p4}.

The coupling of the chiral Lagrangians to external vector ($v_\mu$) and axial vector ($a_\mu$) fields 
is rather straightforward: the ordinary derivative is replaced
by a covariant one,
$
   D_\mu U = \partial_\mu U - i [v_\mu,U] - i \{ a_\mu,U\}
$.
Inserting the photon for the vector field, $v_\mu = e Q A_\mu$ with $Q$ denoting the quark charge matrix,  generates the couplings necessary to calculate, e.g., the electromagnetic
form factor of the pion or pion Compton scattering at leading order.  
Higher orders in addition require the field strength tensor (with appropriate chiral transformation behavior) as a Lagrangian building block.

Now we consider the application of ChPT to the pion vector form factor $\FV(s)$,
\begin{equation}
  \langle \pi^+(p) \pi^-(p') | j_\mu^{\rm em}(0) | 0 \rangle = 
  (p_\mu - p'_\mu) \, \FV(s) \,,
\end{equation}
where $j_\mu^{\rm em} = 2/3 \,\bar u\gamma_\mu u-1/3 \,\bar d\gamma_\mu d \pm\ldots$ denotes the electromagnetic current. 
At $\Order(p^4)$, there are loop diagrams contributing to $\FV(s)$ as well as a tree graph proportional to the low-energy constant $\bar \ell_6$. 
The full one-loop representation is given as~\cite{Gasser:1983yg}
\begin{align}
    \FV(s) &=1+\frac{1}{6F_\pi^2}(s-4m_\pi^2)\bar J(s)+\frac{s}{6}\left(\braket{r^2}_\pi^V+\frac{1}{24\pi^2F_\pi^2}\right) \,,
    \notag\\
    \bar J(s)&=\frac{1}{16\pi^2}\left\{\sigma_\pi\log\frac{\sigma_\pi-1}{\sigma_\pi+1}+2\right\} 
\,, \qquad \sigma_P=\sqrt{1-\frac{4m_P^2}{s}} \,.
\end{align}
The (squared) electromagnetic radius $\langle r^2\rangle_\pi^V$ is defined via
the expansion for small $s$,
\begin{equation}
\FV(s) = 1 + \frac{s}{6} \langle r^2\rangle_\pi^V + \Order(s^2) \,, \quad
\langle r^2\rangle_\pi^V = \frac{1}{16\pi^2 F_\pi^2} ({\bar \ell_6} - 1) \,.\label{eq:piVradius1}
\end{equation}
To understand the role of the LEC $\bar\ell_6$, we consider the contribution of the $\rho$ resonance to this form factor, 
\begin{figure}
\begin{center}
\includegraphics[width=7cm]{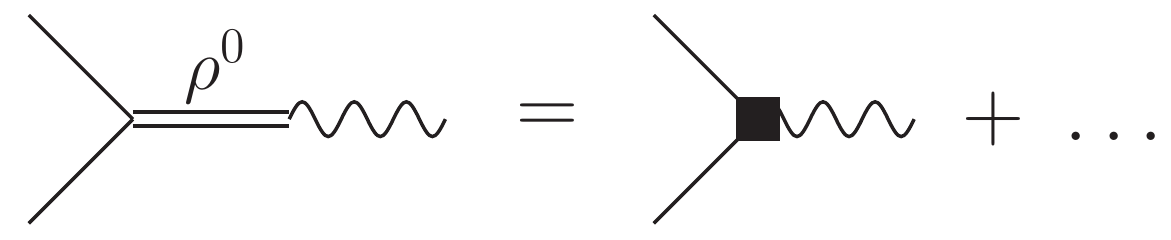}
\caption[Pion vector form factor in ChPT]{The contribution of the $\rho$ resonance to the pion vector form factor
can, at small $t$, be represented by a point-like counterterm.\label{fig:rhoVff}}
\end{center} \vspace{-3mm}
\end{figure}
as shown in Fig.~\ref{fig:rhoVff}. Expanding the $\rho$ propagator for $s \ll m_\rho^2$, 
\begin{equation}
   \frac{s}{m_\rho^2-s} = \frac{s}{m_\rho^2}\biggl(1+\frac{s}{m_\rho^2} + \ldots \biggr) \,,
\end{equation}
we find that identifying the leading term with the $\bar \ell_6$ contribution reproduces the empirical value for $\bar \ell_6$ nicely. This is a modern version of the VMD picture: where allowed by quantum numbers, the numerical values of LECs are dominated by the contributions of vector resonances~\cite{Ecker:1988te,Donoghue:1988ed}.

On the other hand, this connection also neatly demonstrates one of the central limitations of ChPT: it only parameterizes the low-energy tails of resonances in terms of approximate polynomials, but cannot, even at higher orders, reproduce the poles associated with resonant states.  This failure is intimately linked to the fact that ChPT, as a low-energy effective theory, only fulfills unitarity perturbatively:
Eq.~\eqref{eq:PWunitarity} neatly illustrates that the imaginary part of a partial wave is necessarily suppressed relative to the real part within the range of applicability of the chiral power counting scheme, while near a resonance, $\Im t_\ell^I \gg \Re t_\ell^I$.  
The limited range of convergence of the perturbative ChPT expansion is therefore closely related to the violation of exact unitarity.
As a consequence, 
there have been various attempts to 
extend the range of applicability of ChPT to higher energies by invoking unitarization in one way or the other.  These therefore often rely on dispersion-theoretical methods, which are based on the underlying analytic properties of quantum field theories.  

\subsubsection{Dispersion relations; Omn\`es formalism}
We have seen the first dispersion relation already in Eq.~\eqref{eq:DR-FpiV} for the pion vector form factor: in the spirit of the Kramers--Kronig relations from optics or electrodynamics, they allow to express a complex function (and specifically the real part therefrom) in terms of an integral over its imaginary part.  
Dispersion relations rely on the fundamental quantum-field-theoretical properties of analyticity, the mathematical formulation of causality, and unitarity, the consequence of probability conservation
(see Ref.~\cite{Eden:1966dnq} for the classic textbook and, e.g., Ref.~\cite{Oller:2019rej} for a recent one including current applications).  Analyticity allows to express complex functions such as form factors or scattering amplitudes in terms of contour integrals in the complex energy plane according to Cauchy's integral formula. Unitarity dictates that this complex energy plane contains poles and cuts starting from certain branch points, dictated by single- or multi-particle intermediate states, respectively.  The discontinuities along the branch cut are in many cases related to the imaginary parts in a straightforward way (following from Schwarz' reflection principle), $\text{disc}\,f(s) = 2i\,\Im f(s)$.  This results, in the simplest case, in dispersion relations of the form
\begin{equation}
    f(s) = \frac{1}{2\pi i}\int_{s_{\text{thr}}}^\infty \diff s' \frac{\text{disc}\,f(s')}{s'-s} 
    = \frac{1}{\pi}\int_{s_{\text{thr}}}^\infty \diff s' \frac{\Im f(s')}{s'-s} \,. \label{eq:sampleDR}
\end{equation}
Here, $s_{\text{thr}}$ denotes the branch point or outset of the unitarity cut, which in the simplest case---a two-particle intermediate state of equal masses $m$---is simply given by $s_{\text{thr}}=4m^2$.  If the integrals up to infinity in Eq.~\eqref{eq:sampleDR} do not converge, subtractions need to be introduced, which emphasizes the weight of the integrands at low energies (where the imaginary parts are typically known far better, as only few intermediate states are kinematically allowed), at the expense of introducing unknown parameters, the so-called subtraction constants, that need to be fixed from elsewhere.

The unitarity relation for a partial wave was already given in Eq.~\eqref{eq:PWunitarity}.  Of particular interest in the context of this review is a form factor unitarity relation, as it links production (e.g., of a pair of mesons through an electroweak current) to scattering (the final-state interaction of said pair of mesons in the given partial wave): e.g., the pion vector form factor obeys
\begin{equation}
\Im \FV(s) = \FV(s) \sigma_\pi(s) \big[t_1^1(s)\big]^* \theta(s-4m_\pi^2) \,. \label{eq:ImFV}
\end{equation}
In the elastic regime, the partial wave $t_J^I(s)$ can be expressed in terms of the phase shift $\delta_J^I(s)$ according to
\begin{equation}
t_J^I(s) = \frac{e^{i\delta_J^I(s)}\sin\delta_J^I(s)}{\sigma_\pi(s)} \,,
\label{eq:t-phaseshift}
\end{equation}
which leads to the so-called Omn\`es solution already quoted in Eq.~\eqref{eq:Omnes}, as well as Watson's final-state theorem~\cite{Watson:1952ji}: as long as elastic unitarity is fulfilled, the form factor phase $\text{arg}\,\FV(s)$ is identical to the scattering phase shift $\delta_1^1(s)$.  

Dispersion relations for four-point functions, such as scattering amplitudes or three-body systems in production mechanisms, are typically more complicated due to the presence of crossed-channel singularities: cuts in, say, the $t$- and $u$-channels of a scattering amplitude induce, when projected onto $s$-channel partial waves, additional cuts that extend from another branch point to $-\infty$.  The latter necessitate the generalization of dispersion relations as in Eq.~\eqref{eq:sampleDR} to include a second integral over these so-called left-hand cuts.  

\subsubsection{Inverse-amplitude method}\label{sec:IAM}
The partial-wave unitarity relation for $\pi\pi$ scattering~\eqref{eq:PWunitarity}, $\Im t = \sigma_\pi |t|^2$ (omitting for simplicity isospin and angular momentum indices), has the immediate consequence that the imaginary part of the \textit{inverse} partial wave is given by the phase space factor alone:
\begin{equation}
    \Im \frac{1}{t(s)} = -\sigma_\pi \,,
\end{equation}
which leads to
\begin{equation}
    t(s) = \frac{1}{\Re t^{-1}(s) - i \sigma_\pi} \,.
\end{equation}
Applying the next-to-leading-order chiral expansion $t=t^{(2)}+t^{(4)}+\ldots$ to $\Re t^{-1}$ as well as perturbative unitarity in the form $\Im t^{(4)} = \sigma_\pi |t^{(2)}|^2$, one arrives at the expression
\begin{equation}
    t(s) = \frac{\big[t^{(2)}(s)\big]^2}{t^{(2)}(s)-t^{(4)}(s)} \,, \label{eq:IAM-NLO}
\end{equation}
which is the NLO inverse-amplitude method (IAM) expression for the partial wave $t(s)$~\cite{Truong:1988zp,Dobado:1989qm,Truong:1991gv,Dobado:1992ha,Dobado:1996ps,Guerrero:1998ei,GomezNicola:2001as,Nieves:2001de}.  
Equation~\eqref{eq:IAM-NLO} has been extended to next-to-next-to-leading order, employing also the two-loop chiral representation of the $\pi\pi$ scattering amplitude~\cite{Dobado:1996ps,Nieves:2001de}.  
Special care has to be taken for the application of the IAM to $S$-waves to take zeros in the amplitudes into account (so-called Adler zeros)~\cite{GomezNicola:2001as}.

The important observation is that IAM representations of scattering partial waves fulfill unitarity exactly.  In the elastic case, they can be derived rigorously based on dispersion theory, with only the crossed-channel singularities or left-hand cuts approximated according to the chiral expansion~\cite{Dobado:1992ha,Dobado:1996ps}.  In contrast to the strict chiral series, which contains essentially polynomials in the Mandelstam variables in addition to (perturbative) unitarity corrections, it is obvious that the series expansion in the inverse amplitude in Eq.~\eqref{eq:IAM-NLO} allows for the generation of poles for complex values of $s$, hence of resonances.

The generalization of the IAM to coupled channels, which turns Eq.~\eqref{eq:IAM-NLO} into a matrix relation,
underlies the theoretical representation of the various phase shifts and inelasticities~\cite{GomezNicola:2001as} shown in Fig.~\ref{fig:PPphase}.  These clearly show an at least reasonable reproduction of the lowest $S$- and $P$-wave resonances in $\pi\pi$ and $K\pi$ scattering.
For the application of the IAM to form factors and associated production amplitudes, see, e.g., Ref.~\cite{Shi:2020rkz} and references therein.

\subsubsection{Roy equations}\label{sec:Roy}
Meson--meson phase shifts are central objects of interest for this review.
The paradigm case for meson--meson scattering information that is known to excellent precision are the pion--pion phase shifts at low-to-moderate energies.  As a lot of progress in the description of meson decays and production processes are based on the precision with which we nowadays know these, we here briefly recapitulate how these can be constrained rigorously by means of the so-called Roy equations~\cite{Roy:1971tc}.

Roy equations present a coupled system of partial-wave dispersion relations, which make maximal use of analyticity, unitarity, isospin, and crossing symmetry.  The construction is based on a twice-subtracted dispersion relation for fixed Mandelstam variable $t$:
\begin{equation}
T(s,t) = c(t) + \frac{1}{\pi} \int_{4m_\pi^2}^\infty \diff s' \bigg\{
\frac{s^2}{s'^2(s'-s)} + \frac{u^2}{s'^2(s'-u)} \bigg\} \Im T(s',t) \,.
\label{eq:Roy:fixed-t}
\end{equation}
The two integrands correspond to the right- and left-hand cuts, respectively,
and isospin indices are suppressed for simplicity.
The $t$-dependent subtraction function $c(t)$ can be fixed using crossing symmetry.  
Partial-wave expanding the imaginary parts and projecting the resulting amplitude results in the schematic form~\cite{Roy:1971tc}
\begin{equation}
t_J^I(s) = k_J^I(s) + \sum_{I'=0}^2 \sum_{J'=0}^\infty  \int_{4m_\pi^2}^\infty \diff s'
K_{JJ'}^{II'}(s,s') \Im t_{J'}^{I'}(s') \,, \label{eq:Roy:pwsum}
\end{equation}
where we have reinstated the isospin dependence.  The kernels $K_{JJ'}^{II'}(s,s')$ are analytically known; they contain a singular 
Cauchy kernel (diagonal in both isospin and angular momentum) as well as terms resulting from the left-hand cuts.  The subtraction polynomial
$k_J^I(s)$ can be expressed entirely in terms of the two $\pi\pi$ $S$-wave scattering lengths $a_0^{I=0,2}$.  
As long as elastic unitarity is fulfilled, the partial waves $t_J^I(s)$ can be expressed in terms of phase shifts $\delta_J^I(s)$, see Eq.~\eqref{eq:t-phaseshift},
which turns Eq.~\eqref{eq:Roy:pwsum} into a coupled set of integral equations for $\pi\pi$ phase shifts.

Roy equations can only be applied within a limited energy range, e.g., $s \leq s_{\text{max}} = (1.15 \GeV)^2$~\cite{Roy:1990hw}.  The phase shifts for the lowest partial waves (usually $S$ and $P$) are therefore
only solved for at low energies below a certain matching point $s_{\text{m}}$, above which experimental
input is required.  However, the low-energy
$\pi\pi$ scattering phase shifts can still be determined with remarkable accuracy~\cite{Ananthanarayan:2000ht}, and the matching of the 
subtraction constants, i.e., the $S$-wave scattering lengths, to ChPT further strengthens the 
constraints~\cite{Colangelo:2001df}. 
Alternative variants of the Roy equations with different subtraction schemes have also been used in fits to $\pi\pi$ scattering data~\cite{Kaminski:2006qe,GarciaMartin:2011cn}, 
resulting in similarly accurate parameterizations of phase shifts that have been widely used as input to dispersion-theoretical analyses of final-state interactions.

Roy equations can be adapted to other scattering reactions that are less symmetric under crossing.  Such Roy--Steiner equations~\cite{Hite:1973pm}, making use of the relation between different crossed reactions, have been  solved in particular for pion--kaon scattering~\cite{Buettiker:2003pp,Pelaez:2018qny,Pelaez:2020gnd}, which is of high relevance in the context of this review, but also for 
pion Compton~\cite{Hoferichter:2011wk} as well as
pion--nucleon scattering~\cite{Ditsche:2012fv,Hoferichter:2015hva}.

\subsection{Amplitude analyses for three-body decays}\label{sec:DPanalysis}
\begin{figure}
    \centering
    \includegraphics[width=0.95\textwidth]{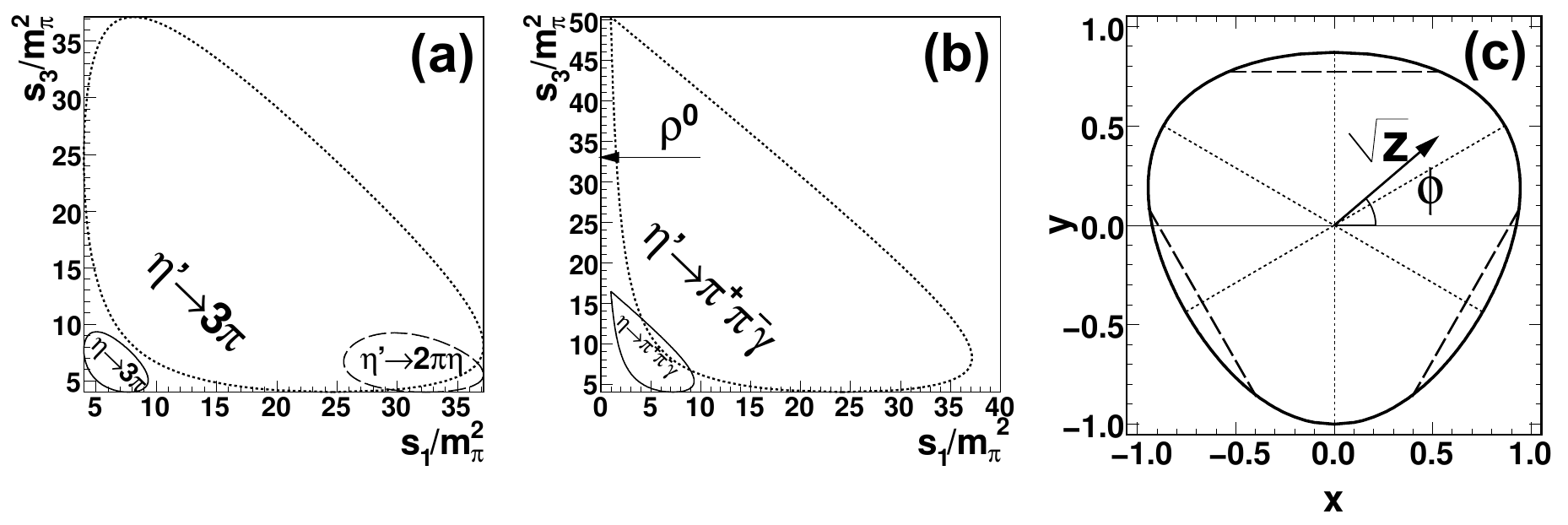}
\caption[Dalitz plots examples]{Dalitz plot boundaries for hadronic and radiative \et\   and \etp\  decays:  (a)~$\eta\to3\pi$  ---   solid  line, $\eta'\to3\pi$  --- dotted  line, and $\eta'\to\eta\pi\pi$  --- dashed line. The variables $s_1$, $s_2$  are expressed in units of the pion mass squared. (b)~$\eta^{(\prime)}\to\pi^+\pi^-\gamma$.  The vertical axis represents the invariant mass squared of the $\pi^+\pi^-$ system.  The position of the $\rho^0$ meson in the dimensionless units is indicated by the arrow. (c)~Example of a normalized Dalitz plot for three identical particles, the kinematic region for the decay $\eta\to3\pi^0$ expressed in $x$ and $y$ variables. The dashed lines represent the thresholds for $\pi^+\pi^-\to\pi^0\pi^0$ rescattering. \label{fig:DPdef}  }
\end{figure}

Amplitude analysis of multibody decays is a tool to study meson interactions gaining more importance in recent years. Often, the knowledge of two-body scattering amplitudes is assumed as a starting point to describe the decay dynamics. In some cases, especially when precise multidimensional data is available, one can try to extract information about meson interactions from such decays. We will illustrate the methods using mainly the simplest case of three-body decays.  A pseudoscalar meson decay to three ground state pseudoscalar mesons is related to meson--meson scattering by crossing symmetry, as discussed in Sec.~\ref{sec:intro}. For a three-body decay $0\to 1+2+3$ the Mandelstam variables are often denoted by
\begin{equation}
s_i\equiv ({p}_0-{p}_i)^2 \,,
\end{equation}
where $i=1,2,3$ and $p_0,{p}_i$ are the four-momenta of the particles.  The relation between the standard notation and  $s_i$ reads: $s\to s_3$, $u\to s_1$, $ t\to s_2$.  Alternatively one can use kinetic energies $T_i$ of the decay particles in the rest frame of the particle ``0'':
\begin{equation}
T_i=\frac{(m_0-m_i)^2 - s_i}{2m_0},
\label{eqn:sT}
\end{equation}
which are linearly related to the $s_i$. Energy conservation imposes the constraints
\begin{equation}
\sum_{i=1}^3T_i=
m_0-\sum_{i=1}^3m_i\equiv Q_0 \quad \text{and} \quad \sum_{i=1}^3s_i=s+t+u=\sum_{i=0}^3m_i^2\,.
\end{equation}
The dynamical information on a three-body decay of scalar particles can be represented by a probability density function drawn in a plane defined by any two of the Mandelstam variables, the so-called Dalitz plot~\cite{Dalitz:1953cp}. Examples of Dalitz plot  boundaries are  shown in  Fig.~\ref{fig:DPdef} using Mandelstam variables  expressed  in units of squared pion masses, where the kinematical ranges for three-body decays of $\eta$ and $\eta'$ mesons are compared.  The density distribution in the Dalitz plot is determined by the quantum numbers of the particles and by the interaction of the produced particles.  The latter can be  influenced by resonances or kinematical thresholds. A two-particle resonance will be seen as a band perpendicular to the corresponding $s_i$ variable in the Dalitz plot if the mass is contained in the plot and the width is significantly less than the kinematical range.  As we will discuss in Sec.~\ref{sec:strong3p}, none of these conditions are fulfilled for $\eta\to3\pi$ and $\eta'\to\eta\pi\pi$  decays presented in Fig.~\ref{fig:DPdef}(a). The example of Fig.~\ref{fig:DPdef}(b) shows the kinematic ranges for radiative decays of $\eta$  and $\eta'$ mesons into $\pi^+\pi^-\gamma$. The $s_3$ variable  corresponds  to  the $\pi^+\pi^-$ invariant  mass squared $M^2(\pi^+\pi^-)$. As we discuss in Sec.~\ref{sec:etapipig}, the $\rho^0$ meson resonance is visible in the $\eta'\to\pi^+\pi^-\gamma$ decay. The peak position in the $s_3$ variable, corresponding to $m_\rho^2/m_\pi^2\approx 33$ is indicated in the figure.

A virtue of the Dalitz plot representation using two of the Mandelstam variables or their linear combinations $T_i$ is that the probability density function is directly proportional to the matrix element
squared $|{\cal M}(s,t,u)|^2$:
\begin{equation}
\diff\Gamma=\frac{1}{(2\pi)^3}\frac{1}{8 m_0}|\overline{\cal M}|^2\diff T_1\,\diff T_2
       =\frac{1}{(2\pi)^3}\frac{1}{32 m_0^3}|\overline{\cal M}|^2\diff s_1\,\diff s_2\,.
\label{eqn:phsp}
\end{equation}
Sometimes it is useful to describe the three-body decay phase space in terms of sequential decays, i.e., processes $0\to (1+2)+3$, using helicity frames. This will, e.g., be the case when we discuss radiative decays into a meson--meson pair. One uses the invariant mass squared of the $(1+2)$ 
system $s$ and the emission angle  $\Omega_3$ of  particle ``3'' in the particle ``0'' rest frame, as well as $\Omega_1^*$, the emission angle of particle  ``1'' in the $(1+2)$ helicity system. The fully differential cross section (not spin averaged) is given by 
\begin{equation}
  \diff\Gamma=\frac{1}{(2\pi)^5}\frac{1}{16 m_0^2}|{\cal M}|^2\frac{|\bf{p}_3||\bf{p}_1^*|}{\sqrt{s}}\diff s\,
  \diff\Omega_3\,\diff\Omega_1^* \,,
\end{equation}
where ${\bf p}_3$ and ${\bf p}_1^*$ are the momenta of particle ``3'' and ``1'' in ``0'' and $(1+2)$ frames, respectively.
If two of the particles are identical or are a particle--antiparticle  pair the $s$ variable  corresponds  to  the  invariant  mass squared  of  the  like particles.   In  this case one can represent the Dalitz plot using the following dimensionless
variables:
\begin{equation}
  x\equiv\frac{\sqrt{3}}{2}\frac{u-t}{m_0 Q_0} \,, \qquad   
y\equiv\frac{3}{2}\frac{(m_0-m_3)^2-s}{m_0 Q_0}-1 \,.\label{eq:DaltzXY}
\end{equation}
As an example the boundary of the Dalitz plot for $\eta\to3\pi^0$ 
in the $x$ and $y$ variables is shown in Fig.~\ref{fig:DPdef}(c).
Decays into three identical particles are conveniently described using polar coordinates $(\sqrt{z},\phi)$ in the $(x,y)$ plane:
\begin{equation}
x=\sqrt{z}\sin\phi \,, \qquad  y=\sqrt{z}\cos\phi \,.
\end{equation}
Such a Dalitz plot has a sextant symmetry and one can limit the range of the $\phi$ angle to $0^\circ\le\phi<60^\circ$. One example is $\eta^{(\prime)}\to3\pi^0$ discussed in Sec.~\ref{sec:strong3p}. The variable $z$ is given by $z=x^2+y^2$ and covers the range $0\le z\le 1$.

\begin{figure}
    \centering
    \includegraphics[width=0.95\linewidth]{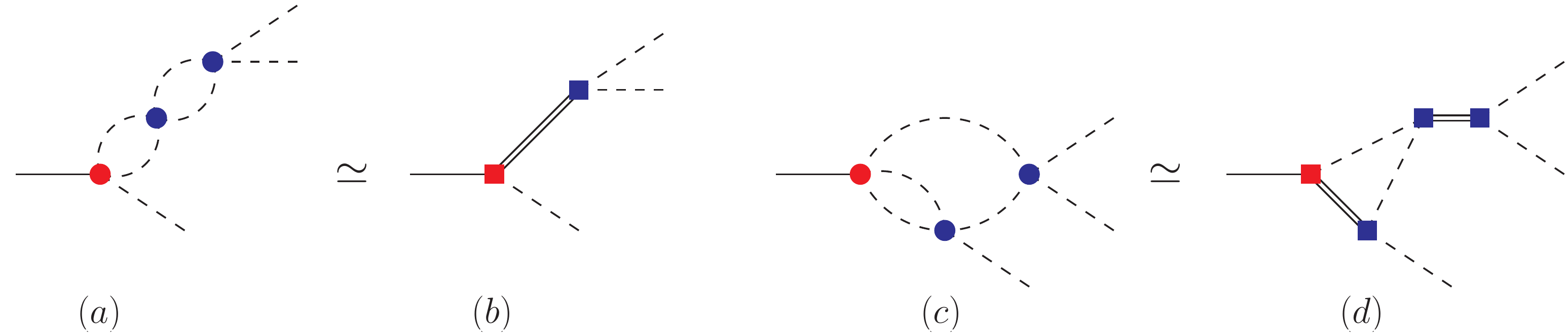}
    \caption{Rescattering effects in three-body decays. Two-body rescattering in the spectator approximation (a) is approximated by resonance (double line) exchange (b). The isobar model in its simplest form does not include rescattering between all three final-state particles (c) or sequential resonance exchanges in different two-body channels (d).}
    \label{fig:3-body-rescattering}
\end{figure}

The most common way to describe the matrix element of a decay is as a sum of interfering decay amplitudes, each proceeding through chains of resonant two-body decays.  This is the so-called isobar model~\cite{Sternheimer:1961zz,Herndon:1973yn}; see Fig.~\ref{fig:3-body-rescattering}.
The diagrams in this model would correspond to the tree contributions in a quantum field theory for the process if all the mass parameters in the propagators are taken to be real. However, in the isobar model the effect of final-state interactions is included by allowing the masses to be complex, i.e., by including finite widths.  In this manner, however, the isobar model neglects more complicated rescattering effects between all three final-state particles; see Fig.~\ref{fig:3-body-rescattering}(c) and~(d).  Below, in the context of reactions or decays of the type $e^+e^-,\,\omega,\,\phi\to\pi^+\pi^-\pi^0$, we will briefly describe the so-called Khuri--Treiman formalism~\cite{Khuri:1960zz} that takes more complete final-state interactions into account.  Other theoretical approaches include the application of Bethe--Salpeter models~\cite{Mai:2017vot,Sadasivan:2020syi} or other formulations explicitly incorporating three-body unitarity~\cite{Jackura:2018xnx,Mikhasenko:2019vhk}, all of which ought to be equivalent within their respective approximations~\cite{Jackura:2019bmu} and describe the same physics.

\section{Experiments at symmetric electron--positron colliders}
\subsection{Experimental facilities}\label{sec:colliders}
The majority of the results presented in this review were obtained  at low-energy symmetric electron--positron colliders in Frascati, Novosibirsk, and Beijing. The facilities and detectors are briefly described in this section. We do not include descriptions of earlier experiments like the CLEOc~\cite{Poling:2006da}, which was a modification of an experiment originally running above
$b\bar b$ threshold to carry out a physics program in the charm region. It had significantly lower luminosity than the facility in Beijing, but still several results persist as world-leading and are discussed in this review. 
Many findings that we will show come from the asymmetric electron--positron colliders PEP-II and KEKB
designed to work in the $b\bar b$ region~\cite{Bevan:2014iga}. The attainable luminosities at these high energies are two orders of magnitude larger and allow the experiments to perform low-energy hadronic studies using higher-order electromagnetic processes such as initial-state radiation and two-photon fusion. These colliders have asymmetric configurations with electron and positron beams of different energies.  In particular many results included in this review stem from the $\sim0.5\ab^{-1}$ of data collected by the BaBar collaboration~\cite{Aubert:2001tu} near the $\Upsilon(4S)$ resonance.

\begin{table}[t]
    \caption[Electron--positron colliders with c.m.\ energy $0.3$--$5.0\GeV$]{Main parameters of the electron--positron colliders with c.m.\ energies in the range $0.3$--$5.0\GeV$. Data taken from Ref.~\cite{PDG}.  
    }
    \label{tab:colliders}
    \begin{center}
    \renewcommand{\arraystretch}{1.3}
    \begin{tabular}{p{5.5cm} rrrrr}
    \toprule
        & VEPP-2M& VEPP-2000& BEPC&BEPC-II&DA$\Phi$NE \\ 
\midrule
Physics start date&1974 &2010 &1989 &2008&1999\\
Physics stop date&2000 &$-$ &2005 &$-$&$-$\\
c.m.\ energy (GeV)& $0.36$--$1.4$&$0.3$--$2.0$ & 5.0&$2.0$--$4.7$&1.020\\
Delivered luminosity (fb$^{-1}$)& 0.1&0.125 &0.11 &20&10\\
Luminosity ($10^{31}$ cm$^{-2}$s$^{-1}$)        &0.5 &4 &12 &100&45\\
Circumference   (m)       &18&24 &240 &238&98\\
Crossing angle (m rad)&0 &0 &0 &22&50\\
 Interactions regions        &2 &2 &1 &1&1\\
 \bottomrule
    \end{tabular}
    \renewcommand{\arraystretch}{1.0}
    \end{center}
\end{table}
\subsubsection[Frascati: DA$\Phi$NE and KLOE]{\boldmath Frascati: DA$\Phi$NE and KLOE}
DA$\Phi$NE at Frascati LNF-INFN is an electron--positron collider optimized to run at the c.m.\ energy corresponding to the $\phi$ meson mass. The visible  $\phi$ peak cross section is about $3\,\mu\text{b}$ and the decay width is $4.249(13)\MeV$, where the main decay modes are into $K^+K^-$ [$49.2(5)\%$] and $K_LK_S$ [$34.0(4)\%$].
During the KLOE experiment's running period from 2001 to 2006, a total collected integrated luminosity of $2.5\fb^{-1}$ at the $\phi$ peak and about $250\pb^{-1}$ off peak were acquired. In 2008, DA$\Phi$NE was upgraded to be the first collider to implement the crab-waist interaction scheme~\cite{Zobov:2010zza}. At the same time, new sub-detectors were added. From 2014 until March 2018, the new KLOE-2 detector collected $5.5\fb^{-1}$ at the $\phi$ resonance. The total KLOE and KLOE-2 data sets correspond to about $2.4\times10^{10}$ $\phi$-meson events. At present (2021), DA$\Phi$NE is used as a worldwide unique low-energy kaon--antikaon source, e.g., for kaonic atoms studies~\cite{Sirghi:2020xtq}.

The KLOE detector consists of a large cylindrical drift chamber (DC)~\cite{Adinolfi:2002uk} surrounded by a hermetic electromagnetic calorimeter (ECAL)~\cite{Adinolfi:2002zx}. The detectors are placed inside a superconducting coil providing an axial magnetic field of $0.52\T$. Having $4\m$ diameter and $3.3\m$ length the DC is the world's largest of this type. The mechanical structure is of carbon fiber composite and consists of 12582 drift cells made of tungsten sense wires arranged in 58 stereo layers, a total of 52140 wires.  A gas mixture of helium (90\%) and isobutane (10\%) is used. The position resolution is $ 150\,\mu \text{m}$ in the transverse and $2\,\text{mm}$ in the longitudinal directions with respect to the beams, and the relative transverse momentum resolution is $<$0.4\% for large-angle tracks.
The ECAL covers $98\%$ of the solid angle and is divided into a barrel and two end caps with a total of 88 modules. Each module is built out of $1\mm$ diameter scintillating fibers embedded in $0.5\mm$ lead foils and the readout is made by photo-multipliers on both sides. The ECAL energy resolution 
is ${\sigma(E)}/{E} = {5.7\%}/{\sqrt{E\,(\text{GeV})}} $
with  excellent timing performance of $\sigma(t) ={ 54\text{ps}}/{\sqrt{E\,(\text{GeV})}} \oplus 140 \,\text{ps}$. The center of gravity of a cluster in the ECAL is measured with a resolution of $1.3\cm$ in the transverse and, by signal time difference,  of $1.2\cm/{\sqrt{E\,(\text{GeV})}}$ in the longitudinal 
direction relative to the fibers.  Around the interaction point (IP), the beam pipe has spherical shape, with a radius of $10\cm$, to allow $K_S$ mesons to decay in vacuum. The beam pipe walls are made of a 60\% beryllium/40\% aluminum alloy $0.5\mm$ thick.

In the KLOE-2 upgrade  a tracker device, a novel Inner Tracker (IT), was added. The IT consists of four concentric cylindrical gas electron multiplier detector (CGEM) layers~\cite{Balla:2013gua} between the IP and the DC to improve the resolution on decay vertices close to the IP. Each layer consists of a triple-CGEM detector with an X-V strip readout. The X strips are placed along the beam axis while the V strips have on average 23$^\circ$ angle with respect to the beam axis. Both X and V strips have a $650\,\mu \text{m}$ pitch. The CGEMs are filled with an argon-based gas mixture. A technology with low material budget (below $2\%$ of the radiation length) was chosen, to minimize multiple scattering of low-momentum tracks, photon conversions, and kaon regeneration.

Reviews of the KLOE experimental results using data collected until 2005, including analyses finished until 2008, are given in Refs.~\cite{Franzini:2006aa,Bossi:2008aa}, while the  physics program for KLOE-2 was laid out in Ref.~\cite{AmelinoCamelia:2010me}.
\begin{figure}[t]
    \centering
    \includegraphics[width=0.99\textwidth]{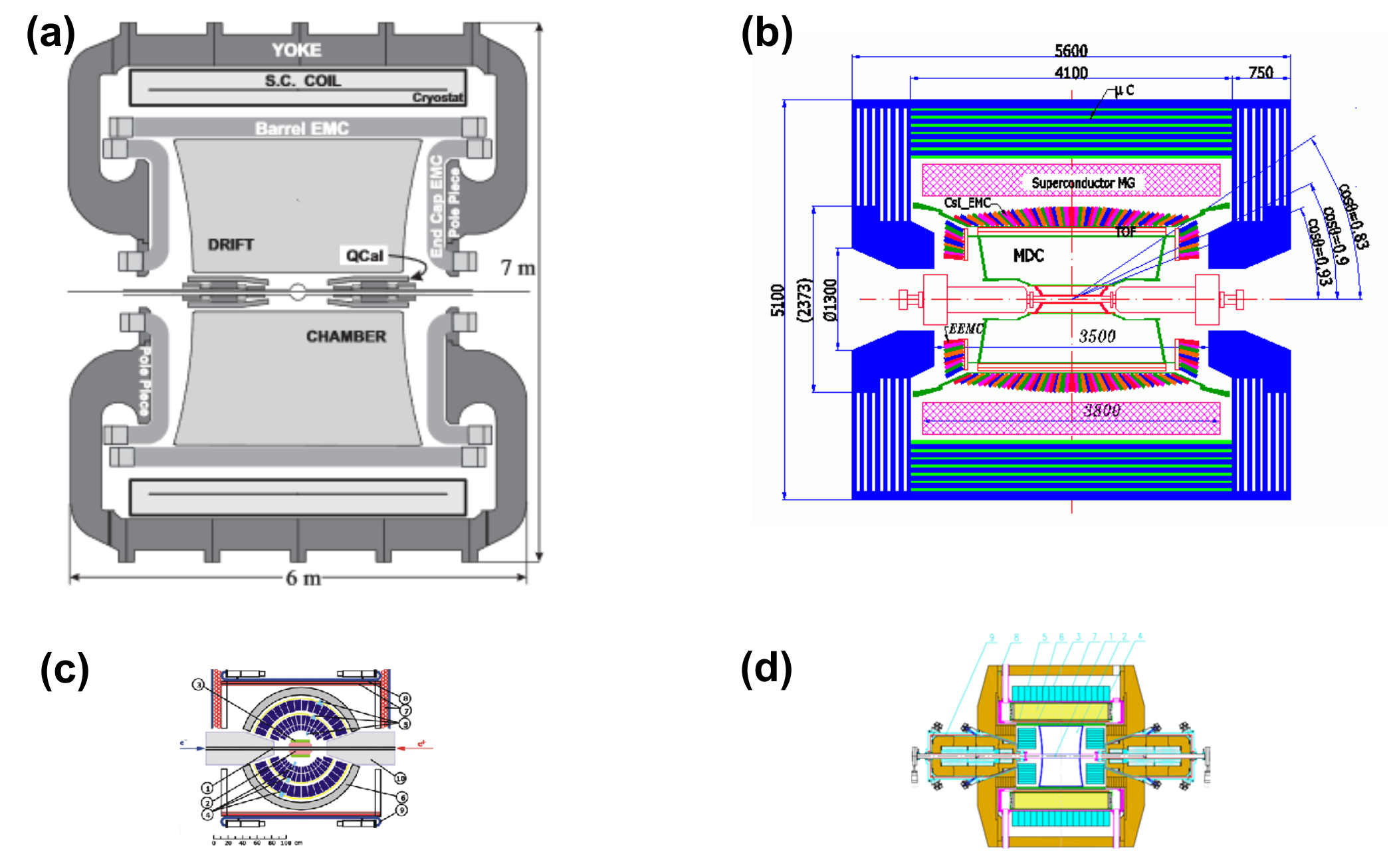}
    \caption[Detectors at the symmetric electron--positron colliders ]{Detectors at the symmetric electron--positron colliders (a) KLOE, (b) BESIII, (c) SND, and (d)~CMD-3. All detectors are drawn approximately in the same scale. We show schematic cross sections along the beam axes.    
    }
    \label{fig:detectors}
\end{figure}
\subsubsection{Beijing: BEPC and BES}
The Beijing Spectrometer (BES) is a large general-purpose solenoidal detector~\cite{Bai:1994zm} at the Beijing Electron--Positron Collider (BEPC) used for \CT\ physics in the c.m.\ energy range of $2$--$5\GeV$. The first version of the detector was completed in 1989 and was in operation for six years.  Afterwards a major upgrade of the detector, called BESII, was installed~\cite{Bai:2001dw} to improve its performance. At the same time BEPC was improved to increase the luminosity by a factor of 1.7--2~\cite{Guo:1996fx}.  Several important results were obtained with BESII, e.g., a precision $R$-value measurement~\cite{Bai:1999pk}, or the observation of the $X(1835)$~\cite{Bai:2003sw}. The results were based on the data samples collected until 2005 at $6+85$ c.m.\ energies between 2 and $5\GeV$, $5.8\times 10^7$ $J/\psi$ events, $1.4\times 10^7$ $\psi'$ events, and $20\pb^{-1}$ data at the peak of the $\psi(3770)$, which underlines the rich physics program in the $\tau$-charm region, including light-hadron spectroscopy, the charmonium spectrum, charm meson decays, and $\tau$ physics. 

To meet the challenge of precision \CT\ physics, a major upgrade of the collider, now called BEPCII, 
was completed in 2008 together with the new  BESIII detector. The collider has separate rings for the electron and positron beams with a single interaction region. The peak luminosity is $10^{33}\,\text{cm}^{-2}s^{-1}$ at $\sqrt s = 3.773\GeV$,  i.e., 100 times larger than its predecessor. The BESIII detector is a large solid-angle magnetic spectrometer schematically shown in Fig.~\ref{fig:detectors}(b) and described in detail in Ref.~\cite{Ablikim:2009aa}. The main components are the superconducting solenoid magnet with a central field of $1\T$, the main drift chamber (MDC), and a cesium iodide (CsI) electromagnetic calorimeter. Up to 2020 the BESIII detector has collected the world's largest samples in the  \CT\ energy range including $10^{10}$ $J/\psi$ events, $4.5\times 10^8$ $\psi'$ events, $2.9\fb^{-1}$ data at the $\psi(3770)$, and more than $15\fb^{-1}$ data above $4\GeV$. Most of the results presented in the review for $J/\psi$ decays use $1.31\times10^9$ events collected in the 2009 and 2012 runs. The physics accomplishments of the BES experiments are reviewed in the recent articles Refs.~\cite{Briere:2016puj,Yuan:2019zfo}.

\subsubsection{Novosibirsk: VEPP and SND, CMD}
The Institute of Nuclear Physics (INP) (currently the Budker Institute of Nuclear Physics) was one of the first research centers to build an  electron--positron collider (see Sec.~\ref{sec:history}). The parameters of the second collider VEPP-2M operating  with several generations of detectors until the end of the 1990s 
are given in Table~\ref{tab:colliders}~\cite{Levichev:2018cvd}. The c.m.\ energies ranged up to $1.4\GeV$ and it
had two interaction points, allowing to collect data simultaneously by two detectors. The collected data on the hadronic cross sections by the two latest experiments CMD-2\cite{Akhmetshin:2000sr,Aulchenko:1993xp} and  SND~\cite{Dolinsky:1991vq,Achasov:1999ju} gives one of the most important contributions to the accuracy for the hadronic contributions to the muon $g-2$.  The purpose of the new successor collider VEPP-2000 is to study light hadrons using variable c.m.\ energies up to 2.0\GeV. The facility reuses the VEPP-2M buildings and its infrastructure. It uses a round beam to suppress beam--beam effects. Initially, until 2013, data were collected using the old  injector system, which limited the luminosity at the highest energies. 
The two new detectors are CMD-3 and the upgraded SND~\cite{Khazin:2008zz} shown schematically in Fig.~\ref{fig:detectors}. 
The detectors are very compact, which is imposed by the size of the ring and the experimental hall.

\subsection{Production of hadronic systems}\label{sec:production}
\begin{figure}
    \centering
\begin{minipage}{0.9\linewidth}
\includegraphics[width=0.40\linewidth]{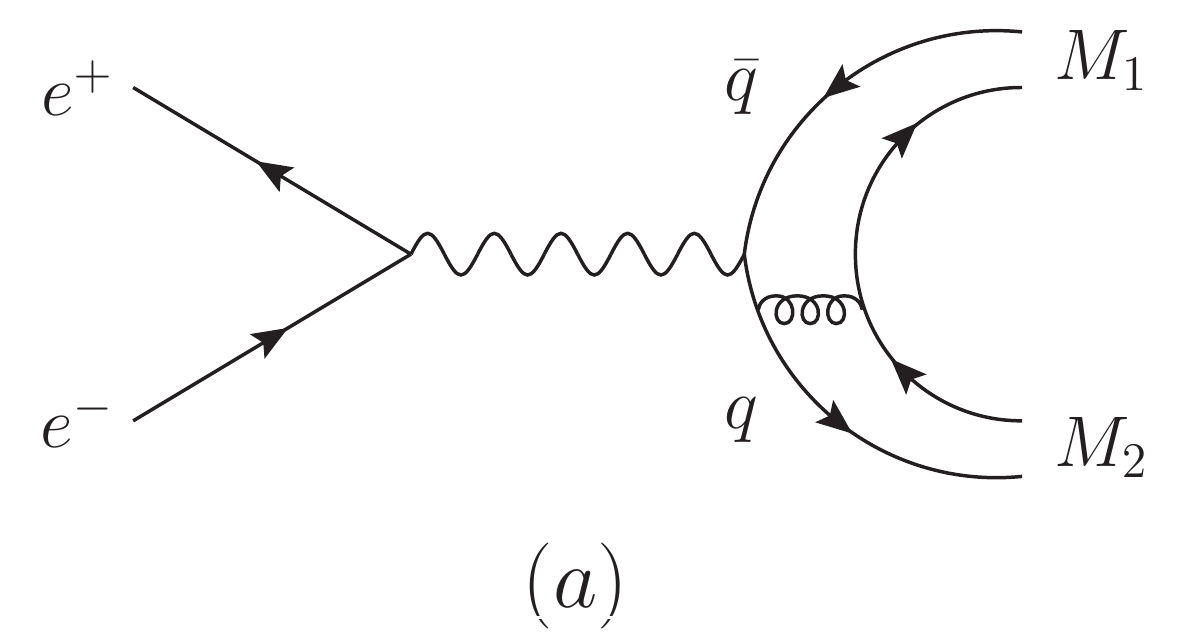} \hfill
\includegraphics[width=0.40\linewidth]{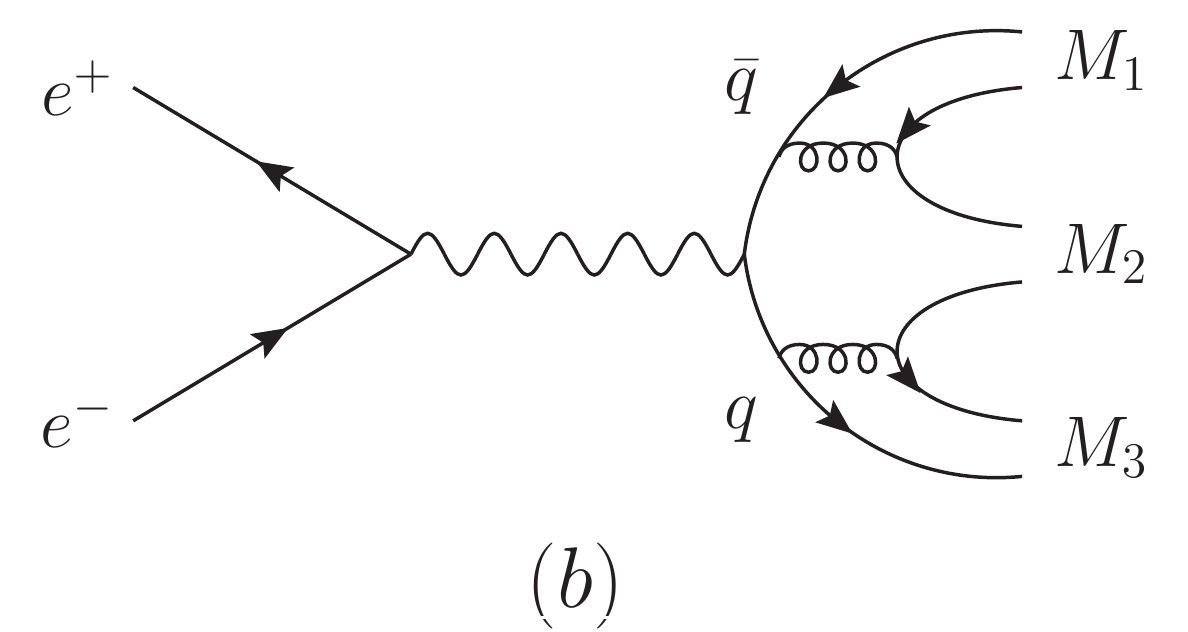}

\vspace*{5mm}
\includegraphics[width=0.38\linewidth]{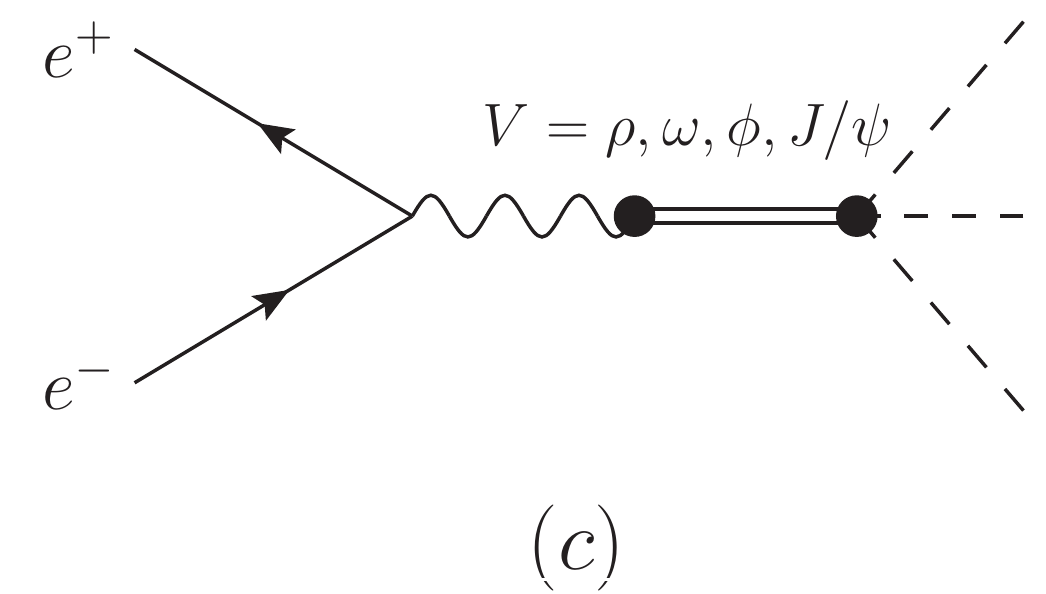} \hfill
\includegraphics[width=0.38\linewidth]{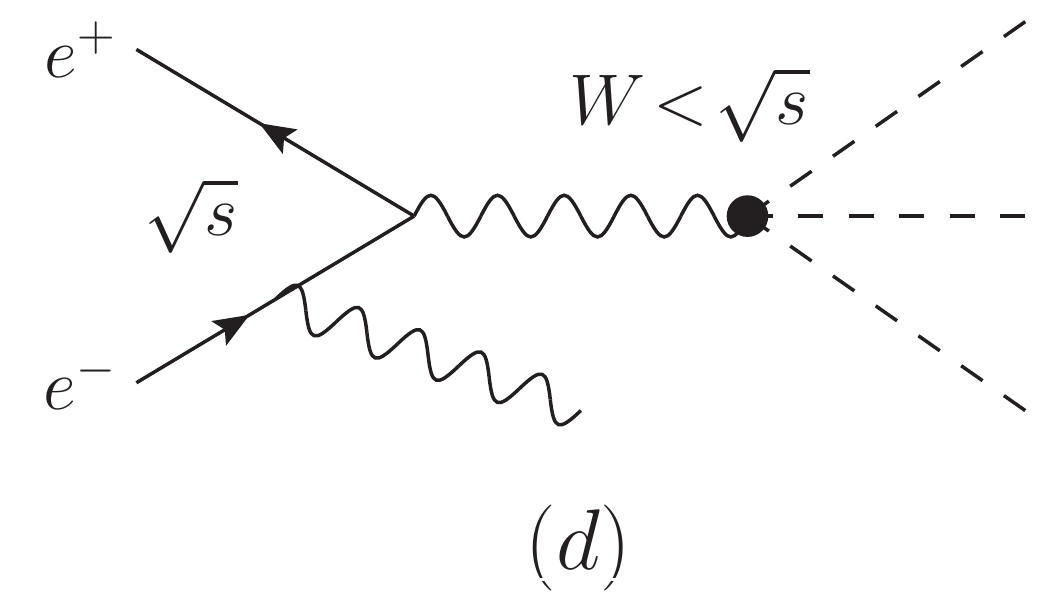}
\end{minipage}

\caption[Mechanisms of hadron production in 
    electron--positron annihilation]{Mechanisms of hadron production in 
    electron--positron annihilation:  (a) a quark diagram for two-meson production;
     (b) a quark diagram for three-meson production;
    (c) a  neutral-vector-meson-dominated process; (d) initial-state radiation production mechanism.  
    \label{fig:mech2}}
\end{figure}

One can describe the ideal setup to study meson--meson interactions as a system without extra hadrons from the production process. The production mechanism should be well understood or easy to parameterize in a model-independent way. In addition, the invariant mass of the meson--meson system should be variable. Such conditions can be achieved, e.g., by changing the c.m.\ energy of the collider or having an extra photon, $\ell\bar\ell$, $\ell\bar\nu_\ell$, or $\nu_\tau$, which carry away part of the initial fixed energy. There are many sources of light-hadron systems at $e^+e^-$ colliders. We consider only processes with sufficiently large cross sections to provide enough events for precision studies. In this section we introduce the main processes where systems of two or three light mesons in the final state can be produced.

\subsubsection{Single-photon annihilation}
\begin{figure}
\centering
\includegraphics[width=0.95\textwidth]{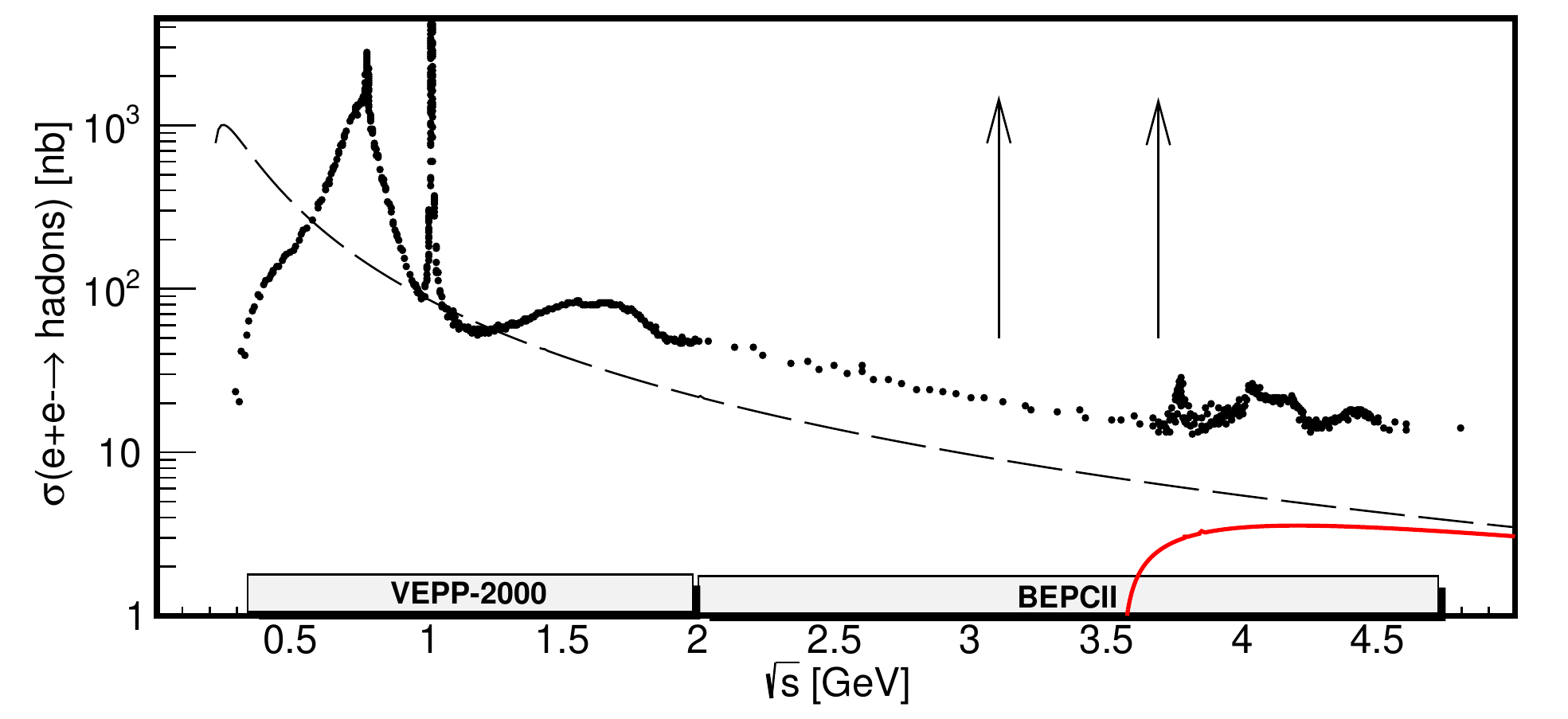}
\caption[Total electron--positron cross section]{Total bare cross section of $e^+e^-\to \text{hadrons}$
  as a function of center-of-mass energy from the compilation of Ref.~\cite{Ezhela:2003pp}. The two vertical arrows indicate the positions of the $J/\psi$ and $\psi'$ peaks. The c.m.\ energy ranges of the VEPP-2000 and BEPCII colliders are also shown. Dashed and solid lines refer to $e^+e^-\to\mu^+\mu^-$ and $e^+e^-\to\tau^+\tau^-$ e.m.\ Born cross sections, respectively. \label{fig:cstot}}
\end{figure}
The direct production of a meson system in the single-photon annihilation process $e^+e^-\to X$ exclusively produces states with quantum numbers $J^{PC}=1^{--}$.  We illustrate such production processes by simplified quark diagrams in Fig.~\ref{fig:mech2}(a) and (b) for meson--meson pairs and three-meson systems, respectively. The measurement of the total cross section for a pseudoscalar meson pair like $\pi^+\pi^-$ as a function of c.m.\ energy is directly related to the meson--meson interactions in the $P$-wave, and allows to extract vector form factors, $\FV(s)$ for the given example. 

\begin{sloppypar}
The differential rate for the production of a hadronic system via a single virtual photon, see Fig.~\ref{fig:mech}(b), can be expressed as the product of leptonic and hadronic tensors, multiplied with the multihadron phase space element: 
\begin{equation}
\diff\sigma = \frac{e^4}{8s^3} \, L_{\mu\nu} H^{\mu\nu}\,  
     \diff\Phi_n(q_1,\dots,q_n) \,, 
\label{eq:sighad}
\end{equation}
where $s=(p_1+p_2)^2$, and \(\diff\Phi_n(q_1,\dots,q_n)\) denotes the $n$-body phase space
 with all statistical factors included.
The leptonic tensor $L_{\mu\nu}=L_{\mu\nu}(p_1,p_2)$  is process-independent
and for unpolarized beams given by the averages over lepton spins,
\begin{equation}
	L_{\nu\mu}(p_1,p_2) =   \frac{1}{4} {\rm Tr}\big[ \gamma_\nu 
	\slashed{p}_1 \gamma_\mu \slashed{p}_2  \big]
	 = p_{1\nu}p_{2\mu}+ p_{2\nu}p_{1\mu} - \frac{1}{2} s\, g_{\nu\mu} \,.
\end{equation}
The leptonic tensor is symmetric, and hence it is only the real
symmetric part of the hadronic tensor $H_{\mu\nu}$ that is relevant for the differential cross section. 
The hadronic tensor is defined via current matrix elements as~\cite{Czyz:2000wh}
\begin{equation}
  H_{\mu\nu}=J^{\rm em}_\mu \big(J^{\rm em}_\nu\big)^* \,,
\end{equation}
where for the production of an $n$-particle hadronic final state, the matrix element of the hadronic current is
\begin{equation}
J^{\rm em}_{\mu} \equiv J^{\rm em}_{\mu}\left(q_1,...,q_n\right) \equiv 
 \bra{h(q_1),...,h(q_n)}j^{\rm em}_{\mu}\ket{0} \,,
\end{equation}
with the electromagnetic current at the quark level given by
\begin{equation}
  \label{eq:defemcurrent}
  j^{\rm em}_\mu = \frac23 \, \bar u \gamma_\mu u - \frac13 \, \bar d \gamma_\mu d - \frac13 \, \bar s \gamma_\mu s + \frac23 \, \bar c \gamma_\mu c + \ldots \, . 
\end{equation}
The quark electromagnetic current can be decomposed into an isospin singlet piece
and a part transforming like the third component of an isospin triplet:
\begin{equation}
 j^{\rm em}_\mu = \frac{1}{\sqrt{2}}\ j^3_\mu + \frac{1}{3\sqrt{2}}\  j^{I=0}_\mu - \frac13 \, \bar s \gamma_\mu s + \frac23 \, \bar c \gamma_\mu c + \ldots \,,\label{eq:emcurrent2}
\end{equation}
where
\begin{equation}
    j^3_\mu=\frac{\bar u \gamma_\mu u - \, \bar d \gamma_\mu d}{\sqrt{2}} \quad {\rm and } \quad 
    j^{I=0}_\mu=\frac{\bar u \gamma_\mu u + \, \bar d \gamma_\mu d}{\sqrt{2}}\,.
\end{equation}
The above expressions for the electromagnetic current give a qualitative picture of the annihilation process to the light-hadron systems. At the lowest energies, below the $K\bar K$ production threshold for open strangeness, the isovector systems should contribute with 90\% to the total cross section, while above $c\bar c$ threshold this contribution is reduced to 45\%. 
As two examples, which we will discuss in this review, we give the expressions for the 
hadronic currents for the dominant contributions to the isovector and the isoscalar processes
$e^+e^-\to \pi^+\pi^-$ and $e^+e^-\to \pi^+\pi^-\pi^0$, respectively.
For $e^+e^-\to \pi^+(q_+)\pi^-(q_-)$, the current matrix element
\begin{equation}
J_{\nu}^{{\rm em},2\pi} 
 = \left(q_{+}-q_{-}\right)_\nu \FV(s) \label{eq:2picurr}
\end{equation}
is determined by only one scalar, dimensionless function, the pion vector form factor $\FV(s)$. We discuss properties of this form factor in Sec.~\ref{sec:eePP}. As we will show, there is a direct connection between $\FV$ and pion--pion interactions in a $P$-wave.
For the three-pion production $e^+e^-\to \pi^+(q_1)\pi^-(q_2)\pi^0(q_3)$, the current matrix element is restricted by negative parity to the form
\begin{equation}
 J_{\nu}^{{\rm em},3\pi} =
 \epsilon_{\nu\alpha\beta\gamma}\ q_{1}^{\alpha}q_{2}^{\beta}q_{3}^{\gamma}
\ F_{3\pi}(s;s_1,s_2,s_3) \,,\label{eq:3picurr}
\end{equation}
where $F_{3\pi}(s;s_1,s_2,s_3)$ is a scalar function of the total  $s=(q_1+q_2+q_3)^2$ and the two-pion invariant masses $s_1=(p_2+p_3)^2$ (plus cyclic permutations). This function depends on three independent Mandelstam variables, as $s_1+s_2+s_3=s+3m_\pi^2$,
and describes the interaction of three pions at a given c.m.\ energy.  We discuss this process more closely in Sec.~\ref{sec:ee-PPP}.
\end{sloppypar}

The basic observable measured at electron--positron colliders is the total cross section obtained by the integration of Eq.~\eqref{eq:sighad} over the kinematic variables defining the final state.
In the case of two-body hadronic production as in $e^+e^-\to \pi^+\pi^-$, it is directly related to the pion dynamics via $|\FV(s)|^2$. An interesting question is to what extent we can infer information about meson--meson dynamics also from the cross section dependence in the case of three and more produced hadrons. A direct connection is established rather by studies of differential distributions via the multidimensional form factors such as $F_{3\pi}(s;s_1,s_2,s_3)$.  However, the cross section is the first and often the only accessible observable as it requires the smallest amount of collected data.
In case there is no interaction, $F_{3\pi}=\mathrm{const.}$ and the cross section should follow three-body phase space or phase space restricted by angular momentum conservation and other symmetries. A strong interaction between a single pair of mesons would lead to the cross section dependence given by the corresponding two-body phase space. The effect is especially important close to threshold and can help to understand the dynamics of the process.  We present examples of cross section measurements for three-body processes in Sec.~\ref{sec:ee-PPP}.

\subsubsection{Strong and radiative decays of vector mesons}\label{sec:prodVmesons}
A single-photon annihilation process into a hadronic system can be mediated by a narrow neutral vector meson [see Fig.~\ref{fig:mech2}(c)], like $\omega$, $\phi$, $J/\psi$, and $\psi'$, strongly enhancing the  hadronic cross section at the c.m.\ energy corresponding to the resonance mass as shown in Fig.~\ref{fig:cstot}.
One can therefore use the vector meson decays as an abundant source of light hadrons for precision studies of meson--meson interactions. 
The  propagator of a neutral vector meson $V$  with momentum $P=p_1+p_2$ takes the form
\begin{equation}
	\frac{i(-g_{\mu\nu}+ P_\mu P_\nu/m_V^2)}{s-\Pi(s)} \,,
	\label{eqn:propagator}
\end{equation}
where $m_V$ is  the  $V$ mass and $\Pi(s)$ is the self-energy of the resonance, $\Pi(s)=m_V^2(s)+im_V(s)\Gamma_V(s)$. 
The absorptive part ${\rm Im}\Pi(s)=m_V(s)\Gamma_V(s)$ is generated by intermediate states that can go on their mass shell, while
the dispersive part ${\rm Re}\Pi(s)=m_V^2(s)$ also obtains contributions from intermediate states that are virtual only (resonance propagation without decay). 
In general the function $\Pi(s)$ is very complicated and for hadronic resonances incalculable in a fundamental way. For a narrow resonance, it can be approximated as a constant $\Pi(s)\simeq m_V^2+im_V\Gamma_V$, where $\Gamma_V$ is the full decay width.  
Simple generalizations to energy-dependent widths take at least the corresponding decay phase spaces into account, thus avoiding imaginary parts below threshold. 
For a resonance produced through an intermediate  photon, the contribution from the $P_\mu P_\nu$ term in the propagator vanishes and the matrix element is structurally identical to the nonresonant annihilation, provided we make the replacement 
\begin{equation}
 \frac{e^2}{s}	\longrightarrow\frac{e_V e_X}{s-m_V^2+im_V\Gamma_V} \,,
 \label{eq:eV}
\end{equation}
where $e_V$ is the resonance coupling to the $e^+e^-$ pair, determined from the  leptonic decay width, 
and $e_X$ is the coupling to the final state $X$. 

In general, two types of experiments for meson--meson interactions are possible using decays of vector mesons. The first is direct production of hadronic systems in strong and radiative decays. To study dynamics in this case, three and more particles should be created, and the radiative decays with just two hadrons are particularly simple. We discuss these in Sec.~\ref{sec:VtoPPg}. The data sets for $\phi$, $J/\psi$, and $\psi'$ mesons collected at KLOE and BESIII are huge, with at the order of $10^{10}$ events (see Sec.~\ref{sec:colliders}), and can be used for detailed studies of the interactions of the hadrons produced in the decays. 
In the second type of experiment a decay into a pair of stable particles (or narrow resonances) is used to tag a hadron to study its further decays. In particular radiative and strong hadronic decays of $\eta$ and $\eta'$ are well-known laboratories for precision studies of pion interactions. Radiative decays of $\phi$ and $J/\psi$ are among the best sources of $\eta$ and $\etap$ mesons since they have large branching fractions ($\BR$) as shown in Table~\ref{tab:eta}. The accompanying monochromatic radiative photon is in most cases well separated from the decay products. Huge numbers of $\phi$ and $J/\psi$ collected at KLOE and BESIII were used for numerous precision studies of  $\eta$ and $\etap$~\cite{Gan:2020aco}. We show some examples in Secs.~\ref{sec:etapipig} and~\ref{sec:strong3p}. Another option is the hadronic two-body processes with a narrow vector meson replacing the radiative photon, e.g., $J/\psi\to\omega\eta$, which allows for competitive study of $\omega$ decays as we discuss in Sec.~\ref{sec:ee-PPP}. Furthermore, at a \CT\ factory production of nonvector narrow charmonium resonances, e.g., $\eta_c$ or $\chi_{cJ}$, allows us to study their decays into 
light-hadron systems. At electron--positron colliders, these charmonia come from radiative decays of
$J/\psi$ and $\psi'$ such as $J/\psi\to\eta_c\gamma$ with $\BR \approx 1\%$ or $\psi'\to\gamma\chi_{cJ}$ with  $\BR \approx 10\%$. 

\begin{table}
 \caption[Data sets of the $\eta/\eta'$ mesons at KLOE and at BESIII]{\label{tab:eta} Estimate of the available $\eta/\eta^\prime$ data sets assuming  $2.4\times10^{10}$ $\phi$ events at KLOE/KLOE-2 and $10^{10}$ $J/\psi$ events at BESIII.}
 \begin{center}
 \renewcommand{\arraystretch}{1.3}
 \begin{tabular}{l c c c }\toprule
        Decay mode    &       $\mathcal{B}$ ($\times 10^{-4}$)~\cite{PDG}& $P_{cm} [\text{MeV}]$  & $\eta/\eta^\prime$ events \\ \midrule
      $\phi\to\gamma\eta$ &$130.3(2.5)$ & 363   &$3.1\times 10^8$ \\ 
          $\phi\rightarrow\gamma\eta'$ &$0.622(21)$ &60 &$1.5\times 10^6$\\  
\midrule
      $J/\psi\rightarrow\gamma\eta^\prime$ &$52.1(1.7)$ &1400&    $5.2\times 10^7$ \\ 
          $J/\psi\rightarrow\gamma\eta$ &$11.08(27)$ &1500&$1.1\times 10^7$\\  
       $J/\psi\rightarrow\phi\eta$ & $7.5(8)$ &1320&    $7.5\times 10^6$ \\ 
          $J/\psi\rightarrow\phi\eta'$ & $4.6(5)$&1192&   $4.6\times 10^6$ \\  
          $J/\psi\rightarrow\omega\eta$ & $17.4(2.0)$ &1394&    $1.7\times 10^7$ \\ \bottomrule
                  \end{tabular}
 \renewcommand{\arraystretch}{1.0}
                  \end{center}
\end{table}
\begin{figure}[t]
\centering
\includegraphics[width=0.90\textwidth]{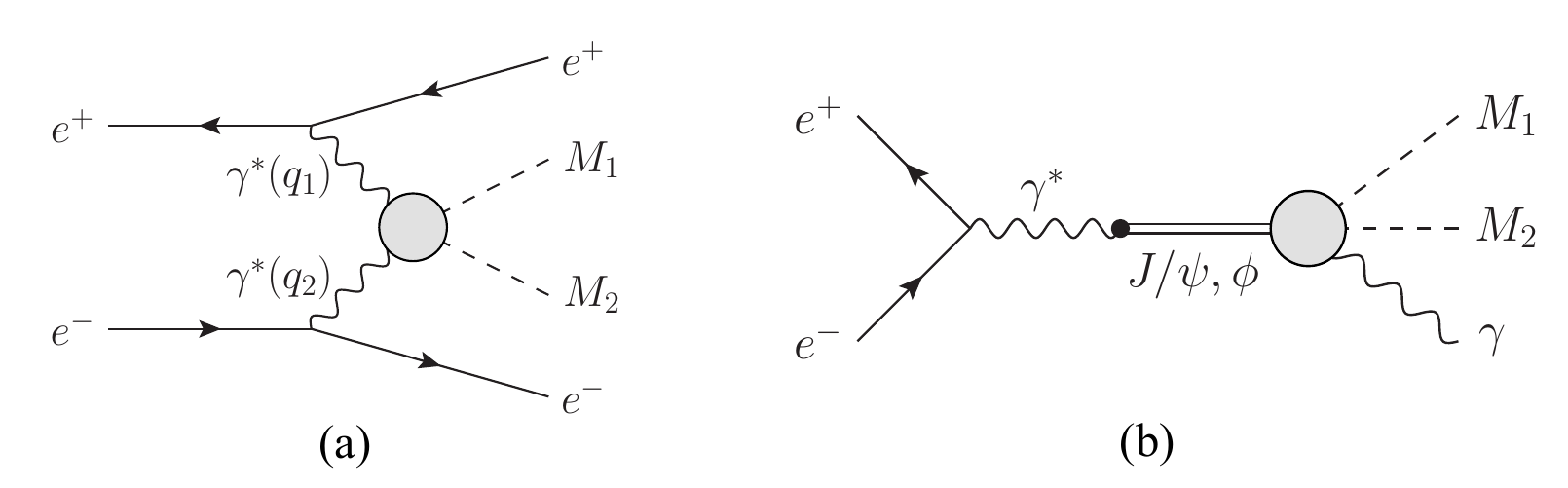} 
\caption[Double radiative processes ]{Production of parity-even meson systems $(M_1M_2)$ in (a) two-photon processes and
(b) radiative decays of $J/\psi$ or $\phi$. 
    \label{fig:twogmech}}
\end{figure}

\subsubsection{Higher-order electromagnetic processes}
With the modern high-luminosity colliders, hadron systems can be studied using higher-order electromagnetic processes like initial-state radiation $e^+e^-\to X\gamma$, shown in  Fig.~\ref{fig:mech2}(d), and two-photon fusion $e^+e^-\to e^+e^-\gamma^*\gamma^*\to e^+e^- X$, Fig.~\ref{fig:twogmech}(a). The invariant mass $W$ of the produced hadronic system $X$ can vary between threshold and the c.m.\ energy $\sqrt{s}$ of the collider. There are recent extensive reviews covering these experimental techniques: Ref.~\cite{Druzhinin:2011qd} for the ISR method
and Ref.~\cite{Danilkin:2019mhd} for the two-photon processes.
In addition, such experiments benefit from
highest  luminosities or/and high energies available at $B$-factories. 

\paragraph{Initial-state radiation processes}
The photon in the $e^+e^-\to X\gamma$ process can be radiated in the initial state from the electron or positron beams as shown in diagram Fig.~\ref{fig:mech2}(d). This corresponds to the ISR mechanism, where the quantum numbers of the hadronic system $X$ are $J^{PC}=1^{--}$. Alternatively the photon can be radiated after annihilation took place, e.g., due to a radiative decay of a vector meson $V$ as in diagram Fig.~\ref{fig:twogmech}(b). This corresponds to final-state radiation (FSR) and the hadronic system is in an even charge conjugation eigenstate $C=+1$, but both spin and parity can have several values. 

The general properties of an ISR process can be derived from QED and are given by the leptonic 
tensor $L_{\rm ISR}^{\mu\nu}$, which is separated from the hadronic part. The main features of the ISR photon distribution are 
the infrared divergence for small energies of the radiated photon and collinear divergences for small emission angles with respect to the electron or positron beams. The ISR process allows one to extract continuum hadronic cross sections for invariant masses $W$ of the hadronic system from threshold up to the  c.m.\ energy  $\sqrt{s}$ of the colliding beams. It effectively converts a fixed energy, high-luminosity collider into a variable-energy machine~\cite{Binner:1999bt}. The ISR technique was first applied  at the KLOE experiment to study $e^+e^-\to\pi^+\pi^-$~\cite{Aloisio:2004bu}. Since then the production cross sections for many multihadron final states were measured using this method. Most of the results were obtained at the BaBar experiment, with accumulated data samples corresponding to the integrated luminosity of $0.5\ab^{-1}$ at the $\Upsilon$ resonances.  In the context of our review, for light-meson dynamics, the application of the method is so far limited to studies of the total cross sections of meson systems. 
This might change in the future when statistics will be sufficient to study multidimensional differential distributions of the produced multihadron states.

\begin{figure}[t]
\begin{center}
\includegraphics[width=0.495\textwidth]{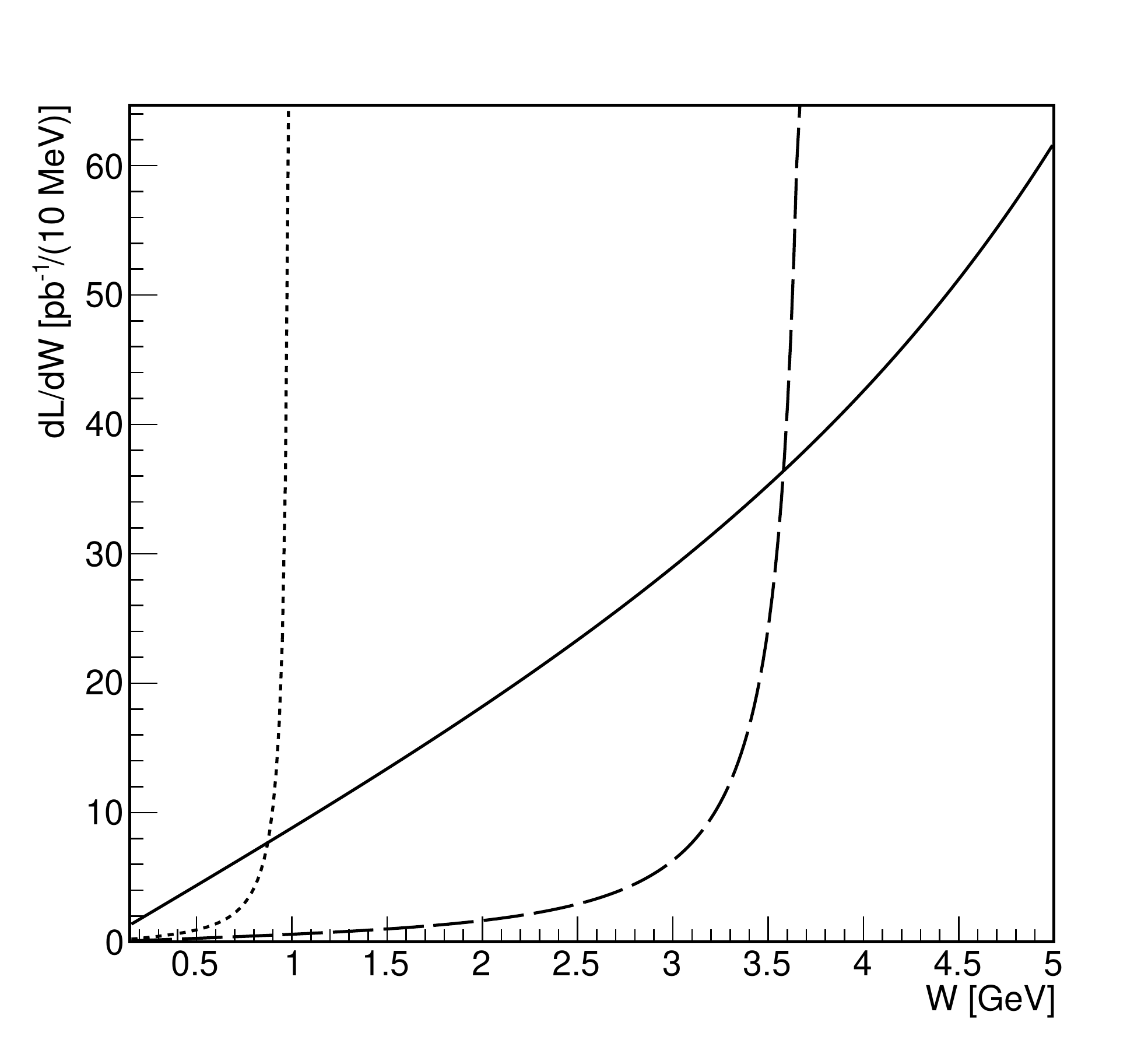}\put(-180,170){\bf\large (a)}
\includegraphics[width=0.495\textwidth]{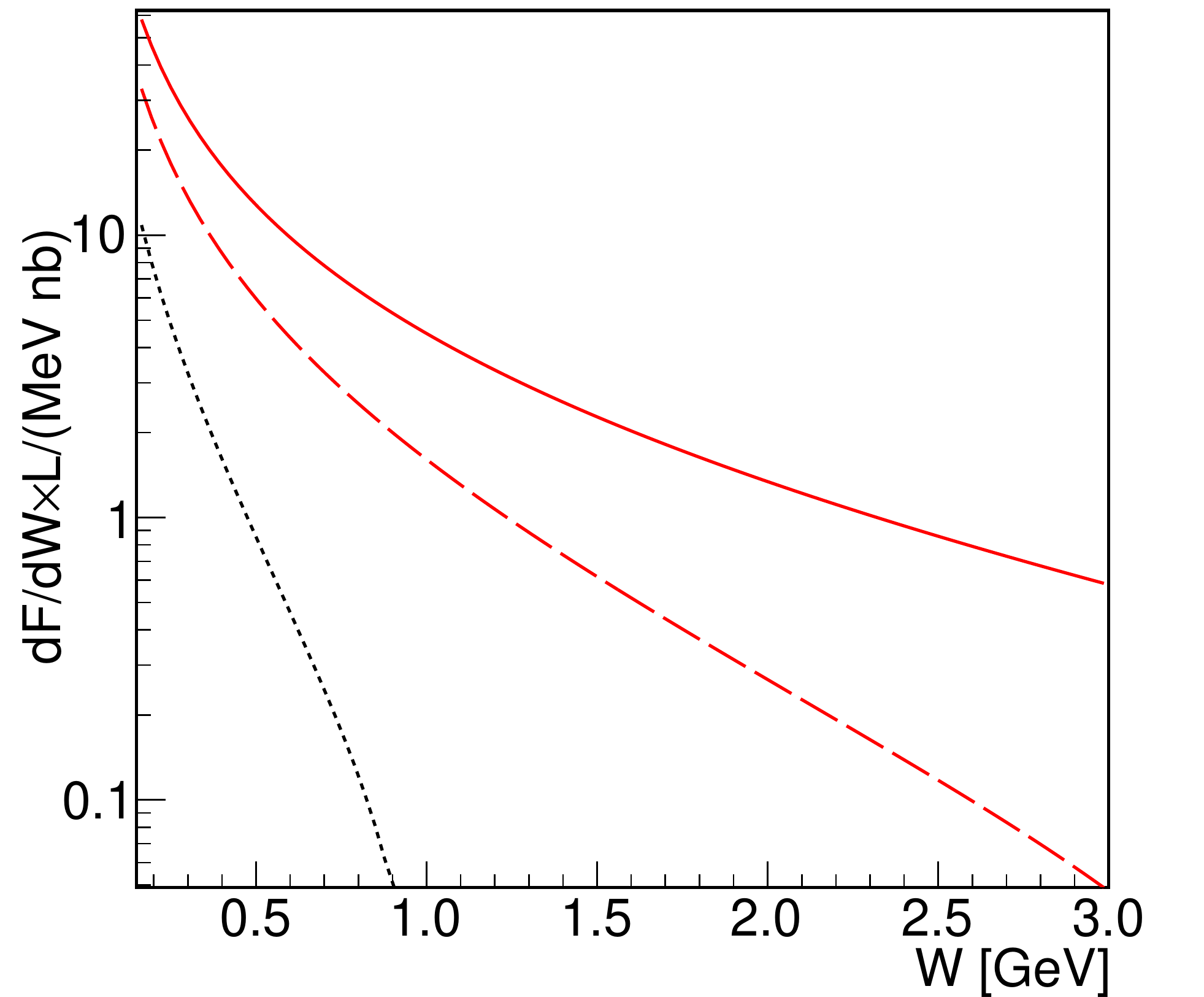}\put(-180,170){\bf\large (b)}
\caption[Gamma-Gamma luminosity]{Indicative differential 
luminosities  as a function of the hadronic system invariant mass $W$
at the $\phi$ meson (dotted line), \CT\ (dashed line), and $B$-factories (solid line)
for (a) ISR processes (for ${\cal L}= 2\fb^{-1}$,
$10\fb^{-1}$, and $1\ab^{-1}$, respectively) and  (b) $\gamma \gamma$ processes (${\cal L}=1\fb^{-1}$).}
\label{fig:ISR-gg}
\end{center}
\end{figure}
The differential cross section for the ISR process $e^+(p_1) + e^-(p_2) \to {\rm hadrons} + \gamma(k_1)$ can be written in terms of the leptonic and hadronic tensors as~\cite{Kuhn:2002xg} 
\begin{equation}
    \diff \sigma=\frac{(4 \pi \alpha)^2}{W^3s^2}L_{\rm ISR}^{\mu\nu}H_{\mu\nu}\frac{s-W^2}{64\pi^3}
    \diff\Phi_n(W;q_1,\ldots,q_n)\diff\Omega\ \diff W \,, \label{eq:ISR1}
\end{equation}
where $\Phi_n(W;q_1,\ldots,q_n)$ denotes the phase space volume of the hadronic system and $\Omega$ the scattering angle of the ISR photon.
 The general form of the ISR leptonic tensor is 
\begin{equation}
L^{\mu \nu}_{\rm ISR} =
a_{00} \, g^{\mu \nu} + a_{11} \, \frac{p_1^{\mu} p_1^{\nu}}{s}
 + a_{22} \, \frac{p_2^{\mu} p_2^{\nu}}{s} 
+ a_{12} \, \frac{p_1^{\mu} p_2^{\nu} + p_2^{\mu} p_1^{\nu}}{s}
+ i \pi \, a_{-1} \; 
\frac{p_1^{\mu} p_2^{\nu} - p_2^{\mu} p_1^{\nu}}{s} \,,
\end{equation}
where the LO coefficients are
\begin{align}
a_{00}^{(0)} &= \frac{2 m^2 z^2(1-z^2)^2}{y_1^2 y_2^2}
- \frac{2 z^2+y_1^2+y_2^2}{y_1 y_2} \,, \qquad
a_{11}^{(0)} = \frac{8 m^2}{y_2^2} - \frac{4z^2}{y_1 y_2} \,, \nonumber\\
a_{22}^{(0)} &= a_{11}^{(0)} (y_1 \leftrightarrow y_2) \,, \qquad 
a_{12}^{(0)} = - \frac{8 m^2}{y_1 y_2} \,, \qquad a_{-1}^{(0)}=0 \,, 
\end{align}
with 
\begin{equation}
y_i = \frac{2 k_1 \cdot p_i}{s}\,, 
\qquad m^2=\frac{m_e^2}{s}\,, \qquad z=\frac{W}{\sqrt{s}}\,.
\end{equation}
The result of the integration of the cross section in  Eq.~\eqref{eq:ISR1} with respect to all variables but the invariant mass of the hadronic system $W$ can be expressed in terms of the ISR luminosity $L$ as 
\begin{equation}
    \frac{\diff\sigma}{\diff W}\approx\frac{\sigma_B(W)}{\cal L}\frac{\diff L}{\diff W} \,,
\end{equation}
where ${\cal L}$ is the integrated luminosity at the primary c.m.\ energy of the colliding electron--positron beams. The term  $\diff L/\diff W$ 
is given by a radiator function as discussed in Sec.~\ref{sec:RadCorr}. The ISR luminosity is compared for colliding beams with c.m.\ energies of $1 \GeV$, $3.77 \GeV$, and $10 \GeV$,
approximately corresponding to the experiments at KLOE-2, BESIII, and $B$-factories, respectively, in Fig.~\ref{fig:ISR-gg}(a). The plot was obtained using indicative values of the integrated luminosities ${\cal L}$ for each facility: $2\fb^{-1}$,
$10\fb^{-1}$, and $1\ab^{-1}$.  The ISR luminosity is a fast increasing function close to the c.m.\ energy of the collider. The cross section measurements using ISR processes have reached such statistical precision that they are often dominated by systematic effects. Therefore the ISR measurements at colliders with different c.m.\ energies provide complementary data sets with different experimental settings. The best example is given by the studies of the $e^+e^-\to\pi^+\pi^-$ process discussed in Sec.~\ref{sec:eePP}.

\begin{sloppypar}
\paragraph{Two-photon processes}
Two-photon production of $C=+1$ hadronic systems in $e^-e^\pm\to e^-e^\pm\gamma^*\gamma^*\to e^-e^\pm X$ is given by the diagram illustrated in Fig.~\ref{fig:twogmech}(a).
The two photons are space-like with squared four-momenta $q^2_1,q^2_2\le0$ and the process can be studied both in electron--positron and electron--electron collisions. The cross section increases as the logarithm of the c.m.\ energy of the colliding beams. 
Since it was realized in the 1970s that significant production rates could be
achieved~\cite{Brodsky:1970vk,Brodsky:1971ud}, the two-photon processes have
been investigated at most of the $e^+e^-$ colliders with the limitations
imposed by the rate of these subleading processes at the luminosity
reached by past-generation colliders.
\end{sloppypar}

Three types of experiments are possible: untagged, where none of the leptons in the final state is recorded; single-tagged, with one lepton measured; and double-tagged, where the two leptons are measured, meaning that the corresponding photon four-momenta are determined. If not tagged, the unmeasured photon 
distributions peak at  $q^2_1(q^2_2)\approx0$.

In the low-energy region the two $\gamma^*$ can be considered quasi-real, so that only $J^{PC}=0^{\pm +}$, $2^{\pm +}$ quantum numbers are allowed~\cite{Yang:1950rg}. If no cut is applied to the final-state leptons, the Weizs\"acker--Williams or equivalent-photon approximation~\cite{Brodsky:1971ud} can be used to understand the main qualitative features. 
Then the event yield, $N_{eeX}$, can be evaluated according to 
\begin{equation}
N_{eeX} = {\cal L}\int\frac{{\rm d} F}
{\diff{W}}\,
\sigma_{\gamma\gamma\to X}(W)\,
{\diff W}\,,
\label{eq:gg-stat2}
\end{equation}
where $W$ is the invariant mass of the produced hadronic system, ${\cal L}$ is the integrated luminosity, and $\diff F/\diff W$ is the $\gamma\gamma$ flux,
\begin{equation}
\frac{{\diff}F}{{\diff}W} = 
\frac{1}{W}\ \left(\frac{2\alpha}{\pi}\right)^2\
\left(\ln\frac{\sqrt{s}}{2m_e}\right)^2 f(z) \,,
\label{eq:gg-stat3}
\end{equation}
where $\sqrt{s}$ is the c.m.\ energy of the collider and 
\begin{equation}
f(z) = -(z^2+2)^2\ \ln{z}
-(1-z^2)\ (3+z^2) \quad {\rm with} \quad 
z = \frac{W}{\sqrt{s}}.
\label{eq:fz}
\end{equation}
Figure~\ref{fig:ISR-gg}(b) shows the flux multiplied by an integrated luminosity ${\cal L}=1\fb^{-1}$, as a function of the $W$ invariant mass for three different c.m.\ energies. This plot demonstrates the feasibility of the measurements of the final states $\pi^+\pi^-$, $\pi^0\pi^0$, or $\pi^0\eta$, whose cross sections are of the order of or larger than $1\nb$,
and of the identification of the resonances produced in these channels. 
Single pseudoscalar ($X=\pi^0$, $\eta$, or $\eta^\prime$) production is also
accessible, which allows for the determination of the two-photon decay widths of these mesons and to measure the transition form factors $F_{X\gamma^*\gamma^*}(q^2_1,q^2_2)$ as a function of the momentum of the virtual photons, $q^2_1$ and $q^2_2$.

In the present review we will give examples of the related radiative processes shown in Fig.~\ref{fig:twogmech}(b). The radiative decays have usually much larger cross sections and precision differential experimental data is available. One can use dispersive methods to relate these processes to the two-photon form factors. A prominent example is the use of information from $e^+e^-\to\pi^+\pi^-$ and $e^+e^-\to\pi^+\pi^-\pi^0$ to deduce the $\pi^0$ transition form factor, i.e., the  $\pi^0\gamma^*\gamma^*$ vertex~\cite{Hoferichter:2014vra,Hoferichter:2018dmo,Hoferichter:2018kwz}.

\subsubsection{Tau lepton decays}
The purely leptonic initial state in $\tau$ decays involving two or three final-state hadrons and the $\tau$ neutrino represents one of the cleanest ways to study mesonic systems. The hadronic systems are produced by elementary currents. The accessible invariant masses of the produced system vary from threshold up to $m_\tau$ due to the energy taken by the emitted $\nu_\tau$. 

Tau leptons are produced in the continuum electroweak process $e^+e^-\to \tau^+{\tau}^-$, see Fig.~\ref{fig:mech}(a).  The threshold $\sqrt{s_{\text{thr}}}=2m_\tau$ is $132\MeV$ below the mass of the $\psi'$ resonance. The differential cross section for a lepton--antilepton pair e.m.\ production process is 
\begin{equation}
    \frac{\diff\sigma}{\diff\Omega}=\frac{\alpha^2}{4s}\sqrt{1-\frac{4m_\ell^2}{s}}\left[\left(1+\frac{4m_\ell^2}{s}\right)+\left(1-\frac{4m_\ell^2}{s}\right)\cos^2\theta\right] \,,
\end{equation}
and the total Born cross section with perturbative photon propagator
\begin{equation}
    \sigma=\frac{4\pi\alpha^2}{3s}\sqrt{1-\frac{4m_\ell^2}{s}}\left(1+\frac{2m_\ell^2}{s}\right)\,.
    \label{eq:eeToll}
\end{equation}
The $e^+e^-\to \tau^+{\tau}^-$ e.m.\ Born cross section is shown in Fig.~\ref{fig:cstot}, where the maximum of $3.8\nb$ is reached at $4.2\GeV$. The cross section in Eq.~\eqref{eq:eeToll} should be multiplied by the lepton form factor $|F_\ell(s)|^2$ accounting for vacuum polarization~\cite{Benayoun:1997ex}.  The effect of the form factor is that the production is enhanced at vector resonances such as the $\psi'$. We will discuss this effect in Sec.~\ref{sec:HVP}.  So far the majority of the experimental studies of the $\tau$ lepton were done at high-energy electron--positron colliders such as LEP or $B$-factories. However, close-to-threshold production offers a cleaner environment. The direct scan at the $\tau^+\tau^-$ production threshold is also the best method to precisely measure the mass of the $\tau$ lepton, an important parameter of the Standard Model. The most precise recent results are from the KEDR~\cite{Anashin:2007zz} and BESIII experiments~\cite{Luo:2015gka}. 

An important feature of the  $e^+e^-\to \tau^+\tau^-$  process is that the $\tau^+\tau^-$ pair is produced in a spin-correlated state. We define $\tau^+$ and $\tau^-$ polarization vectors in their respective rest frames with the same orientations of the spin quantization axes with the $z$ direction along
$\tau^-$ momentum in the overall c.m.\ system. The $y$ direction is given by the vector product of
the incoming electron and the outgoing $\tau^-$ momenta.  The spin correlation 
matrix  depends on the angle $\theta$ between the electron and the $\tau^-$. At c.m.\ energies relevant to \CT\ factories, the electron mass
can be neglected and a single-photon annihilation process dominates.
The spin correlation  matrix for the e.m.\ process in the Born approximation is given as~\cite{Tsai:1971vv} 
\begin{equation}
\frac{1}{1+\eta\cos^2\theta} \left(
\begin{array}{ccc}
  \sin ^2\theta & 0 & \sqrt{1-\eta^2}  \sin \theta \cos \theta  \\
 0 & -\eta  \sin ^2\theta  & 0 \\
  \sqrt{1-\eta^2} \sin \theta  \cos \theta  & 0 & \eta +\cos ^2\theta  
\end{array}
\right) \quad {\rm with} \quad     \eta=\frac{s-4m_\tau^2}{s+4m_\tau^2} \,.
\end{equation}
Close to production threshold, $\eta\approx0$ and the matrix includes $x$--$z$ spin correlation terms. The $\tau^+\tau^-$ spin correlations are reflected in the multidimensional distribution of the daughter particles from the two leptons.

The matrix element for a semileptonic decay $\tau\to X\nu_\tau$ has the form
\begin{equation}
    {\cal M}=\frac{G_F}{\sqrt{2}}\bar u(p_\nu)\gamma^\mu(1-\gamma_5)u(p_\tau)J_\mu\,.
\end{equation}
Since parity is maximally violated in $\tau$ decays, the hadronic current has both vector and axial-vector contributions, $J_\mu=\bra{X}V_\mu-A_\mu\ket{0}$. 
The weak quark current is 
\begin{equation}
j_\mu^-=\bar d\gamma_\mu(1-\gamma_5)u \cos\theta_C+\bar s\gamma_\mu(1-\gamma_5) u\sin\theta_C \,.
\end{equation}
 The general weak hadronic current contains four terms~\cite{Tsai:1971vv}: 
\begin{equation}
    J^\mu=\Big[\big(F_1^\mu+iF_2^\mu\big)-\big(F_1^{5\mu}+iF_2^{5\mu}\big)\Big] V_{ud}+\Big[\big(F_4^\mu+iF_5^\mu\big)-\big(F_4^{5\mu}+iF_5^{5\mu}\big)\Big]V_{us} \,.
\end{equation}
In the limit of exact isospin symmetry, the weak vector current is conserved separately and is directly related to the electromagnetic current~\cite{Cabibbo:1963yz},
i.e.,
the coupling of the $W$ to the $\rho$ meson is given by the replacement $e\to \sqrt{2}g\cos\theta_C$. 
\begin{figure}
    \centering
    \includegraphics[width=0.4\textwidth]{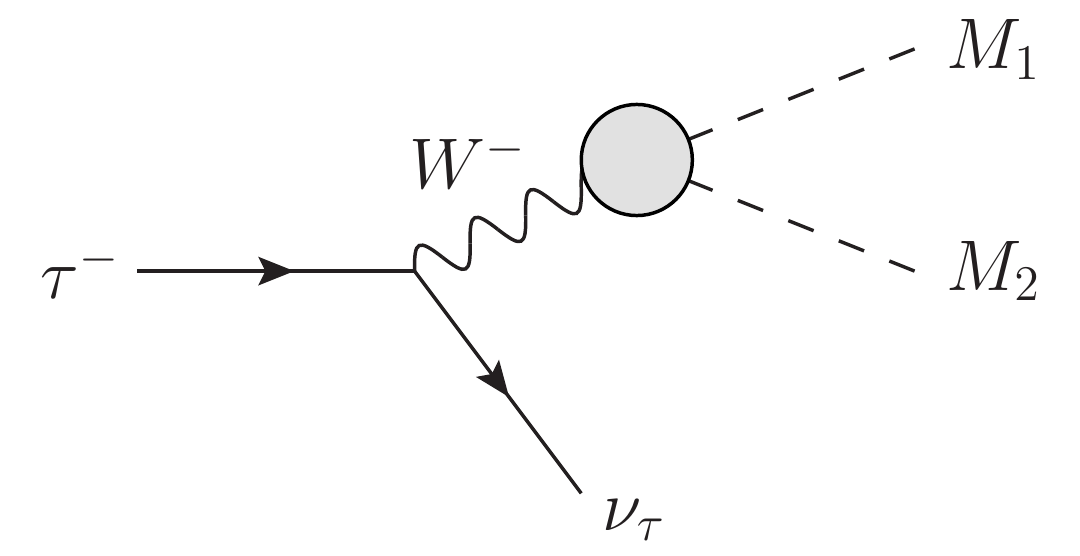} \hspace{1cm}
    \includegraphics[width=0.4\textwidth]{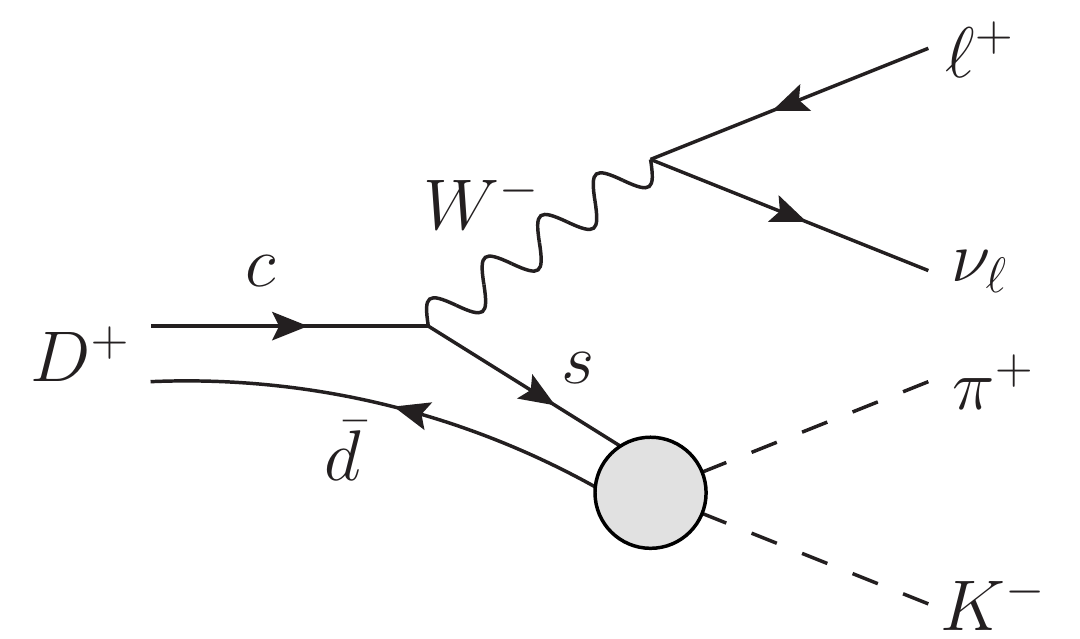}
    \caption[Tau and charmed-meson decays]{Tau and charmed-meson decays as a means to study meson interactions. 
    }
    \label{fig:HadDecays}
\end{figure}
The $\tau$ decays into two pseudoscalar mesons illustrated by the diagram in Fig.~\ref{fig:HadDecays}(left) are particularly clean systems to study meson--meson interactions. Branching fractions for some of such modes are listed in Table~\ref{tab:taudecay2}. In these decays  the hadronic system can have spin-parity quantum numbers $J^P = 0^+$ or $1^-$, and in the isospin limit the conserved-vector-current theorem forbids production of $0^+$ nonstrange states in $\tau$ decays.
Since there are two recent reviews  dedicated to hadronic $\tau$ decays~\cite{Davier:2005xq,Pich:2013lsa} we will discuss only the direct connection between 
$e^+e^-\to\pi^+\pi^-$,  $e^+e^-\to K\bar K$  and $\tau^-\to\pi^-\pi^0\nu_t$, $\tau^-\to K^-\bar K^0\nu_\tau$ in Sec.~\ref{sec:eePP}.
The symmetric $V-A$ structure of the weak current implies that any differences between the two currents are generated by nonperturbative QCD effects. The axial vector current is dominated by
$X=\pi^-\pi^+\pi^-$ with the $a_1(1260)$ resonance decaying (predominantly) to $\rho\pi$, which provides complementary information to the isoscalar vector current in the $e^+e^-\to\pi^-\pi^+\pi^0$ reaction.

\begin{table}[t]
    \caption[]{Branching fractions for example decay modes of the $\tau$ lepton into two light mesons~\cite{PDG}.}
    \renewcommand{\arraystretch}{1.3}
    \begin{center}
    \begin{tabular}{lr}
    \toprule
    Decay mode& $\BR$\\ \midrule
      $\tau\to\pi^-\pi^0\nu_\tau$   &25.49(9)\%  \\ 
      $\tau\to K^-\pi^0\nu_\tau$      & $4.33(15)\times10^{-3}$\\
      $\tau\to K^-K^0\nu_\tau$      & $1.486(34)\times10^{-3}$\\
      $\tau\to\eta K^-\nu_\tau$      & $1.55(8)\times10^{-4}$\\
      $\tau\to\phi\pi^-\nu_\tau$      & $3.4(6)\times10^{-5}$\\
      $\tau\to\phi K^-\nu_\tau$      & $4.4(16)\times10^{-5}$\\
      $\tau\to\omega\pi^-\nu_\tau$      & $1.95(6)\times10^{-2}$\\
      $\tau\to \omega K^-\nu_\tau$      & $4.1(6)\times10^{-4}$\\
      \bottomrule
    \end{tabular}
    \label{tab:taudecay2}
    \end{center}
    \renewcommand{\arraystretch}{1.0}
\end{table}

\subsubsection[Weak decays of $K$ and $D$ mesons]{Weak decays of \boldmath{$K$ and $D$} mesons}\label{sec:weakKD}
The electron--positron colliders are also factories of exclusive quantum-correlated pairs of $K\bar K$ and $D\bar D$ mesons.  The DA$\Phi$NE collider provides 
optimal conditions for the production of entangled kaon pairs close to threshold at the $\phi$ resonance, where the kaon--antikaon modes are the dominant contributions to the decay width: $\BR(\phi\to K^+K^-) = 49.2(5)\%$ and $\BR(\phi\to K_LK_S) = 34.0(4)\%$.
The main experimental advantage is the possibility to select a pure, kinematically well-defined $K_{S}$ beam. Kaons from $\phi$ decays have low momenta  of  $110\MeV$ for neutral kaons.  Due to the beams' crossing angle, there is a transverse momentum component of $15\MeV$ in the detector reference system. The mean decay lengths of $K_L$ and $K_S$ are $3.4\m$ and  $0.59\cm$, respectively. For a detector with a radius of $2\m$ such as  KLOE,  this  gives a geometrical efficiency of ca.\ 30\% for the fiducial volume for the detection of the $K_L$ decay vertex~\cite{Franzini:2006aa}. On the other hand, a fraction of the $K_L$ interacts in the calorimeter, allowing the experiment to efficiently tag the $K_{S}$ beam. However, from $K_S$ decays, with the dominant contribution of two-pion decays, only limited information on meson interactions can be extracted. The more relevant  $K_L$ studies must be carried out at dedicated fixed-target facilities using hadronic production reactions and extracting $K_L$ beams. 

On the contrary, the  potential for meson interaction studies in  $D$ and $D_s$ decays at the  \CT\ colliders is huge. Both semileptonic $D\to X\ell\bar\nu_\ell$ shown  in Fig.~\ref{fig:HadDecays}(right) and  hadronic modes with at least three mesons can be investigated~\cite{Artuso:2008vf}. For $D$ mesons the optimal production process is $e^+e^-\to\psi(3770)\to D\bar{D}$ with cross sections $\sigma(e^+e^-\to D^0 \bar D^0)=3.6\nb$ and $\sigma(e^+e^-\to D^+D^-)=2.8\nb$~\cite{Ablikim:2018tay}. The presently collected data sample by BESIII at the $\psi(3770)$ is $2.9\fb^{-1}$. Also $D_s\bar D_s$ mesons are produced, with the maximal yield within the range available for BESIII at c.m.\ energy of $4170\MeV$, where the cross section is approximately $0.9\nb$~\cite{CroninHennessy:2008yi} and dominated by the $e^+e^-\to D_s^{*+}D_s^-$ intermediate state. The $D_s^{*+}$ decays predominantly to $D^+_s\gamma$, with $\BR(D_s^{*+}\to D_s^+\gamma)=95.5(7)\%$.

The analyses of the $D$-meson decays benefit from a double-tag technique, pioneered by the Mark~III collaboration~\cite{Baltrusaitis:1985iw,Adler:1987as}. With this method one can select and reconstruct the complete kinematics for the data samples of $D^+$ and $D^0$ decays (or charge conjugated modes) with one missing particle. In the recent BESIII analysis~\cite{Ablikim:2018tay} there are three tagging modes for $\bar D^0$ and six for $D^-$ listed in Table~\ref{tab:Dstinfo}.
\begin{figure}
    \centering
    \includegraphics[width=0.7\textwidth]{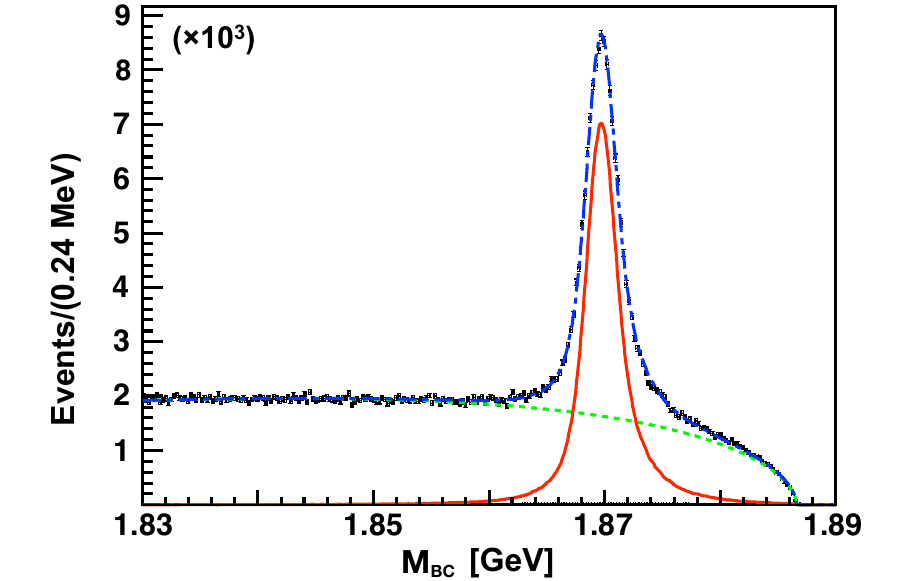}
    \caption[$D$ meson tagging]{Beam-constrained mass $M_{BC}$ distribution  for the single-tag mode $D^+\to K^-\pi^+\pi^+\pi^0$ from the BESIII analysis~\cite{Ablikim:2018tay} (points with error bars). The lines represent the fit:  background contribution (dashed line), signal (solid line), and the sum of the two (dashed--dotted line).}
    \label{fig:DmesonST}
\end{figure}
The criteria for choosing the tagging modes are: large signal yield (large product of $\BR$ and efficiency) and low background. In each event only one $D$ candidate for a given tag mode is selected for $D^+$ and $D^-$ separately. For each tag mode, one selects the signal region from distributions of beam-constrained mass $M_{BC} = \sqrt{E^2_{\text{beam}} - |{\bf p}|^2}$, where ${\bf p}$ is the three-momentum of the tag $D^-$ candidate and $E_{\text{beam}}$ is the beam energy in the c.m.\ system. An example of the  $M_{BC}$ distribution for  $D^+\to K^-\pi^+\pi^+\pi^0$ from BESIII is shown in Fig.~\ref{fig:DmesonST}. One can fit the $M_{BC}$ distributions with Monte-Carlo-based signal shapes, and the background shape is well parameterized with an ARGUS function~\cite{Albrecht:1990am}.
To select the tag, one  requires $1863\MeV<M_{BC}<1877\MeV$. Table~\ref{tab:Dstinfo} shows the number of selected events $N_{\text{tag}}$ and reconstruction efficiencies $\epsilon_{\text{tag}}$ for each tag mode.
\begin{table}[t]
  \caption[$D$ meson tagging at the \CT\ factory]{Channels used for $D$ meson tagging in 
  $e^+e^-\to D\bar D$ at the $\psi(3770)$ resonance. Single-tag efficiencies, $\epsilon_{\text{tag}}$, are given. The efficiencies are corrected for $\BR(K_S^0\to\pi^+\pi^-)$. The number of events is for integrated luminosity of $2.9\fb^{-1}$ and c.c.\ is implied. 
}
\def\1#1#2#3{\multicolumn{#1}{#2}{#3}}
\label{tab:Dstinfo}
\begin{center}
\scalebox{1.0}
{
\renewcommand{\arraystretch}{1.3}
  \begin{tabular}{l r   c }
    \toprule
    \multicolumn{1}{c}{Tag mode}
    & \multicolumn{1}{c}{$N_{\text{tag}}$ ($\times 10^3$)}
    & $\epsilon_{\text{tag}}$ (\%) 
    \\
\midrule
$\bar D^0\to K^+\pi^-$      &   $520$ & $64$\\
$\bar D^0\to K^+\pi^-\pi^-\pi^0$    &      $1080$ & $35$ \\
$\bar D^0\to K^+\pi^+\pi^-\pi^-$                      &  $699$ & $39$ \\
\bottomrule
\phantom{$K_S^0\pi^-\pi^0$}              &  & \\
\phantom{$K_S^0\pi^-\pi^-\pi^+$}     &   &  \\
\phantom{$K^+K^-\pi^-$}                    &  &  \\
\end{tabular}
\renewcommand{\arraystretch}{1.0}
}
\scalebox{1.0}
{
\renewcommand{\arraystretch}{1.3}
  \begin{tabular}{l r   c }
    \toprule
    \multicolumn{1}{c}{Tag mode}
    & \multicolumn{1}{c}{$N_{\text{tag}}$ ($\times 10^3$)}
    & $\epsilon_{\text{tag}}$ (\%) 
    \\
\midrule
$D^-\to K^+\pi^-\pi^-$               &   $798$ & $51$\\
$D^-\to K^+\pi^-\pi^-\pi^0$    &      $245$ & $25$ \\
$D^-\to K_S^0\pi^-$                      &  $93$ & $51$ \\
$D^-\to K_S^0\pi^-\pi^0$              &  $206$ & $26$ \\
$D^-\to K_S^0\pi^-\pi^-\pi^+$     &  $110$ & $27$ \\
$D^-\to K^+K^-\pi^-$                    &  $68$ & $40$ \\
\bottomrule
\end{tabular}
\renewcommand{\arraystretch}{1.0}
}
\end{center}
\end{table}
In total, from the BESIII run at the $\psi(3770)$ more than $2\times10^6$ $D^0/\bar D^0$ and $1.5\times10^6$ $D^+/ D^-$ tagged events are available. This data sample in particular allows studies of semileptonic decays with two pseudoscalar mesons in the final state. Some examples are given in Sec.~\ref{sec:exDtoPPlv}.

\subsection{Radiative corrections}\label{sec:RadCorr}
\subsubsection{Initial-state-radiation corrections}
In scan experiments, the goal is to determine the Born cross section $\sigma_B$ for an exclusive process $e^+e^-\to X$, cf.\ Fig.~\ref{fig:mech}(b). What is directly measured is the visible cross section $\sigma_{\rm vis}$, which includes processes with extra ISR photons radiated, see Fig.~\ref{fig:mech2}(d). The observed cross section depends on the invariant mass  $W<\sqrt{s}$ of the final state, as well as on the emission angle of the radiated photon. One obtains the following relation between $\sigma_B$ and $\sigma_{\rm vis}$ in the case the photon is not observed:
\begin{equation}{\label{eq:radxs}}
\sigma_{\rm vis}(s)=\int_{M_{\rm{th}}}^{\sqrt s} \diff W \frac{2W}{s}F(s,x)\sigma_B(W):=\sigma_B(s)\left\{1+\delta(s)\right\} \,, 
\end{equation}
where $M_{\rm{th}}$ is the sum of the masses of the final particles for the process, $x\equiv1-W^2/s$, and $F(s,x)$ is the radiator function, which can be calculated in QED.
It is necessary to go beyond the first order in $\alpha$ and sum all diagrams with soft multi-photon emission to avoid infrared divergences: both soft multiphoton emission and hard collinear bremsstrahlung in the leading logarithmic approximation have to be taken into account. To calculate the finite-order leading logarithmic correction, the structure function method was implemented in Ref.~\cite{Kuraev:1985hb}. Up to order $\alpha^2$, the radiator function takes the form
\begin{equation}{\label{eqn:secondRad}}
F(s,x)=\beta x^{\beta-1}\Delta-\frac{\beta}{2}(2-x) + \frac{\beta^2} {8}\left\{(2-x)\left[3\ln(1-x)-4\ln x\right]-4\frac{\ln(1-x)}{x}-6+x\right\},
\end{equation}
where
\begin{align}
\beta&=\frac{2\alpha}{\pi}({l}-1) \,,\qquad\Delta=1+\frac{\alpha}{\pi}\left(\frac{3}{2}\;{l}+\frac{1}{3}\pi^2-2\right)+\left(\frac{\alpha}{\pi}\right)^2\delta_2\,,\qquad
{l}=2\ln \frac{\sqrt s}{m_e}  \,,\nonumber\\
\delta_2&=\left(\frac{9}{8}-2\zeta(2)\right){l}^2-\left(\frac{45}{16}-\frac{11}{2}\zeta(2)-3\zeta(3)\right){l}-\frac{6}{5}\zeta(2)^2-\frac{9}{2}\zeta(3)-6\zeta(2)\ln 2+\frac{3}{8}\zeta(2)+\frac{57}{12} \,,\nonumber\\
&\zeta(2)=\frac{\pi^2}{6} \,, \qquad
\zeta(3)\approx1.2020569 \,,
\end{align}
and $\zeta$ is the Riemann zeta function. In most cases 
the above representation of the radiator function up to second order is sufficient to determine the cross section, although contributions from $\Order(\alpha^3)$ are known~\cite{Montagna:1996jv}. A detailed discussion of this important aspect of the cross section measurements at electron--positron colliders is found in the report by the Working Group on Radiative Corrections~\cite{Actis:2010gg}. 

\begin{figure}
    \centering
    \includegraphics[width=0.9\textwidth]{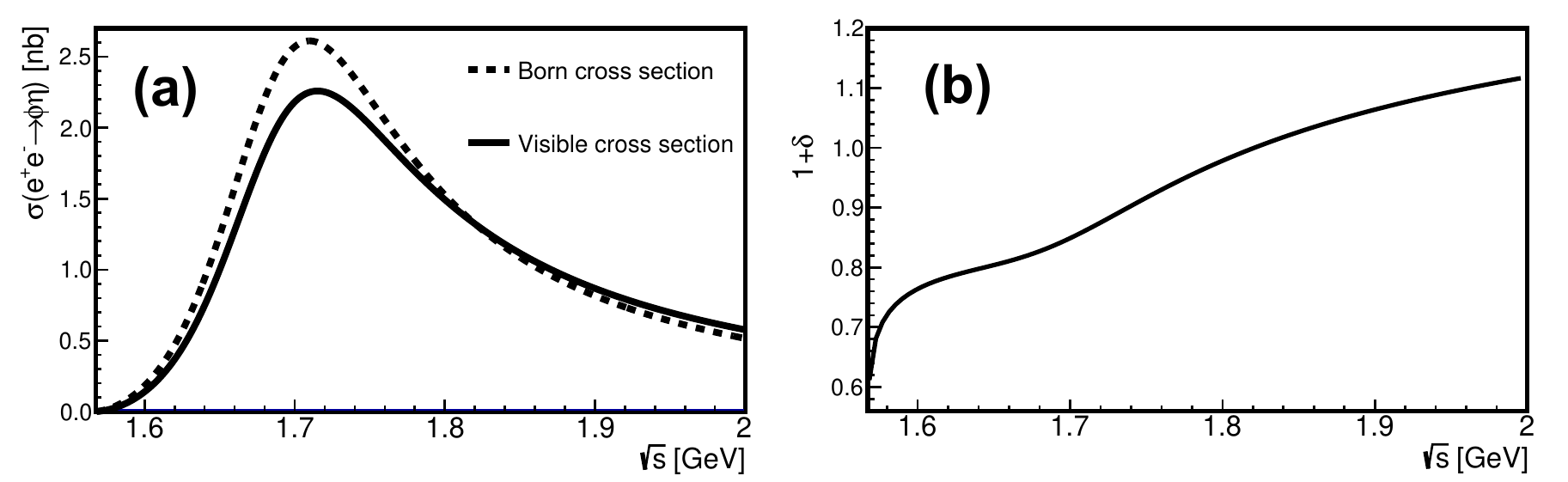}
    \caption[Illustration of the effect of radiative corrections ]{Radiative corrections for the $e^+e^-\to\phi\eta$ process close to threshold. (a) The Born cross section is shown by the dashed line and the visible cross section by the solid line. (b) The corresponding radiative correction term $1+\delta(s)$.}
    \label{fig:radCor}
\end{figure}

To give an illustration of the effect of the radiative corrections, we use a simplified form of the $e^+e^-\to\phi\eta$ reaction close to threshold (at $\sqrt{s}=1.567\GeV$) inspired by the recent CMD-3 analysis using the scan method~\cite{Ivanov:2019crp}. The energy dependence of the Born and visible cross sections is shown in Fig.~\ref{fig:radCor}(a). The process is dominated by the broad $\phi(1680)$ resonance and influenced by the threshold. In this example $\sigma_{\rm vis}(s)<\sigma_B(s)$ for  $\sqrt{s}<1.8\GeV$. This is caused by two effects: the kinematical limit for the ISR photon energy due to the reaction threshold and the initially rising $\sigma_B(s)$. Eventually the contribution of the broad resonance compensates the two effects and $\sigma_{\rm vis}(s)>\sigma_B(s)$. Figure~\ref{fig:radCor}(b) shows the corresponding ISR factor $1+\delta(s)$. The radiative corrections are much larger (up to 20\%) than what one would na\"ively expect for an $\alphaem$ suppressed process.  
Born cross sections measured by energy-scan experiments are obtained in the following way. At each c.m.\ energy point of the collider, $N_i$ events consistent with a signal signature are selected and the background contribution $B_i$ is estimated. The visible cross section is given as    $\sigma_{{\rm vis},i}=(N_i-B_i)/({\cal L}_i\cdot\epsilon_i)$, where ${\cal L}_i$ is the integrated luminosity and $\epsilon_i$ the detection efficiency. If the final state was selected using specific decay modes, the cross section is obtained by dividing by the product of branching fractions of all involved decays. The ISR correction is often performed by an iterative procedure using the measured $\sigma_{{\rm vis},i}$ values directly or with the help of a cross section parameterization. The Born cross section is obtained by solving Eq.~\eqref{eq:radxs} iteratively.  
When dealing with the extracted Born cross sections using this method one should remember that the data points at different energies are correlated. Therefore the necessary practice of scan experiments is to provide  information needed to calculate $\sigma_{{\rm vis},i}$ and to list the determined $1+\delta_i$ factors explicitly. In fact the Born cross sections might be reevaluated when new data sets are available.  For analyses of line shapes close to narrow resonances, often only the visible cross section is shown and both radiative corrections and energy smearing are included in the fit function.  

\subsubsection{Hadronic vacuum polarization \label{sec:HVP}}
Light-meson interactions also play an important role for precise descriptions of purely leptonic processes such as $e^+e^-\to\mu^+\mu^-$, via hadronic vacuum polarization (HVP) in the photon propagator and the contribution to the running of the fine structure constant $\alphaem$. The value of the QED coupling constant $\alphaem$ is determined from the anomalous magnetic moment of the electron with the impressive accuracy of 0.37 parts per billion~\cite{Aoyama:2012wj,Aoyama:2017uqe,Parker:2018vye,Morel:2020dww}. 
However, electromagnetic processes at nonzero momentum transfer squared, $s$, are described by  an effective electromagnetic coupling $\alpha(s)$.  The shift of the effective coupling involves low-energy nonperturbative hadronic effects, which affect the precision. These effects represent the largest uncertainty for electroweak precision tests such as the determination of $\sin^2\theta_W$ at the $Z$ pole or the Standard Model prediction of the muon $g-2$~\cite{Jegerlehner:2001wq}.

\begin{figure}
    \centering
     \includegraphics[width=0.99\textwidth]{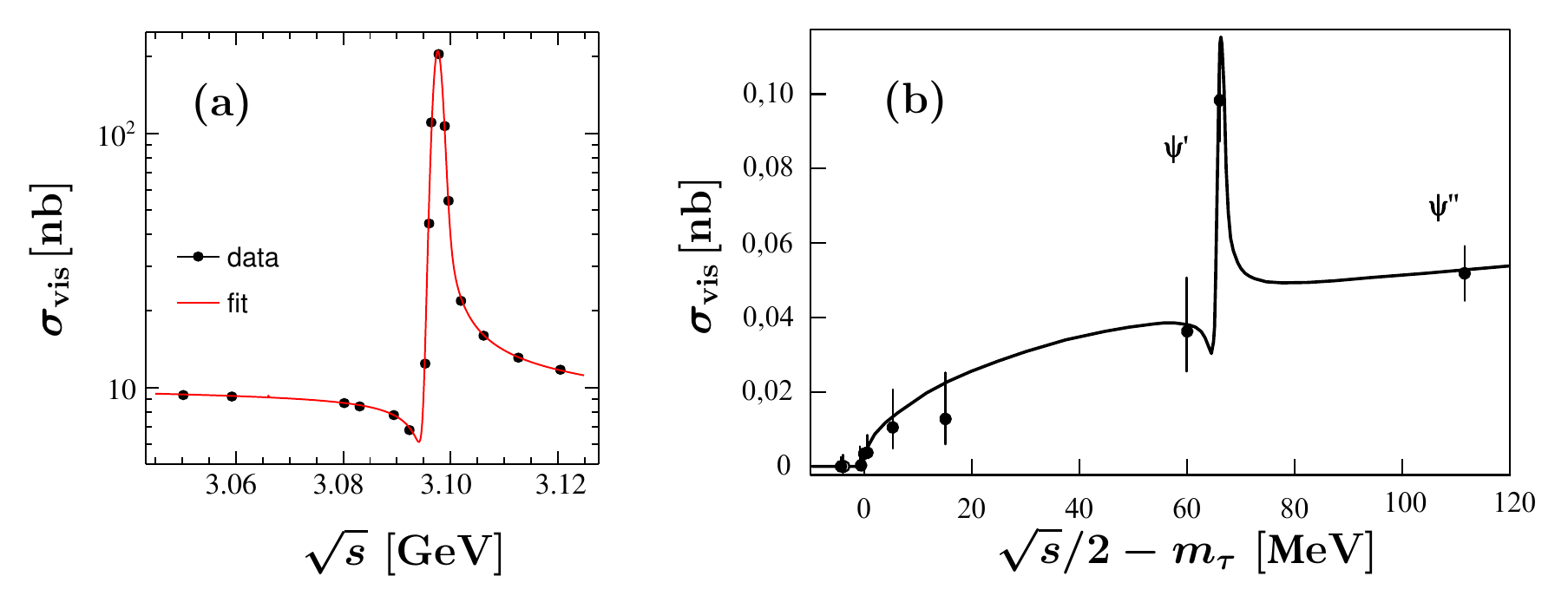}
    \caption[Visible cross section for lepton--antilepton pair production close to resonances]{Visible cross section for lepton--antilepton pair production close to the $J/\psi$ and $\psi'$ resonances. (a) $J/\psi$ scan data for $e^+e^-\to\mu^+\mu^-$ from BESIII~\cite{Ablikim:2018ege}; (b) close-to-threshold $e^+e^-\to\tau^+\tau^-$ data from KEDR~\cite{Anashin:2007zz}. The curves are line shapes  using the Born cross section from Eq.~\eqref{eq:Rcont} convoluted with FSR and the beam energy spread. 
    }
    \label{fig:HVPpsi}
\end{figure}

The effect of a vector meson $V$ in lepton--antilepton pair production is obtained by  combining contributions from the e.m.\  and the $V$ propagators in Eq.~\eqref{eq:eV}. The Born cross section for the annihilation into a muon--antimuon pair $e^+e^-\to V\to\mu^+\mu^-$ close to $m_V$ can be written as~\cite{Ablikim:2018ege}
\begin{equation}
    \sigma=\frac{4\pi\alphaem^2}{s}\left|1+\frac{3s}{\alphaem m_V}\frac{\sqrt{\Gamma_{ee}\Gamma_{\mu\mu}}}{s-m_V^2+im_V\Gamma_V}\right|^2
    \,. \label{eq:Rcont}
\end{equation}
At the peak, the squared modulus in Eq.~\eqref{eq:Rcont} is equal to $|1-3i\, \BR_{\ell\Bar{\ell}}/\alphaem|^2$, where the leptonic branching fractions $\BR_{\ell\bar\ell}$ for narrow vector mesons  are given in Table~\ref{tab:Bee}. One can see that the vector meson propagator dominates for $J/\psi$ and $\psi'$, while for $\omega$ and $\phi$ it is at the percent level and will only be observed in the interference pattern. Figure~\ref{fig:HVPpsi}(a) shows the observed $e^+e^-\to\mu^+\mu^-$ cross section in the vicinity of the $J/\psi$ resonance. The effect is still large for  $e^+e^-\to\tau^+\tau^-$ at the $\psi'$ resonance 
close to $\tau^+\tau^-$ threshold as shown in Fig.~\ref{fig:HVPpsi}(b).
\begin{table}[t]
    \caption[Leptonic decay widths of narrow  vector meson resonances]{Leptonic decay widths of narrow  vector meson resonances (for $\ell=e,\mu$)~\cite{PDG}.}
\renewcommand{\arraystretch}{1.3}
\begin{center}
    \begin{tabular}{cc|ccc}
    \toprule
 $V$   &$\BR_{\ell\bar\ell}$&$V$&$\BR_{\ell\bar\ell}$&$\BR_{\tau\bar\tau}$\\ \midrule
        $\omega$ & $7.4\times10^{-5}$ &$J/\psi$&$6\times10^{-2}$&--\\
        $\phi$ &$3\times10^{-4}$ &$\psi'$&$8\times10^{-3}$&$3\times10^{-3}$\\
        \bottomrule
    \end{tabular}
    \label{tab:Bee}
    \end{center}
\renewcommand{\arraystretch}{1.0}
\end{table}

The modulus in Eq.~\eqref{eq:Rcont} can be treated as the effective e.m.\ running coupling constant $\alpha(s)/\alpha(0)$. 
The most precise method to determine the hadronic part of the $s$-dependence is to use a dispersion integral from a compilation of $e^+e^-\to \text{hadrons}$~\cite{Jegerlehner:2006ju,Jegerlehner:2011mw}.
However,  one can also see the  $\alpha(s)$ variation in an $e^+e^- \to \mu^+\mu^-$ experiment for  $s<1\GeV$, where it is just a small correction. Such a study was performed by KLOE-2~\cite{KLOE-2:2016mgi} using the ISR method. The differential cross section for the process $e^+e^- \rightarrow \mu^+\mu^- \gamma(\gamma)$ was studied in the region $0.600\GeV<W< 0.980\GeV$, where $W$ is the invariant mass of the $\mu^+\mu^-$ pair.  The value of $|\alpha(s)/\alpha(0)|^2$ is extracted from the ratio of the experimental differential cross section to the corresponding cross section obtained from a Monte Carlo simulation, with $\alpha(s)$ set to the constant value of $\alpha(0)$.

\begin{figure}[t]
\includegraphics[width=0.98\textwidth]{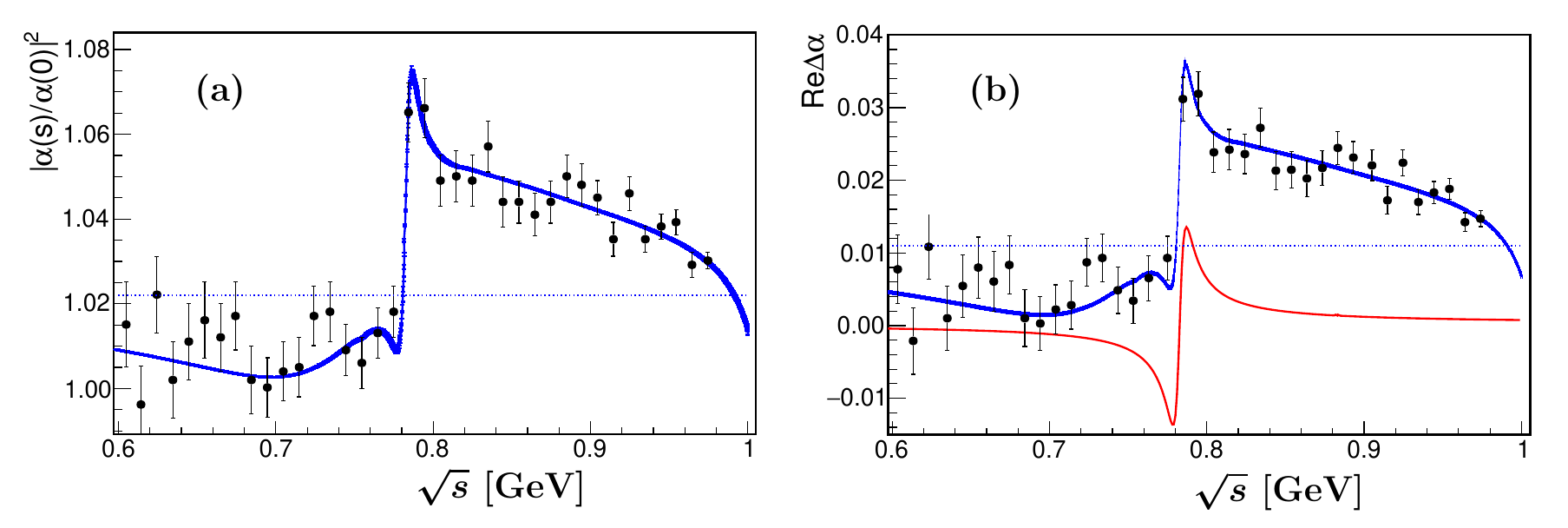}
\caption[Hadronic vacuum polarization in the $\rho$ and $\omega$ mesons mass region]{Hadronic vacuum polarization close to the masses of the $\rho$ and $\omega$ mesons: (a) $|\alpha(s)/\alpha(0)|^2$ and (b) ${\rm
    Re}\,\Delta\alpha$ as determined directly from the KLOE  $e^+e^-\to\mu^+\mu^-$ data
 (points with statistical uncertainties only), compared to the result of a dispersive calculation (blue band) and the leptonic part (dotted line). In addition, the extracted contribution from the $\omega$ meson alone is shown (red solid line).}
\label{fig:HVPrho}
\end{figure}

A photon and two tracks of opposite curvature are required to identify
a $\mu\mu\gamma$ event.  Events are selected with an undetected
photon emitted at small angle, i.e., within a cone of
$\theta_\gamma < 15^\circ$ around the beamline, and the two charged
muons are emitted at large polar angle,
$50^\circ<\theta_\mu<130^\circ$.  The ISR
$\mu^+\mu^-\gamma$ cross section is obtained from the observed number
of events ($N_{\rm obs}$) and the background estimate ($N_{\rm bkg}$) as
\begin{equation}
\frac{\diff\sigma(e^+e^-\rightarrow \mu^+\mu^-\gamma(\gamma))}{\diff{W}}= \frac{N_{\rm obs}-N_{\rm bkg}}{\Delta W}\cdot\frac{1-\delta_{\rm FSR} }{\epsilon({W})\cdot \textit{L}} \,,
\label{mmg}
\end{equation}
where $\Delta W=10\MeV$ is the $W$ bin width, 
$\epsilon$ is the efficiency, ${\cal L}$ is the integrated luminosity, and
$1-\delta_{\rm FSR}$ is the correction applied to remove the final-state
radiation contribution. The extracted ratio $|\alpha(s)/\alpha(0)|^2$ is
shown in Fig.~\ref{fig:HVPrho}(a)
and compared both to the leptonic part from perturbative 
calculations and to the
full combined hadronic and leptonic contributions, where the hadronic part is obtained from a dispersion integral 
over the $e^+e^-\to \text{hadrons}$ data~\cite{Jegerlehner:2011mw}.
The imaginary part of $\Delta\alpha\equiv1-\alpha(0)/\alpha(s)$, which can often be neglected~\cite{Actis:2010gg}, 
is related to the 
total cross section $\sigma(e^+e^-\to\gamma^*\to \text{anything})$:
$\Im\Delta \alpha = - {\alpha}/{3}\,R(s)$,
with $R(s) = \sigma_{\rm tot}/[4\pi\vert\alpha(s)\vert^2/(3s)]$.
$R(s)$ includes leptonic and hadronic contributions:
$R(s)=R_{\rm lep}(s)+R_{\rm had}(s)$.
In the energy region around the $\rho$~meson, the hadronic cross section 
can be approximated by the dominant $2\pi$ contribution:
\begin{equation}
R_{\rm had}(s)= \frac{1}{4} 
\left(1-\frac{4m_\pi^2}{s} \right)^\frac{3}{2} 
\vert \FV(s) \vert ^2\left\lvert \frac{\alpha(0)}{\alpha(s)}\right\rvert^2.
\end{equation}
This information allows one to extract the real part for $\Delta \alpha$ as shown in Fig.~\ref{fig:HVPrho}(b).

The data on $\Re\Delta \alpha$ can be used to determine the
$\omega\to\mu^+\mu^-$ branching ratio. For this purpose we 
 parameterize the hadronic
contributions as a sum of $\rho(770)$, $\omega(782)$, and
$\phi(1020)$, and a nonresonant term~\cite{Actis:2010gg}.
  We use a Breit--Wigner
description for the $\omega$ and $\phi$ resonances and
a Gounaris--Sakurai parameterization~\cite{Gounaris:1968mw} of the pion
form factor for the $\rho$, where the interference with the $\omega$ and
the higher $\rho$ states can be neglected due to the
statistical precision of the data. The dispersion integral 
for $\Delta\alpha_{\rm had}$ reads 
\begin{equation}
 \Delta\alpha_{\rm had}=-\frac{\alpha(0) s}{3\pi}\,{\rm P.V.}\! \!\int_{ 4m_\pi^2}^\infty ds'\, \frac{R_{\rm had}(s')}{s'(s'-s-i\epsilon)},
\end{equation}
where P.V.\ indicates the Cauchy principal value.
The fit of $\Re\Delta\alpha$ with fixed leptonic
and $\phi$ parts shows a clear contribution of the $\rho$ and $\omega$
resonances to the photon propagator, which results in a  
more than $5\sigma$ significance of the hadronic contribution to $\alpha(s)$. This is the strongest direct evidence achieved 
by a single experiment.
The $\omega$ contribution to $\Re\Delta\alpha$ 
obtained from the  fit  could be translated, 
assuming lepton universality, to 
the branching ratio value $\BR(\omega\to\mu^+\mu^-)
=6.6(2.2)\times 10^{-5}$, to be compared to 
the only previous measurement of $9.0(3.1)\times 10^{-5} $
from ALEPH~\cite{Heister:2001kp}.

\section{Example analyses}\label{sec:examples}
\subsection[$e^+e^-\to P\bar{P}$]{\boldmath $e^+e^-\to P\bar{P}$}\label{sec:eePP}
The only reactions in which two ground-state pseudoscalar mesons are produced in electron--positron  annihilation are $e^+e^-\to \pi^+\pi^-$, $K^+K^-$, and $K_LK_S$, but together with $e^+e^-\to \pi^+\pi^-\pi^0$ they dominate the total low-energy cross section. 
The other neutral two-pseudoscalar systems are forbidden by charge conjugation conservation in the electromagnetic interactions. The hadronic current for the process can be written as in Eq.~\eqref{eq:2picurr}:
\begin{equation}
J_{\mu}^{\text{em},P} 
 = \left(q_{1}-q_{2}\right)_\mu F_P^V(s) \,, \label{eq:FFP}
\end{equation}
where $q_1,q_2$ are four-momenta of the produced pseudoscalar mesons and $F_P^V(s)$ is the elastic form factor of the meson $P$. The form factor is a scalar, dimensionless function with the constraint $F_P^V(0)=Q_P$, where $Q_P$ denotes the meson's charge (in units of $e$). The angular cross section is given as
\begin{equation}
    \frac{\diff\sigma}{\diff\Omega}={\alphaem^2}\frac{{q}^3_P}{s^{5/2}}\sin^2\theta|F_P^V(s)|^2=
    \frac{\alphaem^2}{8s}\sigma_P^3\sin^2\theta|F_P^V(s)|^2 \,,
\end{equation}
where ${q_P=\sqrt{s-4m_P^2}/2}$ is the momentum of a pion/kaon in the c.m.\ system and $\sigma_P=\sqrt{1-4m_P^2/s}$ the velocity. The total Born cross section is
\begin{equation}
    {\sigma}=\frac{8\pi\alphaem^2}{3s^{5/2}}{{q}^3_P}|F_P^V(s)|^2=
    \frac{\pi\alphaem^2}{3s}\sigma^3_P|F_P^V(s)|^2\,.
\end{equation}
For the case of a produced pair of charged kaons the cross section close to threshold 
should be multiplied by the Sommerfeld--Gamov--Sakharov factor~\cite{Sommerfeld:1921aaa,Gamow:1928zz,Sakharov:1991pia}
\begin{equation} \label{eq:SGS}
Z(s) =  \frac{\pi\alpha}{\sigma_K}\frac{ 1+{\alpha^2}/{(4\sigma_K^2)}}{1-\exp(-\pi\alpha/\sigma_K)} \,,
\end{equation}
obtained by solving the {Schr\"odinger} equation in a Coulomb potential for a $P$-wave final state.
The normalized leading asymptotic $s$ dependence of the pion and kaon form factors was calculated using  perturbative  QCD methods~\cite{Farrar:1979aw,Lepage:1979zb,Chernyak:1983ej}:
\begin{equation}
    F_P^V(s)_{s\to\pm\infty}=-\frac{64\pi^2{F_P^2}}{{(11-2 n_f/3)}s\ln|s|} \,,
\end{equation}
where $n_f$ is number of flavors and $F_P$ is the pion (kaon) decay constant.

\subsubsection[$e^+e^-\to \pi^+\pi^-$]{\boldmath $e^+e^-\to \pi^+\pi^-$}
The $e^+e^-\to \pi^+\pi^-$ process is the subject of many recent precision studies since it determines the by far most important individual hadronic contribution, both to the value and uncertainty, to the muon anomalous magnetic moment. A precision below 1\% is required to match the expected uncertainty of the new muon $g-2$ experiments~\cite{Aoyama:2020ynm}, which has motivated several measurements of the $e^+e^-\to\pi^+\pi^-$ cross section: 
 CMD2~\cite{Akhmetshin:2003zn,Akhmetshin:2008gz} and SND~\cite{Achasov:2005rg} using the scan technique, and KLOE~\cite{Aloisio:2004bu,Ambrosino:2008aa,Ambrosino:2010bv,Babusci:2012rp}, BaBar~\cite{Aubert:2009ad,Lees:2012cj}, and BESIII~\cite{Ablikim:2015orh} using the ISR method. In addition to these electron--positron results, the  time-like form factor was measured very close to threshold, in the range $ 0.101\GeV^2<s<0.178\GeV^2$, in the fixed-target  NA7 experiment with a 100--$175 \GeV$ positron beam~\cite{Amendolia:1983di}. The pion form factor has been measured also in the space-like region $0.014\GeV^2 < -s < 0.26\GeV^2$ by scattering $300 \GeV$ pions from the electrons of a liquid-hydrogen target.
The long-standing issue with a high-precision evaluation of the HVP contribution of the $\pi^+\pi^-$ channel to $(g-2)_\mu$ has been the fact that two precise individual experimental measurements, KLOE and BaBar, do not agree with each other at a percent level in the most important energy region, that of the $\rho(770)$ resonance, but new experiments are expected to settle this issue soon.   We refrain from discussing this data inconsistency here and instead refer to the recent review Ref.~\cite{Aoyama:2020ynm}, where also the different approaches to account for it in the assessment of the resulting uncertainty in $(g-2)_\mu$ are presented in detail~\cite{Davier:2017zfy,Keshavarzi:2018mgv,Colangelo:2018mtw,Davier:2019can,Keshavarzi:2019abf}.

At low $s$, the process is dominated by the contribution of the $\rho^0$ resonance. In Eq.~\eqref{eq:piVradius1} we have given the LO ChPT expression for $\FV(s)$. In general, the hadronic current can be rewritten by inserting a complete set of intermediate states $X$: 
\begin{equation}
    J_\mu^{\text{em},2\pi}=\sum_X\bra{\pi^+\pi^-}H \ket{X}\bra{X}j_\mu^{\rm em}(0)\ket{0}^* \,, \label{eq:J2pi-X}
\end{equation} 
where the symbolic summation includes the integration over all kinematic variables.
Equation~\eqref{eq:J2pi-X} develops an imaginary part as soon as an intermediate state $X$ can go on its mass shell, which, due to the optical theorem, happens for $s>4m_\pi^2$.
For energies $4m_\pi^2<s<16m_\pi^2$ (below four-pion production threshold), the intermediate $X$ state can be only $\pi^+\pi^-$, and 
we find the elastic unitarity relation for the imaginary part of the pion vector form factor as given in Eq.~\eqref{eq:ImFV}.
We have discussed the ChPT one-loop representation in Sec.~\ref{sec:ChPT}, however, we saw that this necessarily breaks down below the mass of the $\rho$ meson.
Here, we will describe approaches to describe $\FV(s)$ in the range  $\sqrt{s}< 1 \GeV$ with percent precision as required by experiment.

Given the phase shifts $\delta_1^1(s)$, the form factor can be determined as
\begin{equation}
    \FV(s)=R(s)\Omega(s) \,\label{eq:ModInd}
\end{equation}
up to energies where contribution of inelastic channels in $\pi\pi$ $P$-wave scattering become significant.  
Here the ambiguity of the solution allows $R(s)$ to be an arbitrary polynomial in $s$. The Omn\`es function $\Omega(s)$ is given by~\cite{Omnes:1958hv}
\begin{equation}
    \Omega(s)=\exp\left\{\frac{s}{\pi}\int_{4m_\pi^2}^\infty \diff x\frac{\delta_1^1(x)}{x(x-s-i\epsilon)}\right\} \,.
\end{equation}
In modern applications, the function $R(s)$ is generalized to take into account intermediate states beyond two pions; most importantly three pions, which resonate at the mass of the narrow $\omega(782)$ and hence lead to an isospin-violating signal massively enhanced by a small energy denominator, see the discussion of $\rho$--$\omega$ mixing below.  Furthermore, higher inelasticities are typically included in the form of a conformal polynomial.  Such representations have been used to extract the pion--pion $P$-wave phase shift from form factor data with unprecedented precision~\cite{DeTroconiz:2001rip,Leutwyler:2002hm,Colangelo:2003yw,deTroconiz:2004yzs,Colangelo:2018mtw,Colangelo:2020lcg}.

In a VMD picture, one can represent the $\rho^0$ contribution to $\FV(s)$ as
\begin{equation}
    \FV(s)=\frac{m_\rho^2}{m_\rho^2-s-{\Re}{\cal A}(s)-i{\Im}{\cal A}(s)} \,,
\end{equation}
where 
${\cal A}(s)$ represents the $2\pi$ state insertion in the $\rho$-meson propagator~\cite{Bruch:2004py}.  The imaginary part of ${\cal A}(s)$ is normalized at $s=m_\rho^2$ to the $\rho$ decay width $\Gamma_\rho$: ${\Im}{\cal A}(m_\rho^2)=m_\rho\Gamma_\rho$.
The simplest assumption for the ${\cal A}$ amplitude is to take ${\Im}{\cal A}(s)=m_\rho\Gamma_\rho=\text{const.}$ and to neglect the real part setting it to zero.  This corresponds to the standard relativistic Breit--Wigner (BW) formula:
\begin{equation}
  \FV(s)= \frac{m_\rho^2}{m_\rho^2-s-im_\rho\Gamma_\rho}\equiv {\BW}_\rho(s) \,.
  \label{eq:BW}
\end{equation}
This approximation clearly fails for low $s$, as the imaginary part survives below $\pi\pi$ threshold (and even in the space-like region).
A more refined assumption is therefore to consider a variable $\rho$  decay width proportional to the volume of the $P$-wave phase space for the $\rho\to\pi\pi$ decay mode: 
\begin{equation}
  {\Im}{\cal A}(s)= \sqrt s\Gamma_\rho \frac{m_\rho^2}{s}\frac{q^3_\pi(s)}{q^3_\pi(m_\rho^2)}\equiv\sqrt{s}\Gamma_\rho(s) \,. 
\end{equation}
This is equivalent to the assumption that the $\rho\pi\pi$ coupling is constant. There are two commonly used variants of parameterizations using this ansatz for the decay width. One is to neglect the real part as before, leading to the K\"uhn--Santamaria (KS) formula~\cite{Kuhn:1990ad} 
\begin{equation}
    {\BW}_\rho^{KS}(s)=\frac{m_\rho^2}{m_\rho^2-s-i\sqrt{s}\Gamma_\rho(s)}   \,.
\end{equation}
In the Gounaris--Sakurai parameterization (GS)~\cite{Gounaris:1968mw,Kuhn:1990ad}, on the other hand, one 
determines the real part of ${\cal A}(s)$ from a dispersion relation with two subtractions at $s=0$:
\begin{equation}
 {\cal A}(s) =  {\cal A}(0)+s{\cal A}'(0)+\frac{s^2}{\pi}\int_{4m_\pi^2}^{\infty}\frac{\diff x}{x^2}
 \frac{{\Im}{\cal A}(x)}{x-s-i\epsilon} \,.
\end{equation}
This is equivalent to the assumption of
a generalized effective-range formula for $\ell=1$ $\pi\pi$ scattering: 
\begin{align}
   \bigg(\frac{q_\pi^3}{\sqrt{s}}\bigg)\cot\delta_1^1(s)&=q_\pi^2h(s)+a^{1}_1+b^{1}_1q_\pi^2 \,, \notag\\
    h(s)&=\frac{2}{\pi}\frac{q_\pi}{\sqrt{s}}\ln\left(\frac{\sqrt{s}+2q_\pi}{2m_\pi}\right) \quad \xrightarrow[s<4m_\pi^2]{} \quad
   -\frac{1}{\pi}\sqrt{\frac{4m_\pi^2}{{s}}-1}\arccot{\left(\frac{s}{4m_\pi^2-s}\right)^{1/2}}\,,
\end{align}
where $a^{1}_1$ and $b^{1}_1$ are scattering length and effective range parameters\footnote{It is customary in the literature on pion--pion scattering to also refer to the leading terms in the near-threshold expansion of partial waves with nonvanishing angular momentum as ``scattering lengths'' and ``effective ranges,'' even though their dimensions would suggest otherwise.} determined assuming $\rho$ meson dominance:
\begin{equation}
    \left.\cot\delta_1^1\right|_{s=m_\rho^2}=0 \,, \qquad \left.\frac{\diff\cot\delta_1^1}{\diff s}\right|_{s=m_\rho^2}=
    \frac{1}{m_\rho\Gamma_\rho} \,.
\end{equation}
 For  $s< 4m_\pi^2$, $q_\pi$ is defined as
$q_\pi(s)=i\sqrt{4m_\pi^2-s}/2$. 
This ansatz leads to the following form of the form factor:
\begin{equation}
    {\BW}_\rho^{GS}(s)=\frac{m_\rho^2+H(0)}{m_\rho^2-s+H(s)-i\sqrt{s}\Gamma_\rho(s)} \,,
\end{equation}
where
\begin{equation}
    H(s)=\frac{\Gamma_\rho m_\rho^2}{q_\pi^3(m_\rho^2)}\left[q_\pi^2(s)\left(h(s)-h(m_\rho^2)\right)
    +q_\pi^3(m_\rho^2)h'(m_\rho^2)(m_\rho^2-s)\right] \,.
\end{equation}

\begin{figure}[t]
    \centering
    \includegraphics[width=0.49\textwidth]{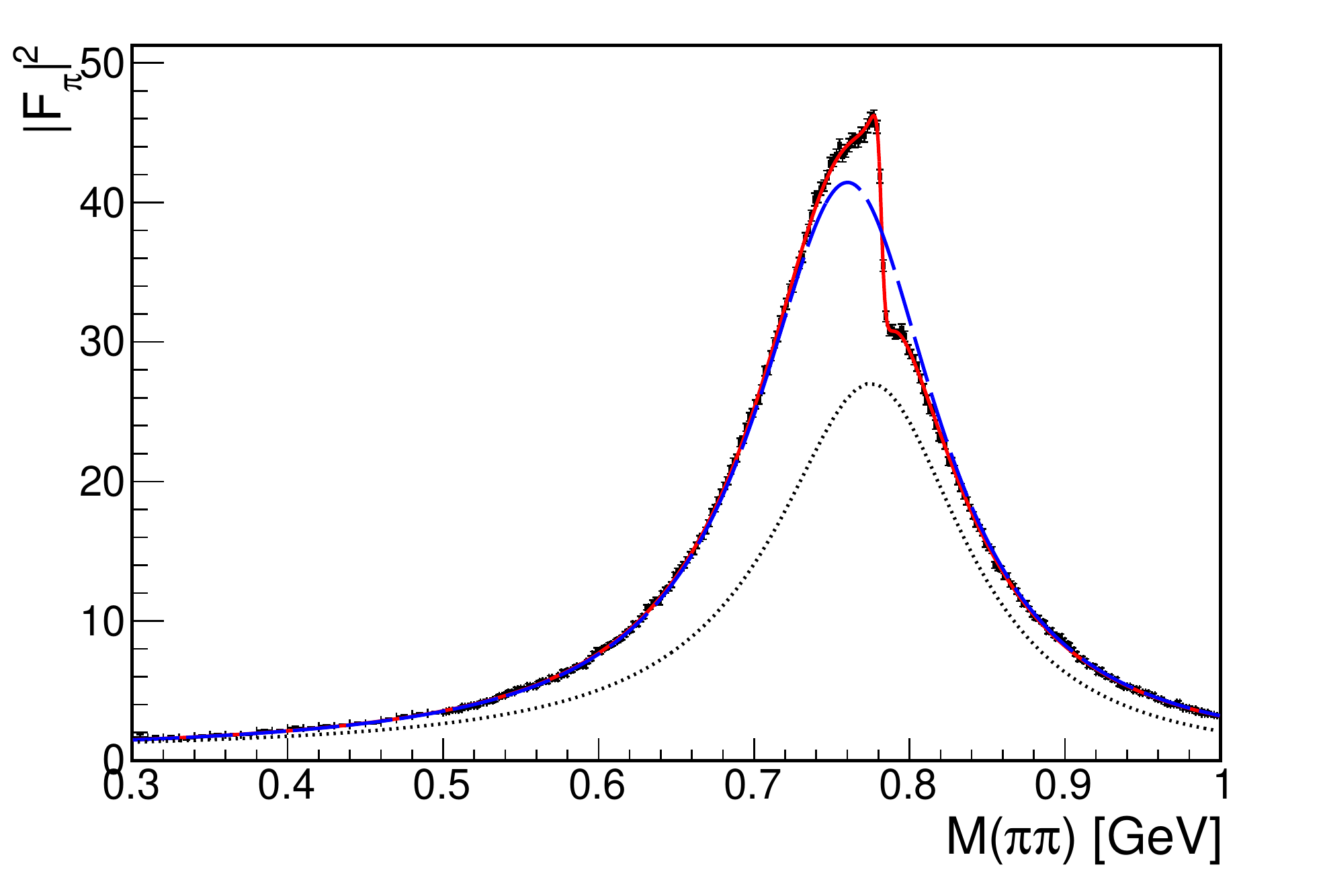}\put(-220,150){\bf (a)}
    \includegraphics[width=0.49\textwidth]{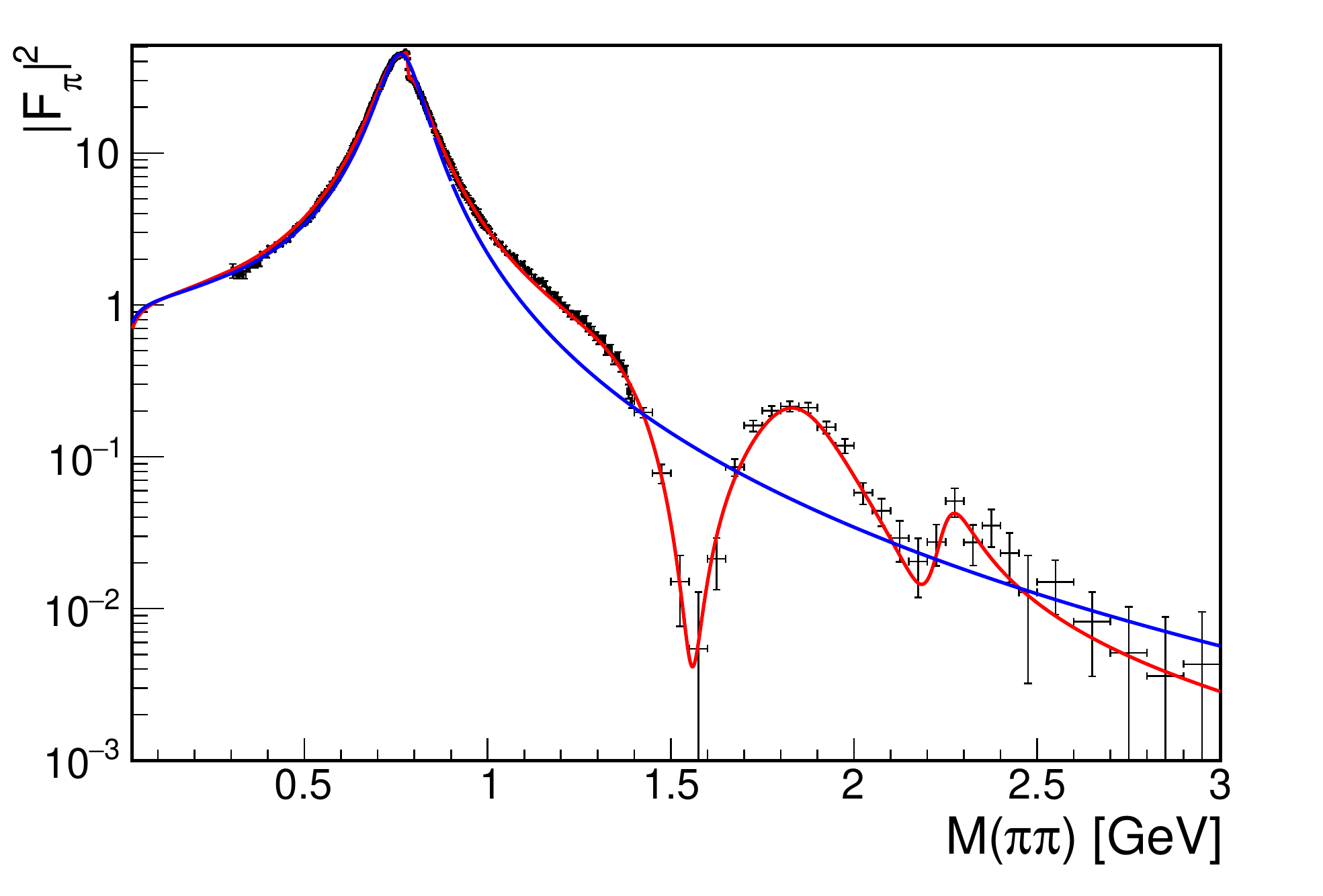}\put(-220,150){\bf (b)}
   \caption[Pion form factor $\FV$:  data and parameterization]{Pion form factor $\FV$ illustrated using BaBar data~\cite{Aubert:2009ad,Lees:2012cj}. (a) Region $\sqrt{s}<1\GeV$ compared to the fixed-width Breit--Wigner Eq.~\eqref{eq:BW} (dotted line), Omn\`es function (blue dashed line), and  
   Omn\`es function multiplied by $R(s)$ including $\rho$--$\omega$ mixing as in Eq.~\eqref{eq:ModIndMix} (red solid line). (b) Range up to $3\GeV$ compared to Gounaris--Sakurai parameterization with only $\rho$ and $\omega$ (blue solid line) and including $\rho'$, $\rho''$ contributions Eq.~\eqref{eq:GS-BaBar} (red solid line). }
    \label{fig:piFV}
\end{figure}
In Fig.~\ref{fig:piFV}(a) the parameterizations are compared to the BaBar data~\cite{Aubert:2009ad,Lees:2012cj} in the range $\sqrt{s}<1\GeV$.
One immediately sees that isospin breaking due to the narrow $\omega$ resonance has to be included to describe the data.  For the description of the data for $s<1\GeV^2$ it is important to include $\rho^0-\omega$ mixing, which in the GS parameterization can be achieved by 
\begin{equation}
    \FV(s)={\BW}^{GS}_{\rho+\omega}(s)\equiv {\BW}^{GS}_\rho(s)\frac{1+c_\omega {\BW}_\omega(s)}{1+c_\omega} \,,
\end{equation}
where for the $\omega$, given its small width, a simple Breit--Wigner function Eq.~\eqref{eq:BW} is used, and $c_\omega$ has to be real to assure that the form factor is real below the two-pion threshold.
In the Omn\`es parameterization, one replaces the polynomial $R(s)$ with a function explicitly including the $\omega$ pole~\cite{Hanhart:2016pcd}
\begin{equation}
    R(s)=1+\alpha_Vs+\lambda s^2+\kappa\frac{ s}{m_\omega^2}\BW_\omega(s)  \,,\label{eq:ModIndMix}
\end{equation}
where the constants $\alpha_V$, $\lambda$, and $\kappa$ are real numbers.

For  energies above $1\GeV$, one has to include the effects of the inelastic resonances $\rho'=\rho(1450)$ and $\rho''=\rho(1700)$~\cite{Barkov:1985ac,Kuhn:1990ad}. In the GS parameterization this reads as
\begin{equation}
    \FV(s)=\frac{{\BW}^{GS}_{\rho+\omega}(s)+ c_{\rho'} {\BW}^{GS}_{\rho'}(s)+ c_{\rho''}  {\BW}^{GS}_{\rho''}(s)}
    {1+c_{\rho'} +c_{\rho''}} \,,\label{eq:GS-BaBar}
\end{equation}
where $c_{\rho'}$ and $c_{\rho''}$ should be real numbers in order to assure the form factor has no imaginary part for $s<4m_\pi^2$. This parameterization describes the $\FV(s)$ data in the range $2m_\pi<\sqrt{s}<3\GeV$ well as illustrated in Fig.~\ref{fig:piFV}(b) for the data collected by BaBar.  
For a unitary and analytic formalism that not only matches the Omn\`es formalism at low energies, but also includes the effects of $\rho'$ and $\rho''$ above to describe the form factor data up to $2\GeV$ equally well, see Ref.~\cite{Hanhart:2012wi}.

The spectral function $v_1(s)$ for the  $\tau^-\to\pi^-\pi^0\nu_\tau$ decay is directly related to 
$\FV(s)$.  It differs from the one measured in $e^+e^-$ collisions by isospin-breaking effects, most prominently the absence of the $\omega$ resonance.  The isospin corrections have been studied theoretically in quite some detail~\cite{Cirigliano:2001er,Cirigliano:2002pv} and subsequently applied with the goal to use $\tau$-decay data to broaden the data basis for the two-pion HVP evaluation~\cite{Alemany:1997tn}.  This has lead to the puzzling observation that both do not seem to agree with each other very well, and that the significance of the $(g-2)_\mu$ discrepancy depends crucially on whether $\tau$ data are employed for the extraction of the pion form factor or not; see, e.g., Ref.~\cite{Davier:2009ag}.  While a phenomenological analysis based on a model for $\rho$--$\gamma$ mixing has been able to reconcile $e^+e^-$ and $\tau$-decay data to a large extent~\cite{Jegerlehner:2011ti}, the current consensus seems to be that isospin breaking in the pion vector form factor is not yet sufficiently well understood to include $\tau$ data in the evaluation of $(g-2)_\mu$ reliably~\cite{Aoyama:2020ynm}.

\begin{figure}[t]
    \centering
    \includegraphics[width=0.4\textwidth]{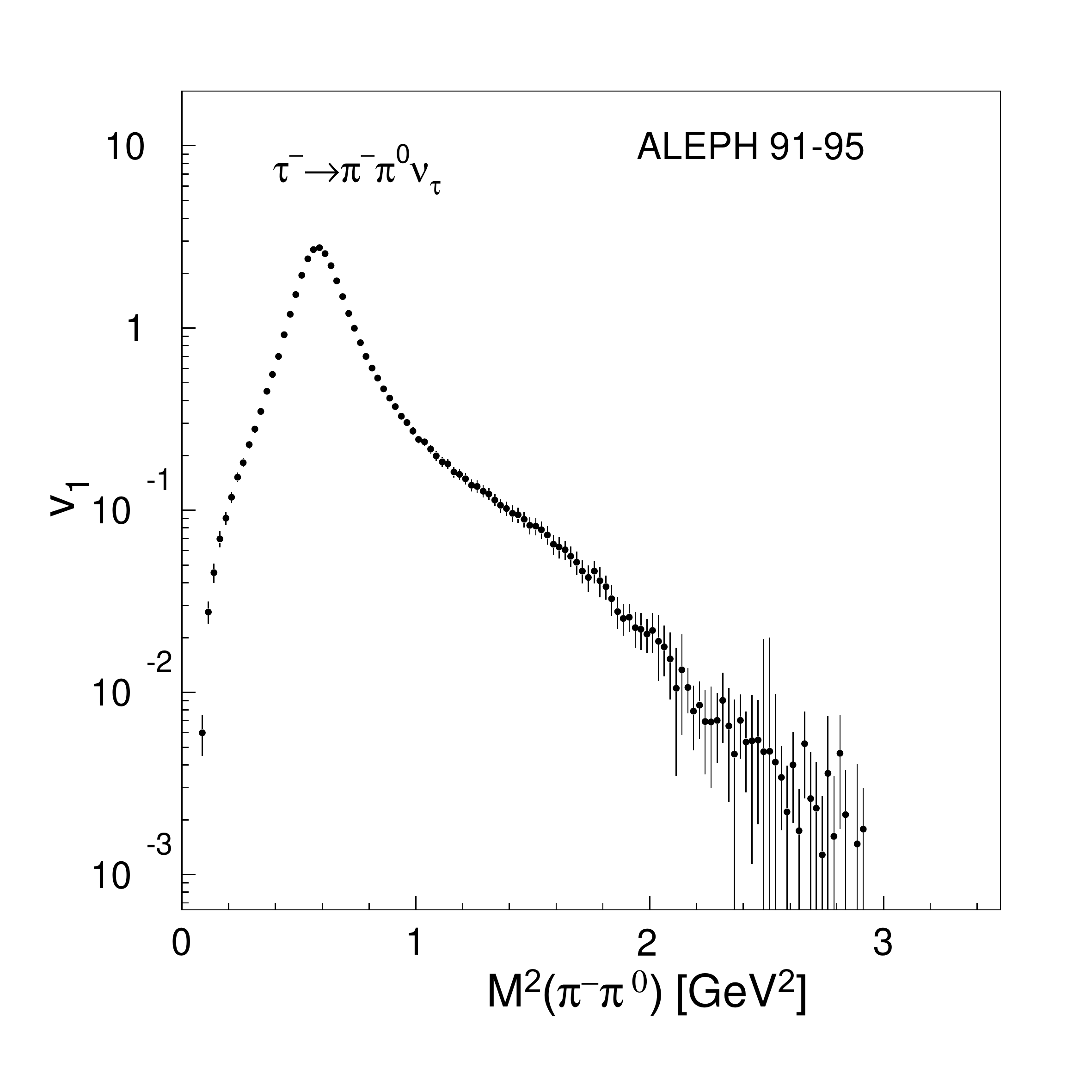}\put(-175,160){\Large\bf  (a)} \hfill
\includegraphics[width=0.56\textwidth]{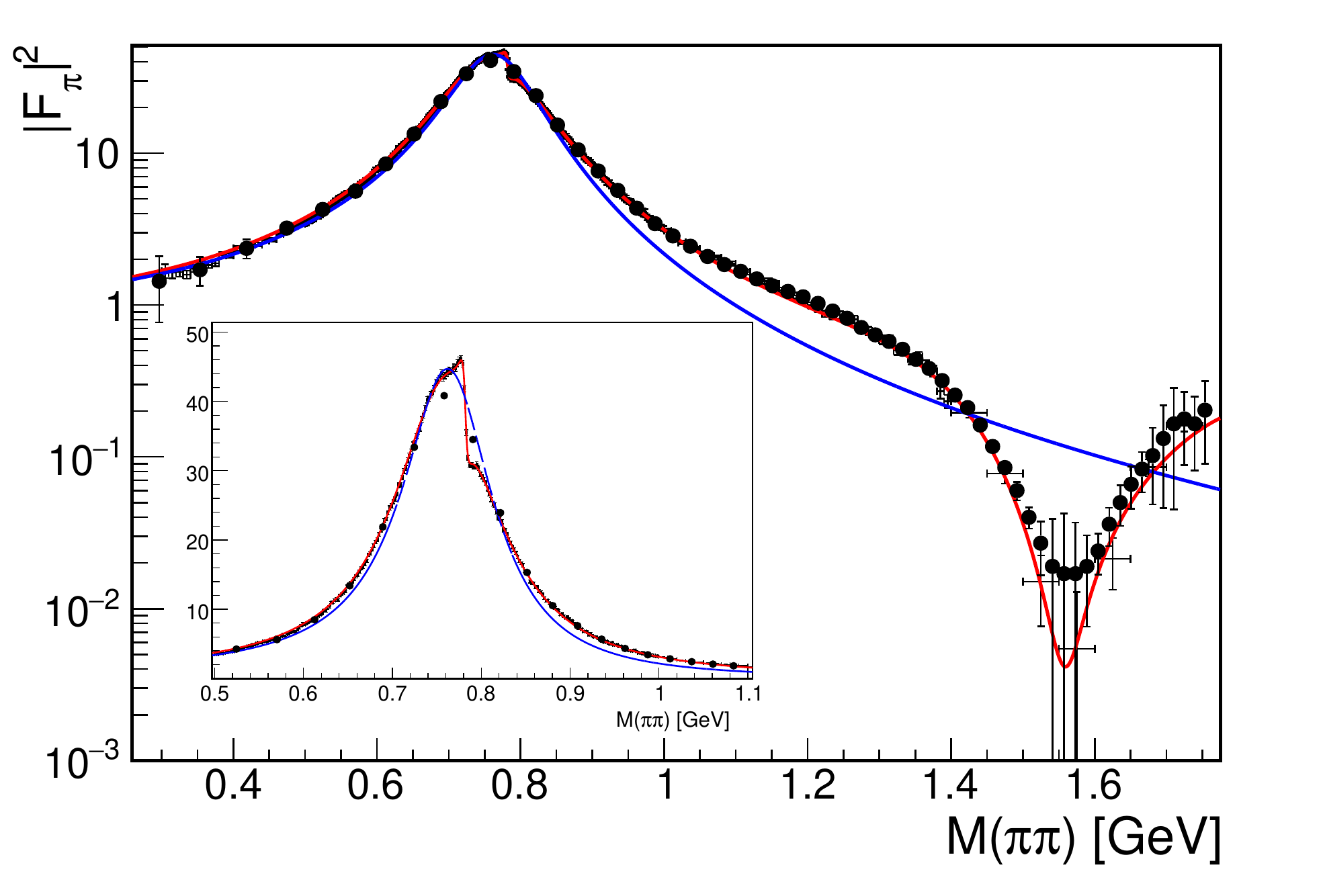}\put(-260,160){\Large\bf  (b)}
    \caption[Spectral functions for pion pairs]{(a) Spectral function $v_1$ for $\tau^-\to\pi^-\pi^0\nu_\tau$ from ALEPH~\cite{Schael:2005am}. (b) Comparison of pion vector form factor extracted from $e^+e^-$ [BaBar data, see caption of Fig.~\ref{fig:piFV}(b)] and from $\tau$ decays data from Belle~\cite{Fujikawa:2008ma} (black circles). The insert shows the comparison close to the $\rho$ peak.
    \label{fig:tauV1}}
\end{figure}

The spectral functions are obtained from the differential decay rate 
$\diff\Gamma/\diff s$, which up to electroweak radiative 
corrections can be written
as~\cite{Tsai:1971vv,Anderson:1999ui} 
\begin{equation}
   \frac{\diff\Gamma(\tau^-\to\pi^-\pi^0\nu_\tau)}{\diff s} = 
   \frac{G_F^2\, |V_{ud}|^2\,}
        {32\pi^2\, m_\tau^3}\,
   (m_\tau^2-s)^2 \,
   (m_\tau^2 + 2 s ) \,
         v_1(s) \,, 
   \label{eq:dVtau} 
\end{equation}
where $s$ is the invariant mass squared of the $\pi^-\pi^0$ system,
and $v_1(s)$ is the vector spectral function for the $\pi^-\pi^0$ system. 
The conserved vector current relates  $v_1(s)$ to the isovector part of the pion vector form factor $\FV(s)$:
\begin{equation}
     v_1(s)=\frac{1}{4\pi}\frac{\sigma_\pi^3}{3}|\FV{}^{(I=1)}(s)|^2  \,. \label{eq:tauV1}
\end{equation}
In Fig.~\ref{fig:tauV1}(a) 
the spectral function $v_1$ from ALEPH~\cite{Schael:2005am} is shown. In the panel (b) of this figure the extracted pion form factor from  Belle~\cite{Fujikawa:2008ma} is shown and compared to $\FV$ from the BaBar experiment displayed in Fig.~\ref{fig:piFV}(b).

\subsubsection[$e^+e^-\to K^+K^-$ and $e^+e^-\to K_LK_S$]{\boldmath $e^+e^-\to K^+K^-$ and $e^+e^-\to K_LK_S$}
The thresholds for the $e^+e^-\to K^+K^-$ and $e^+e^-\to K_LK_S$ reactions are just $32\MeV$ and $24\MeV$ below the $\phi$ mass, respectively.  
The dynamics of the process is described by the elastic form factors defined in Eq.~\eqref{eq:FFP}. 
In the static limit, $s=0$, they represent the electric charges:  $F_{K^+}^V(s=0)=1$ and $F_{K^0}^V(s=0)=0$. The slope of the charged kaon form factor in the static limit, related to the electromagnetic radius $\braket{r^2}_{K^+}^V$ defined in  Eq.~\eqref{eq:piVradius1}, was determined using the space-like kaon form factor from kaon--electron scattering experiments~\cite{Dally:1980dj,Amendolia:1986ui}.
Since the $K\bar K$ system is not an eigenstate of isospin,
both form factors are the sum of isoscalar and isovector components, where~\cite{Bruch:2004py}
\begin{equation} 
F_{K^+}^{V(I=0)}=F_{K^0}^{V(I=0)} \quad \text{and} \quad F_{K^+}^{V(I=1)}=-F_{K^0}^{V(I=1)}\label{eq:KKiso} \,.
\end{equation}
In the $SU(3)$ flavor symmetry limit, $F_{K^+}^V(s)=\FV(s)$ and $F_{K^0}^V(s)=0$. However, the near-threshold cross sections  of $K_SK_L$ and $K^+K^-$ in the $\phi$ meson region are almost the same, implying the isovector contribution to be very small there.

Precision scans of the cross sections around the $\phi$ peak were performed in Novosibirsk. The latest results 
are from CMD-3 for both $K_{S}K_{L}$~\cite{Kozyrev:2016raz} and $K^+K^-$~\cite{Kozyrev:2017agm} in the energy range 
$1.0$--$1.06\GeV$.
\begin{figure} 
\centering
\includegraphics[width=0.98\textwidth]{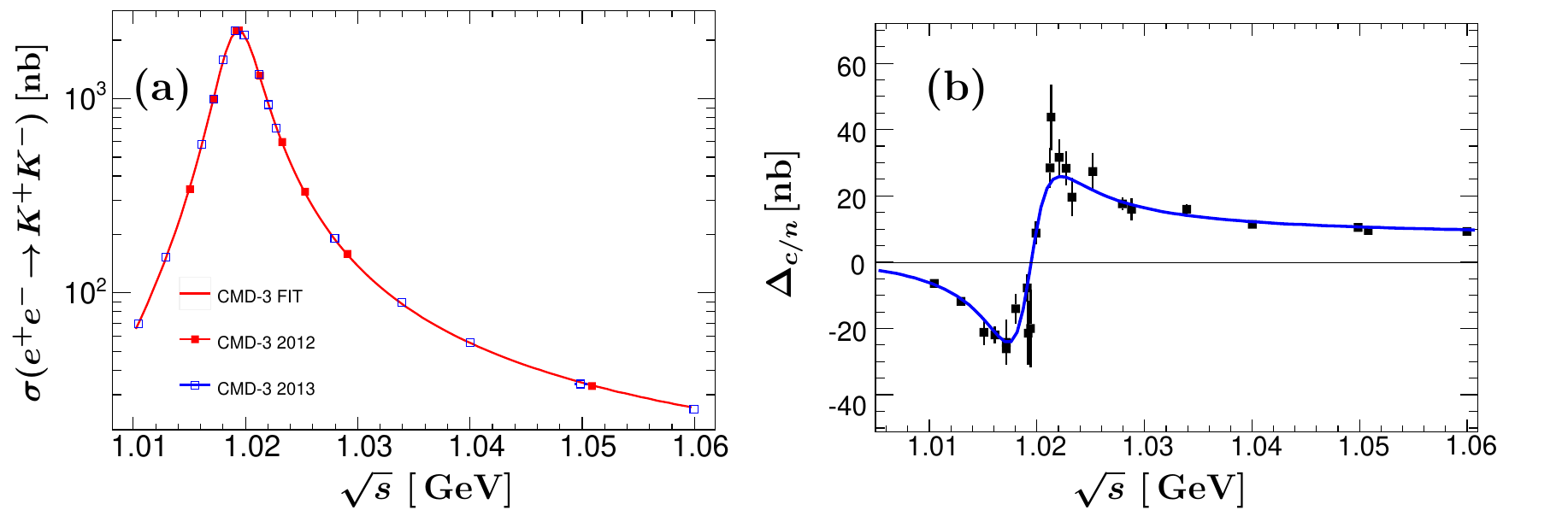}
  \caption[$e^+e^-\to K^+ K^-$ cross section close to the $\phi$ meson ]{ \label{fig:KKcross_phi} 
  (a) Cross section for $e^+e^-\to K\bar K$ in the vicinity of the $\phi$ meson.
(b) The normalized difference of the charged and neutral cross sections $\Delta_{c/n}$ defined in Eq.~\eqref{eq:KKdiff_phi}.
Only statistical uncertainties are shown. Data and fits (solid lines) are from the CMD-3 experiment~\cite{Kozyrev:2017agm}.
}
\end{figure}
The Born cross section for $e^+e^-\to K^+K^-$ from the CMD-3 experiment 
is shown in Fig.~\ref{fig:KKcross_phi}(a) with the peak value of $2.23(2)\,\mu\text{b}$.  This cross section exceeds the prior CMD-2~\cite{Akhmetshin:2008gz} result by about 10\%, which is explained by a not-accounted-for systematic effect in the CMD-2 analysis. However, there is still a tension~\cite{Aoyama:2020ynm} with the BaBar measurement~\cite{Lees:2013gzt} obtained using the ISR method. 

The issue of isospin corrections for the two channels has received quite some theoretical attention~\cite{Bramon:2000qe,FloresBaez:2008jp,Benayoun:2011mm}. 
Their cross sections are compared by plotting the following corrected difference:
\begin{equation}
\Delta_{c/n}(s) = \frac{q_{K^0}^3}{q_{K^{\pm}}^3}  \frac{1}{Z(s)}\sigma{(e^+e^- \to K^{+} K^{-})}  -   \sigma{(e^+ e^- \to K^0_{S}K^0_{L})} \,, \label{eq:KKdiff_phi}
\end{equation}
where $Z(s)$ is the
Sommerfeld--Gamov--Sakharov factor defined in Eq.~\eqref{eq:SGS}.
Equation~\eqref{eq:KKdiff_phi} is defined such that most of the common uncertainties cancel.
The experimental value of $\Delta_{c/n}$ is shown in Fig.~\ref{fig:KKcross_phi}(b), where both cross section data sets are taken from CMD-3~\cite{Kozyrev:2017agm}.
The isovector contributions (nonresonant in the physical region) are dominated by the $\rho$ and have opposite sign, see Eq.~\eqref{eq:KKiso}. The  difference $\Delta_{c/n}$ probes interference of such contributions with the dominant $\phi$ pole mechanism. The magnitude of $\Delta_{c/n}$ in Eq.~\eqref{eq:KKdiff_phi} is proportional to the ratio ${g_{\rho KK} g_{\phi KK}}/{D_{\phi}(s){D_{\rho}(s)}}$ and allows us to determine the $\rho$ contribution to $K$-meson production experimentally.

\begin{figure}[t]
    \centering
\includegraphics[width=0.98\textwidth]{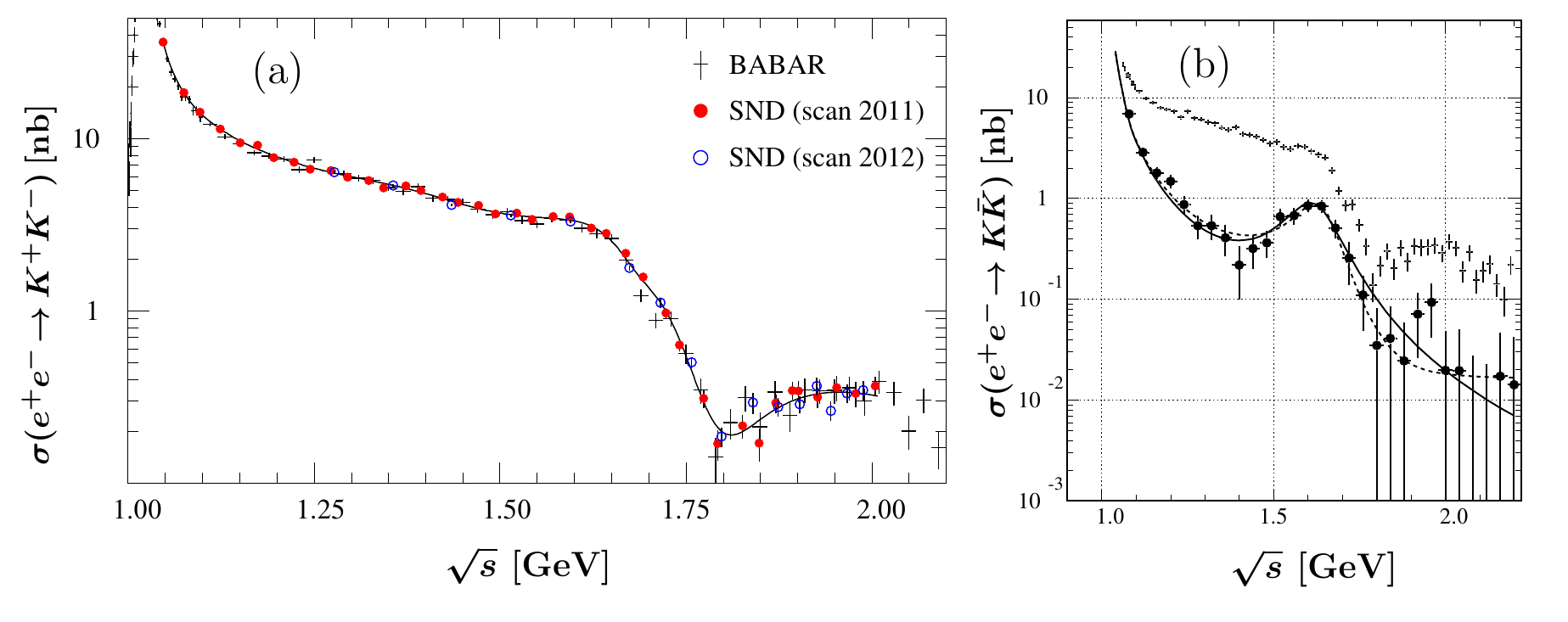}
    \caption[Born cross section for  $e^+e^- \to K\bar K$ below $2\GeV$]{Born cross section for  $e^+e^- \to K\bar K$. (a)  $e^+e^- \to K^+K^-$ data from SND~\cite{Achasov:2016lbc} and BaBar~\cite{Lees:2013gzt}. (b) Comparison of the  $e^+e^-\to K^+K^-$ (crosses)~\cite{Lees:2013gzt} and  $e^+e^-\to K_SK_L$ (black circles and lines representing VMD model parameterizations)~\cite{Lees:2014xsh} cross sections from BaBar. }
    \label{fig:KKff}
\end{figure}
At slightly higher energies, the cross sections start to deviate significantly as shown in Fig.~\ref{fig:KKff}(b) for the $\sqrt{s}$ region below 2\GeV.  In particular, while 
the $\phi(1680)$ contribution is seen as a peak for $F_{K^0}^V$, for the $F_{K^+}^V$ it is seen only as a shoulder. This difference indicates strong interference between isoscalar and isovector amplitudes.

Measurements of the form factors in this energy range were performed by BaBar both for $K^+K^-$~\cite{Lees:2013gzt} and  $K_{S}K_{L}$~\cite{Lees:2014xsh}, using the ISR method with the energy range from threshold up to  $5\GeV$. SND measured $e^+e^-\to K^+K^-$ in 2007~\cite{Achasov:2007kg} in the range $1.04$--$1.38\GeV$, and in 2016~\cite{Achasov:2016lbc} using two separate energy scans for c.m.\ energies $1.05$--$2.00\GeV$. The results are consistent with the BaBar experiment and have a comparable or better accuracy, see Fig.~\ref{fig:KKff}(a). For  $e^+e^-\to K_SK_L$, there are CMD-2~\cite{Akhmetshin:2002vj} and SND~\cite{Achasov:2006bv} measurements of the cross section in the energy range $\sqrt{s}=1.04$--$1.38\GeV$. For energies below $1.2\GeV$, the cross section exceeds VMD model predictions when only $\rho(770)$, $\omega(782)$, and $\phi(1020)$ are taken into account. The measured cross section agrees well with BaBar.

Within the VMD model, a possible ansatz for the kaon form factors to cover the energy region up to $2\GeV$ is given by~\cite{Bruch:2004py}
\begin{align}
    F_{K^{+}}^{V(I=1)}(s)&=\frac{1}{2}\left\{c_\rho\BW_\rho^{KS}(s)+c_{\rho'}\BW_{\rho'}^{KS}(s)+c_{\rho''}\BW_{\rho''}^{KS}(s)\right\} \,,\notag\\
    F_{K^{+}}^{V(I=0)}(s)&=\frac{1}{6}\left\{c_\omega\BW_\omega(s)+c_{\omega'}\BW_{\omega'}(s)+c_{\omega''}\BW_{\omega''}(s)\right\} 
    +\frac{1}{3}\left\{c_\phi\BW_\phi(s)+c_{\phi'}\BW_{\phi'}^{KS}(s)\right\} \,.
\end{align}
See, e.g., Ref.~\cite{Beloborodov:2019fmw} for a recent application of such a model in a combined fit to $e^+e^-\to K^+K^-$ and $e^+e^-\to K_SK_L$ together with the $\tau$ decay $\tau^-\to K^-K_S\nu_\tau$, see below.

\begin{figure}[t]
    \centering
\includegraphics[width=0.6\textwidth]{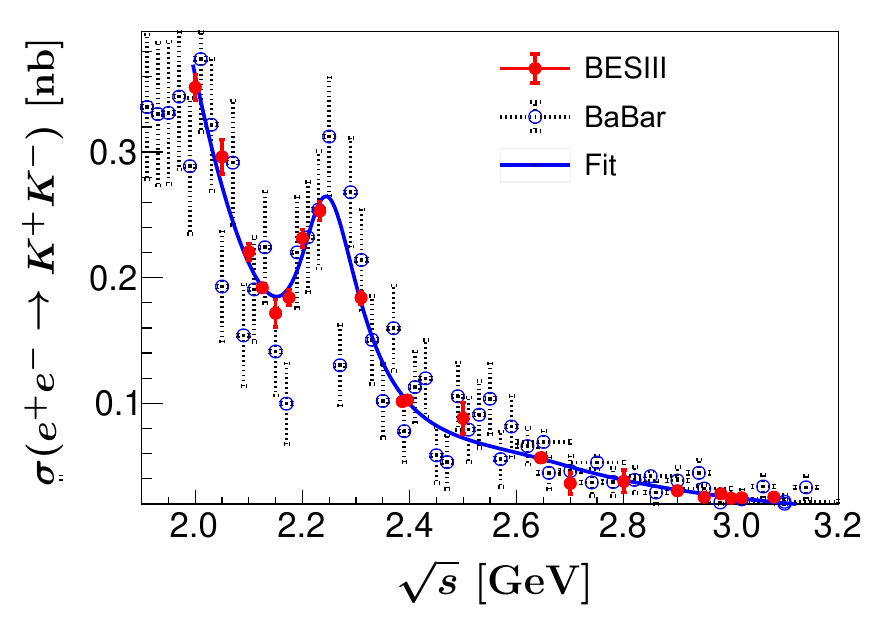}
\caption[Born cross section for  $e^+e^- \to K^+ K^-$ for 2--$3\GeV$]{Born cross section for  $e^+e^- \to K^+ K^-$ in the 2--$3\GeV$ range. Data is from BESIII~\cite{Ablikim:2018iyx} and BaBar~\cite{Lees:2013gzt}.}
    \label{fig:KKff2}
\end{figure}

\begin{figure}[t!]
    \centering
    \includegraphics[width=0.4\textwidth]{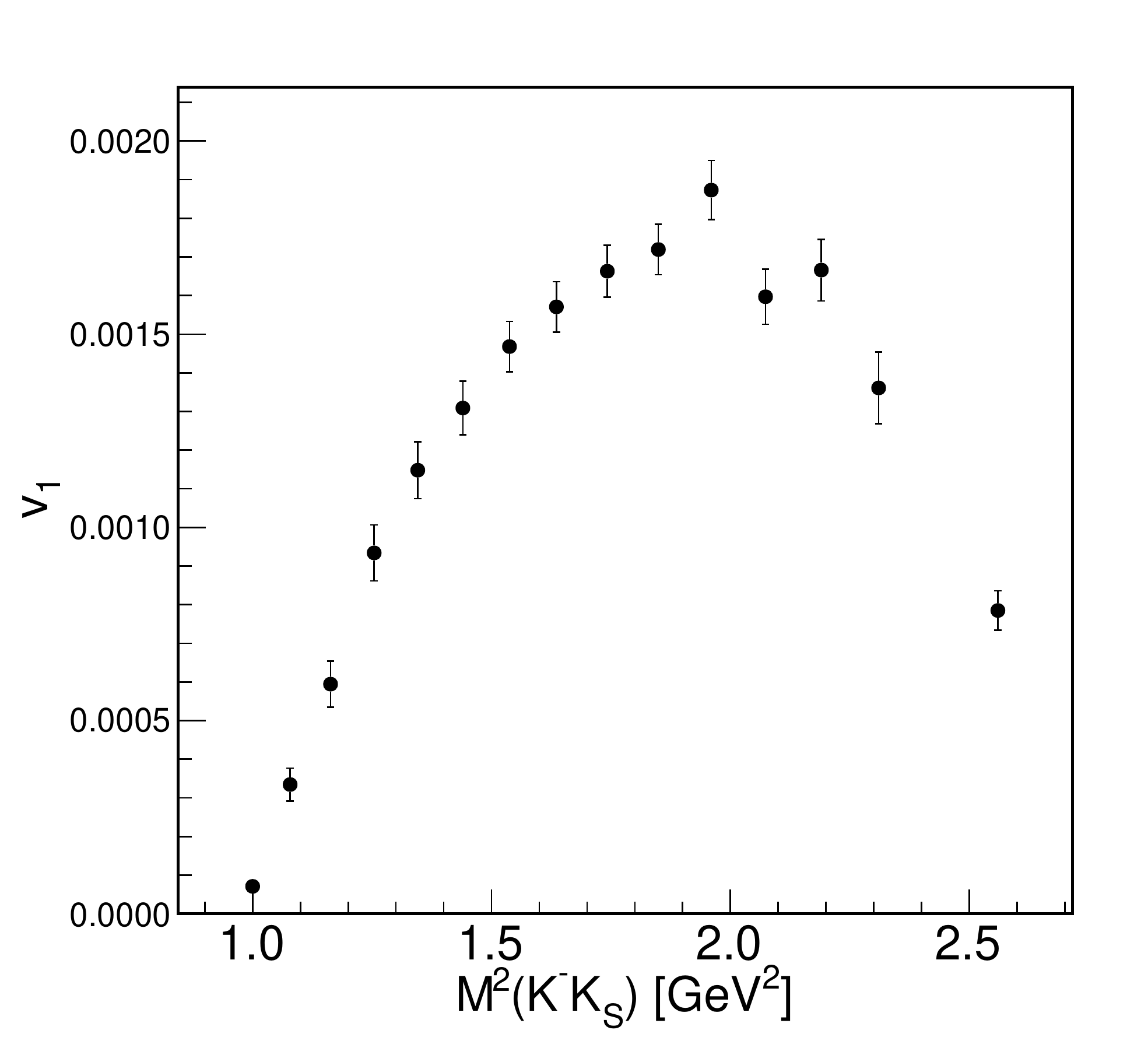}\put(-170,150){\Large\bf  (a)}
    \includegraphics[width=0.59\textwidth]{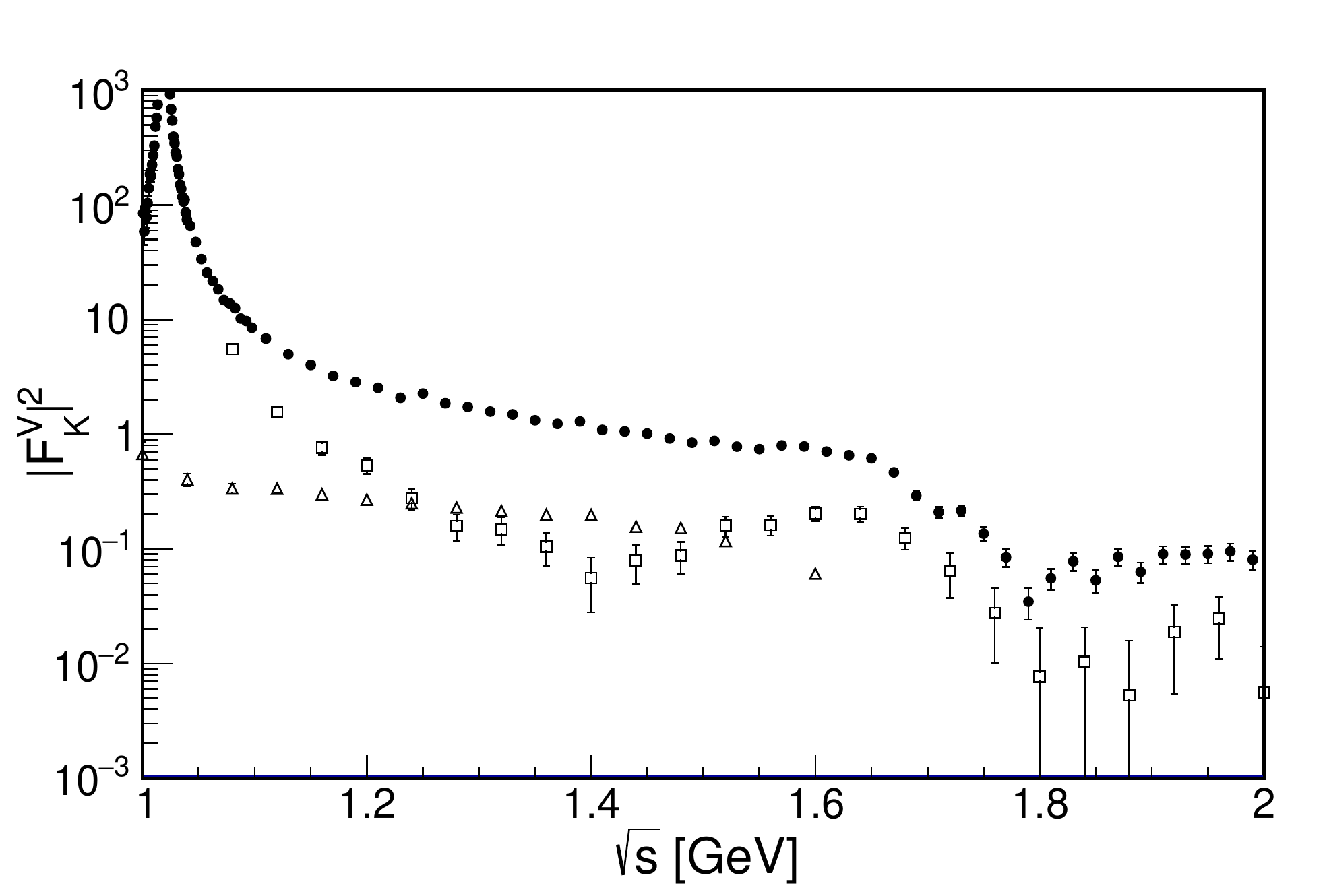}\put(-230,150){\Large\bf  (b)}
    \caption[Spectral functions $v_1$ and $F^V_K$]{(a) Spectral function $v_1$ for  $\tau^-\to K^-K_S\nu_\tau$ from BaBar~\cite{BaBar:2018qry}. (b) Comparison of the extracted kaon form factor $|F_K^{V(I=1)}(s)|^2$ (open triangles) with $|F_{K^0}^{V}(s)|^2$ from $e^+e^-\to K_SK_L$~\cite{Lees:2014xsh} (open squares) and $|F_{K^+}^{V}(s)|^2$ from $e^+e^-\to K^+K^-$~\cite{Lees:2013gzt} (black circles). 
    }
    \label{fig:KKtau}
\end{figure}

Above $2 \GeV$,
BESIII has measured the $e^+e^-\to K^+K^-$ cross section in a scan experiment at 22 energies between $2\GeV$ and $3.08\GeV$~\cite{Ablikim:2018iyx}, see Fig.~\ref{fig:KKff2}(a). A resonant structure around $2.2\GeV$ is observed in the cross section line shape. A Breit--Wigner fit yields a mass of $M=2239(13)\MeV$ and a width of $\Gamma=140(21)\MeV$. The structure seems to be distinct from the $\phi(2170)$ resonance. Since the kaon form factors include both isovector and isoscalar contributions, it is difficult to interpret the observed structure from a $e^+e^-\to K^+K^-$ cross section measurement only. For this reason, BaBar has carried out a more comprehensive analysis of several final states in Ref.~\cite{BABAR:2019oes}, and the  average $e^+e^-\to K_SK_L$ cross section in the energy range from 1.98 to $2.54\GeV$ is  
$4(7)\pb$, consistent with zero. The interference patterns observed in $K^+K^-$ as well as $\pi^+\pi^-$, $\pi^+\pi^-\eta$, and $\omega\pi\pi$ channels in the resonance energy range, suggest an evidence for an isovector resonance $\rho(2230)$.
---
The measured  $K^+K^-$ form factor at $3\GeV$ and $7\GeV$ is 5.3 and 2.6 times larger than the LO asymptotic values, respectively~\cite{Lees:2015iba}, while the measured slope is consistent with the prediction.

Similarly to the case of $\FV(s)$, complementary information on the isovector part of the kaon form factor for $\sqrt{s}$ below $m_\tau$ is obtained from $\tau$ decays, since the spectral function $v_1(s)$ in  $\tau^{-}\to K^{-}K_S\nu_{\tau}$ is related in the same way to $|F_K^{V(I=1)}(s)|^2$ as given in Eq.~\eqref{eq:tauV1} for the pion form factor.
In Fig.~\ref{fig:KKtau}, recent data on the spectral function $v_1$ from the BaBar experiment~\cite{BaBar:2018qry} is plotted and compared to the kaon form factor from the $e^+e^-\to K\bar K $ data~\cite{Lees:2013gzt,Lees:2014xsh}, 
using $4.3\times10^8$ $e^+e^-\to \tau^+\tau^-$ events produced at a c.m.\ energy around $10.6\GeV$. 
As an example for a theoretical analysis dedicated specifically to the isovector kaon form factor, see Ref.~\cite{Gonzalez-Solis:2019iod}. 

\subsection[$e^+e^-\to PV$]{\boldmath $e^+e^-\to PV$}\label{sec:eePV}
The next most important contribution to electron--positron annihilation is given by the 
final states with a vector and a pseudoscalar meson, $e^+e^-\to VP$, where $V$ and $P$ are members of the ground state nonets. We treat two of the vector resonances $\omega$ and $\phi$ separately as quasi-stable states and in this section focus on  $e^+e^-\to \omega P$, $\phi P$, where the pseudoscalar meson has no electric charge. The remaining cases, with contributions from the broad nonisoscalar vector mesons $\rho(770)$ or $K^*(892)$,  are discussed in the context of three-pseudoscalar production mechanisms in Sec.~\ref{sec:ee-PPP}.

\begin{figure}[t]
    \centering
\includegraphics[width=\linewidth]{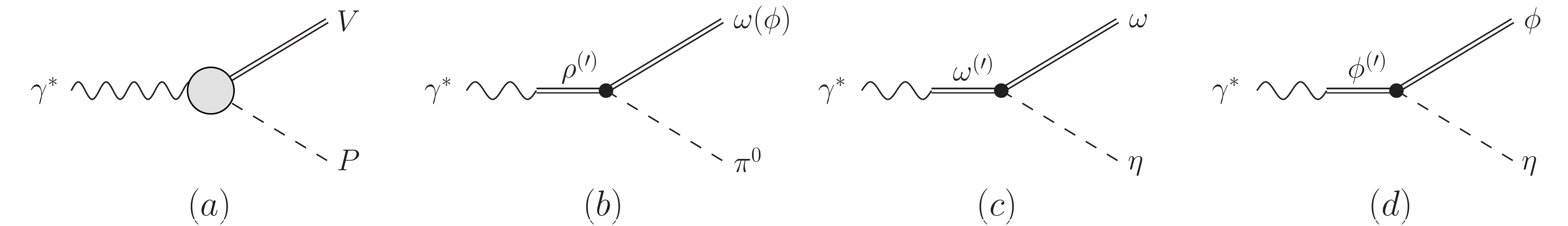}
    \caption[$\gamma^*\to VP$ vertex ]{(a) Form factor $F_{VP}(s)$, as well as VMD-type contributions to the specific final states
    (b)~$\omega(\phi)\pi^0$, (c)~$\omega\eta$, and (d)~$\phi\eta$.
    The final state $\phi\pi^0$ is OZI-suppressed; we also disregard the OZI-suppressed intermediate vector mesons for the $\omega\eta$ and $\phi\eta$ form factors.  Note that we concentrate on \textit{narrow} vector mesons in the final state and disregard those with a $\rho$, which are discussed separately as three-body continuum channels ($\pi^+\pi^-\pi^0$ and $\pi^+\pi^-\eta$) in Sec.~\ref{sec:ee-PPP}.}
    \label{fig:VMD-VPP}
\end{figure}
The Born cross section for a two-body production process  $e^+e^-\to VP$ with narrow vector and pseudoscalar mesons is related to the transition form factor  $F_{VP}(s)$  that represents the dynamics of the $\gamma^*VP$ vertex shown in Fig.~\ref{fig:VMD-VPP}(a) and defined as~\cite{Landsberg:1986fd,Pacetti:2008hd}
\begin{equation}
    J_\mu^{\text{em},VP}=\epsilon_{\mu\nu\rho\sigma}\varepsilon^\nu q_1^\rho q_2^\sigma F_{VP}(s) \,,
\end{equation}
where $\varepsilon^\nu$ is the polarization vector of the corresponding vector meson. 

Renewed and enforced interest in these transition form factors is not least motivated by their impact on the transition form factors of the light, flavor-neutral pseudoscalars $P=\pi^0$, $\eta$, $\eta'$, describing the couplings $P\to\gamma^*\gamma^*$: at low energies, their isoscalar spectral functions are entirely dominated by the narrow $\omega$ and $\phi$ resonances, and hence there is a close relation between $F_{VP}(s)$ and the pseudoscalar transition form factors with one virtuality fixed at the vector mass, $F_{P\gamma^*\gamma^*}(s,m_V^2)$, which has been exploited variously in dispersive analyses connected to light-by-light scattering~\cite{Hanhart:2013vba,Colangelo:2014pva,Hoferichter:2014vra}.

Contrary to the elastic form factors such as $\FV(s)$, the transition form factors $F_{VP}(s)$ have dimension $\GeV^{-1}$; they yield nontrivial information on 
the strength of the radiative transitions that is not constrained by gauge invariance. The value of the form factor at $s=0$ is directly related to the decay width $\Gamma(V\to P\gamma)$ or  $\Gamma(P\to V\gamma)$, depending on the sign of $m_V-m_P$:
\begin{equation}
    \Gamma(V\to P\gamma)=\frac{\alpha}{3} E_\gamma^3 |F_{VP}(0)|^2 \, , \qquad  
    \Gamma(P\to V\gamma)=\alpha E_\gamma^3 |F_{VP}(0)|^2 \, ,
\end{equation}
where $E_\gamma$ is the radiative photon energy.
\begin{table}[t!]
    \caption[Transition form factors for the $VP\gamma^*$ processes]{Transition form factor strengths for the $VP\gamma^*$ processes given by $F_{VP}(0)$, as extracted from the respective radiative decays. The remaining columns quote the upper limits of the decay regions and the thresholds for the production processes.}
    \renewcommand{\arraystretch}{1.3}
    \begin{center}
    \begin{tabular}{lllcc}
    \toprule
         $VP$&Radiative& $F_{VP}(0)$&Decay region& Production threshold\\
         &decay&$[\text{GeV}^{-1}]$ &$|m_V-m_P|~[\text{GeV}]$&$m_V+m_P~[\text{GeV}]$\\
         \midrule
         $\omega\pi^0$&$\omega\to\pi^0\gamma$     & 2.31(3)&0.648&0.918\\
    $\omega\eta$&$\omega\to\eta\gamma$     & 0.445(2)&0.235&1.331\\
    $\omega\eta'$&$\eta'\to\omega\gamma$     &0.41(1) &0.175&1.740\\
    $\phi\pi^0$& $\phi\to\pi^0\gamma$    & 0.134(2)&0.884&1.154 \\
    $\phi\eta$& $\phi\to\eta\gamma$    & 0.691(7) &0.472&1.567\\
    $\phi\eta'$& $\phi\to\eta'\gamma$    & 0.71(1) &0.062&1.977\\
    \bottomrule
    \end{tabular}
    \end{center}
    \renewcommand{\arraystretch}{1.0}
    \label{tab:PVdata}
\end{table}
In Table~\ref{tab:PVdata}, the values for the various $F_{VP}(0)$ are given, as well as the kinematic ranges for the time-like transition form factors discussed in this section.
The cross section for  $e^+e^-\to VP$ reads
\begin{equation}
    \sigma(e^+e^-\to VP)=\frac{4\pi\alpha^2}{3s^{3/2}}|F_{VP}(s)|^2P_f(s) \,,
\end{equation}
where the factor $P_f(s)$ describes the reaction energy dependence of the phase space of the final state. In the narrow-width approximation, $P_f(s)=p_V^3$, where $p_V$ is the c.m.\ momentum in the final state. For $V=\omega$ or $\phi$, the approximation is not valid close to  threshold, where the phase space volume corrections for the specific $V$ decays modes  
$V\to f$ should be included. 
The form factor
can be studied in two separate kinematic regions. The $e^+e^-\to VP$ reaction covers the $\sqrt{s}>m_V+m_P$ region
of the form factor. If $m_V>m_P$, the decay $V\to \ell^+\ell^-P$ covers the $2m_\ell<\sqrt{s}<m_V-m_P$ region, where $\sqrt{s}$ is the invariant mass of the $\ell^+\ell^-$ system. In the contrary case, if $m_P>m_V$ the  $P\to \ell^+\ell^-V$ decay covers the $2m_\ell<\sqrt{s}<m_P-m_V$ region. 

Depending on the final state of the reaction, the virtual-photon contribution is either isovector
or isoscalar, and the form factors will be dominated by the corresponding (excited) vector mesons.
Figures~\ref{fig:VMD-VPP}(b)-(d) show the cases with narrow vector mesons that we will discuss in the present section. For illustration purposes the data will be compared with parameterizations using
a na\"ive VMD  model (photon couplings {\it solely} through vector mesons). The form
factor  $F_{VP}(s)$ is then given by~\cite{Landsberg:1986fd}
\begin{equation}
   F_{VP}(s)=   \sum_{V'} \frac{g_{PV'V}}{g_{V'}} 
{\BW}_{V'}(s) \,.
\label{eq:FVMDo}
\end{equation}
The sum extends over the neutral vector mesons ($Q=S=0$): $\rho^0$, $\omega$, $\phi$, \ldots; $g_{PVV'}$ and  $g_{V}$ are their
couplings to the $VP$ meson pair and to the photon, respectively. The functions ${\BW}_V(s)$ are given by the Breit--Wigner formula with fixed width from  Eq.~\eqref{eq:BW}, for all vector resonances except for the $\rho(770)$. The contributing vector resonances are selected by isospin and the OZI rule for the specific reaction channel. 
The form factor has to be a real function below inelastic threshold, which is equal to $\sqrt{s_{in}}=2m_\pi$ and  $\sqrt{s_{in}}=3m_\pi$ for the isovector and isoscalar virtual-photon contributions, respectively. 
Several studies that improve on simple VMD models have been performed using chiral Lagrangians including vector mesons, thus necessarily concentrating on decay kinematics~\cite{Terschluesen:2010ik,Terschlusen:2012xw,Chen:2013nna,Chen:2014yta}.

\begin{figure}[t!]
    \centering
    \includegraphics[width=0.46\textwidth]{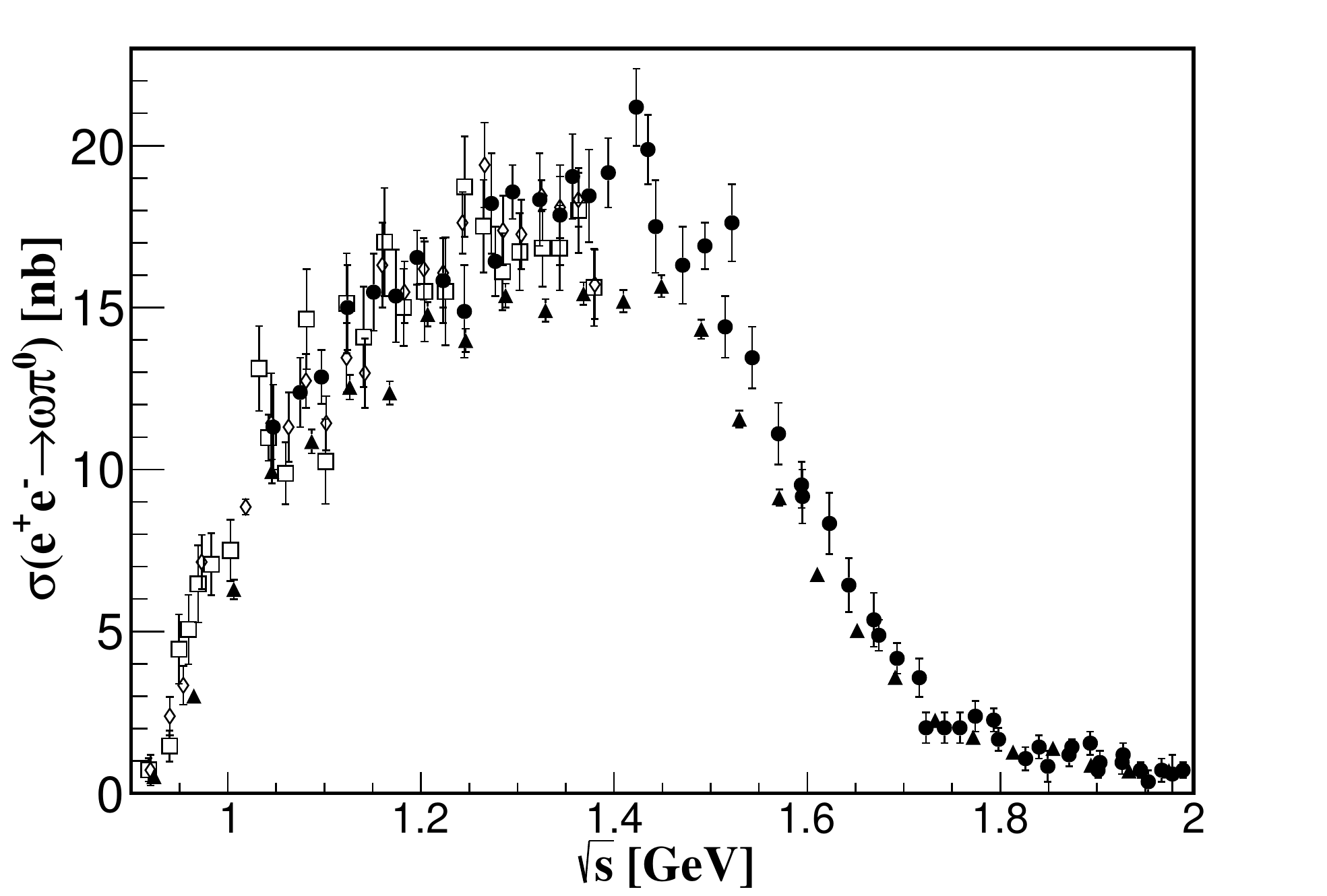}\put(-200,130){\Large\bf  (a)}
    \includegraphics[width=0.53\textwidth]{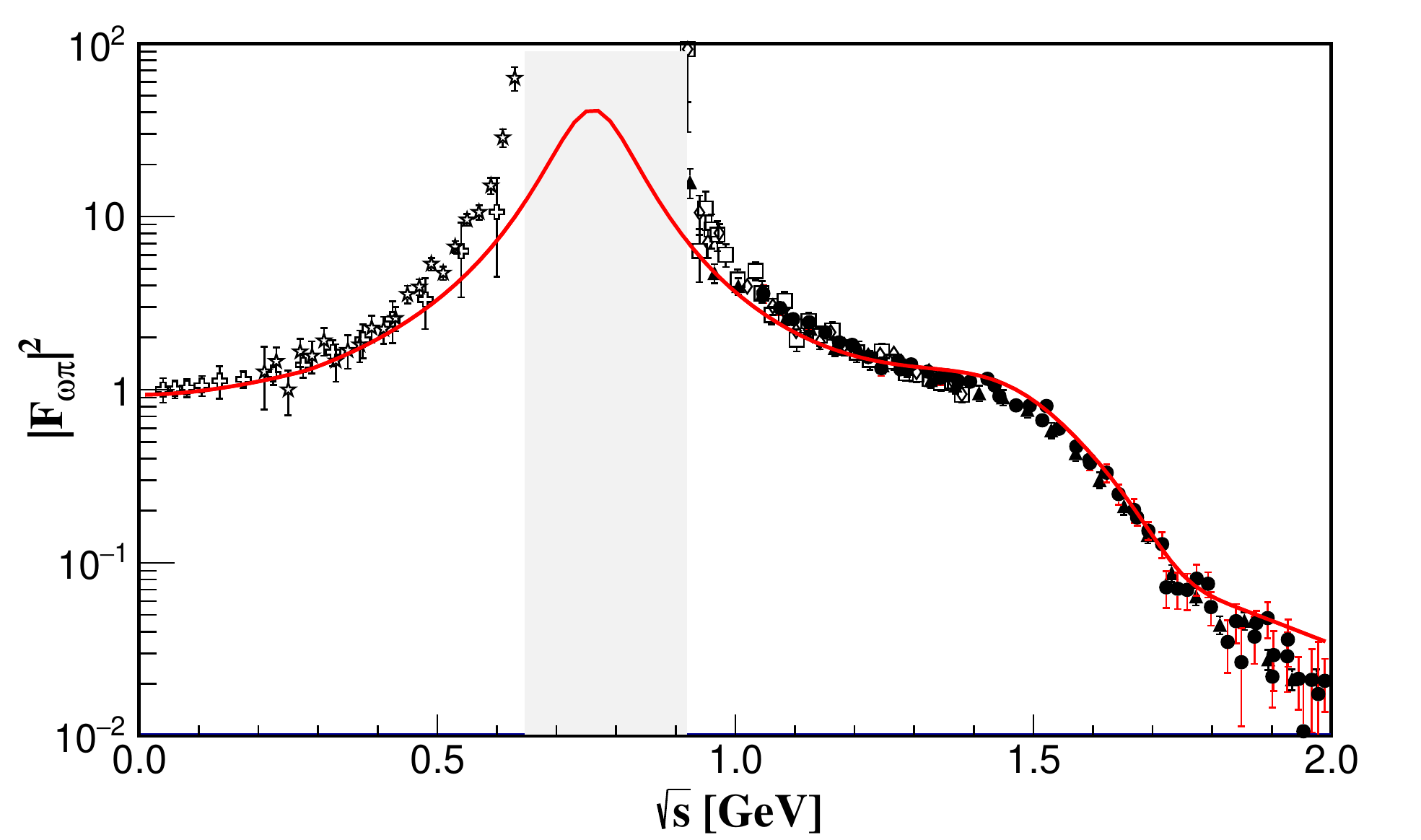}\put(-230,130){\Large\bf  (b)}
    \caption[ $e^+e^-\to \omega\pi^0$ and $|F_{\omega\pi}|^2$]{(a) Born cross section $e^+e^-\to \omega\pi^0$, data from SND~\cite{Achasov:2000wy,Achasov:2016zvn} (diamonds and circles), CMD2~\cite{Akhmetshin:2003ag} (squares), and BaBar~\cite{TheBaBar:2017vzo} (triangles). (b) Form factor $|F_{\omega\pi}|^2$ in the two experimentally accessible kinematic regions $4m_e^2<s<(m_\omega-m_\pi)^2$ and $s>(m_\omega+m_\pi)^2$.
    Form factor data in the decay region is taken from NA60~\cite{Arnaldi:2016pzu} (stars) and A2~\cite{Adlarson:2016hpp} (crosses). The line is the VMD parameterization from Eq.~\eqref{eq:Fompi}.}
    \label{fig:eeOmPiFF}
\end{figure}

\paragraph{\boldmath $e^+e^-\to \omega\pi^0$}
The process $e^+e^-\to \omega\pi^0$ was studied from threshold up to $\sqrt{s}=2\GeV$ by the SND~\cite{Achasov:2000wy,Achasov:2012zza,Achasov:2013btb,Achasov:2016zvn} and CMD-2~\cite{Dolinsky:1986kj,Akhmetshin:2003ag} experiments, as well as by BaBar~\cite{TheBaBar:2017vzo}. 
The measured cross section is shown in Fig.~\ref{fig:eeOmPiFF}(a). The $F_{\omega\pi}(s)$  form factor should be dominated by a coherent contribution of isovector $\rho$ resonances and is parameterized in VMD as
\begin{equation}
    F_{\omega\pi}(s)={\frac{g_{\rho\omega\pi}}{g_\rho}}\left({\BW}^{GS}_\rho(s)+c_1{\BW}_{\rho'}(s) +c_2{\BW}_{\rho''}(s)+\ldots\right) \,,\label{eq:Fompi}
\end{equation}
where for the $\rho(770)$ contribution a Gounari--Sakurai parameterization is used, while for $\rho'$ and $\rho''$ fixed widths are assumed. Often the fits to the experimental data allow the coefficients $c_1,c_2,\ldots$ to be complex numbers.  However, the nonzero relative phases violate unitarity relations for the form factor, i.e., they have to be real below the inelastic threshold. 
The most recent and most precise data on $e^+e^-\to \omega\pi^0$ is from the SND
experiment~\cite{Achasov:2016zvn}, from BaBar~\cite{TheBaBar:2017vzo}, and from BESIII above $2 \GeV$~\cite{Ablikim:2020das}. The description of the data in the c.m.\ energy range $1.05$--$2.00\GeV$ requires three $\rho$-like states. The extracted form factor is shown in Fig.~\ref{fig:eeOmPiFF}(b). The data in the  $2m_\ell<\sqrt{s}<m_\omega-m_\pi$ region for the form factor is obtained from $\omega\to \pi^0\ell^+\ell^-$ decays. The two most precise results from the  NA60~\cite{Arnaldi:2016pzu} and A2~\cite{Adlarson:2016hpp} experiments are included. In NA60, the $\omega$ mesons are produced in $pp$ or $p$--nucleus collisions and the $\omega\to \pi^0\mu^+\mu^-$ decay is extracted from inclusive dimuon invariant-mass distributions. In  the A2 experiment, $\omega$ mesons stem from the exclusive photoproduction process $\gamma p\to p\omega$, and the decay mode $\omega \to \pi^0 e^+e^-$ is studied. The NA60 result for the form factor seems to be inconsistent with the VMD parameterization and the other data sets. In particular, more refined dispersion-theoretical analyses of the $\omega\to\pi^0\gamma^*$ transition form factor based on input for $\omega\to3\pi$ and $\FV$ fail to accommodate such a steep rise in the form factor towards the upper end of the decay region~\cite{Koepp:1974da,Schneider:2012ez,Danilkin:2014cra,Albaladejo:2020smb}, and very general limits based on analyticity and unitarity exclude at least the NA60 data points at highest energies~\cite{Ananthanarayan:2014pta}, in particular when combined with the information above $\omega\pi^0$ production threshold~\cite{Caprini:2015wja}.

\begin{figure}[t!]
    \centering
    \includegraphics[width=0.6\textwidth]{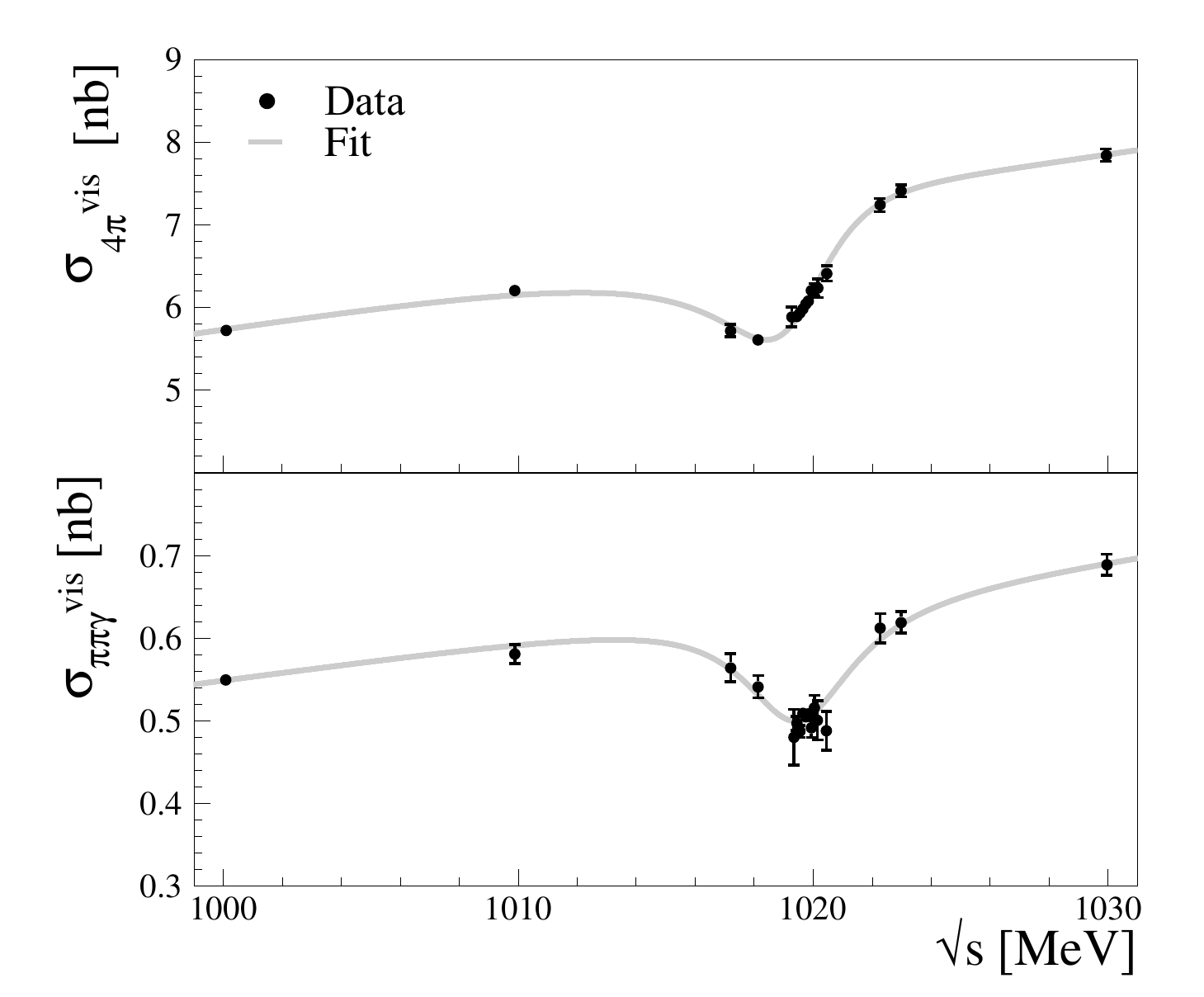}
    \caption[Visible cross section $\sigma(e^+e^-\to\pi^0\omega)$ close to the $\phi$ resonance]{Visible cross section $\sigma(e^+e^-\to\pi^0\omega)$ close to the $\phi$ resonance as observed in $\omega\to\pi^0\pi^+\pi^-$ (upper) and $\omega\to\pi^0\gamma$ (lower panel) channels by the KLOE experiment~\cite{Ambrosino:2008gb}.}
    \label{fig:eeOmPiKLOE}
\end{figure}
Dedicated studies of interference effects in $e^+e^-\to \omega\pi^0$ in the energy region close to  $\sqrt{s}=m_\phi$ allow one to determine $\BR(\phi\to\omega\pi^0)$, which is suppressed by
the  Okubo--Zweig--Iizuka (OZI) rule and $G$-parity. Such measurements were first performed at SND~\cite{Aulchenko:2000zq} and subsequently at KLOE~\cite{Ambrosino:2008gb}. The most precise KLOE data relies on $600 \pb^{-1}$ collected at a few energy points between 1.00 and $1.03 \GeV$ to extract the cross sections for $e^+e^-\to\pi^0\pi^0\pi^+\pi^-$ and $e^+e^-\to\pi^0\pi^0\gamma$. The dependence of the cross sections near the $\phi$ can be parameterized as
\begin{equation}
    \sigma(s)=\sigma_0+A\cdot(\sqrt{s}-m_\phi)\left|1-{z}\frac{\Gamma_\phi}{m_\phi}\cdot
    {{\BW}_\phi(s)}\right|^2 \,,
\end{equation}
where $z$ is a complex parameter describing the interference between the $\phi$ decay amplitude and nonresonant processes. 
In Fig.~\ref{fig:eeOmPiKLOE} the data points with the superimposed fit functions are shown for both channels. By fitting the interference pattern for both final states, a ratio $\Gamma(\omega\to\pi^0\gamma)/\Gamma(\omega\to\pi^+\pi^-\pi^0)=0.0897(16)$ was determined. This result improves the uncertainty by nearly a factor of four with respect to the previous measurement from SND~\cite{Aulchenko:2000zq}, but the value is $1.9 \sigma$ lower. 
The parameters $\sigma_0$ and $z$ describing the $e^+e^-\to\pi^0\pi^0\pi^+\pi^-$
reaction allowed the experiment to extract  $\BR(\phi\to\omega\pi^0)=4.4(6)\times10^{-5}$.

\begin{figure}[t!]
    \centering
    \includegraphics[width=0.4\textwidth]{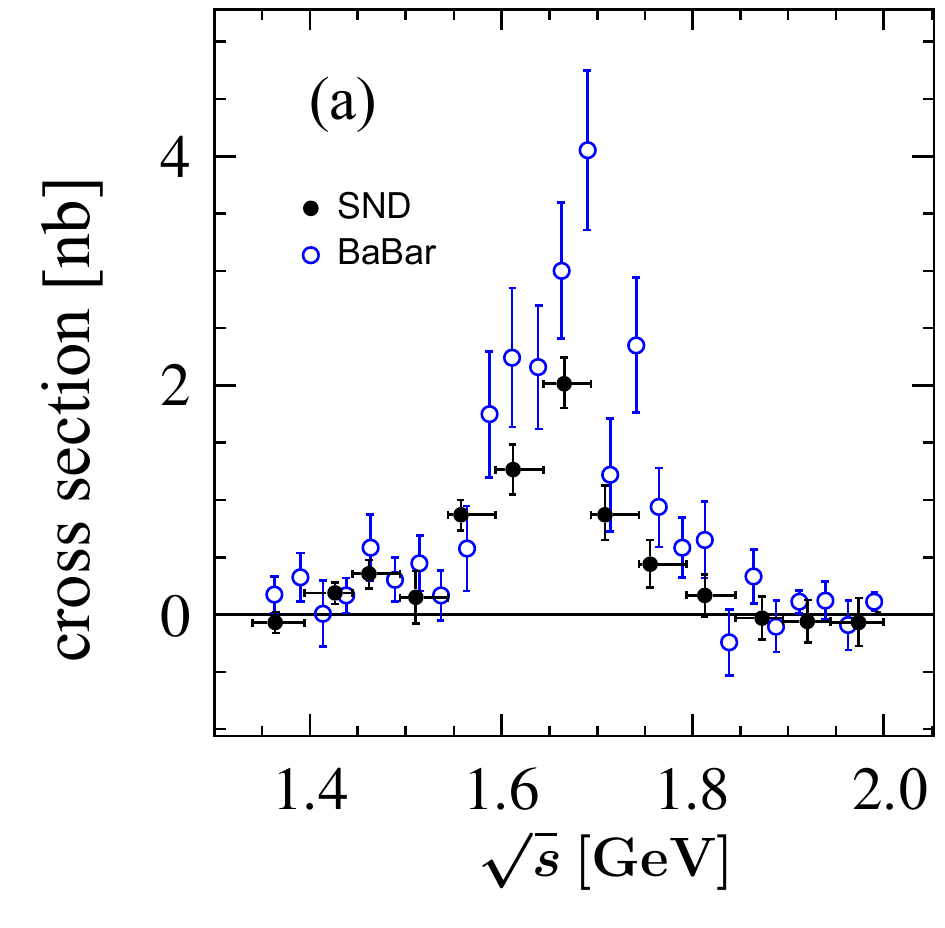} \hfill
    \includegraphics[width=0.525\textwidth]{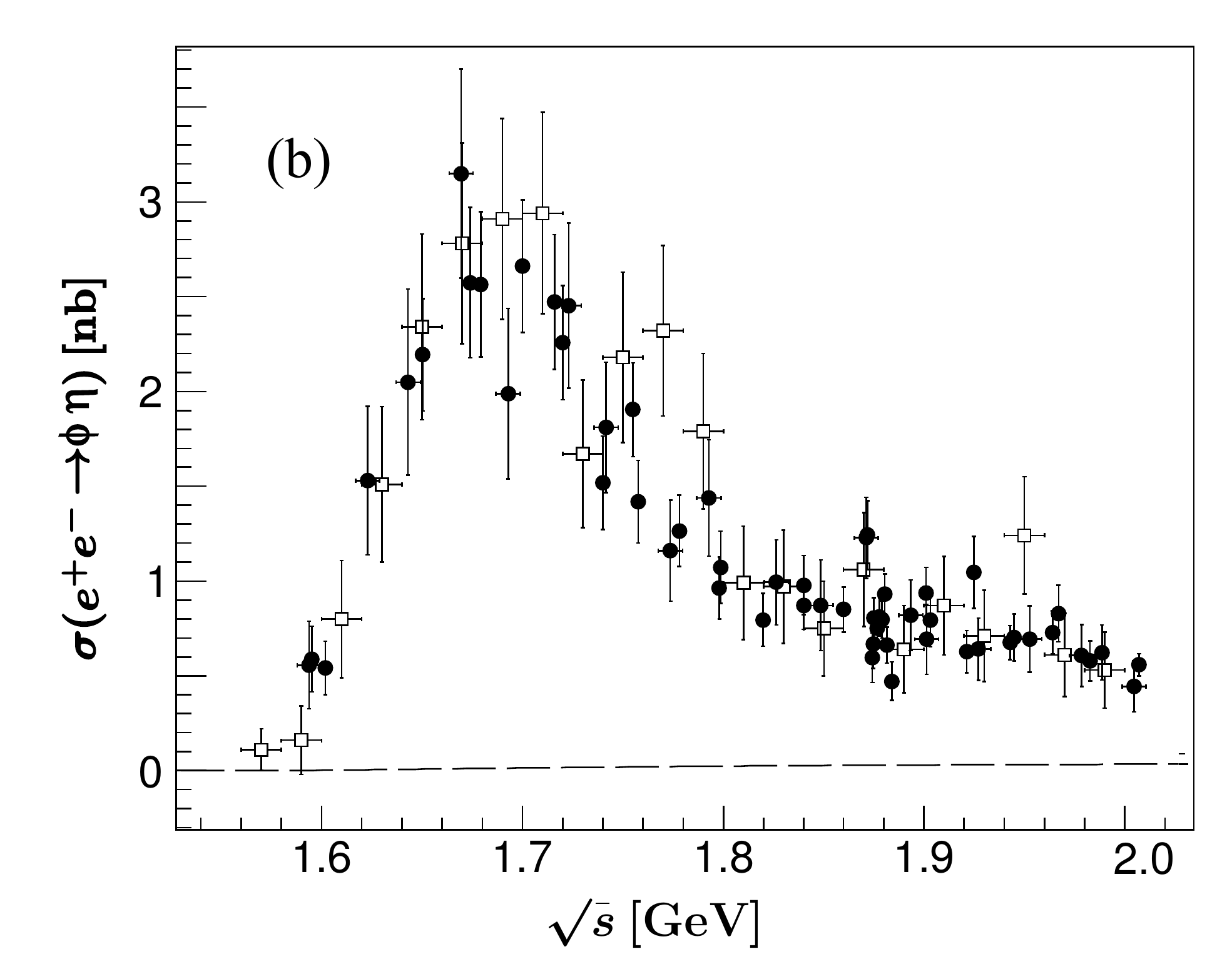}
    \caption[$e^+e^- \to \omega\eta$, $e^+e^- \to \phi\eta$]{ (a) The $e^+e^- \to \omega\eta$ cross sections from SND~\cite{Achasov:2016qvd} and BaBar~\cite{Aubert:2006jq}. (b) The $e^+e^- \to \phi\eta$ cross section from BaBar~\cite{Aubert:2007ym} (open boxes) and CMD-3~\cite{Ivanov:2019crp} (filled circles). }
    \label{fig:eeOmEta}
\end{figure}
\paragraph{\boldmath $e^+e^- \to \omega\eta$}
The most precise cross section measurements for the process $e^+e^- \to \omega\eta$  below $2.00\GeV$ are from SND~\cite{Achasov:2016qvd} and CMD-3~\cite{CMD-3:2017tgb}, extracted using the $\omega\to \pi^+\pi^-\pi^0$ decay mode. 
There is a significant discrepancy between these results and the previous BaBar~\cite{Aubert:2006jq} measurement; see Fig.~\ref{fig:eeOmEta}(a). The process is dominated by the effective contributions of $\omega(1650)$ and $\phi(1680)$, the fitted effective width of $110(20)\MeV$ is smaller than 
the PDG value for the $\omega(1650)$ of $315(35)\MeV$, but consistent with $150(50)\MeV$ 
for the $\phi(1680)$ resonance. The contribution of the $\omega(782)$ pole is neglected in the fits. The contribution from the $\omega(1420)$ is small, but necessary to describe the asymmetry in the peak in the measured distribution. In this channel, the recent BESIII measurement~\cite{Ablikim:2020das} indicates a resonance in the range $2\GeV<\sqrt{s}<2.5\GeV$.

\begin{figure}[t!]
    \centering
    \includegraphics[width=0.55\textwidth]{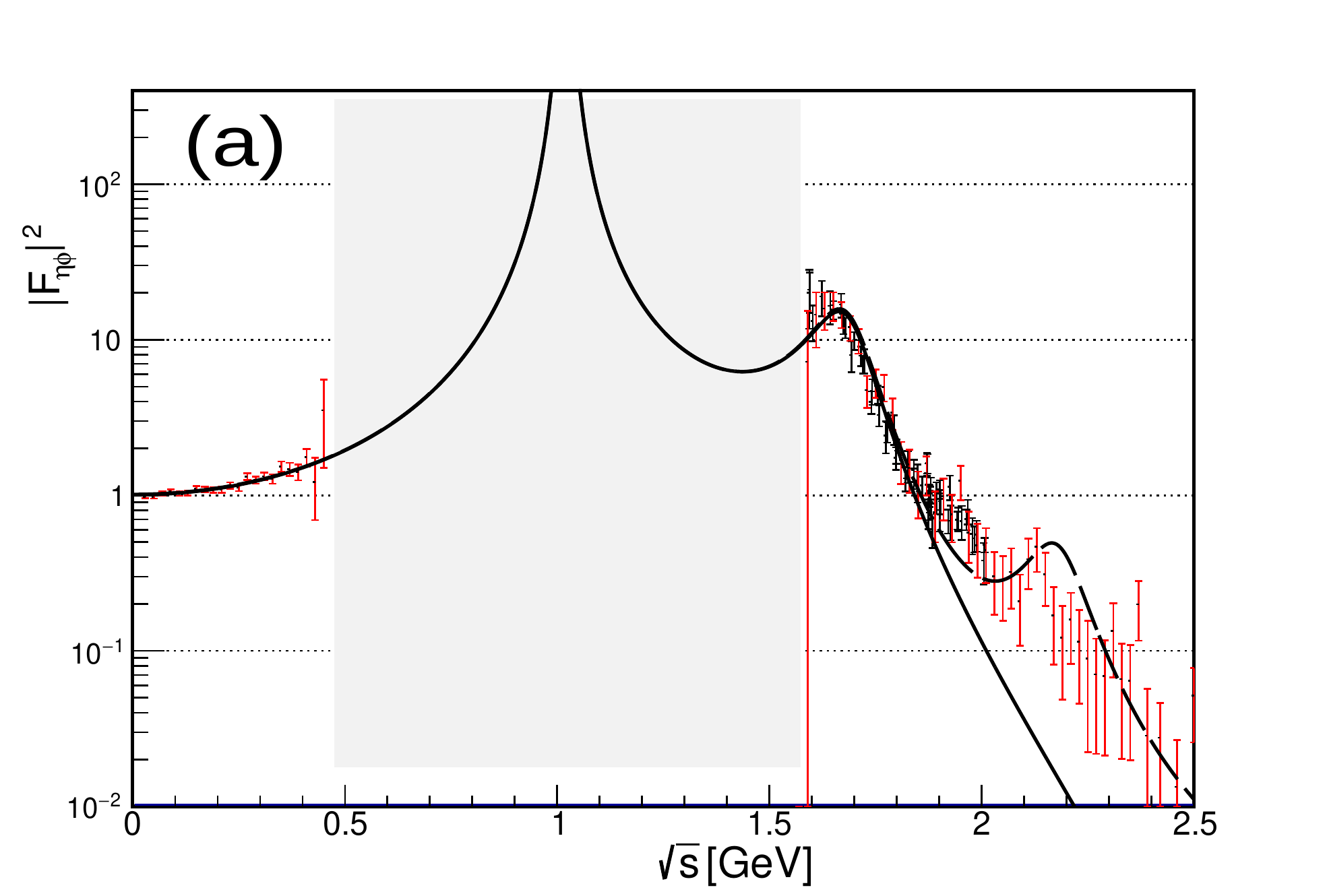}
    \includegraphics[width=0.44\textwidth]{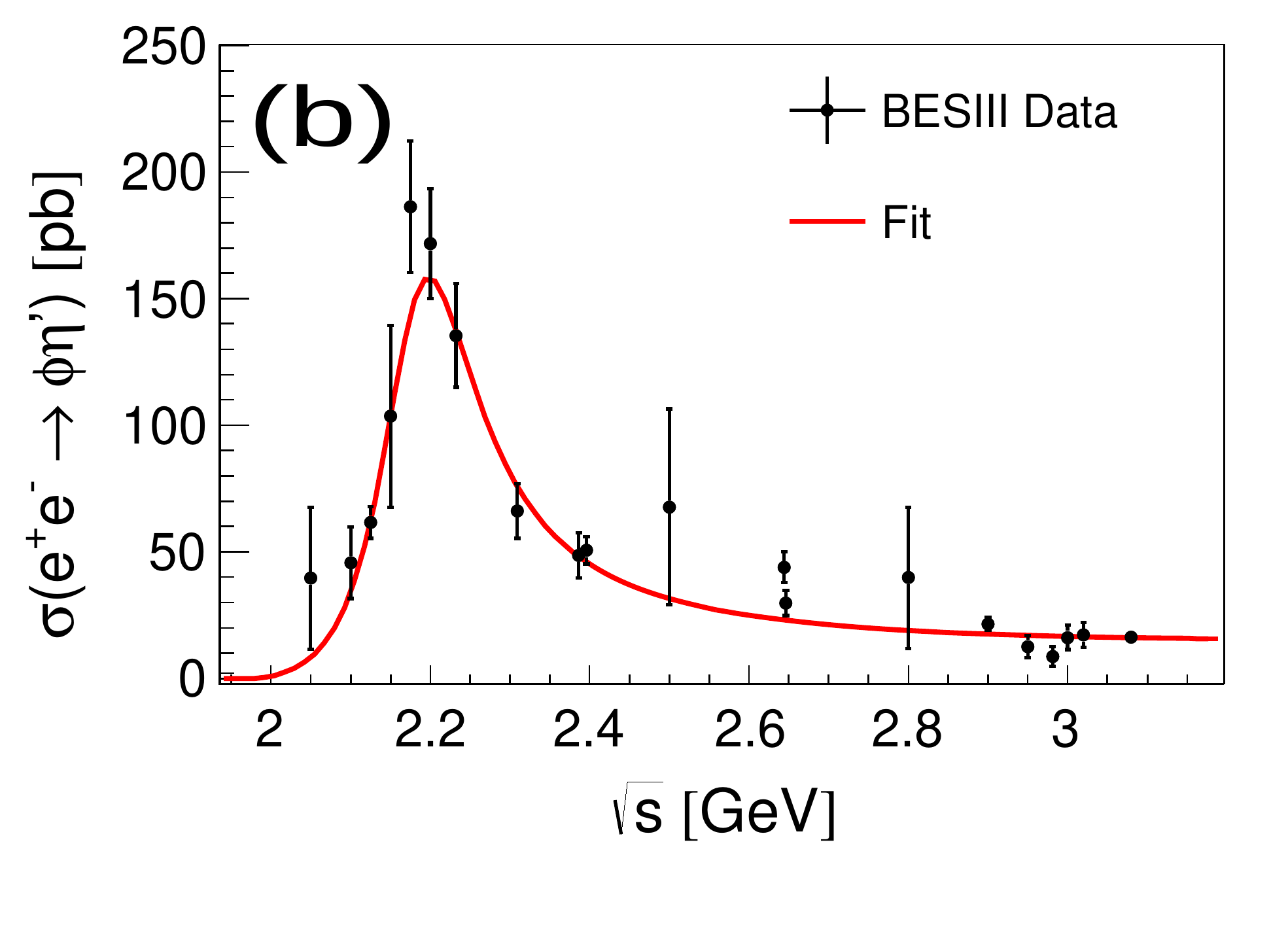}
    \caption[Form factor $|F_{\phi\eta}|^2$]{(a) Form factor $|F_{\phi\eta}|^2$ in two accessible kinematic regions (nonshaded areas). The data in the $\phi\to\eta e^+e^-$ region is from KLOE~\cite{Babusci:2014ldz}. The data in the production region  $e^+e^- \to \phi\eta$ is from BaBar~\cite{Aubert:2007ym} (red error bars) and CMD-3~\cite{Ivanov:2019crp} (black error bars in the range up to 2\GeV). 
    The curves are the VMD model parameterization Eq.~\eqref{eq:phietVMD} with the $\phi$ and $\phi'$ contributions (solid line) and additionally with the $\phi(2170)$ state (dashed line).
    (b) Cross section for $e^+e^-\to\phi\eta'$ measured at BESIII~\cite{Ablikim:2020coo}. 
    \label{fig:PhiEtap}}
    \label{fig:FFphieta}
\end{figure}
\paragraph{\boldmath $e^+e^- \to \phi\eta$ and $e^+e^- \to \phi\eta'$}
The process $e^+e^- \to \phi\eta$ was studied at BaBar using $K^+K^-\gamma\gamma$~\cite{Aubert:2007ym} and $K_LK_S\gamma\gamma$~\cite{TheBABAR:2017vgl} final states for c.m.\ energies up to $4\GeV$. Below $2\GeV$ it was studied at SND~\cite{Achasov:2018ygm} and    CMD-3~\cite{Ivanov:2019crp} with $\phi{\to}K^+K^-$. The resulting distributions from the three experiments are consistent with each other, as shown in Fig.~\ref{fig:eeOmEta}(b). The close-to-threshold cross section is dominated by
the $\phi(1680)$ resonance and well described with the simple ansatz
\begin{equation}
    F_{\phi\eta}(s)=\frac{g_{\phi\phi\eta}}{g_{\phi}}{\BW}_{\phi}(s)+\frac{g_{\phi'\phi\eta}}{g_{\phi'}}{\BW}_{\phi'}(s)+\ldots,\label{eq:phietVMD}
\end{equation}
where $\phi'$ denotes the $\phi(1680)$. Complementary information about the form factor is given by the KLOE measurement of the Dalitz decay $\phi\to\eta e^+e^-$~\cite{Babusci:2014ldz}. The extracted form factor is shown in Fig.~\ref{fig:FFphieta}(a). The line assumes $\phi$ and $\phi'$  meson contributions. 

The first measurement of  $e^+e^- \to \phi\eta'$ by BESIII~\cite{Ablikim:2020coo} shown in Fig.~\ref{fig:PhiEtap}(b) has a similar qualitative behavior of the cross section  close to threshold 
and is dominated by the apparent resonance with mass $2.180(20)\GeV$ and width $149(18)\MeV$ at c.m.\ energies below $3.08 \GeV$. This is likely 
the discussed strangeonium candidate,  the $\phi(2170)$,  observed in the ISR process $\rightarrow \gamma\phi f_0(980)$ at BaBar~\cite{Aubert:2006bu,Lees:2011zi} and
Belle~\cite{Shen:2009zze}. At  BESII~\cite{Ablikim:2007ab} and BESIII~\cite{Ablikim:2014pfc} it was observed in the $J/\psi\to\eta\phi f_0(980)$ decay. 
The discovery of the $\phi(2170)$ has triggered speculations that it might be an $s$-quark counterpart of the $Y(4260)$. 

\begin{figure}[t]
    \centering
    \includegraphics[width=0.54\textwidth]{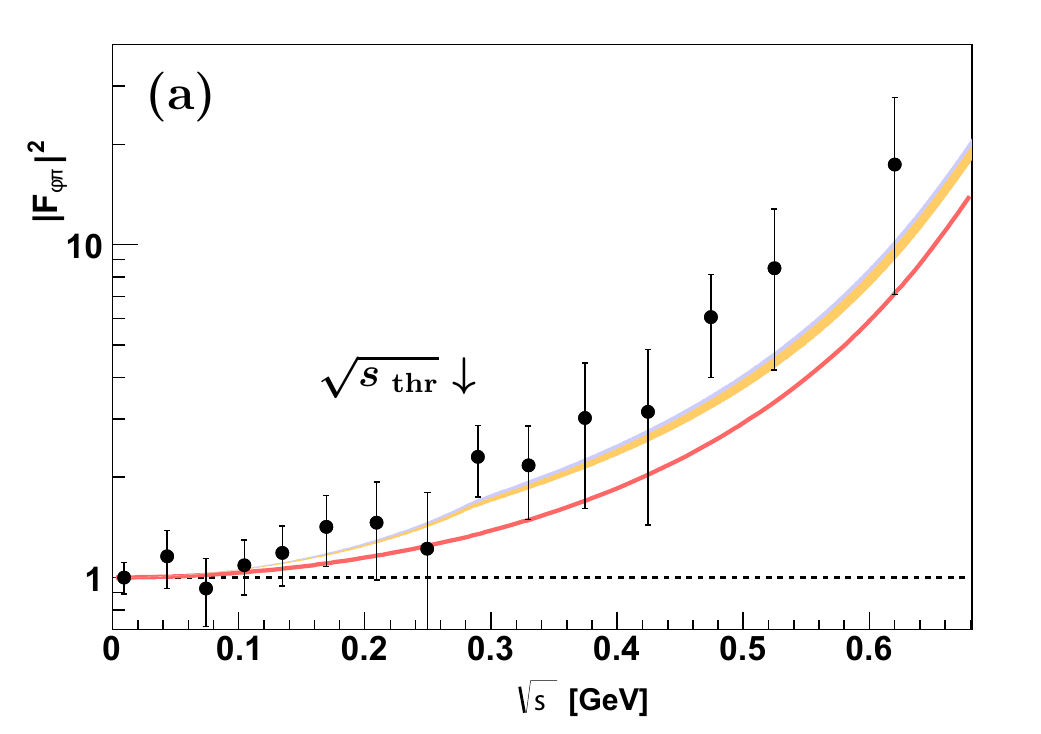}
    \includegraphics[width=0.40\textwidth]{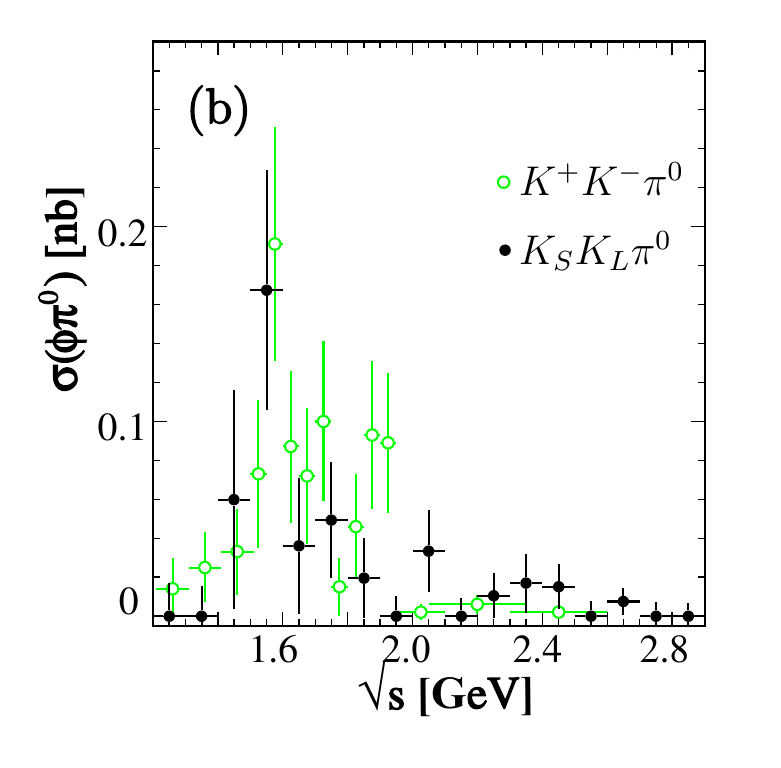}
     \caption[$e^+e^-\to\phi\pi^0$]{(a) Form factor $|F_{\phi\pi}|^2$ determined from the $\phi\to e^+e^-\pi^0$ decay at KLOE~\cite{Anastasi:2016qga} (points), VMD predictions (solid line), and dispersive calculations~\cite{Schneider:2012ez,Danilkin:2014cra} (bands). (b) Born cross section for the $e^+e^-\to\phi\pi^0$ process, measured at BaBar
   using $K^+K^-\pi^0$~\cite{Aubert:2007ym} and $K_SK_L\pi^0$~\cite{TheBABAR:2017vgl} final states.}
    \label{fig:PhiPi0}
\end{figure}

\paragraph{\boldmath $e^+e^- \to \phi\pi^0$}
The $e^+e^- \to \phi\pi^0$ process was measured at BaBar in $K^+K^-\pi^0$~\cite{Aubert:2007ym} and $K_SK_L\pi^0$~\cite{TheBABAR:2017vgl}, using the two main 
  $\phi$ decay channels; cf.\ also Ref.~\cite{Pacetti:2009pg} for a theoretical analysis. The cross section is shown in Fig.~\ref{fig:PhiPi0}(b). The transition form factor $|F_{\phi\pi}|^2$ was also measured in the decay $\phi\to \pi^0 e^+e^-$ at KLOE~\cite{Anastasi:2016qga}.  The decay region (see the upper range in Table~\ref{tab:PVdata}) is dominated by the $\rho^0$ contribution, which should be visible as a peak. 
  More refined theoretical predictions have been obtained using dispersion relations in analogy to $F_{\omega\pi}$~\cite{Schneider:2012ez,Danilkin:2014cra}, based on an analysis of $\phi\to3\pi$ Dalitz plot data~\cite{Aloisio:2003ur,Niecknig:2012sj}.
  However, in the KLOE measurement, due to background from $\phi\to\pi\rho$, $\rho\to\pi\pi$ processes, it was only possible to study the form factor up to electron--positron invariant masses of $625 \MeV$. The KLOE result for the form factor is shown in 
Fig.~\ref{fig:PhiPi0}(a). The curves correspond to the predictions from VMD and dispersive calculations~\cite{Schneider:2012ez,Danilkin:2014cra}. 
\subsection[$V\to P_1P_2\gamma$]{\boldmath $V\to P_1P_2\gamma$}\label{sec:VtoPPg}
Amplitude analyses of two pseudoscalar mesons produced in radiative $\phi$ and $J/\psi$ decays are particularly simple. Possible hadronic states have even angular momentum values with positive parity and charge conjugation quantum numbers ($J^{PC}=0^{++}, 2^{++}, 4^{++},$ etc.). Systems such as $\pi^0\pi^0$, $\pi^+\pi^-$, $K^+K^-$, $K_S^0K_S^0$,  $\eta^{(\prime)}\eta^{(\prime)}$, and $\eta^{(\prime)}\pi^0$ can be produced. The branching fractions for the processes are listed in Table~\ref{tab:VtgPP}. The $K_LK_L$ final state is not included since due to the long lifetime, the $K_L$ is difficult to register in the detectors used at electron--positron colliders. Also, no partial-wave analysis (PWA) of the
processes $J/\psi\to\gamma\eta\pi^0$ has been performed yet since it was recently observed at BESIII~\cite{Ablikim:2016exh}. 
\begin{table}[t]
      \caption{Branching fractions of radiative decays to a pair of pseudoscalar mesons.\label{tab:VtgPP}}
\begin{center}
\renewcommand{\arraystretch}{1.3}
  \begin{tabular}{rlrlr}
  \toprule
    &Final state&$\BR$& &Ref.\\ \midrule
    $\phi\to\gamma P_1P_2$&$\gamma\pi^+\pi^-$& $4.1(1.3)\times10^{-4}$&&\cite{Akhmetshin:1999dh}\\
    &$\gamma\pi^0\pi^0$& $1.09(6)\times10^{-4}$&&\cite{Aloisio:2002bt}\\
    &$\gamma\pi^0\eta$& $7.06(22)\times10^{-5}$&&\cite{Ambrosino:2009py}\\
    \midrule
    $J/\psi\to\gamma P_1P_2$&$\gamma\pi^0\pi^0$& $1.15(5)\times10^{-3}$&&\cite{Ablikim:2015umt}\\
    &$\gamma\pi^+\pi^-$& $2\BR(J/\psi\to\gamma\pi^0\pi^0)$ && 
    inferred from Ref.~\cite{Ablikim:2006db}
    \\
    &$\gamma\eta\pi^0$& $2.14(31)\times10^{-5}$&& \cite{Ablikim:2016exh} \\
    &$\gamma\eta\eta$& $1.47(2)\times10^{-4}$&&  excl. $\phi\eta$~\cite{Ablikim:2013hq}\\
    &$\gamma K_S^0K_S^0$& $8.1(4)\times10^{-4}$&&  \cite{Ablikim:2018izx} \\
    &$\gamma K^+K^-$& --&&  \\ 
    \bottomrule
  \end{tabular}
  \renewcommand{\arraystretch}{1.0}
\end{center}
\end{table}
 
The amplitude of a radiative decay $V\to P_1P_2\gamma$ is often presented assuming factorization~\cite{Ablikim:2015umt}, 
\begin{equation}
    {\cal M}=\sum_X\bra{P_1P_2}H_{S} \ket{X}\bra{X\gamma}H_{EM}\ket{V} \,, \label{eq:VtoPPg}
\end{equation}
where  $H_{EM}$ and  $H_{S}$ represent the electromagnetic transition $V\to\gamma X$ and the strong final-state interaction, respectively.
However, since the strength of the two transitions might be comparable, this representation is only approximate, e.g., for $\phi\to\pi^0\pi^0\gamma$ there is a contribution from the $\phi\to\rho^0\pi^0$  intermediate state with subsequent $\rho^0\to\pi^0\gamma$ decay.
In helicity reference systems, the $e^+e^-\to V\to \gamma(X\to P_1P_2)$ process is described by
the mass squared $s$ of the $X$ system, radiative photon emission angle $\theta$ with respect to the incoming positron in the reaction's c.m.\ system, and the angle $\theta_1$ of the meson $P_1$ in the $P_1P_2$ helicity frame. The amplitude depends on the angular momentum $j$  and helicity $\lambda$ of the produced $P_1P_2$  system as well as the helicity of the initial state $m=\pm 1$.
Denoting the photon helicity by $\lambda_\gamma=\pm1$, the amplitude for the radiative process at fixed invariant mass squared $s$ of the $P_1P_2$ system is 
\begin{equation}
\braket{s,\theta,\phi: \lambda,\lambda_\gamma\vert H\vert 1 m}=
 \sqrt{\frac{2j+1}{4\pi}}{\cal D}^{1 *}_{m,\lambda-\lambda_\gamma}(\phi,\theta,0)
d_{\lambda,0}^j(\theta_1)H_{\lambda,\lambda_\gamma}^j(s) \,,
    \label{eq:VtoPPgHel}
\end{equation}
where ${\cal D}^{J}_{\lambda,\lambda'}(\phi,\theta,0)$ and $d_{\lambda,0}^j(\theta_1)$ are Wigner rotation functions. The hadronic system's helicity amplitudes
$H_{\lambda,\lambda_\gamma}^j(s)$ depend on $s$ and helicities $\lambda$ and $\lambda_\gamma$.  The amplitude can also be written in the radiative multipole basis: 
\begin{equation}
H_{\lambda,\lambda_\gamma}^j(s)=\sqrt{\frac{2J_\gamma+1}{4\pi}}
\braket{J_\gamma\ -\lambda_\gamma;j\ \lambda\vert 1\ \lambda\ -\lambda_\gamma}
\frac{\delta_{\lambda_\gamma,1}+\delta_{\lambda_\gamma,-1}P\, (-1)^{J_\gamma-1}}{\sqrt{2}}T_{J_\gamma j}(s) \,,
      \label{eq:VtoPPgMult}  
\end{equation}
where $J_\gamma$ is the photon angular momentum and $P$ the parity of the meson pair.
The amplitude $j=0$ has only one component (E1) and $|H|^2$ from Eq.~\eqref{eq:VtoPPgHel} can be written as
\begin{equation}
|H(\theta,s)|^2\propto (1+\cos^2\theta) |T(s)|^2\,.
\end{equation}
Any amplitude with $j>0$ has three components, e.g., the $2^{++}$ amplitude has components related to E1, M2, and E3 radiative transitions.

\subsubsection[Radiative $\phi$ decays]{\boldmath Radiative $\phi$ decays}
First results for the $\phi$ radiative decays into pairs of pseudoscalar mesons were obtained at the SND and CMD-2 experiments: observations of $\phi\to \pi^0 \pi^0 \gamma$~\cite{Aulchenko:1998xy},  $\phi\to \pi^+ \pi^- \gamma$~\cite{Akhmetshin:1999dh}, and evidence for $\phi\to \eta \pi^0 \gamma$~\cite{Achasov:1998cc}. Further studies from the Novosibirsk experiments include Refs.~\cite{Achasov:2000ym,Akhmetshin:1999di}.
Here we focus on the most recent high-statistics data from KLOE. 

\begin{figure}[t]
    \centering
    \includegraphics[width=0.48\textwidth]{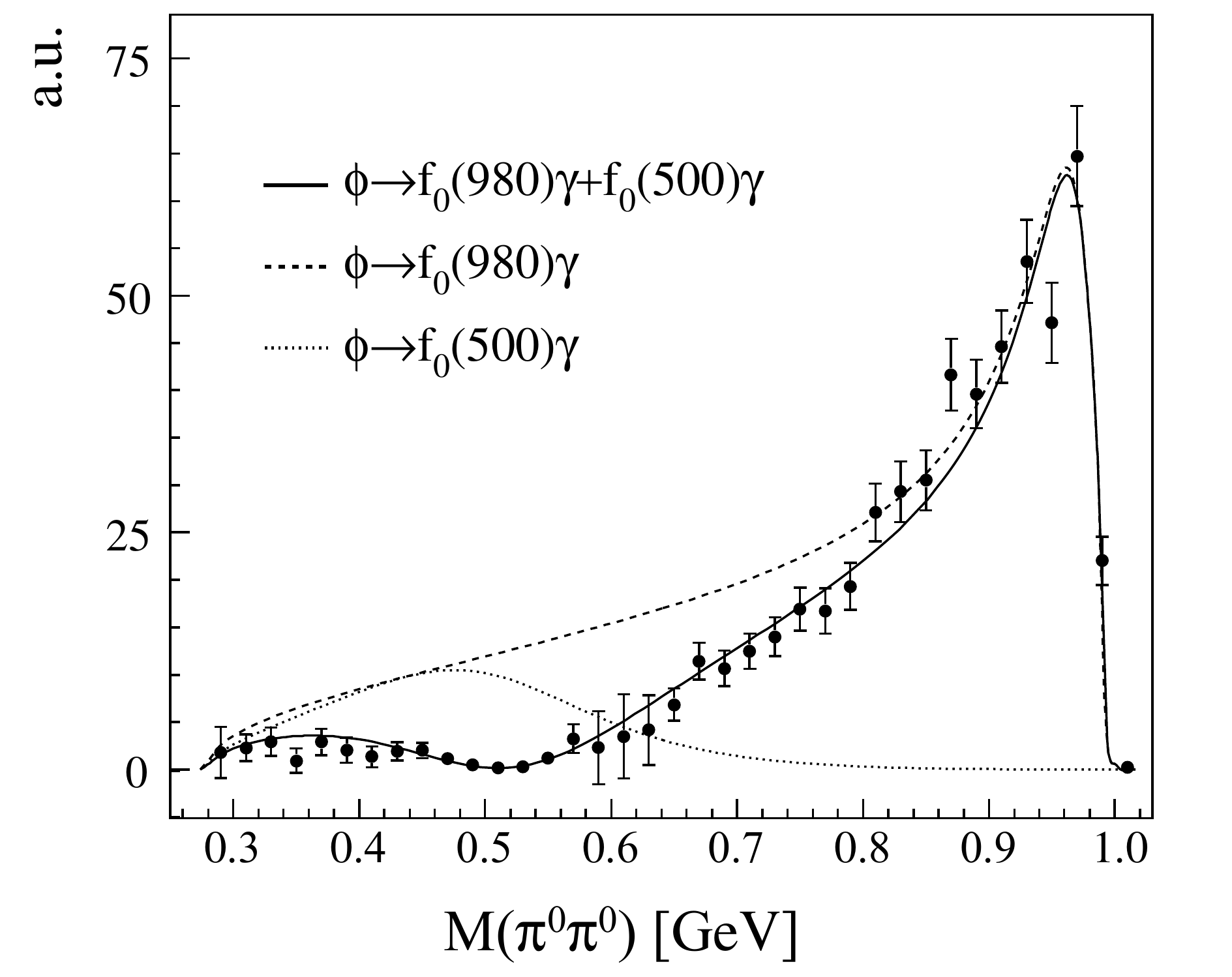}\put(-200,180){\Large\bf  (a)} \hfill
    \includegraphics[width=0.51\textwidth]{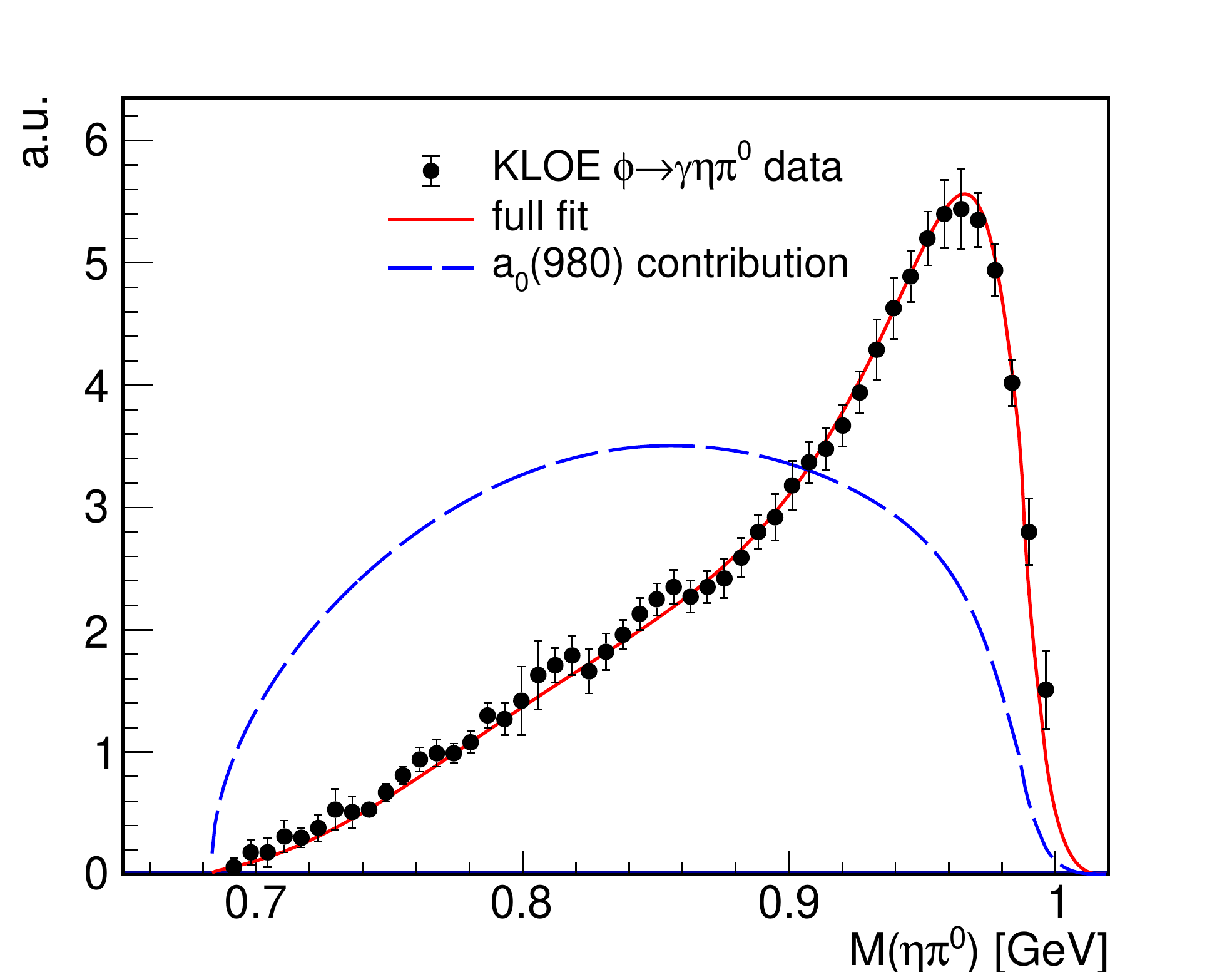}\put(-220,170){\Large\bf  (b)}
    \caption[Radiative decays of $\phi$ to $\pi^0\pi^0$ and $\eta\pi^0$ from KLOE]{(a) $M(\pi^0\pi^0)$ spectrum from $\phi\to\po\po\gamma$~\cite{Aloisio:2002bt}. (b) $M(\eta\pi^0)$ spectrum from $\phi\to\eta\po\gamma$~\cite{Ambrosino:2009py}. Both are unfolded and acceptance corrected distributions. The curves represent interfering contributions for the scalar amplitudes with structureless $\phi$--scalar-resonance--photon couplings~\cite{Isidori:2006we}.
    \label{fig:KLOEp0p0g}}
\end{figure}
The process $\phi\to\pi^0\pi^0\gamma$ was studied at KLOE~\cite{Aloisio:2002bt} using a signal data sample of  $2438(61)$ events. Later, a full Dalitz plot analysis was performed using a much larger data sample of 128530(660) events~\cite{Ambrosino:2006hb}. The $M(\pi^0\pi^0)$ spectrum from the first measurement is shown in Fig.~\ref{fig:KLOEp0p0g}(a). 
The dominant contribution to the processes is from the kaon-loop mechanism~\cite{Achasov:1987ts,Bramon:1992ki,Achasov:1997ih,Achasov:2005hm}, where the main ingredients are the large direct coupling of the $\phi$ to the charged-kaon pair together with final-state radiation and the strong $K^+K^-\to\pi\pi$ rescattering. However, the decay distributions can also be described assuming structureless couplings of the $\phi$ meson to a scalar resonance and a photon~\cite{Isidori:2006we}, as the coherent sum of the $\phi\to f_0(500)\gamma$ and $\phi\to f_0(980)\gamma$ processes.  The interference term between $f_0(500)$ and $f_0(980)$ amplitudes is destructive. The contribution of the crossed channel $\phi\to\rho^0\pi^0$ with $\rho^0\to\pi^0\gamma$ is small and can be fixed from the known branching fractions. 

\begin{figure}
    \centering
    \includegraphics[width=0.95\textwidth]{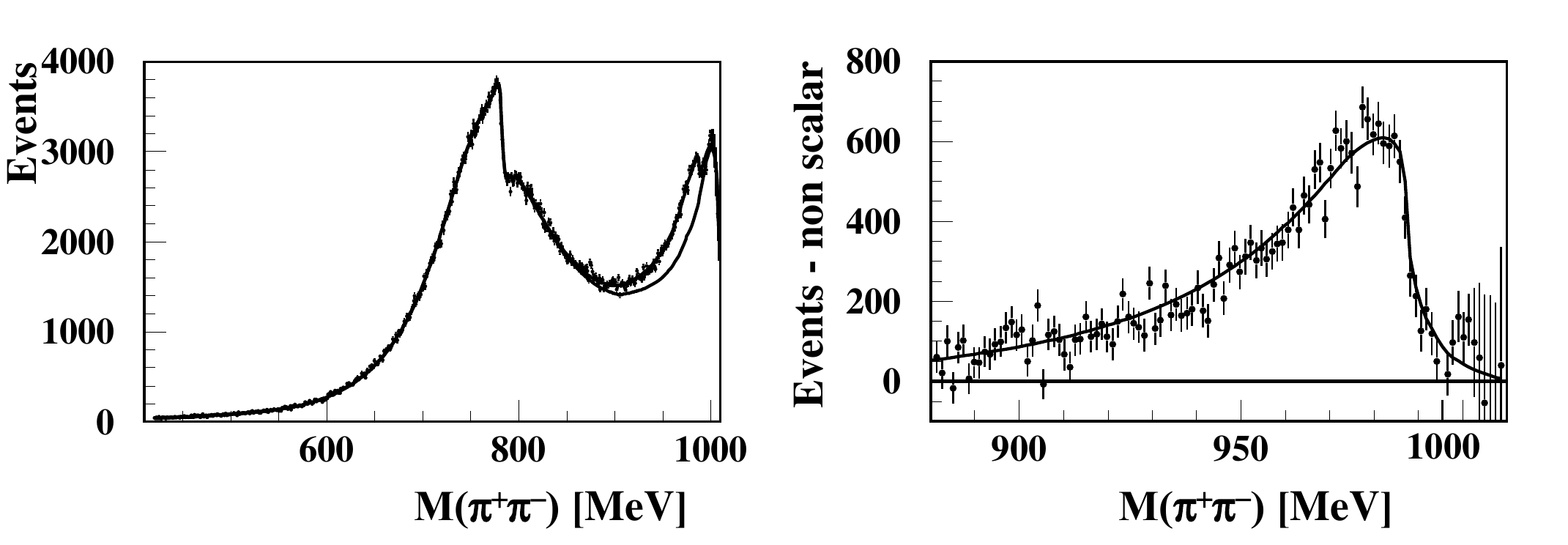}\put(-440,130){\Large\bf  (a)}\put(-190,130){\Large\bf  (b)}
    \caption[Radiative decays of $\phi$ to $\pi^+\pi^-$ from KLOE ]{Invariant-mass distribution 
    from $\phi\to\gamma\pi^+\pi^-$ decays. (a) Data spectrum compared with the fitting spectrum (upper curve following the data points)
and with the estimated nonscalar part therein (lower curve); (b) the fitting function compared to the spectrum
obtained subtracting 
the nonscalar part 
in the $f_0(980)$ region.}
    \label{fig:KLOEf0pippim}
\end{figure}

Studies of the  $e^+e^- \to \pi^+ \pi^- \gamma$ reaction close to the $\phi$ peak require dedicated  methods since it is dominated by the  ISR mechanism, with the photon emitted at small angles with respect to the beams. The analysis at KLOE is based on $6.7\times10^5$  events collected at a c.m.\ energy around the $\phi$ mass~\cite{Ambrosino:2005wk}. The photon is not reconstructed but the ISR contribution is suppressed by requiring the polar angle $\theta_\gamma$ of the $\pi^+\pi^-$ momentum to be greater than $45^\circ$. In the selected sample, still the main contribution is from continuum $e^+e^-\to\pi^{+}\pi^{-}\gamma$ events with ISR photon, but a clear contribution from the intermediate process $\phi\to f_0(980)\gamma$ is observed in the  $M(\pi^+\pi^-)$ spectrum shown in Fig.~\ref{fig:KLOEf0pippim}(a). An additional small contribution comes from the decay $\phi\rightarrow\rho^{\pm}\pi^{\mp}$ with $\rho^{\pm}\rightarrow\pi^{\pm}\gamma$. It contributes to the low-mass region, $400\MeV<M(\pi^+\pi^-)<600\MeV$, with total branching fraction 
$\BR(\phi\rightarrow\rho^{\pm}\pi^{\mp})\times\BR(\rho^{\pm}\to\pi^{\pm}\gamma)\approx4\times 10^{-5}$. Finally, the possibility to observe the decay chain $\phi\rightarrow f_0(500)\gamma\rightarrow\pi^+\pi^-\gamma$ should be considered. Each mechanism contributes to the $M(\pi^+\pi^-)$ invariant-mass spectrum in a characteristic way. 
The $\pi^+\pi^-$ pair is in a $J^{PC}=0^{++}$ state for FSR and the scalar-mesons decays, while ISR produces them in $J^{PC}=1^{--}$ quantum numbers. A sizable interference effect between FSR and $f_0$ decays is expected in the invariant-mass spectrum. On the other hand, interference terms between states with opposite $C$ 
change sign when $\pi^+$ and $\pi^-$ are interchanged and cancel in the $M(\pi^+\pi^-)$ distributions if symmetric cuts on $\theta_{\gamma}$ are used. However, a sizable forward--backward asymmetry for the $\pi^+$ angle $\theta_{\pi^+}$ defined with respect to the electron beam is expected and seen in the data.

The number of events $N_i$ in a bin of the invariant-mass spectrum was described by the function
\begin{equation}
N_i={\cal L}\epsilon_i\left(\Delta{\sigma_{\rm ISR}}+
      \Delta{\sigma_{\rm FSR}}+\Delta{\sigma_{\rho\pi}}+
      \Delta{\sigma_{\rm scal}}\pm
      \Delta{\sigma^{\rm INT}_{\rm scal+FSR}}\right) \,,
\end{equation}
where ${\cal L}$ is the integrated luminosity, $\epsilon_i$ is the selection efficiency, 
and the notation $\Delta\sigma$ represents the differential cross section $\diff\sigma/\diff m$ integrated over the $i$th bin range. The analytic expressions for the first and second terms, ISR and FSR, were obtained in  Ref.~\cite{Achasov:1997gb},
while the $\rho\pi$ term is given in Ref.~\cite{Achasov:1997ih}. 
For the ISR term, $\FV$ was included in the KS parameterization~\cite{Kuhn:1990ad}. 
Figure~\ref{fig:KLOEf0pippim}(b) shows the invariant-mass distribution with nonscalar (ISR, FSR, and $\rho\pi$) contributions subtracted. The data suggests destructive
interference between the scalar--gamma and FSR amplitudes, which damps the low-mass tail
of the $f_0(980)$ and no improvement in the quality of the kaon-loop fit is observed when the $f_0(500)$
is included. 

The decay $\phi \to \eta \pi^0 \gamma$ is dominated by $\phi\to a_0(980) \gamma$. Two analyses of this process were performed by KLOE. The first from 2002~\cite{Aloisio:2002bsa}, based on the same integrated luminosity as for $\phi\to\pi^0\pi^0\gamma$, yielded 802 candidate event. The second analysis with more than an order of magnitude larger statistics~\cite{Ambrosino:2009py} used two $\eta$ decay modes, $\eta\to\gamma\gamma$ with 13270(190) and $\eta\to\pi^+\pi^-\pi^0$ with $3602(70)$ events after background subtraction. The unfolded $M(\eta\pi^0)$ distribution is shown in Fig.~\ref{fig:KLOEp0p0g}(b). The solid curve represents the result of the fit of the $a_0(980)$ contribution interfering with other broad scalars represented using structureless $\phi$--scalar-resonance--photon couplings~\cite{Isidori:2006we}. The dashed line is the $a_0$ meson contribution. 

\subsubsection[Radiative  $J/\psi$ decays]{\boldmath Radiative  $J/\psi$ decays}
The branching fraction for $J/\psi\to\gamma g g$ processes is relatively large, close to 10\%~\cite{Besson:2008pr}, which makes radiative $J/\psi$ decays a perfect place to study hadronic final states produced by two-gluon fusion. The radiative $J/\psi$ decays to pion and kaon pairs were hence primarily motivated by the search for light glueballs~\cite{Becker:1986zt,Augustin:1987fa,Bai:1996wm,Bai:1998tx}.

\begin{figure}[t!]
    \centering
    \includegraphics[width=0.9\textwidth]{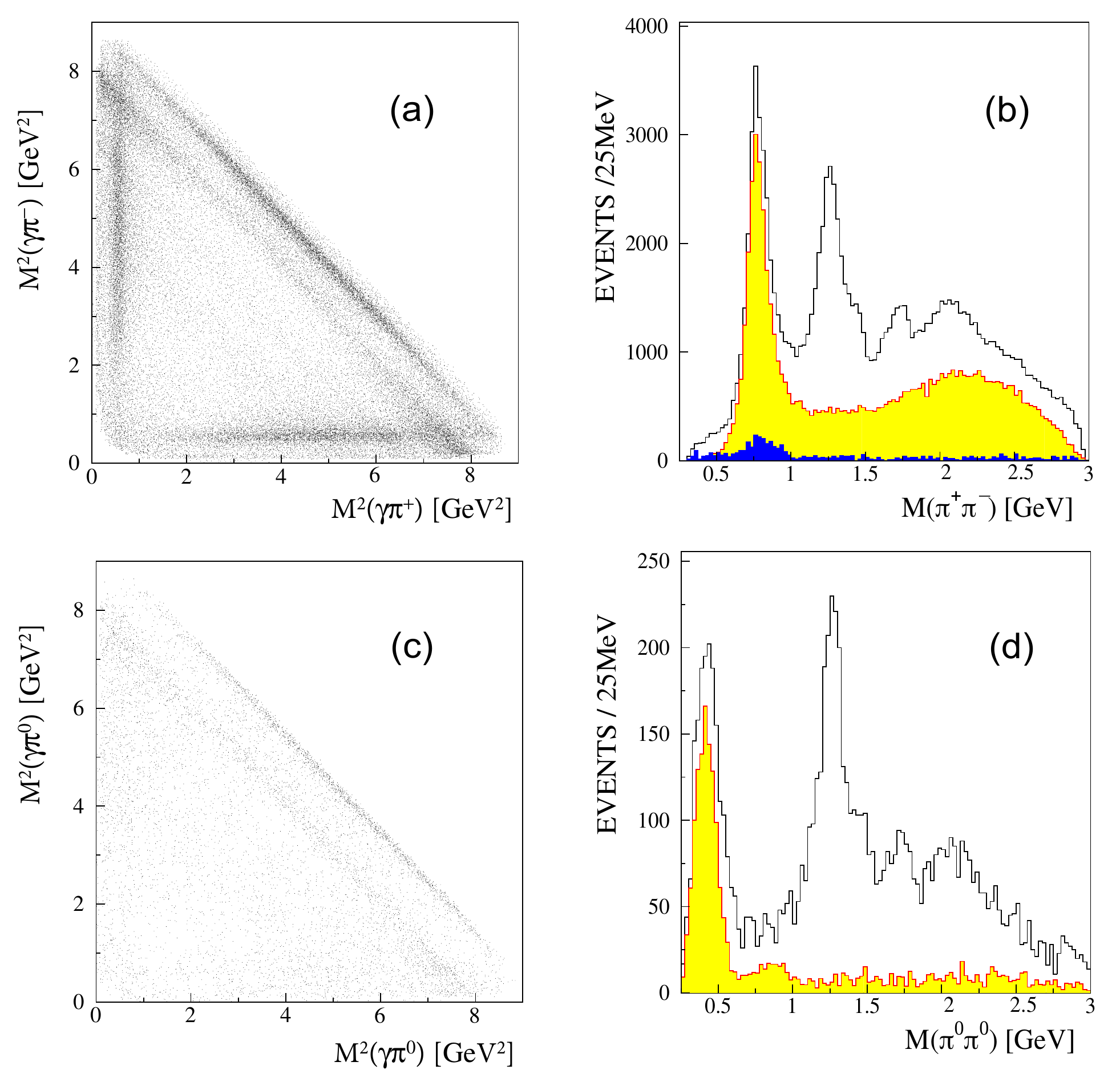}
    \caption[Radiative $J/\psi$ decays to  $\pi^{+}\pi^{-}$ and  $\pi^{0}\pi^{0}$ from the BESII experiment]{Radiative decays of $J/\psi$ to  $\pi^{+}\pi^{-}$ and  $\pi^{0}\pi^{0}$ from the BESII experiment~\cite{Ablikim:2006db}.  We show both Dalitz plots (left) 
    and two-pion invariant-mass distributions (right)
    for $J/\psi\to\pi^{+}\pi^{-}\gamma$ (top) and $J/\psi\to\pi^{0}\pi^{0}\gamma$ (bottom).}
    \label{fig:BESIIPsiPPg}
\end{figure}

Using high-statistics data from BESIII, it is possible to extract information on meson systems in a model-independent way. First attempts of such analyses were performed at BESII, where radiative $J/\psi$ decays to  $\pi^{+}\pi^{-}$ and  $\pi^{0}\pi^{0}$ were studied~\cite{Ablikim:2006db}. The Dalitz plots and invariant-mass spectra are shown in Fig.~\ref{fig:BESIIPsiPPg}.
The prominent features include the $f_{0}(1500)$,  $f_{0}(1710)$, and  $f_{2}(1270)$ resonances. The Dalitz plot for the $\pi^{+}\pi^{-}$ channel shows significant contributions of the $\rho\pi$ background.
For the  $J/\psi $ radiative decays to $K^+K^-$ and $K^0_SK^0_S$, a PWA was carried out in the 1--$2\GeV$ mass range based on the same data sample, using both a mass-independent and a global analysis~\cite{Bai:2003ww}. The process involves production of the $f'_2(1525)$ and $f_0(1710)$ resonances. The latter peaks at a mass of $1740(20)\MeV$ with a width of $166(15)\MeV$. The scalar components dominate the spectrum. A comprehensive study of the two-pseudoscalar meson spectra from radiative $J/\psi$ and $\psi^\prime$ decays was also performed using a $53\pb^{-1}$ data sample  collected by CLEO-c~\cite{Dobbs:2015dwa} at c.m.\ energy $\sqrt{s}=3.686\GeV$. The aim of the analysis was the search for glueball states, and no full amplitude analysis was implemented.

With a much larger $J/\psi$ data sample, the BESIII collaboration has performed two amplitude analyses for the channels $\pi^0\pi^0$~\cite{Ablikim:2015umt}, $4.3\times10^5$ events, and $K_{S}K_{S}$~\cite{Ablikim:2018izx}, $1.6\times10^5$ events.
The neutral channels provide clean samples due to the lack of sizable backgrounds like $J/\psi\to\rho\pi$ for  $\pi^{+}\pi^{-}$~\cite{Ablikim:2006db} and  $J/\psi\rightarrow K^{+}K^{-}\pi^{0}$ for $K^{+}K^{-}$.
Since the initial studies suggest that contributions from $j\ge 4$ amplitudes are negligible, the goal is to extract the scalar and tensor components of the $\pi^0\pi^0$  and $K_{S}K_{S}$ systems as functions of the invariant masses squared $s$, while making minimal assumptions about the properties or number of poles in the amplitudes.  Such model-independent descriptions allow one to integrate these results with other related analyses from complementary reactions, and to develop phenomenological models that can be used to directly fit experimental data to obtain parameters of interest.  The amplitude components for $\pi\pi$ and $K_SK_S$ radiative decays are determined independently for many small ranges of the $s$ variable. This allows one to reconstruct complex amplitudes $T_{J_\gamma j}(s)$ independently for each region as the intensity $|T_{J_\gamma j}|$ and the relative phase. Such a procedure makes minimal assumptions about the $s$-dependence of the $\pi\pi$ and  $K_SK_S$ interactions, reduces potential systematic biases due to the amplitude model, and minimizes the risk for experimental artifacts. These results are easy to combine with those of similar reactions for a more comprehensive study of the light scalar and tensor meson spectra. 
These mass-independent analyses use the following general assumptions: the intensity and the phase difference for each amplitude are continuous functions of $s$, and each set of the $\pi^0\pi^0$ amplitudes with the same quantum numbers is constrained to have the same phase below the $K\bar K$ threshold (the vertex factors associated with the production process are purely real numbers unless additional channels are available). However, rescattering effects may become significant  above the  $K\bar K$ threshold and  generate phase differences between the amplitudes of the same $j^{PC}$.

One should be aware of some drawbacks of this approach. First, in order to resolve the contribution of potentially narrow resonances, a large number of bins is needed and the result is a set of about a thousand parameters that describe the amplitudes, with no single parameter tied to an individual resonance of interest. Second, mathematical ambiguities result in multiple sets of optimal parameters in each bin. With only $j = 0$ and $j = 2$ resonances significant, there are two ambiguous solutions. In general, if one includes $j\ge 4$, the number of ambiguous solutions increases. Finally, in order to make the results manageable for subsequent analyses, an arbitrary assumption of Gaussian errors must be made. 
 
\begin{figure}[t]
    \centering
    \includegraphics[width=0.49\textwidth]{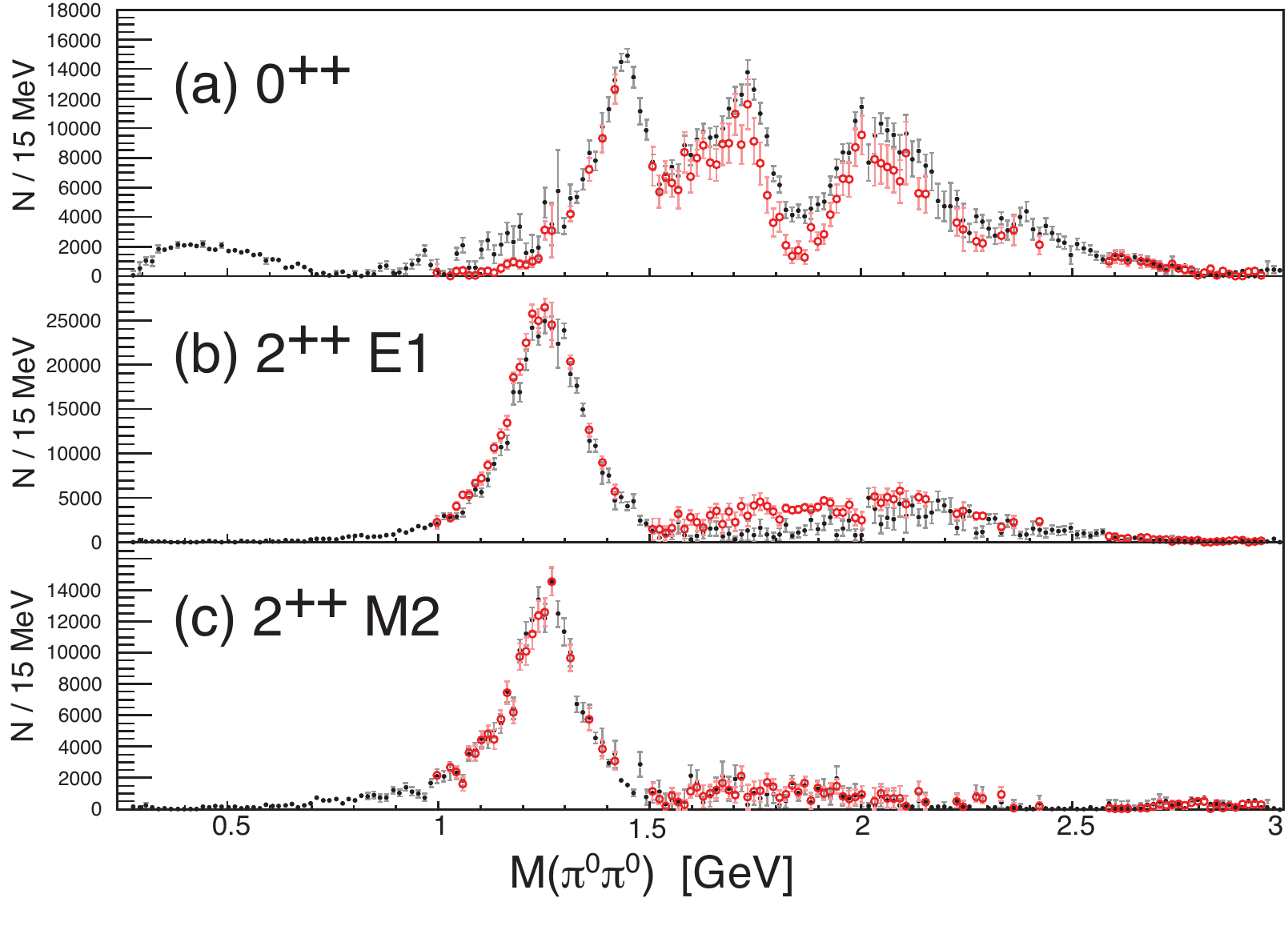}
    \includegraphics[width=0.49\textwidth]{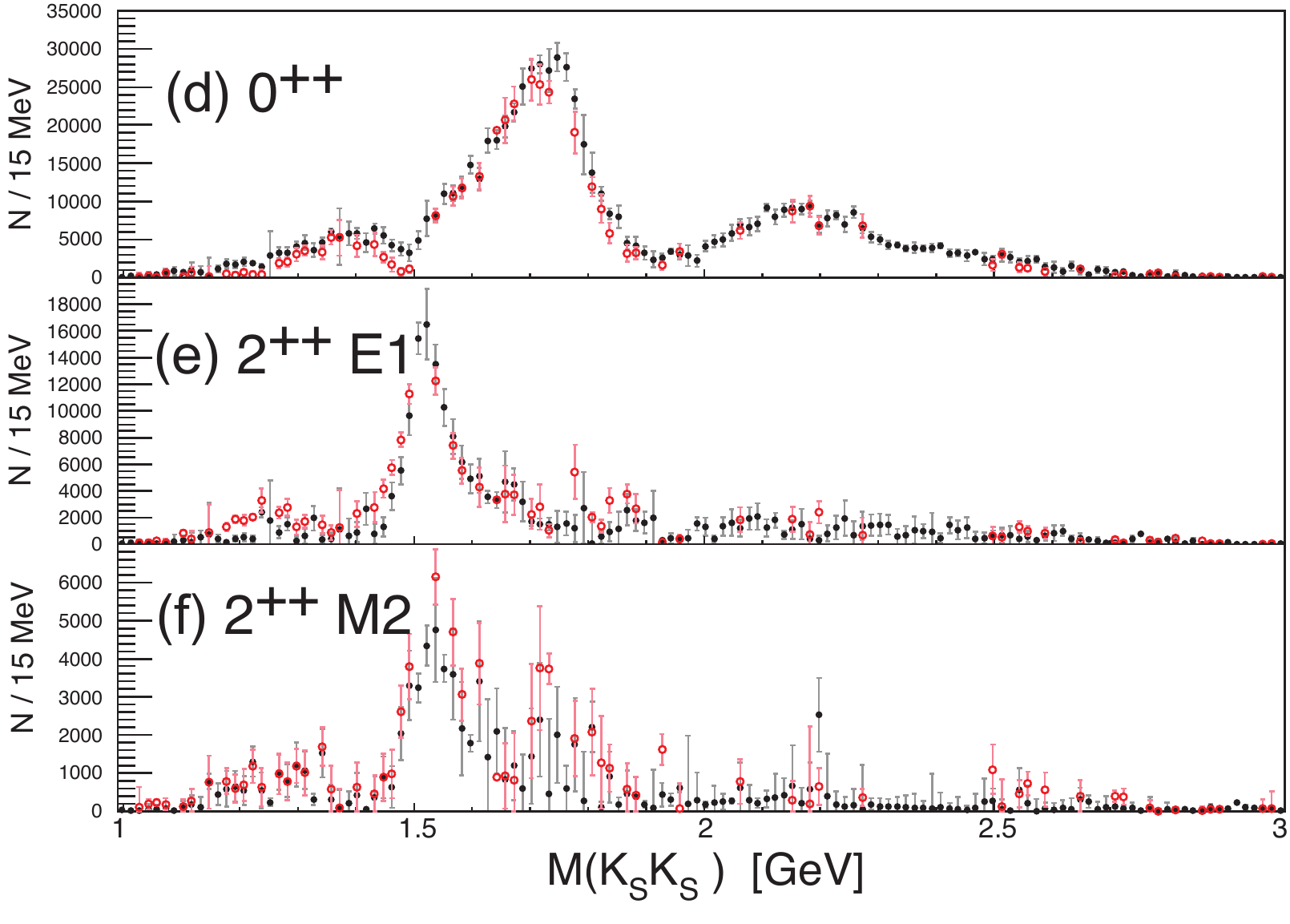}
    \caption[Radiative $J/\psi$ decays to $\pi^0\pi^0$ and $K_SK_S$ at BESIII]{Intensities of the three dominating partial waves $0^{++}$ and $2^{++}(E1,M2)$ for (a)--(c) $J/\psi\to\pi^0\pi^0\gamma$~\cite{Ablikim:2015umt} and (d)--(f)  $J/\psi\to K_S K_S\gamma$~\cite{Ablikim:2018izx} from BESIII. The solid and open markers represent the two possible solutions of the mass-independent fits. }
    \label{fig:VtoPPg1}
\end{figure}

The extracted intensities are shown in Fig.~\ref{fig:VtoPPg1}. The three dominating partial waves $0^{++}$ and $2^{++}(E1,M2)$ are shown for both reactions. The complex amplitudes thus determined include the relative phases with respect to the $2^{++}E1$ wave. For $\pi^0\pi^0$ there are four significant structures: below $0.6\GeV$ likely $f_0(500)$; below $1.5 \GeV$ and near $1.7 \GeV$, where one might expect $f_0(1370)$, $f_0(1500)$, and $f_0(1710)$;  and near $2.0 \GeV$, which could be attributed to the $f_0(2020)$. 
In the $2^{++}$ amplitude, a dominant contribution of a structure consistent with the $f_2(1270)$ is seen.
The remaining broad contribution is significantly different from the one assumed in the previous model-dependent analyses.
The  structures in the $K_SK_S$ channel are consistent with the $f'_2(1525)$ and $f_0(1710)$ resonances. The scalar contribution is stronger than the tensor one.

For  the $K_{S}K_{S}$ system, in addition a model-dependent amplitude analysis was performed where the invariant-mass spectrum was parameterized as a coherent sum of Breit--Wigner line shapes, with the goal of extracting the resonance parameters of intermediate states. The mass-independent results are consistent with those of the model-dependent analysis.

\begin{figure}[t]
    \centering
    \includegraphics[width=0.8\textwidth]{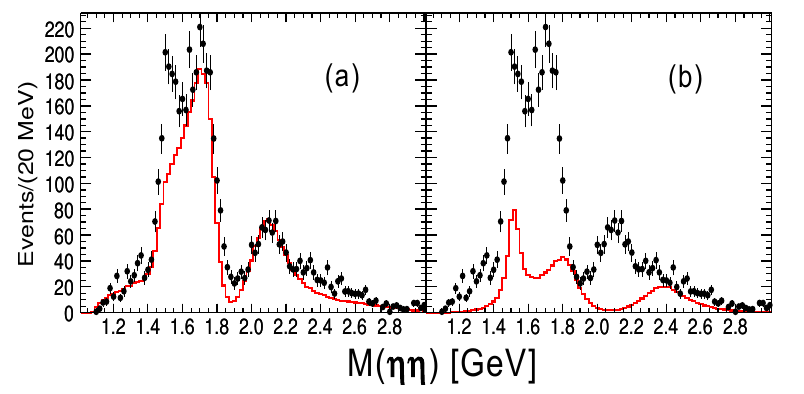}
    \caption[Invariant mass of the $M(\eta\eta)$ system from $J/\psi$ radiative decay]{Invariant mass of the $M(\eta\eta)$ system from the analysis of $J/\psi$ radiative decays at BESIII~\cite{Ablikim:2013hq}. The figure shows the experimental data as well as extracted (a) scalar and (b) tensor amplitudes from the partial-wave analysis.}
    \label{fig:B3etaetag}
\end{figure}

\paragraph{\boldmath $J/\psi \to \gamma \eta \eta$}
A mass-dependent amplitude analysis of radiative $J/\psi$ decays into $\eta\eta$, using a data sample of 5460 candidate events, was performed at BESIII using the relativistic covariant tensor amplitude method~\cite{Ablikim:2013hq}. The scalar and tensor components from the fit to the data are shown in Fig.~\ref{fig:B3etaetag}. 
The scalar
spectrum is described by the contributions from the $f_{0}(1500)$, $f_{0}(1710)$, and $f_{0}(2100)$
states, while the tensor spectrum is dominated by the $f_{2}^\prime(1525)$, $f_{2}(1810)$, and $f_{2}(2340)$ resonances. 

\subsection[$P\to\pi^+\pi^-\gamma$]{\boldmath $P\to\pi^+\pi^-\gamma$}\label{sec:etapipig}
\begin{figure}[t]
    \centering
    \includegraphics[width=0.75\textwidth]{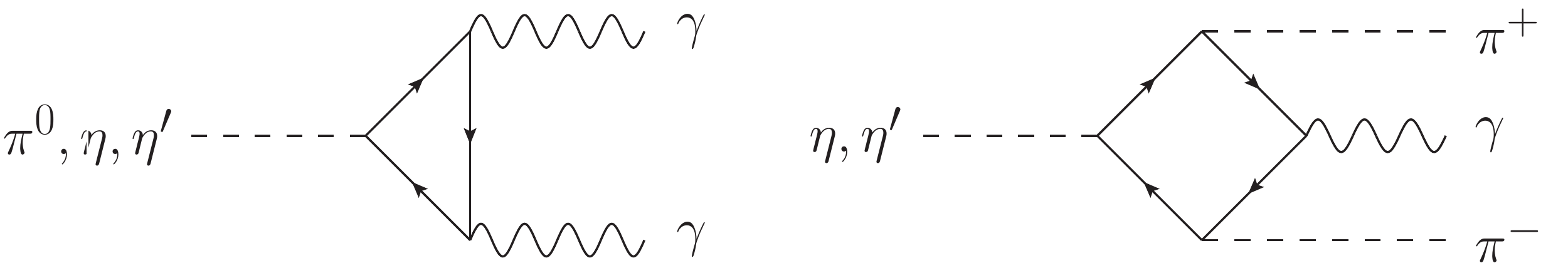}
    \caption[WZW anomalies]{Triangle (left) and box (right) anomaly from the WZW Lagrangian.}
    \label{fig:WZWbox}
\end{figure}

\begin{figure}[t]
\begin{center}
\includegraphics[width=0.99\textwidth]{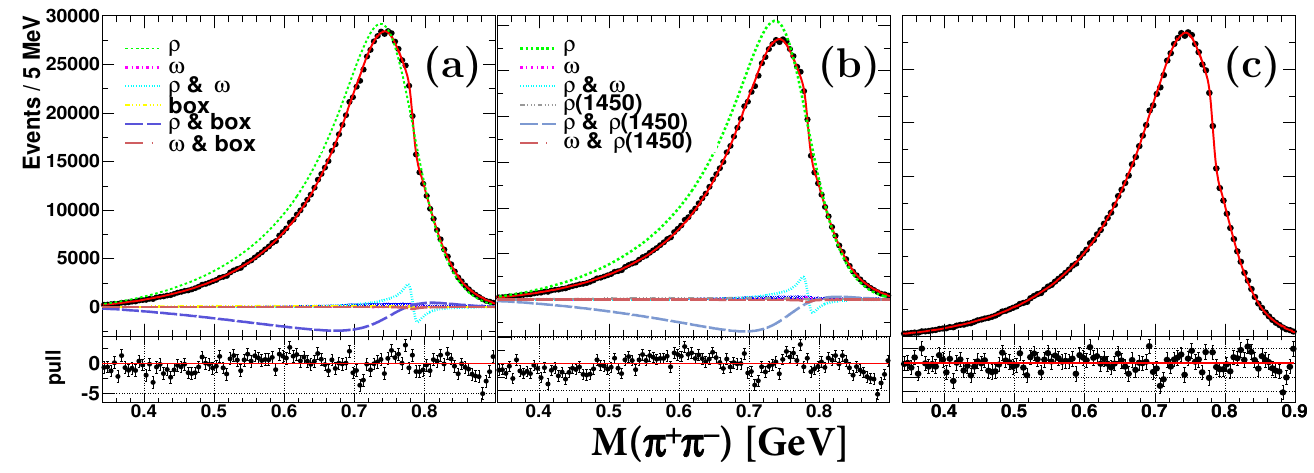}
      \caption[$M(\pip\pim)$ spectrum for $\eta'\to\pi^+\pi^-\gamma$ from BESIII]{Two-pion invariant-mass distribution $M(\pip\pim)$ from  $\eta'\to\pi^+\pi^-\gamma$ BESIII data~\cite{Ablikim:2017fll}. The figure shows 
      model-dependent fits to the $M(\pip\pim)$ spectrum with combinations of (a) $\rho$, $\omega$, and box anomaly, as well as (b)  $\rho$, $\omega$, and $\rho'$.   
      The results of the model-independent fit including $\rho^0$--$\omega$ interference is shown in panel (c).}
  \label{fig:etap-invidata}
\end{center}
\end{figure}

Radiative production of $\pi^+\pi^-$ pairs is one of the main decay modes of the $\eta$ and $\eta'$ mesons, with  branching fractions  of $4.60(16)\%$ and $29.1(5) \%$, respectively. 
Due to its odd intrinsic parity, the decay amplitude for the radiative decay of a pseudoscalar meson $P=\eta,\eta',\eta_c$ 
can be written in terms of a scalar amplitude ${\cal F}(s,t,u)$  as~\cite{Kubis:2015sga}
\begin{equation}
    \left<\pi^+(q_+)\pi^-(q_-)\gamma(k,\epsilon)|H|P\right>=i\epsilon_{\mu\nu\alpha\beta}\epsilon^\mu(k)q_{+}^\nu q_{-}^\alpha k^\beta
    {\cal F}(s,t,u) \,,
\end{equation}
where $s=(q_{+}+ q_{-})^2$,  $t=(q_{-}+ k)^2$, and  $u=(q_{+}+ k)^2$. 
One can express the amplitude in terms of two independent variables: $s$ and the $\pi^+$ polar angle in the $\pi^+\pi^-$ helicity system,
\begin{equation}
    z\equiv\cos\theta_\pi=\frac{t-u}{\sigma_\pi(m_P^2-s)}\,.
\end{equation}
The partial-wave expansion for the $\pi^+\pi^-$ system is given by
\begin{equation}
   {\cal F}(s,t,u)=\sum_{{\rm odd}\ l} P_l'(z)f_l(s)=f_1(s)+\sum_{{\rm odd}\ l\ge3} P_l'(z)f_l(s) \,,
\end{equation}
where only odd partial waves contribute due to $C$ conservation. The decays of $\eta$ and $\eta'$ are completely dominated by the $P$-waves. In the limit of vanishing momenta and quark masses, the amplitude is fixed by the chiral 
anomaly of QCD, which has been incorporated in the  Wess--Zumino--Witten (WZW) effective action~\cite{Wess:1971yu,Witten:1983tw}. 
The relevant contribution of the Lagrangian to $\eta\to\pi^+\pi^-\gamma$ in the chiral limit is the box term describing the direct coupling of three pseudoscalar mesons and a photon, see Fig.~\ref{fig:WZWbox}(right).
On the other hand in the VMD model the $\eta/\eta' \to \pi^+ \pi^- \gamma$ decay is described by the $\rho^0\gamma$ intermediate state~\cite{GellMann:1962jt}:
\begin{equation}
    f_1(s)=c_{\rho}\;{\BW}_{\rho+\omega}^{GS}(s)+c_{\rho'}\;{\BW}_{\rho'}^{GS}(s)+\dots\,.
\end{equation}
In the past, the dipion invariant-mass distribution for $\eta'\to\pi^+\pi^-\gamma$ was studied and interpreted within the VMD model by several experiments~\cite{Bartel:1982qd,Behrend:1982ze,Berger:1984xk,Althoff:1984jq,Aihara:1986sp,Albrecht:1987ed}. The distribution is given as 
\begin{equation}
     \sqrt{s}\ \frac{\diff\Gamma}{\diff {s}}=
     \frac{1}{12 (8\pi m_{\eta'})^3}(m_{\eta'}^2-s)^3(s-4m_\pi^2)^{3/2}|f_1(s)|^2 \,.
\end{equation}
The general conclusion was that the $\rho^0$ peak was shifted by about $+20\MeV$~\cite{Bityukov:1990db}. This
discrepancy was attributed to the WZW box anomaly contribution, which was included as an
extra nonresonant term in the decay amplitude, and it was suggested that fits to the dipion distribution would allow one to determine the fraction of the box contribution~\cite{Benayoun:1992ty}.
Evidence for the box anomaly using such a method was reported in 1997 by the Crystal Barrel experiment~\cite{Abele:1997yi}, using a sample of 7490(180) events, but this observation was not confirmed by the subsequent measurement by the L3 collaboration~\cite{Acciarri:1997yx} with 2123(53) events.  
In the recent BESIII analysis~\cite{Ablikim:2017fll}, a low-background data sample of $9.7\times10^{5}$ $\etap\to\gamma\pi^+\pi^-$ decays candidates was selected. The distribution of the dipion invariant mass, $M(\pi^+\pi^-)$,
is displayed in Fig.~\ref{fig:etap-invidata}.  
The $\rho^0-\omega$ interference is clearly  seen for the first time in this decay. However, the data cannot be described by the ${\BW}_{\rho+\omega}^{GS}(s)$ function alone. The fit result is much improved by including a contact term from the box anomaly, see Fig.~\ref{fig:etap-invidata}(a). However, a data description of similar quality can be obtained by replacing the box anomaly with a $\rho^0(1450)$ contribution, see Fig.~\ref{fig:etap-invidata}(b). 

Ultimately, the notion to separate additive contributions of a contact term/the box anomaly and resonance exchanges is misleading, as the universality of final-state interactions dictates all $P$-wave-produced
pion pairs to undergo rescattering as encoded in an Omn\`es function $\Omega(s)$ built on the corresponding $\pi\pi$ phase shift $\delta_1^1$.
A model-independent description of  $\eta/\eta' \to \pi^+ \pi^- \gamma$ decays therefore uses an approach closely resembling the one introduced in Eq.~\eqref{eq:ModInd} for the pion vector form factor: $f_1(s)=P(s) \Omega(s)$~\cite{Stollenwerk:2011zz}. For the $\eta'$ decay, the process-specific part denoted as $P(s)$ is given by~\cite{Hanhart:2016pcd}
\begin{equation}
    P(s) = A(1 + \kappa s + \lambda s^2)+ \frac{\xi}{m_\omega^2} \cdot {\BW}_{\omega}(s) \,,
\end{equation}
where the normalization $A$ can be matched to the chiral-limit value given by the WZW anomaly.
The reaction-specific term $P(s)$ includes $\rho^0$--$\omega$ mixing in first order in isospin violation as in the e.m.\ pion vector form factor $\FV(s)$. The fit to the BESIII $\eta'\to\pip\pim\gamma$  data is shown  in Fig.~\ref{fig:etap-invidata}(c), yielding $\kappa=0.992(39)\GeV^{-2}$, $\lambda=-0.523(39)\GeV^{-4}$, and  $\xi=6.7(2)\times10^{-3}\GeV^{-1}$.  Since the fit to 
$|\FV(s)|^2$ results rather in linear slope parameters of the order of $\kappa \sim 0.1$, it is clear that the shape of the dipion distribution is different. 
The presence of a significant quadratic term is consistent with calculations including  $\pi^\pm a_2^\mp$ crossed-channel contributions~\cite{Kubis:2015sga}.

\begin{figure}[t]
    \centering
    \includegraphics[width=0.49\textwidth]{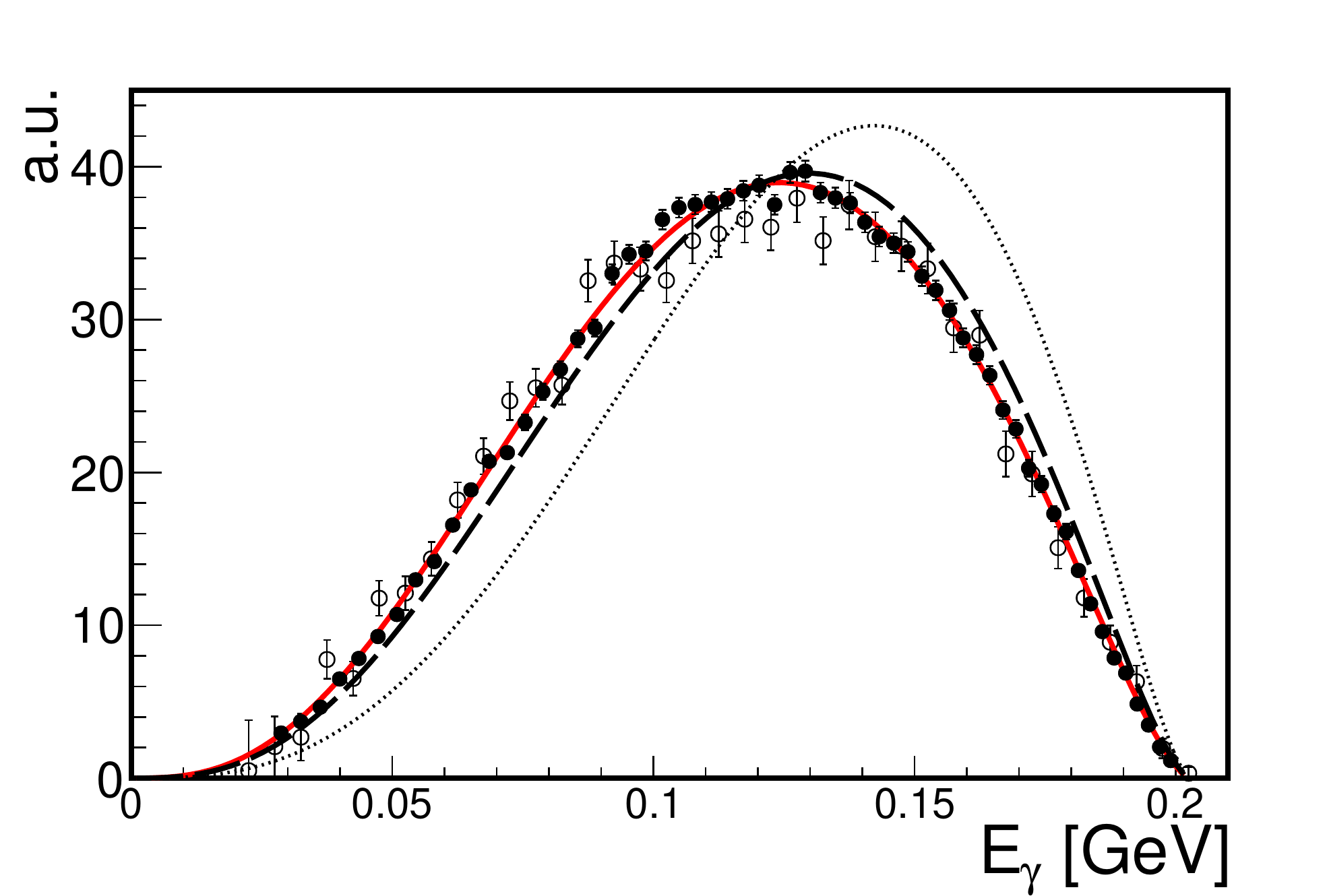}\put(-200,120){\Large\bf  (a)}
    \includegraphics[width=0.49\textwidth]{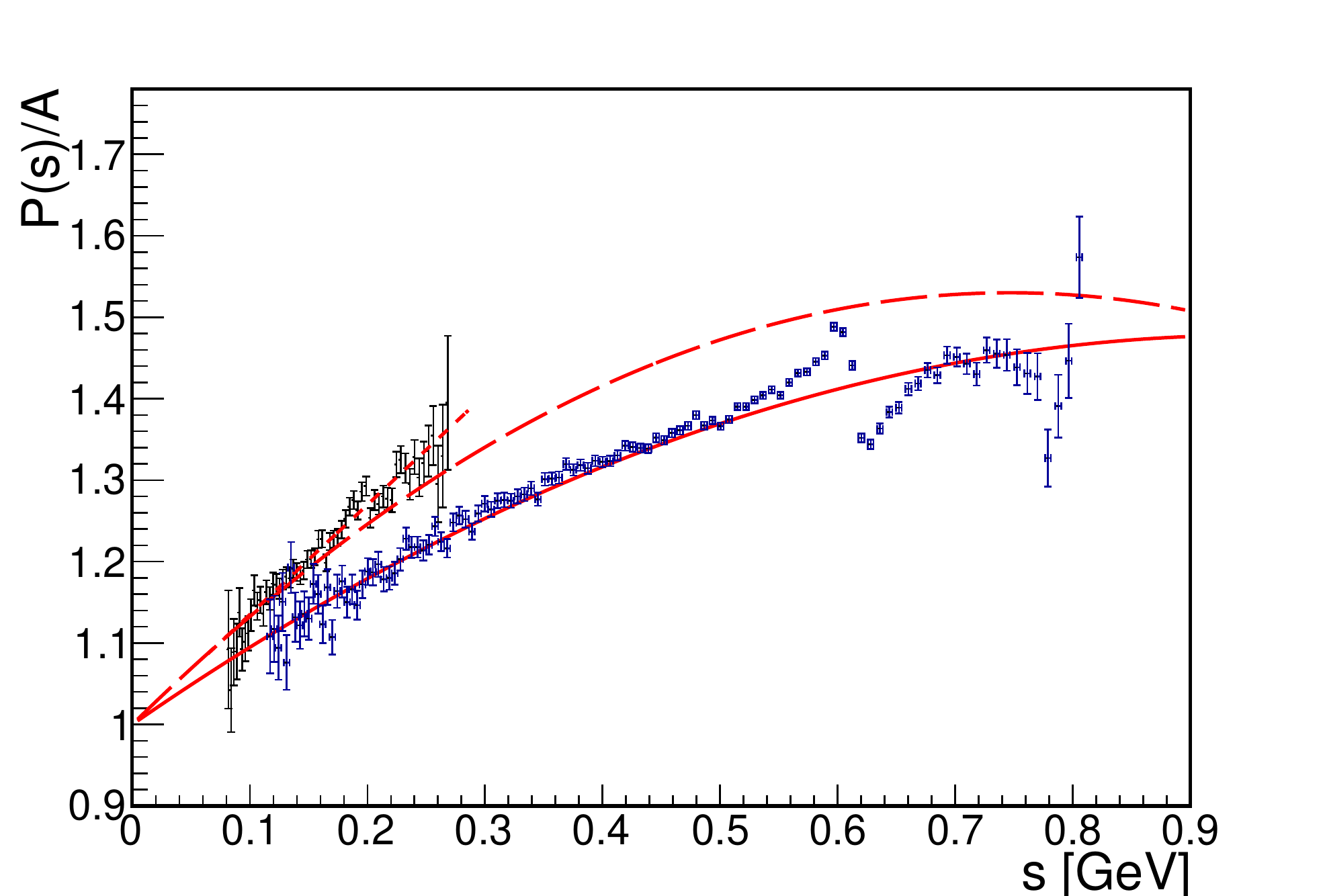}\put(-200,120){\Large\bf  (b)}
    \caption[Differential distribution for $\eta/\eta'\to\pi^+\pi^-\gamma$]{(a) The radiative photon energy distribution for $\eta\to\pi^+\pi^-\gamma$. Black data points are from KLOE~\cite{Babusci:2012ft} and open data points from WASA-at-COSY~\cite{Adlarson:2011xb}. The dotted line is a pure $P$-wave matrix element ${\cal F}(s)\equiv 1$, the dashed line is an Omn\`es function only ($P(s)=1$), and the (red) solid line is  the fit with $\kappa=1.32\GeV^{-2}$. (b) The data points show the extracted reaction-specific part $P(s)/A$ for $\eta\to\pi^+\pi^-\gamma$ from KLOE-2 and  $\eta'\to\pi^+\pi^-\gamma$ from BESIII. The dotted and solid lines are the polynomial part of the fits  $1 + \kappa s + \lambda s^2$. In addition, the dashed line shows calculations including $\pi^\pm a_2^\mp$ crossed-channel contributions~\cite{Kubis:2015sga}.
   }
    \label{fig:Keta2pig}
\end{figure}

The reaction-specific part can be compared to the  $\eta \to \pi^+ \pi^- \gamma$ decay, which has a much smaller kinematic range and necessarily stays closer to the chiral limit, see Fig.~\ref{fig:DPdef}(b). The parameters were determined experimentally by WASA-at-COSY~\cite{Adlarson:2011xb} and KLOE~\cite{Babusci:2012ft}. The differential distributions in Fig.~\ref{fig:Keta2pig}(a) are expressed in terms of the radiative photon energy $E_\gamma$ in the $\eta$ rest frame, 
\begin{equation}
    E_\gamma=\frac{m_\eta^2-s}{2 m_\eta} \,,
\end{equation}
which is a linear function of the Mandelstam variable $s$. The distributions are compared to the pure $P$-wave matrix element and to an unmodified Omn\`es function. 
For the reaction-specific part, only the term linear in $s$ is necessary, $\kappa=1.32(13)\GeV^{-2}$, although crossed-channel contributions will likely induce curvature here, too, that only remains unnoticed due to the smaller kinematic range~\cite{Kubis:2015sga}.
In Ref.~\cite{Hanhart:2013vba} it was hypothesized that the reaction-specific part could be similar for the $\eta$ and $\etap$ decays; the precise data for $\eta$ and $\eta'$ shows, however, that the linear coefficients differ. The reaction-specific parts for both processes are shown in Fig.~\ref{fig:Keta2pig}(b).

\subsection[$D\to P_1P_2\ell^+\nu_\ell$]{\boldmath $D\to P_1P_2\ell^+\nu_\ell$}\label{sec:exDtoPPlv}

\begin{figure}[t]
\centering
\includegraphics{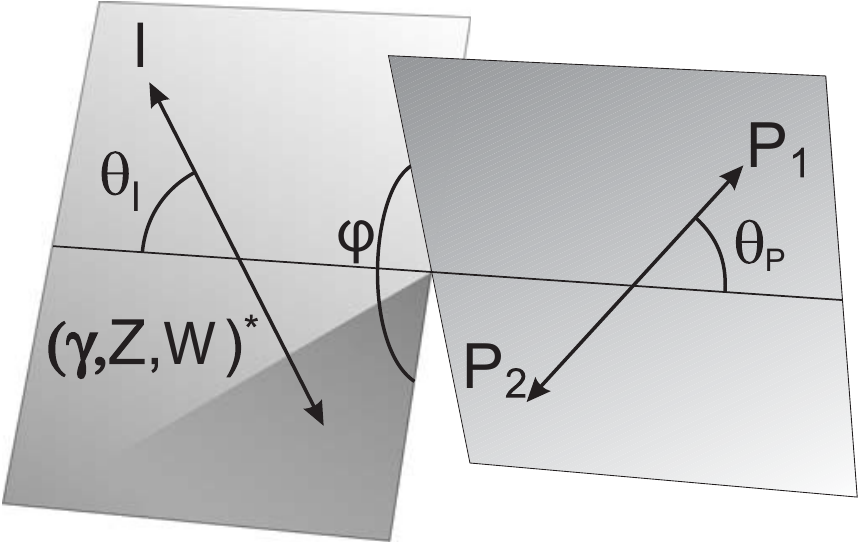}
\caption[Decay plane for semileptonic  weak decays]{Definition of the kinematic variables used for the description of semileptonic  weak decays and e.m.\ conversion decays.}
\label{fig:AsymmIll}
\end{figure}

Semileptonic (SL) decays of charmed mesons are a valuable source of information for both the weak and strong interactions~\cite{Antonelli:2009ws,Artuso:2008vf}.  The partial decay width is proportional to the square of the Cabibbo--Kobayashi--Maskawa (CKM) matrix elements $|V_{cs}|$ or $|V_{cd}|$, and the effect of the
strong interaction in the initial and final hadrons is described by form factors.
The SL decays into a meson pair $D\to P_1P_2 \ell^+ \nu_\ell$ are potentially one of the best places  to study isolated $P_1P_2$ systems at \CT\ factories. The form factors involved are  related to $P_1P_2$  scattering phases and inelasticities. 
\begin{table}[t]
\caption[]{Branching fractions $\BR$ for the main semileptonic $D$ and $D_s$ decays into pseudoscalar meson pairs.  Unless marked otherwise, the branching fractions are taken from Ref.~\cite{PDG}.
\label{tab:DtoPPev}
}
\begin{center}
\renewcommand{\arraystretch}{1.3}
  \begin{tabular}{lrll}
  \toprule
    Decay&$\BR$& & Ref.\\ \midrule
   $D^0\to K^-\pi^0e^+\nu_e$& $1.6\big({}^{+1.3}_{-0.5}\big)\times10^{-2}$&&\\
   $D^0\to \bar K^0\pi^-e^+\nu_e$& $1.44(4)\times10^{-2}$&&\cite{Ablikim:2018lmn}\\
   $D^0\to \pi^0\pi^-e^+\nu_e$& $1.45(7)\times10^{-3}$&&\cite{Ablikim:2018qzz}\\
   \quad $D^0\to \rho^-e^+\nu_e$& $1.45(7)\times10^{-3}$&&\cite{Ablikim:2018qzz}\\ \midrule
   $D^+\to K^-\pi^+e^+\nu_e$& $4.02(18)\times10^{-2}$&&\\
    $D^+\to K^-\pi^+\mu^+\nu_\mu$& $3.65(34)\times10^{-2}$&&\\
    $D^+\to \eta\pi^0e^+\nu_e$& $1.7\big({}^{+0.8}_{-0.7}\big)\times10^{-4}$&&\\
    $D^+\to \pi^+\pi^-e^+\nu_e$& $2.45(10)\times10^{-3}$&&\cite{Ablikim:2018qzz}\\
    \quad $D^+\to \rho^0e^+\nu_e$& $1.86(9)\times10^{-3}$&&\cite{Ablikim:2018qzz}\\
   \bottomrule
  \end{tabular}
\renewcommand{\arraystretch}{1.0}
\end{center}
\end{table}
As seen in Table~\ref{tab:DtoPPev}, there is a variety of possible two-pseudoscalar-meson systems with relatively large branching ratios, allowing experiments to collect sufficient data samples. We have presented the double-tagging technique and examples of the available data sets at BESIII in Sec.~\ref{sec:production}.
The matrix elements of $D\to P_1P_2 \ell^+ \nu_\ell$ have the following form: 
\begin{equation}
 \bra{P_1P_2} \bar q \gamma_\mu(1-\gamma_5) c\ket{D} \bar u_\nu\gamma^\mu(1-\gamma_5)v_\ell\,=
 \sum_J\braket{P_1P_2|X^J}\bra{X^J} \bar q \gamma_\mu (1-\gamma_5)c\ket{D}   
 \bar u_\nu\gamma^\mu(1-\gamma_5)v_\ell\,.
\end{equation}
Compared to strong and electromagnetic processes, there is no background from potential on-shell crossed channels. The goal is to extract the final-state interaction term $\braket{P_1P_2|X^J}$, possibly in a model independent way, in the range of the accessible invariant masses squared $s$. 
However, one should understand the hadronic initial-state interaction (ISI) term, which depends on  the invariant mass squared $q^2$ of the $\ell^+\nu_\ell$ system.
The remaining three kinematic variables describing the decay are  the angle between $P_1$ and the $D$ direction in the $X$ rest frame ($\theta_P$),  the angle between the $\nu_{\ell}$ and the $D$ direction in the $\ell^+\nu_\ell$ rest frame ($\theta_\ell$), and the angle between the two decay planes ($\phi$); see Fig.~\ref{fig:AsymmIll}. The nontrivial dynamics depends only on three kinematic variables $s$, $q^2$, and $\theta_{P}$.

The formalism for the fully differential decay distribution is based on the description of $K\to\pi\pi\ell\nu_\ell$ ($K_{\ell 4}$) decays by Pais and Treiman~\cite{Pais:1968zza}. 
The hadronic current in a semileptonic decay into a vector state $X^1\to P_1P_2$ 
involves four form factors $V$, $A_0$, $A_1$, and $A_2$~\cite{Wirbel:1985ji}:
\begin{align}
\bra{X^1(p_X,\varepsilon)} \bar q \gamma_\mu c\ket{D} &=
2\frac{V(q^2)}{m_D+m_{X}} \epsilon_{\mu\nu\alpha\beta}
p_D^\nu p_{X}^\alpha \varepsilon^{\beta},
\nonumber \\
\langle X^1(p_X,\varepsilon)| \bar q \gamma_\mu \gamma_5 c | D \rangle &=
i \left(m_D + m_{X}\right)
\left(\varepsilon_\mu - \frac{\varepsilon\cdot q}{q^2} q_\mu \right) A_1(q^2)
\\
&- i \frac{\varepsilon\cdot q}{m_D+m_{X}}
\left((p_D+p_X)^\mu-\frac{m_D^2-m_{X}^2}{q^2} q^\mu \right) A_2(q^2)
\nonumber \\
&+ 2i m_X \frac{\varepsilon\cdot q}{q^2} q_\mu ~A_0(q^2),
\nonumber
\end{align}
where $\varepsilon$ is the polarization of the intermediate state $X^1$. The vector form factor $V$ is dominated by vector-meson-resonance exchanges, $A_0$ is dominated by pseudoscalar-meson exchanges, and $A_1$ and $A_2$ are dominated by axial-meson exchanges. 
For decays into a scalar state $X^0$ one has 
\begin{align}
 \bra{X^0(p_X)} \bar q \gamma_\mu \gamma_5 c\ket{D} =   -iF_+(q^2)\left((p_D+p_X)^\mu-\frac{m_D^2-m_{X}^2}{q^2} q^\mu \right)-iF_0(q^2)\frac{m_D^2-m_{X}^2}{q^2} q^\mu \,.
\end{align}
The differential decay width of $D\to P_1P_2 e^+\nu_{e}$ can be expressed as~\cite{Lee:1992ih} 
\begin{equation}
\diff^5\Gamma=\frac{G^2_F|V_{cq}|^2}{(4\pi)^6m^2_{D}}p_X\sigma \,{\cal I}(s, q^2, \theta_{P}, \theta_\ell, \phi) \diff s\, \diff q^2\, \diff\cos\theta_{P}\, \diff\cos\theta_\ell\, \diff\phi \,,
\end{equation}
where $\sigma=2p^{*}/m_X$, $p_X$ is the momentum of the $X$ system in the $D$ rest frame, and $p^*$ is the momentum of $P_1$ in the $X$ rest frame.
The dependence of the decay density $\mathcal{I}$ for the cases with electrons/positrons, i.e., neglecting terms with the lepton mass, is given by
\begin{align}
\mathcal{I}&=\mathcal{I}_1
+\mathcal{I}_2\,{\cos 2}\theta_\ell
+\mathcal{I}_3\,{\sin}^2\theta_\ell{\cos}2\phi
+\mathcal{I}_4\,{\sin}2\theta_\ell{\cos}\phi  
+\mathcal{I}_5\,{\sin}\theta_\ell{\cos}\phi
\nonumber\\&
+\mathcal{I}_6\,{\cos}\theta_\ell
+\mathcal{I}_7\,{\sin}\theta_\ell{\sin}\phi +\mathcal{I}_8\,{\sin}2\theta_\ell{\sin}\phi
+\mathcal{I}_9\,{\sin}^2\theta_\ell{ \sin}2\phi,
\label{eq:Ifunc}
\end{align}
where the functions $\mathcal{I}_{1,\ldots,9}$ depend on $s$, $q^2$, and $\theta_{P}$ and can be expressed in terms of three form factors, $\mathcal{F}_{1,2,3}$. 
The form factors can be expanded into partial waves including $S$-wave ($\mathcal{F}_{10}$), $P$-wave ($\mathcal{F}_{i1}$), and $D$-wave ($\mathcal{F}_{i2}$), to show their explicit dependencies on $\theta_{P}$.
So far, high-statistics analyses of the semileptonic decays (like $D^+\rightarrow K^+\pi^-e^+\nu_e$)  do not require a $D$-wave component and the form factors can be written as
\begin{equation}
\mathcal{F}_1=\mathcal{F}_{10}+\mathcal{F}_{11}\cos\theta_{P} \,, \qquad
\mathcal{F}_2=\frac{1}{\sqrt{2}}\mathcal{F}_{21} \,, \qquad
\mathcal{F}_3=\frac{1}{\sqrt{2}}\mathcal{F}_{31} \,,
\label{eq:F1}
\end{equation}
where $\mathcal{F}_{11}$, $\mathcal{F}_{21}$, and $\mathcal{F}_{31}$ are related to the helicity basis form factors $H_{0,\pm}(q^2)$~\cite{Richman:1995wm,Lee:1992ih}.
The helicity form factors can in turn be related to the two axial-vector form factors, $A_1(q^2)$ and $A_2(q^2)$, as well as the vector form factor $V(q^2)$.
The $A_{1,2}(q^2)$ and $V(q^2)$ are all taken as the simple pole form
$A_i(q^2)=A_{1,2}(0)/(1-q^2/m^2_A)$ and $V(q^2)=V(0)/(1-q^2/m^2_V)$, with pole masses $m_V=m_{D_s^*(1^-)}=2.1121\GeV$ and $m_A=m_{D_s^*(1^+)}=2.4595\GeV$.
The form factor $A_1(q^2)$ is common to all three helicity amplitudes. Therefore, it is natural to define two form factor ratios as $r_V=V(0)/A_1(0)$ and $r_2=A_2(0)/A_1(0)$
at the dilepton momentum squared $q^2=0$.

The experimental determination of these ISI form factors is performed through the study of the differential decay width $\diff\Gamma /\diff q^2$. Dispersive representations~\cite{Boyd:1997kz} allow one to place constraints on the shapes of the form factors by exploiting their analytic properties, leading to polynomial expansions in certain conformal variables (instead of the Mandelstam variables themselves) that account for the lowest branch cuts.  Specifically for the charm sector, only the $D\to\pi$ form factors have been discussed in detail so far~\cite{Ananthanarayan:2011uc,Grinstein:2015wqa,Caprini:2017ins}.
The most common model-dependent parameterization has been a single-pole form factor, where the pole is the lowest-mass resonance formed by the initial- and final-state hadron. The ISI form factor influences mainly the low $M(P_1P_2)$ region. In addition, there are already first calculations of the  ISI form factors in lattice QCD~\cite{Lubicz:2017syv}. 
A technique developed by the FOCUS experiment allows for  the nonparametric determination of the form factors in $D^+\to K^- \pi^+e^+ \nu_{e}$~\cite{Link:2005dp}, later adopted also by the CLEO-c experiment~\cite{Shepherd:2006tw}, thus providing model-independent input for theoretical interpretation~\cite{Fajfer:2006uy}. 

The SL decays are dominated by $X^1$ states of the meson pair. For instance, the semileptonic decays into $K\pi$ pairs are dominated by the $K^*(892)$, and those into $\pi\pi$ pairs by the $\rho(770)$ resonances. The unique possibility is to study the interference of parity-even amplitudes with these dominant and well-known contributions. 
The FOCUS experiment~\cite{Link:2002ev} was the first to report evidence for an even $K^-\pi^+$ amplitude interfering with the dominant $\bar{K}^{\star 0}$ component in the decay $D^+\to K^-\pi^+\mu^+ \nu_{\mu}$. CLEO-c~\cite{Shepherd:2006tw} has seen the
same effect in $D^+\to K^- \pi^+e^+ \nu_{e}$. These observations open prospects for exclusive charm semileptonic decays as a tool to measure phase-shift differences between the $P$- and $S$-waves~\cite{Pelaez:2016tgi}.
Below, we give some examples using BESIII analyses that are based on $2.93\fb^{-1}$ of data collected at the c.m.\ energy of $3.773\GeV$. 

\begin{sloppypar}
\paragraph{\boldmath $K\pi$ system: $D^{+} \to K^{-} \pi^+ e^+ \nu_e$ and $D^0\to \bar{K}^0\pi^-e^+\nu_e$} 
Measurements of $\bar{K}\pi$ resonant and nonresonant amplitudes in the decay $D^+\rightarrow K^-\pi^+e^+\nu_e$ have been reported by
the CLEO~\cite{Shepherd:2006tw,Briere:2010zc}, BaBar~\cite{delAmoSanchez:2010fd}, and BESIII~\cite{Ablikim:2015mjo} collaborations. In these studies, a nontrival $S$-wave component is observed along with the dominant $P$-wave one.
The BESIII analysis is based on  a nearly background-free sample of 18262 double-tagged events (see Sec.~\ref{sec:weakKD}). 
\begin{figure}[t]
    \centering
    \includegraphics[width=0.49\textwidth]{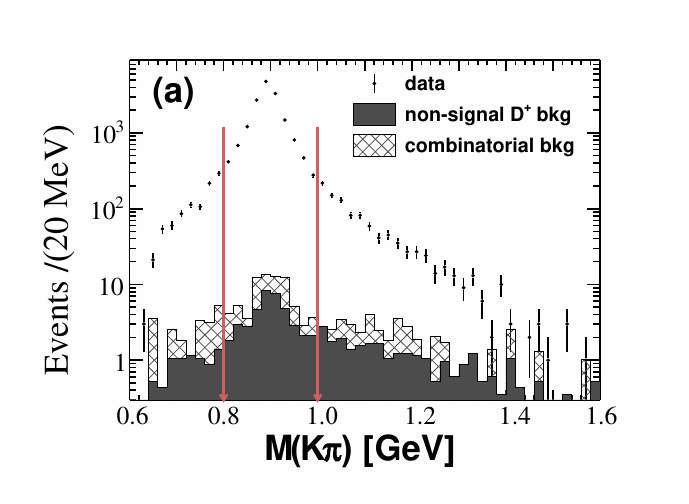}
    \includegraphics[width=0.49\textwidth]{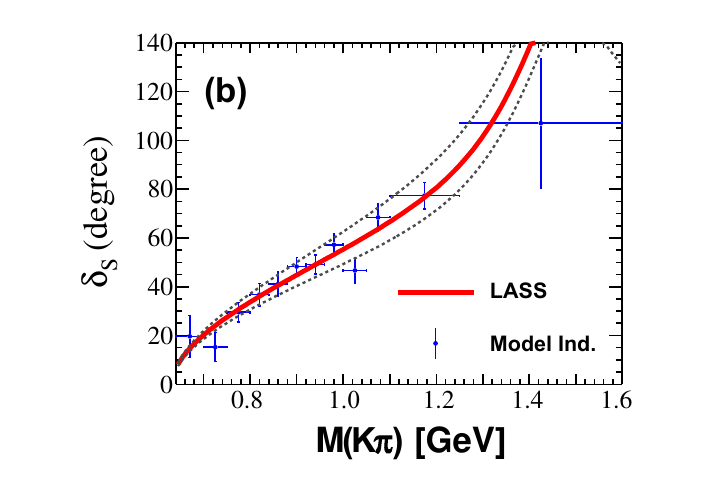}
    \caption[ $D^+\to K^-\pi^+e^+\nu_e$] {Study of the $K\pi$ system at BESIII in  $D^+\to K^-\pi^+e^+\nu_e$~\cite{Ablikim:2015mjo}. (a) Distribution of the selected candidates with respect to the $K\pi$ invariant mass $M({K\pi})$. The range between the arrows corresponds to the $K^*$-dominated region. The dots with error bars represent data, the shadowed histogram and the hatched area show background contributions.   (b) Dependence of the $S$-wave phase on $M({K\pi})$, assuming that the signal is composed of the $S$-wave and the $\bar K^{*}(892)^{0}$. The points with error bars correspond to the model-independent extraction; the solid line corresponds to the result based 
  on the LASS parameterization; the dotted line shows the $1\sigma$ confidence band.}
    \label{fig:DKpi}
\end{figure}
In Fig.~\ref{fig:DKpi}(a) the invariant-mass distribution  $M({K\pi})$
is shown and the $K^*$-dominated region is indicated.  A PWA shows that this dominant component is accompanied by an $S$-wave contribution accounting for $6.1(3)\%$ of the total rate, while other components are negligible.  The helicity form factors of the $\bar K^{*}(892)^{0}$ were studied in a model-independent way.
The $S$-wave phase as a function of $M({K\pi})$, determined 
assuming that the signal is composed of the $S$-wave and the $\bar K^{*}(892)^{0}$, is shown in Fig.~\ref{fig:DKpi}(b). The points with error bars correspond to the model-independent extraction and are compared to a parameterization of the $K\pi$ elastic scattering phase $\delta_0^{1/2}$ from the LASS experiment~\cite{Aston:1987ir} shown in Fig.~\ref{fig:PPphase}(h).
\end{sloppypar}

\begin{figure}[t!]
    \centering
    \includegraphics[width=0.98\textwidth]{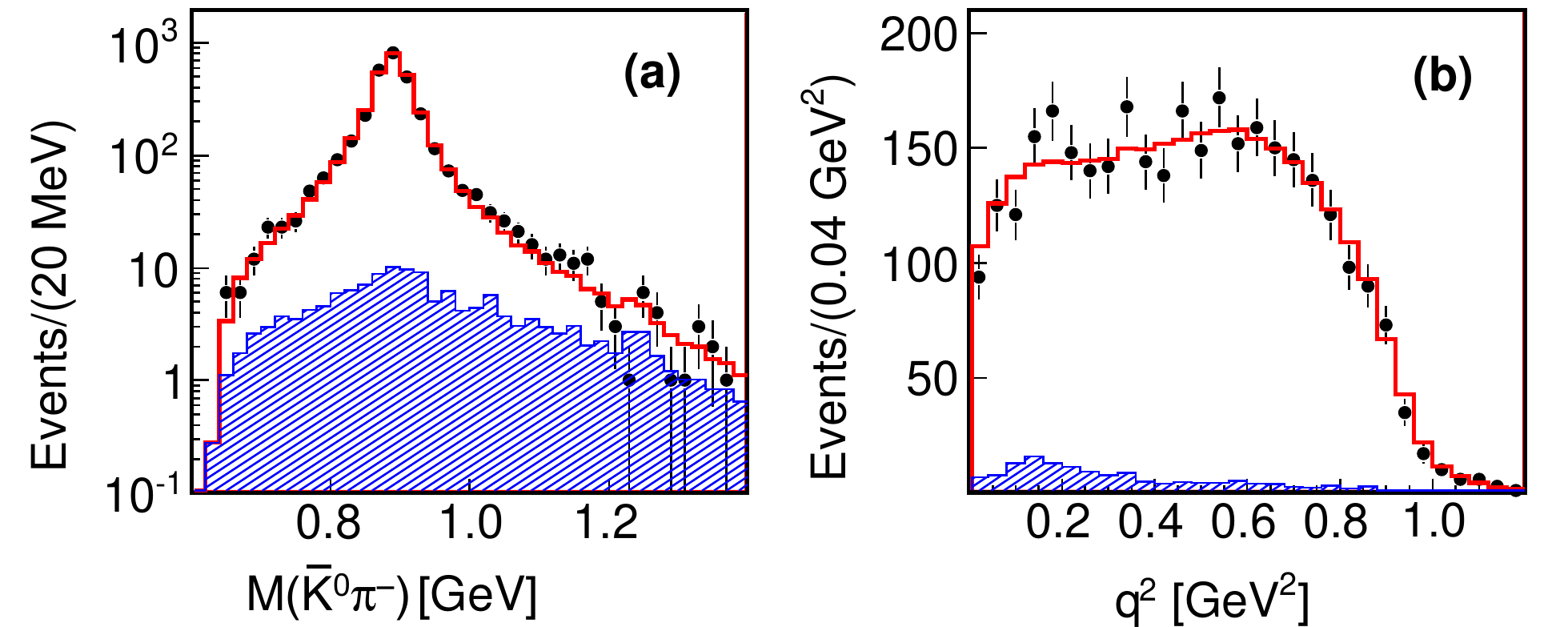}
    \caption[ $D^0\to \bar K^0\pi^-e^+\nu_e$] {Study of the $K\pi$ system at BESIII in  $D^0\to \bar K^0\pi^-e^+\nu_e$~\cite{Ablikim:2018lmn}. (a) Distribution of the selected candidates with respect to the $K\pi$ invariant mass $M({K\pi})$. 
    (b) Distribution with respect to dilepton invariant mass squared $q^2$. 
    The dots with error bars represent data and the hatched area shows background contributions. 
}
    \label{fig:DKpiM}
\end{figure}

A study of the  isospin-related mode $D^0\to \bar{K}^0\pi^-e^+\nu_e$ provides complementary information on the $\bar{K}\pi$ system. Furthermore, the form factors in the $D\rightarrow Ve^+\nu_{e}$ transition, where $V$ refers to a vector meson, have
been measured in decays of $D^+\rightarrow \bar{K}^{*0}e^+\nu_e$ and $D\rightarrow \rho e^+\nu_e$~\cite{CLEO:2011ab}, while no form factor  in $D^0\rightarrow K^{*}(892)^-e^+\nu_e$ has been studied yet.
Therefore, the study of the dynamics in the decay $D^0\rightarrow K^{*}(892)^- e^+\nu_e$ provides essential new information on the family of $D\rightarrow V e^+\nu_e$ decays. A study of the decay $D^0 \rightarrow \bar{K}^0\pi^-e^+\nu_{e}$ was carried out at BESIII using $3112(64)$ double-tagged events~\cite{Ablikim:2018lmn}. This decay is dominated by the vector $K^{*}(892)^-$ contribution, as seen in the $M(K\pi)$ distribution in Fig.~\ref{fig:DKpiM}(a), and the determined $\bar{K}^0\pi^-$ $S$-wave component accounts for $5.5({1.2})\%$ of the total decay rate. 
The hadronic form factor ratios $r_{V}=1.46(7)$ and $r_{2}=0.67(6)$ were extracted from the distribution of the dilepton invariant mass squared shown in Fig.~\ref{fig:DKpiM}(b).

\begin{figure}[t]
    \centering
    \includegraphics[width=0.8\textwidth]{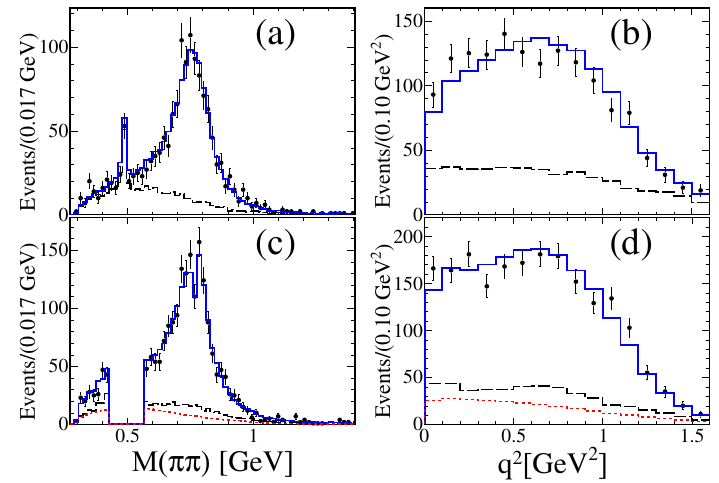}
    \caption[$D\to \pi\pi e^+\nu_e$] {
    Dipion (left) and dilepton (right) invariant-mass distributions for 
    ${D}^0\to\pi^-\pi^0 e^+{\nu}_e$ (top) and
    $D^+\to\pi^-\pi^+ e^+\nu_e$ (bottom), respectively.
    The lines are results of simultaneous PWA fit. 
    The solid lines represent the fits, the dashed lines show the corresponding backgrounds, while the dotted lines in (c) and (d) show the scalar $D^+\to f_0(500)e^+\nu_e$ component.
}
    \label{fig:Dpipi}
\end{figure}

\paragraph{\boldmath $\pi\pi$ system: ${D}^0\to\pi^-\pi^0 e^+{\nu}_e$ and $D^+\to\pi^-\pi^+ e^+\nu_e$}
The two-pion SL decays are Cabibbo-suppressed by $|V_{cd}|^2$.
BESIIII has performed a partial-wave analysis of  $1102(45)$  events of ${D}^0\to\pi^-\pi^0 e^+{\nu}_e$ and $1667(50)$  events of $D^+\to\pi^-\pi^+ e^+\nu_e$~\cite{Ablikim:2018qzz}. 
The pion invariant-mass spectra are dominated by the $\rho$ meson contribution as shown in  Fig.~\ref{fig:Dpipi}(a) and (c). The narrow peak in the $M(\pi^-\pi^0)$ distribution in panel (a) and the removed region in the $M(\pi^+\pi^-)$ distribution in panel (c) correspond to backgrounds with kaons, where the $K_S$ contribution from ${D}^+\to K_S e^+{\nu}_e$ is particularly severe. An interesting feature is a hint at $\rho^0$--$\omega$ interference seen as a narrow peak in  the $M(\pi^+\pi^-)$ distribution close to the $\omega$ mass. The interference pattern is completely different from what we have seen in $\FV$, which can be understood from the different isospin structure of the transition that has the pion pair emerge effectively from a $\bar d\gamma_\mu d$ source; cf.\ the discussion in Ref.~\cite{Daub:2015xja} for a related case. However, the most important conclusion of this analysis is the observation of a large $\pi^+\pi^-$ $S$-wave contribution of $26(2)$\% in the  $D^+\to\pi^-\pi^+ e^+\nu_e$ decay, which is required to describe the data.  
The corresponding dilepton spectra related to the form factors are shown in Fig.~\ref{fig:Dpipi}(b) and (d).

\begin{figure}[t]
    \centering
    \includegraphics[width=0.9\textwidth]{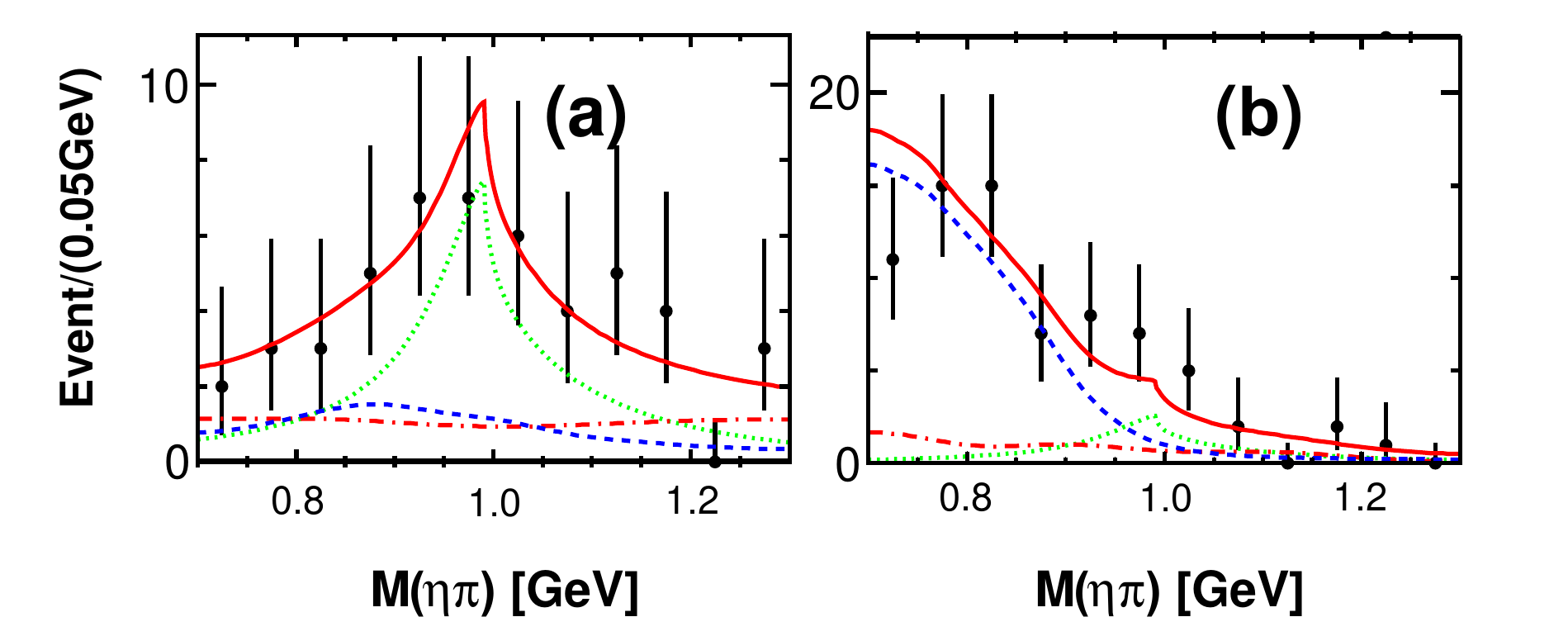}
    \caption[$D\to \eta\pi^0e^+\nu_e$ ] {Invariant mass $M({\eta\pi})$ for (a)  $D^0 \to a_0(980)^- e^+ \nu_e$
    and (b)~$D^+ \to a_0(980)^0 e^+ \nu_e~$. 
    The (red) solid curves are the overall fits, the dashed and dotted--dashed lines 
    denote backgrounds, and the (green) dotted lines show the fitted signal shape.}
    \label{fig:Detapi}
\end{figure}

\paragraph{\boldmath $\eta\pi$ system: $D\to \eta\pi e^+\nu_e$}  
The $D^0 \to \eta\pi^- e^+ \nu_e$ and   $D^+ \to\eta\pi^0 e^+ \nu_e~$ decays were studied at 
BESIII~\cite{Ablikim:2018ffp}. Since  the  isovector $\eta\pi$ system cannot originate from the vector current with positive $G$ parity (as long as isospin is conserved), the reaction allows to study a clean signal of the $a_0(980)$ scalar contribution. The invariant mass $M(\eta\pi)$ spectra are shown in Fig.~\ref{fig:Detapi}.

\subsection[$V\to V' P_1P_2$]{\boldmath $V\to V' P_1P_2$}\label{sec:VtoPPV}
In the context of our review, the  $J/\psi$ and $\phi'$ decays to a narrow vector meson $V'$ and a system of two pseudoscalar mesons $P_1P_2$ are tools to learn about the interaction of the pseudoscalar pair. 
The presence of the narrow vector meson in the final state of the strong decay is used as a spin/isospin/$SU(3)_{\rm flavor}$ filter, and the hope is it can be assumed to be a passive spectator in the reaction.  We expect to be able to write the $V\to V' P_1P_2$
amplitude in an analogous way as Eq.~\eqref{eq:VtoPPg}:
\begin{equation}
    {\cal M}=\sum_X\bra{P_1P_2}H_{2} \ket{X}\bra{X V'}H_{1}\ket{V} \,, \label{eq:VtoPPV}
\end{equation}
where the two steps of the reaction proceed via strong interactions $H_1$, $H_2$.
\begin{figure}
\begin{center}
\includegraphics[width=0.49\textwidth]{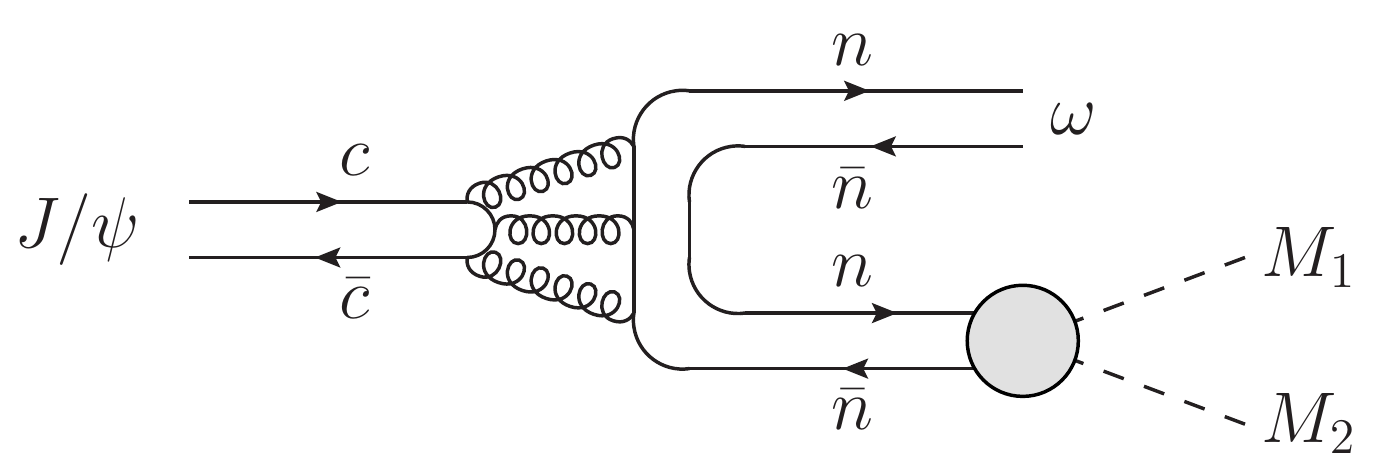}
\put(-110,-20){(a)}
\hfill
\includegraphics[width=0.49\textwidth]{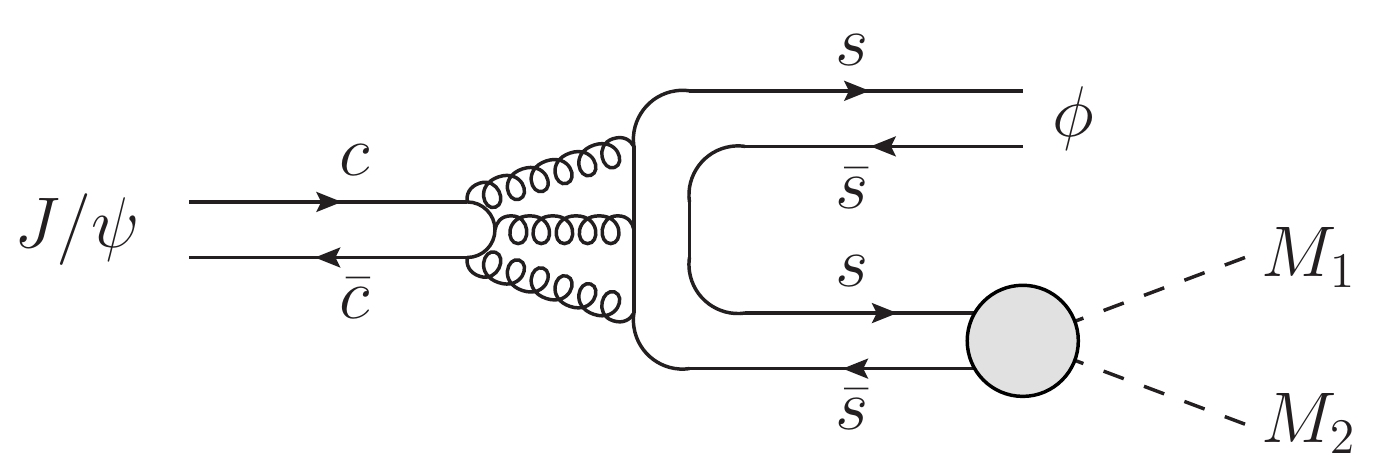} 
\put(-110,-20){(b)}
\caption[Flavor filtering in $V\to V'M_1M_2$ decays]{Production of parity-even meson systems $(M_1M_2)$ in
(a) decay of $J/\psi$ into $\omega(M_1M_2)$ and (b)
$\phi(M_1M_2)$.  The two processes select $\bar nn=(\bar uu+\bar dd)/2$ or $\bar ss$ $(M_1M_2)$ 
meson systems. 
    \label{fig:VtVPPa}}
\end{center}
\end{figure}
In the case of isoscalar vector mesons $V$ and $V'$ the produced   meson systems $X$ will be isoscalar.
Flavor filtering is illustrated in Fig.~\ref{fig:VtVPPa}.  Processes without a significant $s\bar s$ quark contribution in the meson state $(M_1M_2)$ are suppressed in decays to $\phi (M_1M_2)$ due to double OZI suppression, since they require additional disconnected (i.e., producing separate hadrons) quark lines in the final state. This can be seen, e.g., from the comparison of the branching ratios 
\begin{equation}
    \frac{\BR(J/\psi\to\omega\pi^+\pi^-)}{\BR(J/\psi\to\omega K\bar K)}\approx 3.8\,, \qquad
 \frac{\BR(J/\psi\to\phi\pi^+\pi^-)}{\BR(J/\psi\to\phi K\bar K)}\approx0.53\,.
\end{equation}
In some cases the assumption that the vector meson is a spectator is well founded, e.g., the $\phi\pi$ interaction is suppressed by the OZI rule in $J/\psi\to\phi\pi\pi$. In other cases, like $J/\psi\to\phi K\bar K$, the situation is not so clear as we will discuss later in this section.
The high-statistics data of charmonium decays provides a new opportunity to extract information of meson--meson scattering, in particular scalar mesons. Nowadays  the $f_0(500)$ is a well-established example of a resonance with its pole deep in the complex plane, which requires ChPT and dispersive methods to describe its properties~\cite{Pelaez:2015qba}. The $f_0(500)$ state was first observed in $\pi\pi$ scattering and also in some production processes.
It is needed in order to explain the $\pi\pi$ scattering phase shift data.
The observation of clear bumps in charmonia decays at BESII such as in $J/\psi\to\omega\pi^+\pi^-$ 
was an important step towards general acceptance of the $f_0(500)$ resonance.

\begin{table}[t]
      \caption{Branching fractions for $V\to V'P_1P_2$ decays. Unless marked otherwise, the branching fractions are taken from Ref.~\cite{PDG}. \label{tab:VtoVPP}}
\begin{center}
\renewcommand{\arraystretch}{1.3}
  \begin{tabular}{llrll}
  \toprule
    &Final state&$\BR$& & Ref.\\ \midrule
    $J/\psi\to V^0 P_1P_2$&$\omega\pi^0\pi^0$& $3.4(8)\times10^{-3}$&&\\
    &$\omega\pi^+\pi^-$& $7.2(1.0)\times10^{-3}$&&  \\
    &$\omega\pi^0\eta$& $3.4(1.7)\times10^{-4}$&& \cite{Lees:2018dnv} \\
    &$\omega K\bar K$& $1.9(4)\times10^{-3}$&&  \\ \midrule
     &$\phi\pi^0\pi^0$& $5.0(1.0)\times10^{-4}$&&  \\
     &$\phi\pi^+\pi^-$& $9.4(1.5)\times10^{-4}$&&  \\
    &$\phi\eta\eta'$&$2.32(17)\times{10^{-4}}$&&\cite{Ablikim:2018xuz}\\
    &$\phi K\bar K$& $1.77(16)\times10^{-3}$&&  \\ \midrule
    $\psi'\to J/\psi P_1P_2$&$J/\psi\pi^+\pi^+$& $0.3468(30)$&&\\
    &$J/\psi\pi^0\pi^0$& $0.1824(31)$&&\\ 
 \bottomrule
  \end{tabular}
\renewcommand{\arraystretch}{1.0}
  \end{center}
\end{table}

\subsubsection[$\psi'\to J/\psi\pi\pi$]{\boldmath $\psi'\to J/\psi\pi\pi$}
We start from the two $\psi' \to J/\psi\pi\pi$ processes, the main $\psi'$ decay modes, see Table~\ref{tab:VtoVPP}. Here, the assumptions leading to Eq.~\eqref{eq:VtoPPV} should be well fulfilled, since crossed-channel $J/\psi \pi$ rescattering should be very weak, and no resonances are expected in the quark model. 
The advantages of this reaction to study quasi-isolated dipion systems in the low-mass region, $M(\pi\pi)<m_{\psi'}-m_{J/\psi}$, were pointed out already in 1975~\cite{Brown:1975dz,Cahn:1975ts}. 

\begin{figure}[t]
    \centering
    \includegraphics[width=0.49\textwidth]{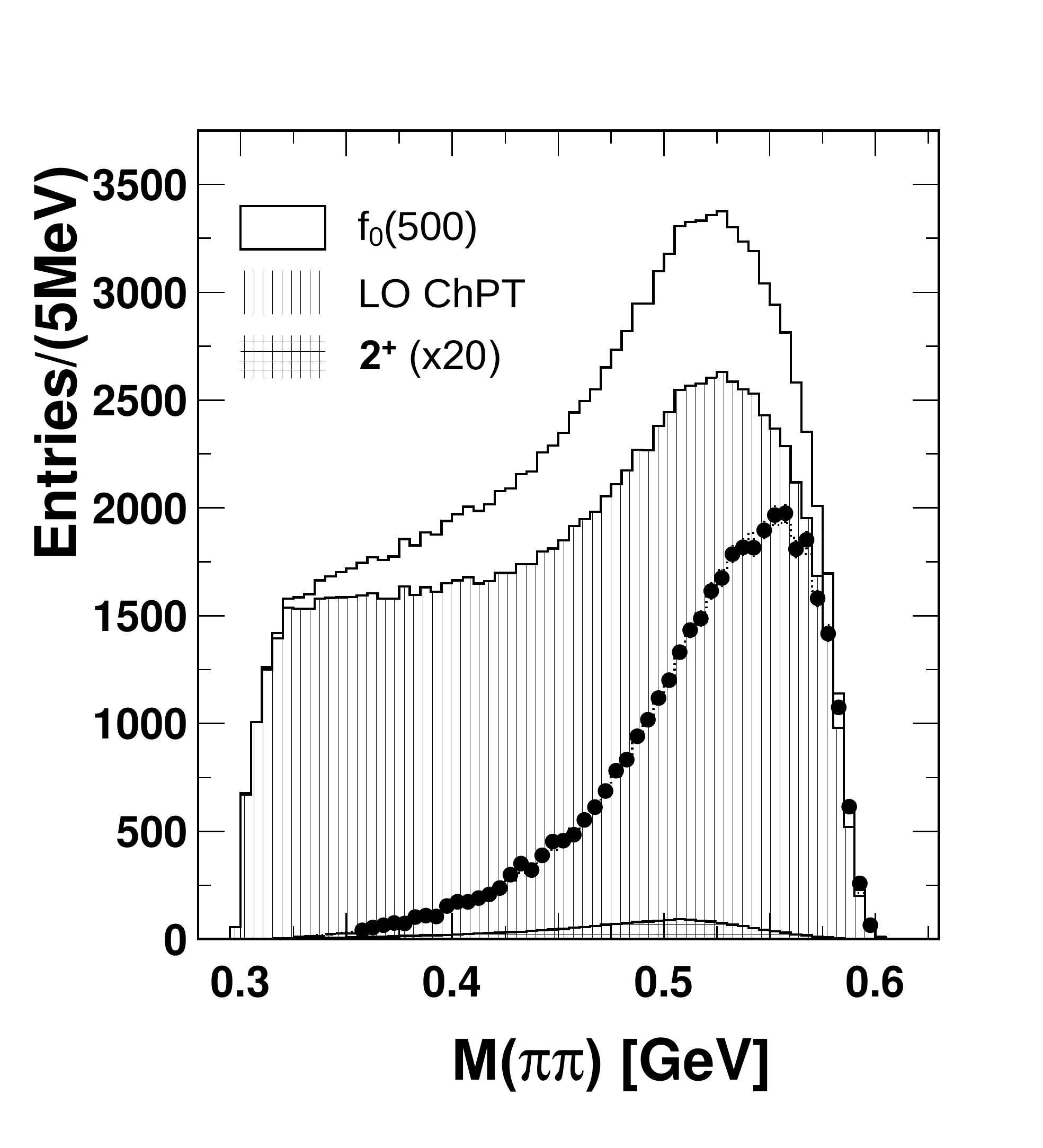}\put(-60,230){\Large\bf  (a)}
    \includegraphics[width=0.49\textwidth]{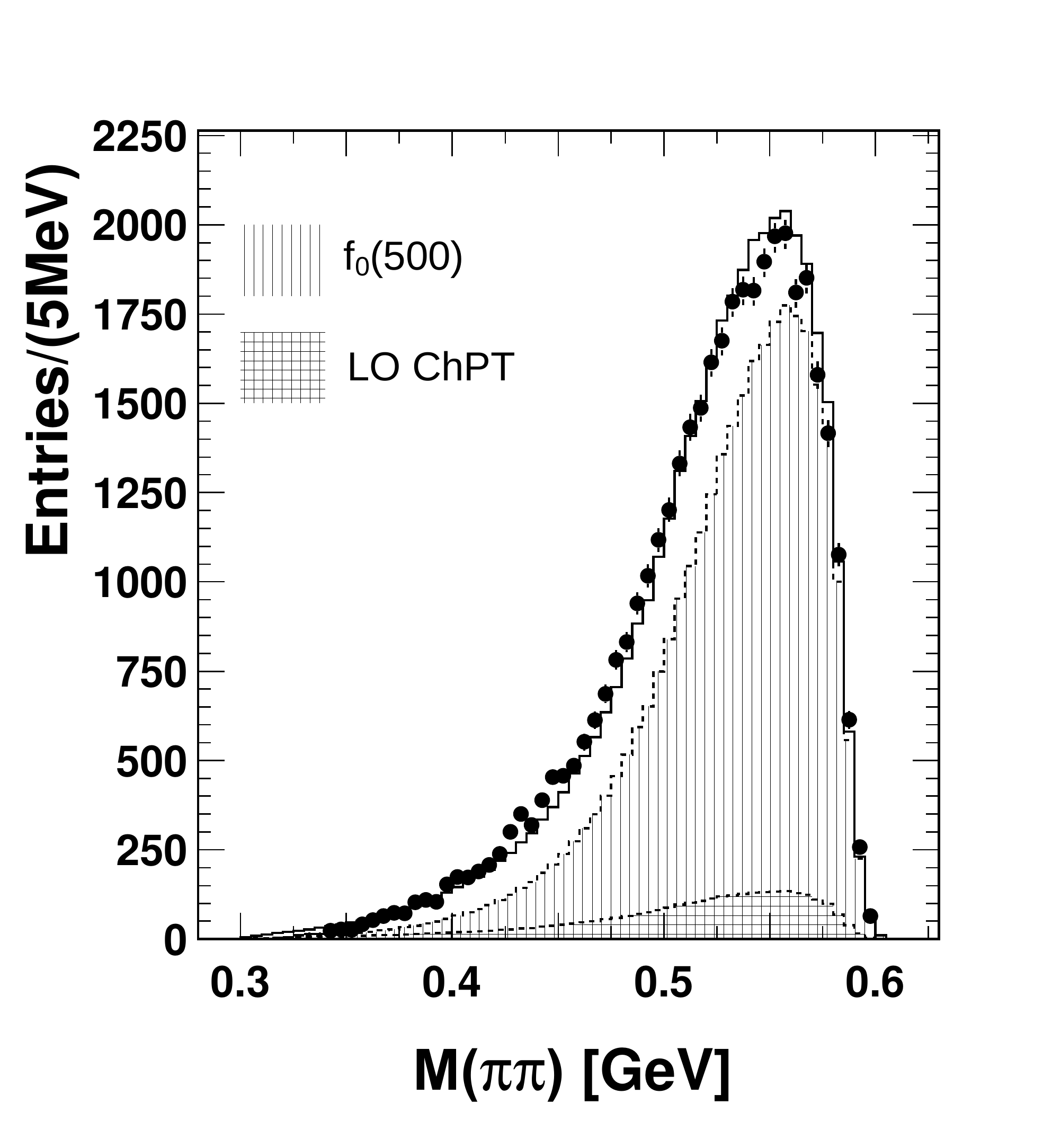}\put(-60,230){\Large\bf  (b)}
    \caption[$M(\pi^+\pi^-)$ distribution from $\psi'\to J/\psi\pi^+\pi^-$]{The $\pi^+\pi^-$ invariant-mass distribution in $\psi'\to J/\psi\pi^+\pi^-$ from  BESII~\cite{Ablikim:2006bz}. Panel (a) shows an isobar model fit, which includes the contributions from $f_0(500)$, $D$-wave term (enlarged by a factor of 20 in the figure), and contact term.
The coherent sum of the contributions describes the data well. Panel (b) shows the invariant mass fitted by the chiral unitarity approach~\cite{Guo:2004dt}. Dots with error bars are data, and the histograms are the fit results.
    \label{fig:Psi2ToJPpp}}
\end{figure}

The dipion invariant-mass distribution in this process is strongly peaked towards its higher end, in contrast to what is expected from phase space alone, and the angular distribution favors an $S$-wave state for the dipion system~\cite{Bai:1999mj,Ablikim:2006bz}. Within a standard PWA approach, three main contributions have been considered: the $f_0(500)$ resonance, a $D$-wave term, and a contact term required by ChPT~\cite{Ishida:2002uf}: $A=A_0+A_2+A_{\rm contact}$. The VPP production vertex is taken as a constant and the amplitudes $A_{0,2}$ are taken from the isobar model.
The $M(\pi\pi)$ distribution from the BESII analysis~\cite{Ablikim:2006bz}, using a clean data sample of 40000 $\psi'\to\pi^+\pi^-J/\psi$ events with  $J/\psi\to\mu^+\mu^-$, is shown in Fig.~\ref{fig:Psi2ToJPpp}(a). It is compared to the results of the PWA with the above ansatz. Four different Breit--Wigner parameterizations for the mass dependence of the $f_0(500)$ resonance were tried. All fit the data well, but the solutions have strong destructive interference with the contact term, especially in the low $\pi\pi$ invariant-mass region. This effect is anticipated by chiral symmetry~\cite{Brown:1975dz,Ishida:2002uf}. The $D$-wave contribution amounts to only 0.3 to 1\%. The extracted $f_0(500)$ pole parameters using different parameterizations are consistent with each other, and the average is $550(95)-i 230(77)$. An alternative fit shown in Fig.~\ref{fig:Psi2ToJPpp}(b) describes the VPP vertex using an effective Lagrangian and includes universal $\pi\pi$ $S$-wave rescattering in the final state~\cite{Guo:2004dt}, thus overcoming the violation of unitarity by na\"ively employing contact term and $f_0(500)$ resonance (as an approximation to rescattering effects) additively. Here, the $f_0(500)$-pole parameters are taken from Ref.~\cite{Oller:1997ti}, where they were determined from the $\pi\pi$ scattering data. 
The contribution of the contact term (LO ChPT) and the interference is small in this approach.
The model does not include a $D$-wave, which might explain the slightly worse fit quality.

\begin{figure}[t!]
    \centering
    \includegraphics[width=0.85\textwidth]{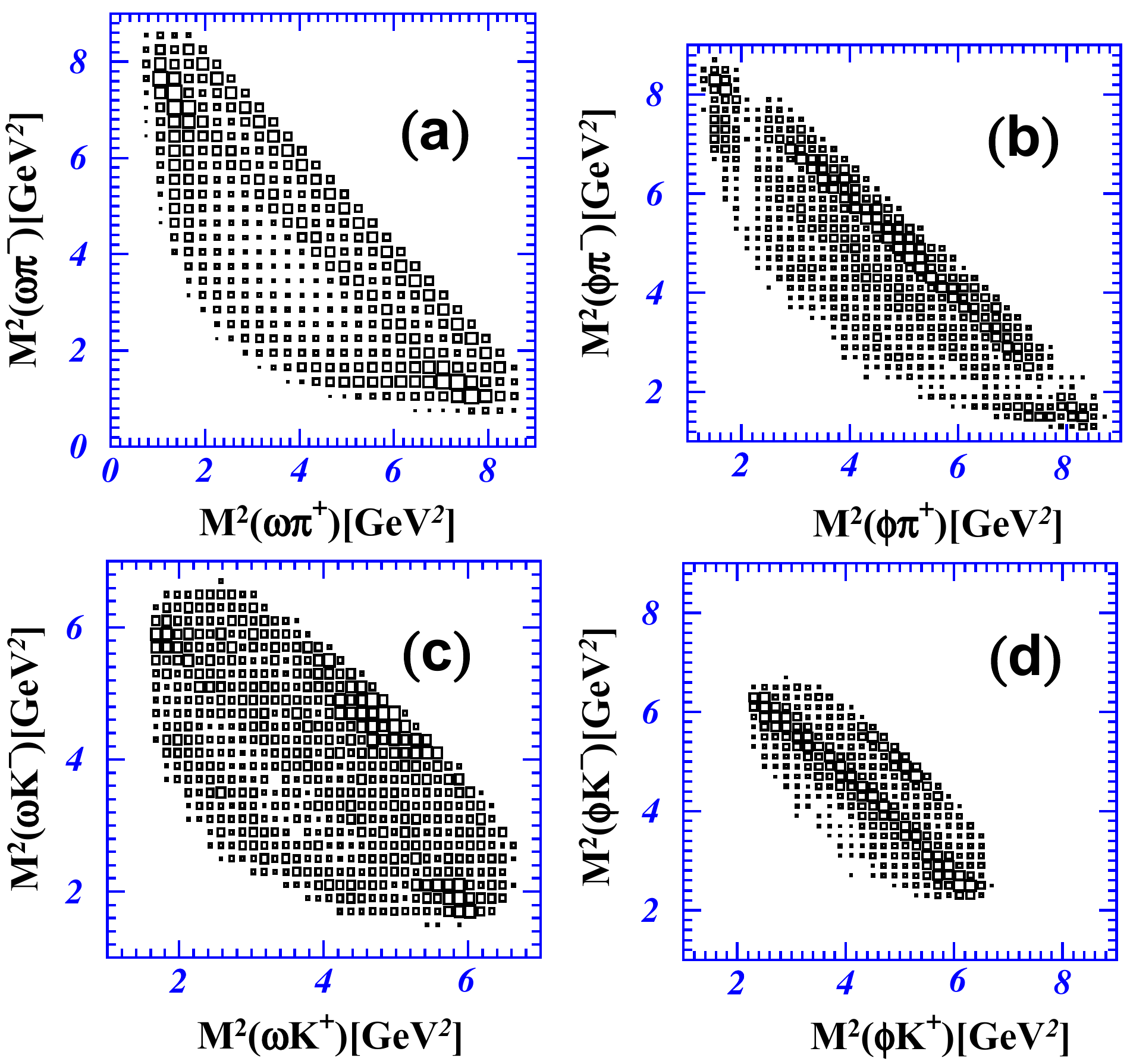}
    \caption[Dalitz plots for $J/\psi\to VPP$]{Dalitz plots for (a) $J/\psi \to \omega \pi^+ \pi^-$~\cite{Ablikim:2004qna}, (b) $J/\psi \to \phi\pi^+ \pi^-$ (with removed $K^*(890)K\pi$ background)~\cite{Ablikim:2004wn}, (c) $J/\psi \to \omega K^+ K^-$~\cite{Ablikim:2004st}, and (d) $J/\psi \to \phi K^+K^-$~\cite{Ablikim:2004wn}.}
    \label{fig:DP-VPP}
\end{figure}

\subsubsection[$J/\psi \to \omega (\pi^+ \pi^-, K^+ K^-)$ and $J/\psi\to\phi (\pi^+ \pi^-, K^+ K^-)$]{\boldmath $J/\psi \to \omega (\pi^+ \pi^-, K^+ K^-)$ and $J/\psi\to\phi (\pi^+ \pi^-, K^+ K^-)$}

\begin{figure}[t!]
    \centering
    \includegraphics[width=0.8\textwidth]{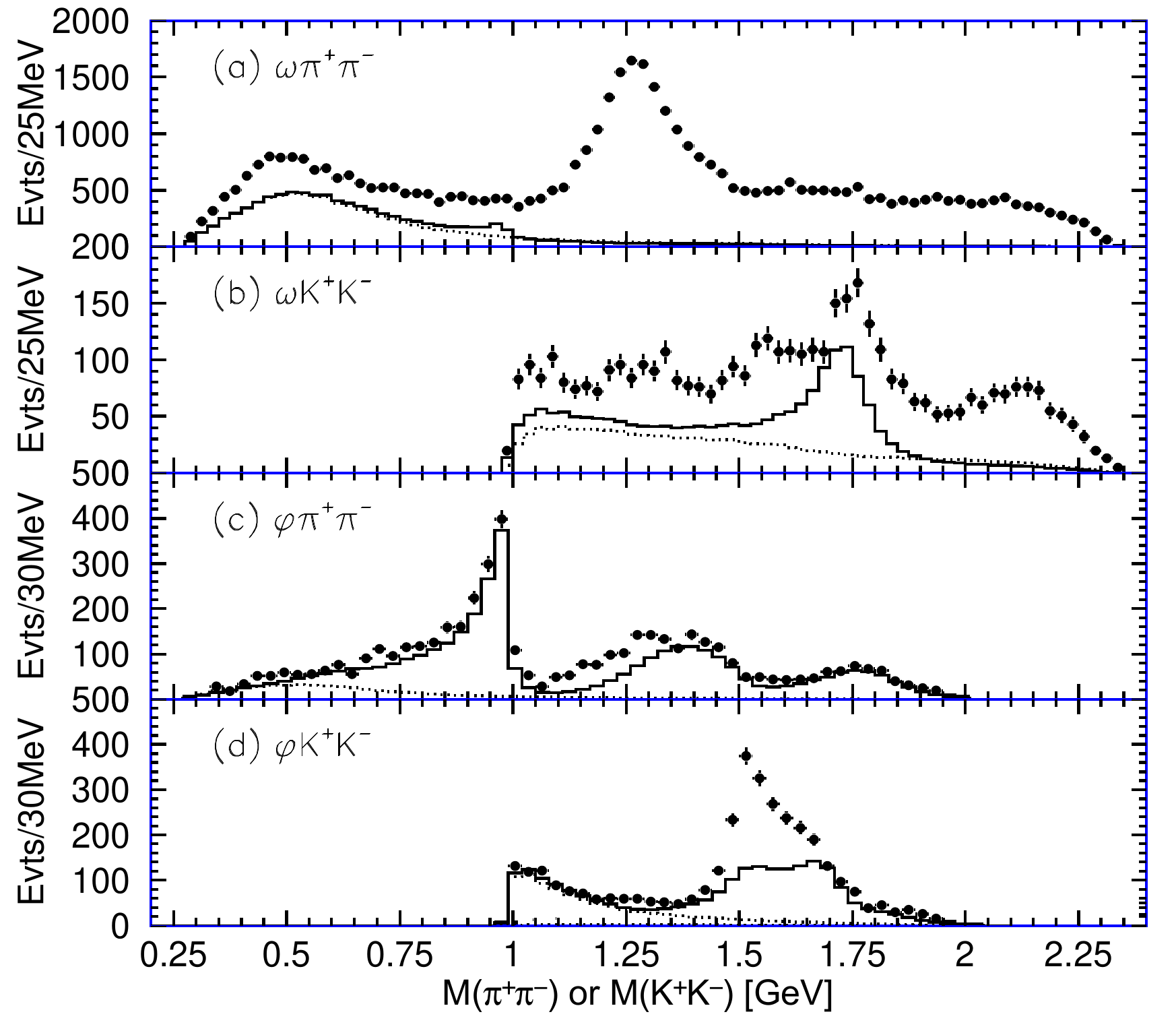}
    \caption[$J/\psi\to VPP$]{Invariant-mass distributions (uncorrected for acceptance) of pseudoscalar meson pairs recoiling against $\omega$ or $\phi$ in $J/\psi$ decays measured at BESII. 
    The solid-line histograms are the scalar contributions from the mass-dependent PWA. The dotted lines in (a) and (c)
    ($\pi^+\pi^-$ channels) denote the $f_0(500)$, and in (b) and (d)
    ($K^+K^-$ channels) the $f_0(980)$. 
     }
    \label{fig:psiPPV}
\end{figure}

In the BESII experiment, a sample of 58 million $J/\psi$ events was used to study $J/\psi \to \omega \pi^+ \pi^-$~\cite{Ablikim:2004qna}, $J/\psi \to \omega K^+ K^-$~\cite{Ablikim:2004st}, and $J/\psi \to \phi \pi ^+\pi ^-$, $\phi K^+K^-$ decays~\cite{Ablikim:2004wn}. The vector mesons are reconstructed via their main decay modes $\omega\to\pi^+\pi^-\pi^0$ and $\phi\to K^+K^-$. Dalitz plots for all these channels are shown in Fig.~\ref{fig:DP-VPP}. The prominent role of the resonances in the two-pseudoscalar-meson systems is seen from the predominantly anti-diagonal pattern of the plots, which indicates that the spectator assumption might be a reasonable approximation.

The standard approach to analyses of such reactions is a PWA using, e.g., the covariant tensor formalism~\cite{Zou:2002ar}. In Ref.~\cite{Ablikim:2004qna},  in addition to the $\omega$ spectator scenario of Eq.~\eqref{eq:VtoPPV} two crossed channels were considered for  $J/\psi \to \omega \pi^+ \pi^-$: $\rho(1450)\pi$ and $b_1(1235)\pi$ (the $b_1(1235)$ is a $J^{PC}=1^{+-}$ state, see Table~\ref{tab:pdgMesons}).
However, the analysis does not include the sub-threshold channel $\rho(770)\pi$.
In the spectator scenario, two isoscalar $\pi\pi$ waves were included: scalar $0^{++}$ and tensor $2^{++}$. They were explicitly represented by the  contributions from $f_0(500)$, $f_0(980)$, as well as $f_2(1565)$, $f_2(2240)$ resonances, respectively.   Four different parameterizations of the $f_0(500)$ pole were tried in the analysis. The $f_0(500)$ contribution dominates the low-mass region of the $\pi\pi$ invariant mass,
see Fig.~\ref{fig:psiPPV}(a), as a clear broad enhancement. The Monte Carlo (MC) study excluded a background and acceptance effects as the explanation. 
The extracted mass and width of the $f_0(500)$ resonance  vary significantly depending on the parameterization. The variation was taken as a measure of the systematic uncertainty. The combined result for the pole position of the $f_0(500)$ is $541(39)-i\,252(42)\MeV$.

\begin{figure}
    \centering
    \includegraphics[width=0.9\textwidth]{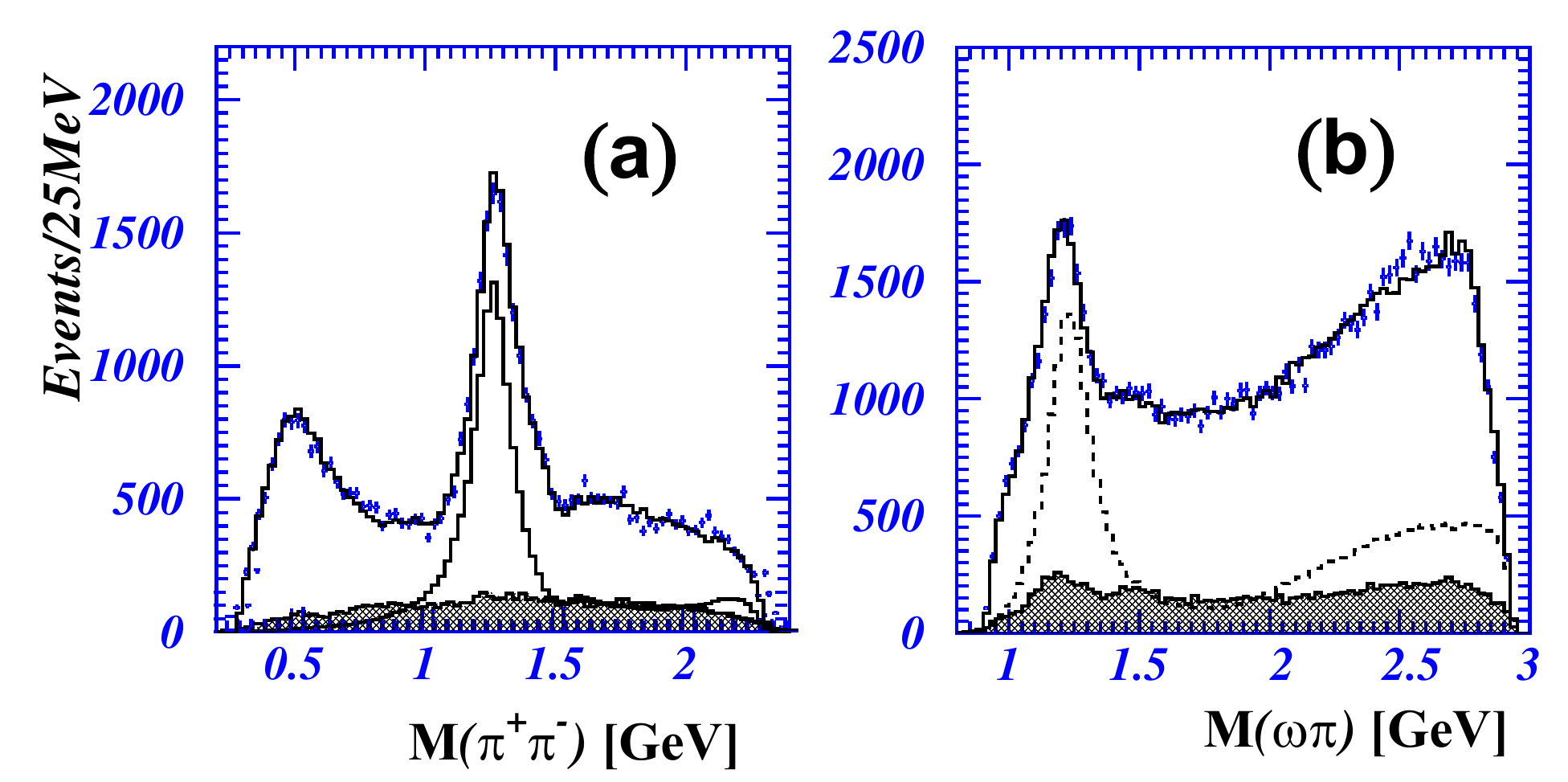}
    \caption[$J/\psi \to \omega \pi^+ \pi^-$ fit]{Invariant-mass distributions for $J/\psi \to \omega \pi^+ \pi^-$ from BESIII~\cite{Ablikim:2004qna}: (a) $M(\pi\pi)$ and  (b) $M(\pi\omega)$. 
    In panel~(a) the PWA fit contribution from 
    $f_2$ states is shown, while in panel~(b) the dashed line shows the $b_1(1235)\pi$ contribution. The shaded areas represent the background estimated from sidebands. }
    \label{fig:psiwpipi}
\end{figure}

The $2^{++}$ wave is dominated by the $f_2(1270)$ as shown in Fig.~\ref{fig:psiwpipi}(a).
The crossed-channel contribution, clearly visible as horizontal and vertical bands  along the low-mass edges of the Dalitz plot in Fig.~\ref{fig:DP-VPP}(a), is dominated by $b_1(1235)\pi$, see Fig.~\ref{fig:psiwpipi}(b). 
Finally, in the $0^{++}$ wave, in addition to the $f_0(500)$, a small (1\%) but significant contribution from the $f_0(980)$ is needed, see Fig.~\ref{fig:psiPPV}(a).

This picture is opposite for  $J/\psi\rightarrow\phi \pi\pi$, where the $f_0(980)$ is the dominating contribution to $M(\pi\pi)$ below $1 \GeV$. The distorted line shape of the $f_0(980)$ is clearly seen in the $\pi\pi$ mass spectrum recoiling against the $\phi$, which is shown in Fig.~\ref{fig:psiPPV}(c).
The $f_0(980)$ is observed in both kaon--kaon channels, the $J/\psi \to \omega K^+  K^-$ and $J/\psi \to \phi K^+K^-$ data sets (the dotted lines in Fig.~\ref{fig:psiPPV}(b) and \ref{fig:psiPPV}(d)).
In Ref.~\cite{Ablikim:2004wn}, the $\phi \pi ^+\pi ^-$ and $\phi K^+K^-$ data are fitted simultaneously assuming common resonance masses in the two channels. Only the channels with spectator $\phi$ are considered.
In the PWA, the $f_0(980)$ resonance is parameterized by the Flatt\'e formula Eq.~\eqref{eq:Flatte} for the $\pi\pi$ and $K\bar K$ channels. The clean signal in the $\pi\pi$ mass spectrum offers a unique opportunity to determine the ratio of the couplings to the two channels, which is found to be $g_{KK}/g_{\pi\pi}=4.21(33)$. 

An additional state in the $\pi \pi$ system with mass $1790(35)\MeV$ and width $270(45) \MeV$ has likely spin~0, and we denote it as $f_0(1790)$. It is distinct from the $f_0(1710)$, since the branching fraction ratio $K\bar K/\pi\pi$ for the $f_0(1790)$ is an order of magnitude lower than for the $f_0(1710)$~\cite{PDG}.
This implies the existence of two states: the $f_0(1790)$ decaying dominantly into $\pi\pi$,
and the $f_0(1710)$ dominantly into $K\bar K$. The $f_0(1790)$ is a natural candidate for the radial excitation of the $f_0(1370)$. 

The $\phi K\bar K$ data contains a strong peak due to the $f_2'(1525)$. The shoulder on its upper side may be fitted assuming interference between $f_0(1500)$ and $f_0(1710)$. 
The data indicates a possible small contribution from the $f_0(1790)$ interfering with the $f_0(1500)$.

The $\phi \pi \pi$ data exhibits a broad peak in the $M(\pi\pi)$ spectrum centered at $1.33\GeV$. It may be fitted with the $f_2(1270)$ and a dominant $0^+$ signal made from $f_0(1370)$, interfering with a smaller $f_0(1500)$ component. There is evidence that the $f_0(1370)$ signal is resonant, from interference with the $f_2(1270)$. 

Now we turn to alternative descriptions, motivated by unitarity, of the above four decay 
processes. At first we consider decay processes  $J/\psi \to \phi \pi ^+\pi ^-$, $\phi K^+K^-$ 
in the $\phi$ spectator approximation, where the meson pairs are in an $S$-wave. The two amplitudes  are related to the $\pi\pi\leftrightarrow K\bar K$ coupled-channel meson--meson scattering amplitudes~\cite{Morgan:1993rn}:
\begin{align}
    F(J/\psi\to\phi\pi^+\pi^-)&=\sqrt{\frac{2}{3}}\left[\alpha_1(s)t_{\pi\pi\to\pi\pi}(s)+
    \alpha_2(s)t_{K\bar K\to\pi\pi}(s)\right] \,,\\
    F(J/\psi\to\phi K^+K^-)&=\sqrt{\frac{1}{2}}\left[\alpha_1(s)t_{\pi\pi\to K\bar K}(s)+\alpha_2(s)t_{K\bar K\to K\bar K}(s)\right] \,,
\end{align}
where $\alpha_i=\kappa_i/(s-\lambda_i)+\gamma_{i0}+\gamma_{i1} s$ for $i=1,2$
are real functions of the couplings of $J/\psi$ to the corresponding channels and are free parameters. This is an example of the $K$-matrix method in the zero-range approximation~\cite{Dalitz:1960du}.

The $S$-waves in the $J/\psi$ decays into $\omega$ and $\phi$, accompanied by $\pi\pi$ and $K\bar K$ pairs, have also been analyzed in the framework of a chiral unitary approach~\cite{Meissner:2000bc,Lahde:2006wr}.  Here, the decay amplitudes are assumed to be proportional to appropriate mixtures of nonstrange and strange scalar form factors for pions and kaons; the latter, including various chiral low-energy constants, are extracted from a fit to data.  While the gross features of the spectra up to slightly above $1\GeV$, such as the contributions of $f_0(500)$ and $f_0(980)$, are reproduced, some properties of the extracted form factors are in conflict with constraints from low-energy QCD.  This is likely due to the neglect of crossed-channel dynamics, which have been included in a similar framework in Refs.~\cite{Roca:2004uc,Liu:2009ub}

\subsubsection[Observation of $a_0^0(980)-f_0(980)$ mixing in $J/\psi\to\phi\eta\pi^0$]{\boldmath Observation of $a_0^0(980)-f_0(980)$ mixing in $J/\psi\to\phi\eta\pi^0$}

The neutral isovector  $a^{0}_{0}(980)$ and the isoscalar $f_{0}(980)$ $0^{++}$ 
resonances can mix. The leading contribution to the isospin-violating
transition amplitudes for $f_{0}\leftrightarrow a^{0}_{0}$ can be shown to be due to the
difference of the unitarity-cut contributions of charged and neutral $K\bar{K}$ pairs, arising due to the mass difference
between the two.  While away from thresholds, isospin breaking is expected to scale as a polynomial in the light quark mass difference or, accordingly, as $M_{K^\pm}^2-M_{K^0}^2$, in the vicinity of both cuts this scaling is enhanced to $\sqrt{M_{K^\pm}^2-M_{K^0}^2}$, and hence nonanalytic in the quark mass difference. 
As a consequence, a strong 
enhancement of the mixing signal is predicted between the charged and
neutral $K\bar{K}$ thresholds,  with the width of $8\MeV$.
The mixing amplitudes strongly depend on the couplings of $a_0^0$
and $f_0$ to $K\bar{K}$. Precise measurements of the mixing transitions are
therefore important probes of the properties of these two
scalar states and can give new insights into their nature~\cite{Achasov:1979xc,Kerbikov:2000pu,Close:2000ah,Achasov:2002hg,Achasov:2003se,Wu:2007jh,Hanhart:2007bd,Wu:2008hx}.  
Two kinds of mixing intensities $\xi_{fa}$ and
$\xi_{af}$ for the $f_{0}\to a^{0}_{0}$ and
$a^{0}_{0}\to f_{0}$ transitions, respectively, are accessible in charmonia
decays~\cite{Wu:2007jh,Wu:2008hx}:
\begin{equation}
\xi_{fa} = \frac {\BR\left[J/\psi \to \phi (f_{0} \to 
a^{0}_{0} \to \eta\pi^{0})\right]} {\BR\left[J/\psi\to\phi
(f_{0}\to\pi\pi)\right]} \,, \qquad
\xi_{af}  = \frac {\BR\left[\chi_{c1} \to \pi^{0}
(a^{0}_{0}\to f_{0} \to \pi^{+}\pi^{-})\right]}
{\BR\left[\chi_{c1}\to\pi^{0}( a^{0}_{0}\to\pi^{0}\eta)\right]} \,.
\end{equation}
 The branching ratio of the chain $J/\psi\to\phi f_{0}(980)\to\phi a^{0}_{0}(980)\to\phi\eta\pi^{0}$ is expected to be about $O(10^{-6})$ similar to the estimated total amount from two background reactions: the e.m.\ transition $J/\psi\to \gamma^*\to \phi a_0$ and $J/\psi \to K^* \bar K + c.c. \to \phi a_0$. However, the peak width from the $a_0(980)$--$f_0(980)$ mixing is about $8 \MeV$, much smaller than that from other mechanisms~\cite{Wu:2007jh}.  

\begin{figure}[t]
    \centering
    \includegraphics[width=0.9\textwidth]{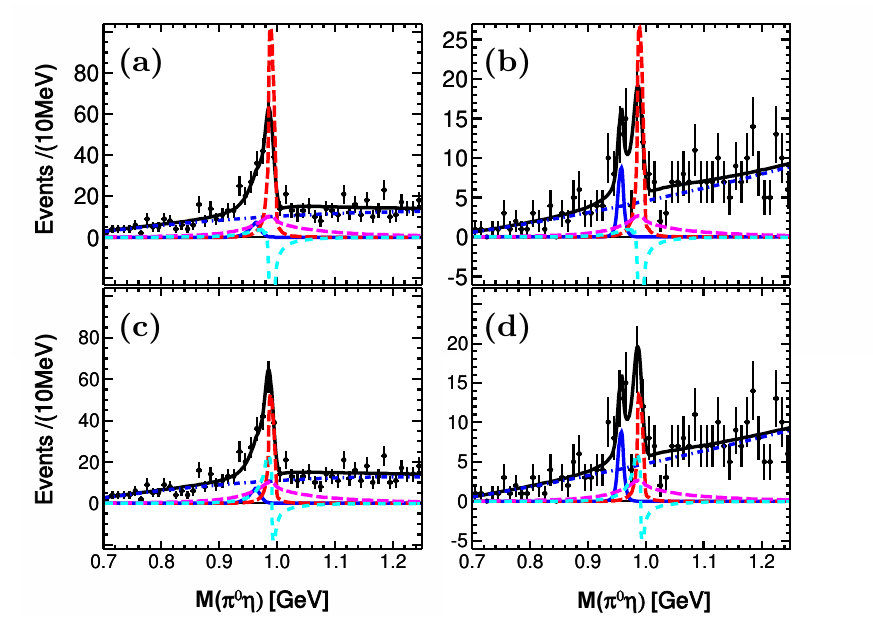}
    \caption[$a_0-f_0$ mixing]{Fits to the $M(\pi^0\eta)$ spectra of the
    $J/\psi\to\phi\eta\pi^0$ decay for destructive (upper panels)  and constructive
(lower panels) interference between e.m.\ and $a_0-f_0$ mixing contributions.
The $\eta$ meson is reconstructed via two decay channels: panels (a), (c) are for $\eta\to\gamma\gamma$ and (b), (d) for  $\eta\to\pi^+\pi^-\pi^0$. The points are data, the (black) solid lines represent the fit result, the (red) dashed curves represent the $a_0-f_0$ mixing signal, the (pink)
dashed lines represent the e.m.\ processes, the (light-blue) dotted curves represent the interference, the (blue) solid lines show the $\eta'$ background,
and the (blue) dotted--dashed lines represent the continuum
background. }
    \label{fig:a0f0mix}
\end{figure}
The two processes were observed for the first time at BESIII
and the two mixing parameters were extracted~\cite{Ablikim:2018pik}. For $J/\psi\to\phi\eta\pi^{0}$ there is a significant electromagnetic contribution and  $\xi_{fa}$ is determined from the interference signal observed in the $M(\pi^0\eta)$ distributions shown in Fig.~\ref{fig:a0f0mix}.
Two  solutions are possible:
$\xi_{fa}=0.99(35)\%$ and   $\xi_{fa}=0.41(25)\%$.
For $\chi_{c1}\to\pi^{0}\pi^{+}\pi^{-}$, the electromagnetic contribution is negligible and a signal of 42(7) events is observed, leading to $\xi_{af}=0.40(17)\%$.
In addition one can study the dependence of the mixing parameters $\xi_{fa}$ and $\xi_{af}$ on the recoil mass to $\phi$ and $\pi^0$, respectively~\cite{Wu:2008hx}.  This allows one to determine bounds on the $f_{0}$ and $a^{0}_{0}$ couplings to $K\bar K$. The predicted value of the couplings is different for the quark model, multi-quark, and hadronic-molecule scenarios.

\subsubsection[Search for isovector strangeonium-like states $Z_s$]{\boldmath Search for isovector strangeonium-like states $Z_s$}

The observation of isovector charmonium-like structures, like $Z_c^{\pm,0}(3900)$ in the $\pi J/\psi$ final states~\cite{Ablikim:2013mio,Liu:2013dau,Xiao:2013iha,Ablikim:2015tbp}, has triggered extensive discussions of their nature, including  interpretations as tetraquark, molecular, or hadroquarkonium states.
This observation challenges also our initial assumption that $J/\psi$ or $\phi$ act as spectators in the decays discussed in this section.
It is possible to conceive a similar $Z_s$ structure, where the $c\bar c$ pair is replaced by an $s\bar s$ pair. A plausible strategy to search for a $Z_s$ is to use the $\phi(2170)\to\phi\pi\pi$ decay 
as the production process and to search for a peak in the $\phi\pi$ system.
It is based on the analogy to $Y(4260)\to Z_c(3900)\pi$ since the $\phi(2170)$ is regarded as the strangeonium equivalent of the $Y(4260)$. Furthermore, as the narrow $Z_c(3900)$ state is located close to the $D^*\bar D$ threshold, the interesting range to search for a $Z_s$ is around the $K^*\bar K$ threshold of $1.4 \GeV$~\cite{Chen:2011cj}, obviously complicated by the fact that the $K^*$ width is nonnegligible.

\begin{figure}
    \centering
    \includegraphics[height=0.23\textheight]{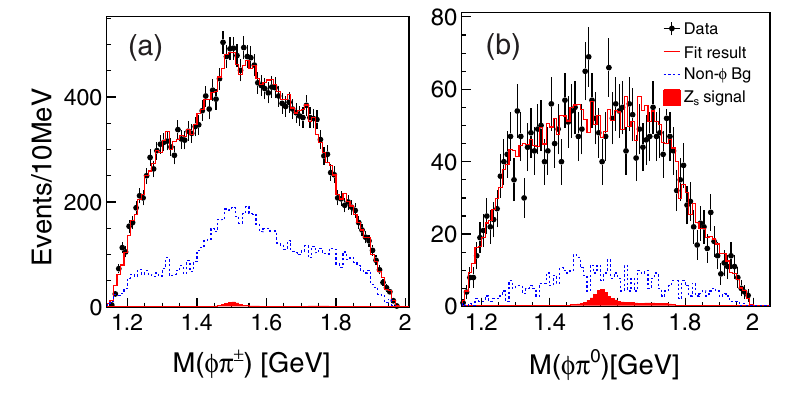} \hfill
    \includegraphics[height=0.23\textheight]{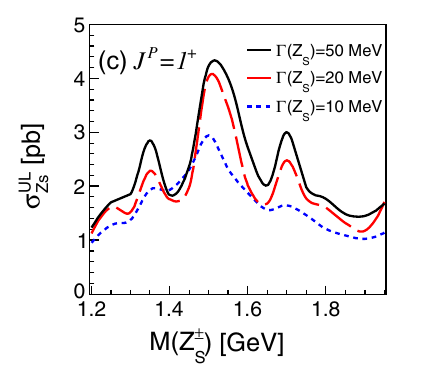}
    \caption[Search for $Z_s$]{Invariant-mass distribution of the (a) $\phi\pi^\pm$ (b)  $\phi\pi^0$  system. Panel (c) shows the determined upper limits for $Z_c^\pm$ assuming $J^{P}=1^+$ and different widths. }
    \label{fig:Zs}
\end{figure}

Using a  $108\pb^{-1}$ data sample taken at the center-of-mass energy of $2.125 \GeV$, a search for the $Z_s$ in the process $e^+e^-\to\phi\pi\pi$ was carried out at BESIII~\cite{Ablikim:2018ofc}. No signal of a narrow $Z_s$ is observed in the $\phi\pi$ invariant-mass spectra shown in Fig.~\ref{fig:Zs}(a)--(b).  The corresponding upper limits on the $Z_s$ production cross section  at the 90\% C.L.\ were
determined for different mass and width hypotheses, as shown in Fig.~\ref{fig:Zs}(c) for the $1^+$ spin-parity assignment. More data is needed to explore lower production cross sections or larger 
widths of the possible $Z_s$ structure. A complementary analysis  path is a complete amplitude analysis of the $J/\psi\to\phi\pi\pi$ process. Such an analysis is ongoing at BESIII, but the presence of many resonant structures in the $\pi\pi$ mass spectra poses a significant challenge. 

\subsection[$e^+e^-\to P_1P_2P_3$]{\boldmath $e^+e^-\to P_1P_2P_3$}\label{sec:ee-PPP}
\subsubsection{Continuum cross sections}\label{sec:ee-PPP-cont}
Production of three pseudoscalar mesons via  single-photon $e^+e^-$ annihilation for c.m.\ energies below $2\GeV$ is dominated by quasi-two-body processes involving a pair of one ground-state pseudoscalar and one vector meson. These processes include $e^+e^-\to\pi^+\pi^-\pi^0$ (dominated by the 
$\rho\pi$ intermediate state),  $e^+e^-\to\pi^+\pi^-\eta$ ($\rho\eta$),  $e^+e^-\to K\bar K\pi$ ($K\bar K^*$), and $e^+e^-\to K\bar K\eta$ (where the cross section is saturated by $\phi\eta$~\cite{Aubert:2007ym}). They could be described approximately by the methods discussed in Sec.~\ref{sec:eePV}, using the VMD model of Eq.~\eqref{eq:FVMDo}.
However, the presence of broad vector resonances (as opposed to $\omega$ or $\phi$) requires the inclusion of the proper phase space factors, and probably of a more elaborate understanding of the reaction dynamics. 

\paragraph{\boldmath $e^+e^-\to \pi^+\pi^-\pi^0$}

\begin{figure}[t!]
    \centering
    \includegraphics[width=0.8\textwidth]{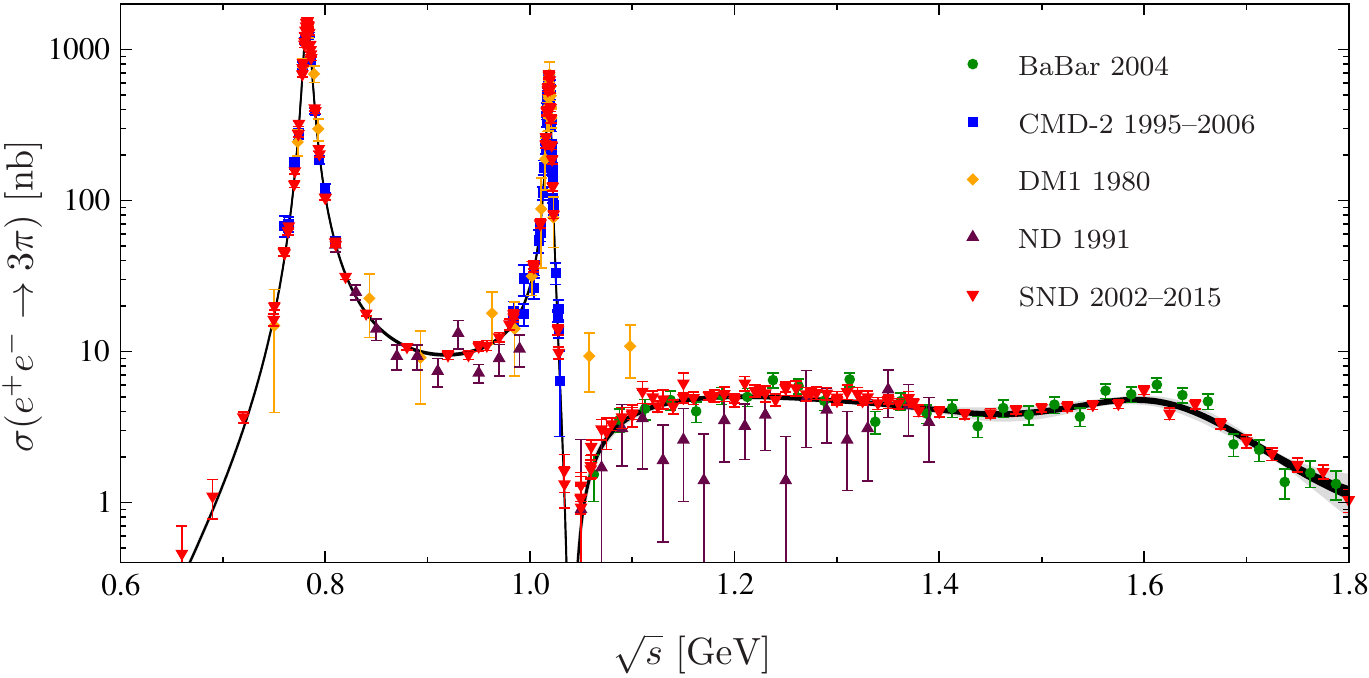}
    \caption[Cross section $e^+e^-\to\pi^+\pi^-\pi^0$ ]{ Compilation of $e^+e^-\to\pi^+\pi^-\pi^0$ data with dispersive $\gamma^*\to\pi^+\pi^-\pi^0$ amplitude parameterization~\cite{Hoferichter:2019gzf}.}
    \label{fig:eeto3pi}
\end{figure}

A compilation of the cross section measurements  for the most important process  $e^+e^-\to \pi^+\pi^-\pi^0$ process is shown in Fig.~\ref{fig:eeto3pi}. 
The low-energy region is totally dominated by contributions from narrow isoscalar resonances 
$\omega$ and $\phi$. Above the $\phi$, due to destructive interference with the nonresonant contribution
the cross section has a narrow dip,
reducing the cross section to almost zero: 
\begin{equation}
    \sigma(e^+e^-\to3\pi)(s)=\frac{4\pi\alphaem}{(s)^{3/2}}{P_f(s)}\left|{\BW}_\omega(s)+c_1{\BW}_{\phi}(s) +\ldots\right|^2 \,,
\end{equation}
where $P_f(s)$ is a generalized phase space factor that depends on the final-state interactions in the three-pion system.
The data using energy scans is from the
SND(VEPP-2M)~\cite{Achasov:2000am,Achasov:2002ud,Achasov:2003ir}, SND(VEPP-2000)~\cite{Aulchenko:2015mwt}, and CMD-2~\cite{Akhmetshin:1995vz,Akhmetshin:1998se,Akhmetshin:2003zn,Akhmetshin:2006sc} experiments.
For completeness, also earlier data from DM1~\cite{Cordier:1979qg}, DM2~\cite{Antonelli:1992jx}, and ND~\cite{Dolinsky:1991vq} are shown. The ISR method data is from BaBar~\cite{Aubert:2004kj}, and most recently from BESIII~\cite{Ablikim:2019sjw}.

The starting point for a more elaborate description of three-pion final-state interactions is a dispersive representation of the $\gamma^*\to 3\pi$ amplitude $F_{3\pi}(s;s_1,s_2,s_3)$ defined in Eq.~\eqref{eq:3picurr},
which depends both on the two-pion invariant masses $s_{1-3}$ and on the dilepton total energy squared $s$.  The cross section is obtained therefrom via
\begin{equation}
\sigma_{e^+ e^- \to 3\pi}(s) = \alpha^2\int_{s_1^\text{min}}^{s_1^\text{max}} \diff s_1 \int_{s_2^\text{min}}^{s_2^\text{max}} \diff s_2 \,
\frac{(s_1-4m_\pi^2)\,\lambda(s,m_\pi^2,s_1)\sin^2\theta_{s_1}}{768 \, \pi \, s^3}  \, |F_{3\pi}(s;s_1,s_2,s_3)|^2 \,; 
\end{equation}
see Ref.~\cite{Hoferichter:2019gzf} for details on the necessary kinematical relations.  The amplitude $F_{3\pi}(s;s_1,s_2,s_3)$ has been constructed and studied in detail in the context of the $\pi^0$ transition form factor~\cite{Hoferichter:2014vra,Hoferichter:2018dmo,Hoferichter:2018kwz}.  It is decomposed into symmetrized $P$-wave amplitudes that are calculated in a Khuri--Treiman framework~\cite{Khuri:1960zz}, which consistently incorporates arbitrary pairwise rescattering to all orders; see also Sec.~\ref{sec:3pidynamics} below for additional details.  The $s$-dependence is subsequently parameterized in terms of dispersively improved $\omega$, $\phi$, $\omega'(1420)$, and $\omega''(1650)$ resonances, supplemented by a conformal polynomial taking further inelasticities into account.  As a side product, resonance pole parameters for $\omega$ and $\phi$ have been extracted from the cross section data fits with improved precision~\cite{Hoid:2020xjs}.

\paragraph{\boldmath $e^+e^-\to \pi^+\pi^-\eta$}

Recent results on the $e^+e^-\to \pi^+\pi^-\eta$ reaction include several Novosibirsk scan measurements: CMD-2~\cite{Akhmetshin:2000wv} with $\eta\to\pi^+\pi^-\pi^0$, SND(VEPP-2M)~\cite{Achasov:2010zzd} with 
 $\eta\to\gamma\gamma$, SND(VEPP-2000) with $\eta\to\gamma\gamma$~\cite{Aulchenko:2014vkn} and $\eta\to3\pi^0$~\cite{Achasov:2017kqm}, and CMD-3 with $\eta\to\gamma\gamma$~\cite{Gribanov:2019qgw}. In addition the process was measured using the ISR method at BaBar with $\eta\to\pi^+\pi^-\pi^0$~\cite{Aubert:2007ef}, $\eta\to\gamma\gamma$~\cite{TheBABAR:2018vvb}, and $\eta\to3\pi^0$~\cite{Lees:2018dnv}.
\begin{figure}[t]
    \centering
    \includegraphics[width=0.35\textwidth]{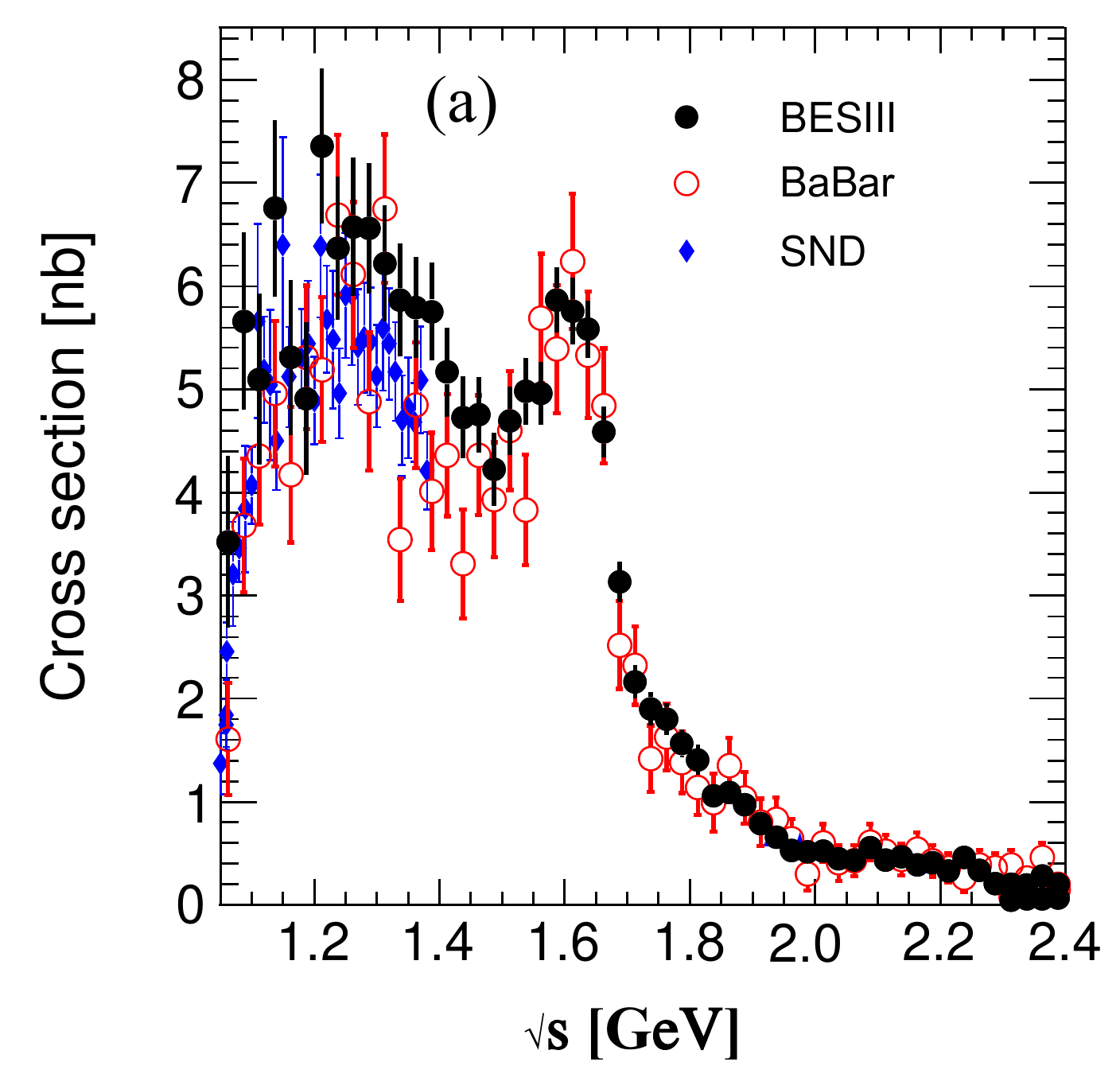}\hspace{1cm}
    \includegraphics[width=0.49\textwidth]{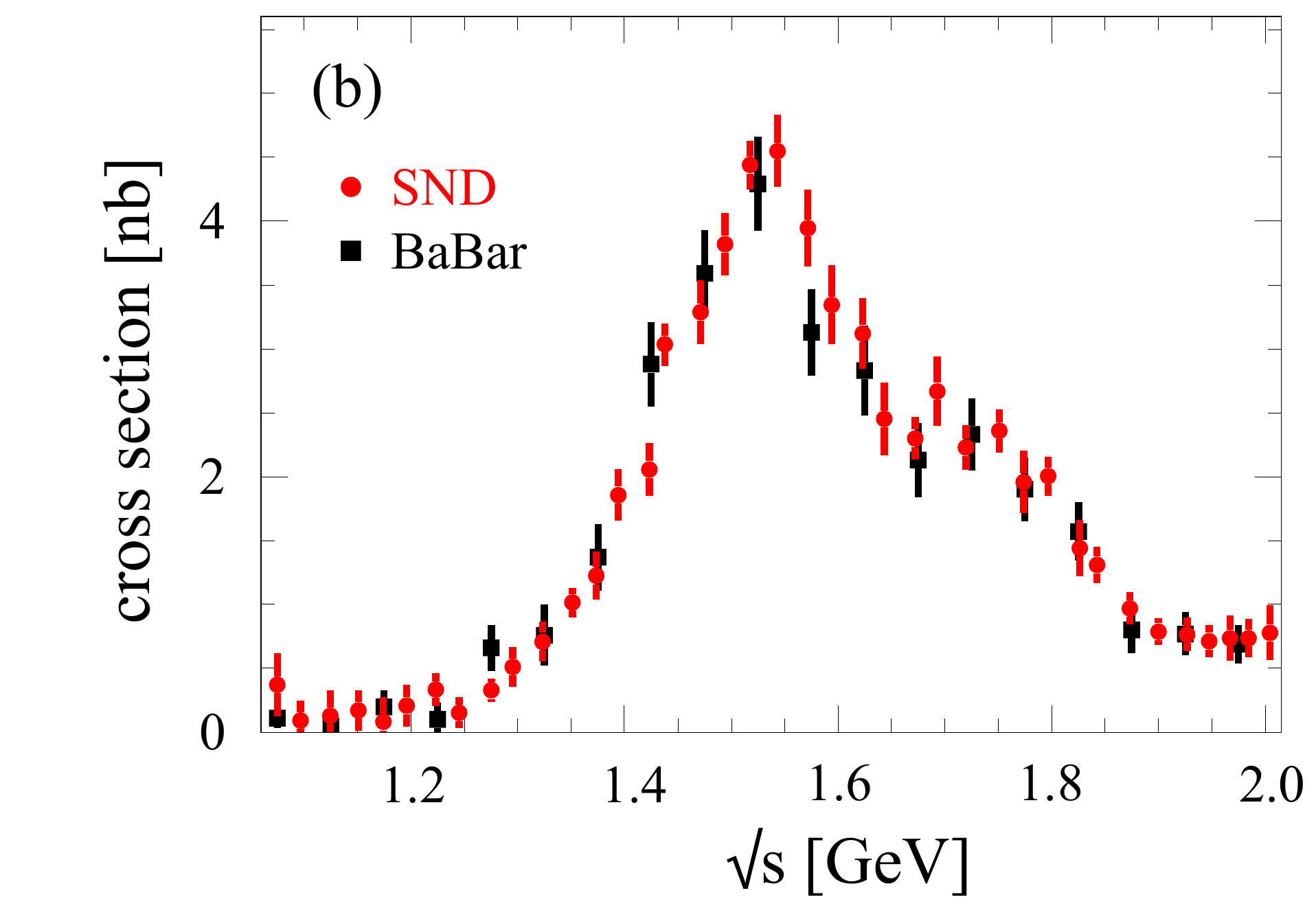}
    \caption[Cross section for $e^+e^-\to\pi^0\pi^+\pi^-$ and $\eta\to\eta\pi^+\pi^-$]{Data on three-pseudoscalar production cross sections in electron--positron annihilation for $\sqrt{s}>1\GeV$:  (a)  $e^+e^-\to\pi^0\pi^+\pi^-$~\cite{Achasov:2002ud,Aubert:2004kj,Ablikim:2019sjw}, (b) $e^+e^-\to\eta\pi^+\pi^-$~\cite{Achasov:2017kqm,Aubert:2007ef}.} 
    \label{fig:eeto3P}
\end{figure}
A compilation of results in Fig.~\ref{fig:eeto3P}(b) shows that the data sets are largely consistent with each other.
The energy dependence of the $e^+e^-\to\eta\pi^+\pi^-$ cross section has been fitted with a VMD model,
where the main contribution comes from the $\eta\rho^0(770)$ intermediate state~\cite{Achasov:1984ru}, and the total cross section has a large contribution from the $\rho(1450)$.
In the narrow-width approximation for the $\rho(770)$, 
the cross section can be expressed in terms of a transition form factor $F_{\rho\eta}(s)$ for the $\gamma^*\rho\eta$ vertex, in a similar way as discussed in Sec.~\ref{sec:eePV}. Due to the large width of the $\rho$ meson, the $P$-wave $\pi^+\pi^-$ two-body phase space has to be integrated out for each $M(\pi^+\pi^-)$ in the range
$2m_\pi<M(\pi^+\pi^-)<\sqrt{s}-m_\eta$~\cite{Achasov:1984ru,Aulchenko:2014vkn}.  A more refined analysis of the $\pi\pi$ invariant-mass distribution and the influence of crossed-channel ($\pi\eta$) dynamics is of high interest~\cite{Xiao:2015uva}, in particular with regard to its impact on dispersion-theoretical analyses of the $\eta$ transition form factor~\cite{Stollenwerk:2011zz,Hanhart:2013vba,Kubis:2015sga}.
Theoretical analyses using resonance chiral theory~\cite{Dai:2013joa,Qin:2020udp} are constrained to isovector vector resonances ($\rho(770)$ and its radial excitations), but contain neither $\pi\eta$ interactions nor nontrivial three-body rescattering effects.

The obtained cross section data has also been used to test the conserved-vector-current hypothesis and make predictions for the corresponding $\tau$ decay $\tau^-\to\pi^-\pi^0\eta\nu_\tau$~\cite{Gilman:1987my,Eidelman:1990pb}.  The latter have similarly been studied in resonance chiral theory~\cite{Dumm:2012vb}.

\begin{figure}[t]
    \centering
    \includegraphics[height=0.3\textwidth]{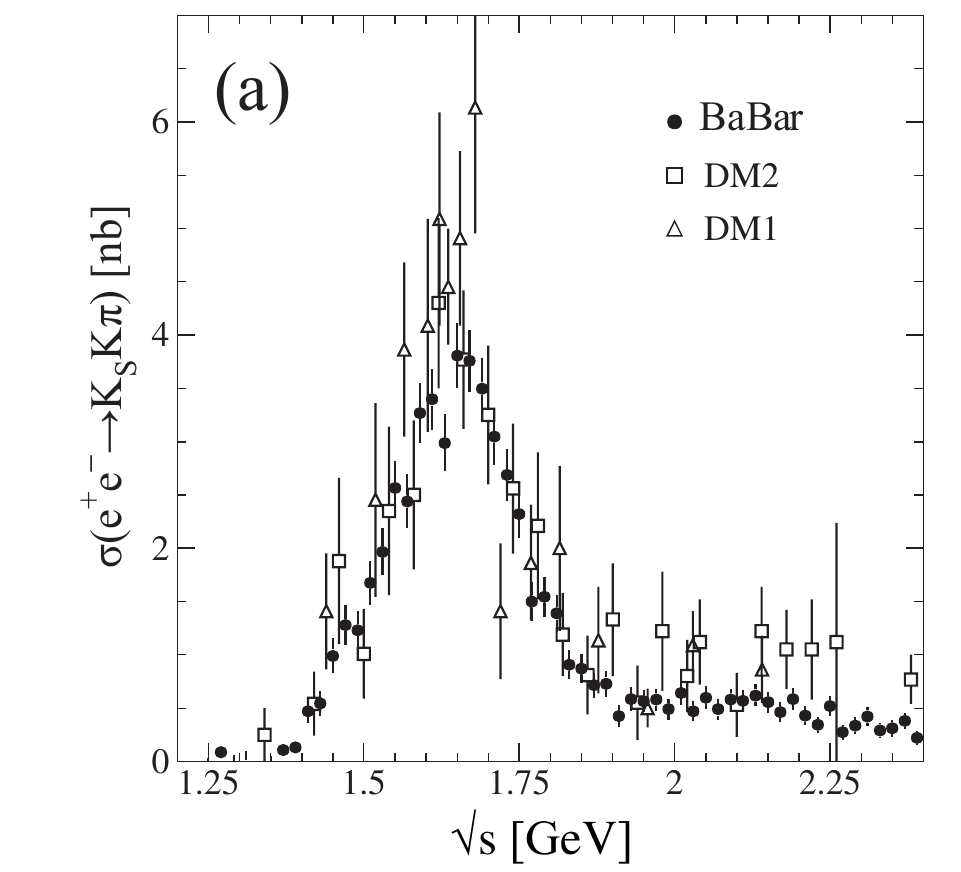} \hfill
    \includegraphics[height=0.3\textwidth]{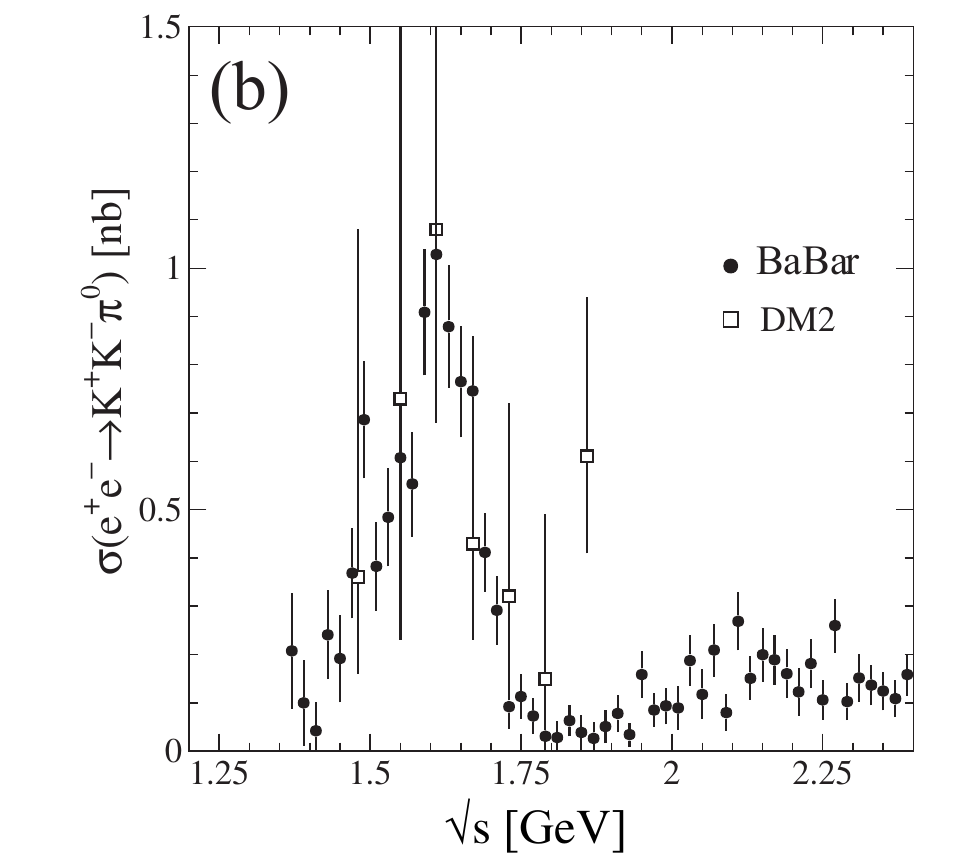} \hfill
    \includegraphics[height=0.3\textwidth]{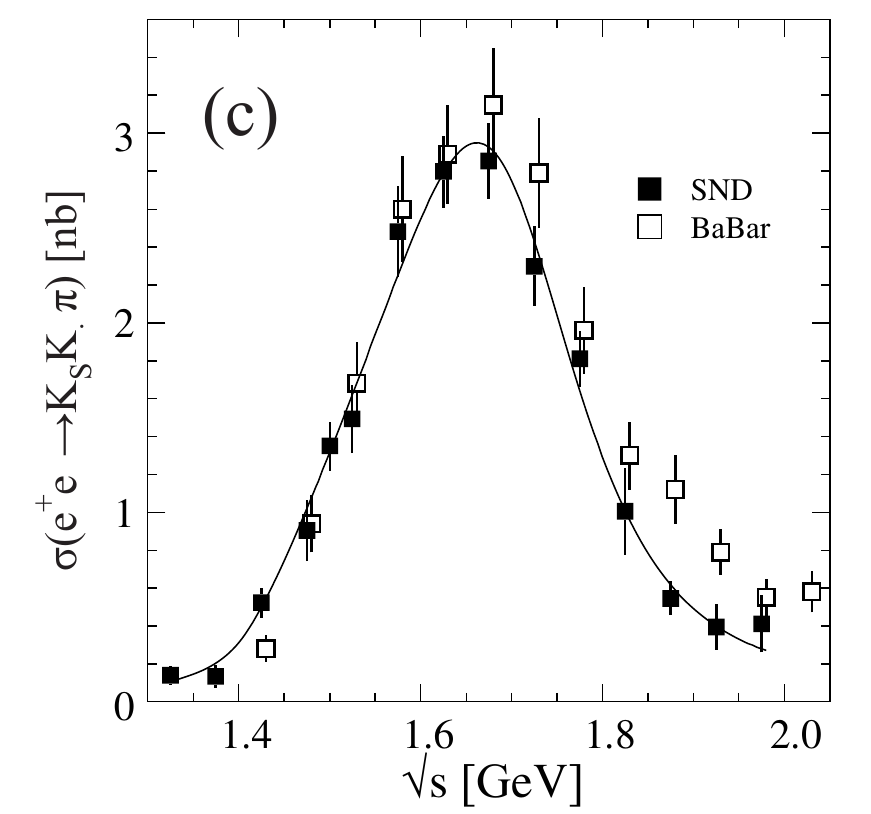}
    \caption[Cross section  $e^+e^-\to K\bar K\pi$]{Cross sections of  $e^+e^-\to K\bar K\pi$ processes:  (a) $K_S K^\pm \pi^\mp$~\cite{Buon:1982sb,Bisello:1991kd,Aubert:2007ym}, (b) $K^+ K^- \pi^0$~\cite{Bisello:1991kd,Aubert:2007ym}, (c)  $K_S K_L \pi^0$~\cite{TheBABAR:2017vgl,Achasov:2017vaq}.}
    \label{fig:eetoKKpi}
\end{figure}
\paragraph{\boldmath ${e^+e^- \rightarrow K \bar K\pi}$}
For $e^+e^- \to K \bar K\pi$, there is data for three charge modes: $K^{+} K^{-} \pi^0$, $K_{S} K^\pm \pi^\mp$, and $e^+e^- \to K_S K_L \pi^0$. 
They give sizable contributions to the total annihilation cross section into hadrons in the c.m.\ range between 1.5 and $1.8 \GeV$.
The processes are also important for the spectroscopy of excited $\phi$ mesons, since their main decay modes are expected to include kaon pairs. For example, the main decay mode for the $\phi(1680)$ is $K^*(890)\bar K$. In the analysis of the two $K \bar K \pi^0$ channels, the OZI-suppressed intermediate state $\phi\pi^0$ is analysed separately and we have discussed it in Sec.~\ref{sec:eePV}. 

Measurements of all three channels were performed at BaBar using the ISR technique~\cite{Aubert:2007ym,TheBABAR:2017vgl}. In addition, SND at VEPP-2000 has measured the $e^+e^- \to K_S K_L \pi^0$ cross section by scanning the c.m.\ energy range $\sqrt{s}=1.3$--$2.0\GeV$~\cite{Achasov:2017vaq}. The results for the three channels are shown in Fig.~\ref{fig:eetoKKpi}.
The cross sections are dominated by $K^*(892)\bar K$ intermediate states. The fraction of the $\phi\pi^0$ contribution in the neutral-pion channels at $1.7 \GeV$ is about 1\%~\cite{Aubert:2007ym}.

\subsubsection{Dynamics of three-pion final states}\label{sec:3pidynamics}
\begin{table}[t]
      \caption[]{Branching fractions of neutral-vector-mesons decays into three pseudoscalar mesons. Unless marked otherwise, the branching fractions are taken from Ref.~\cite{PDG}. 
      \label{tab:VtghPP}}
\begin{center}
\renewcommand{\arraystretch}{1.3}
  \begin{tabular}{rlrlr}
  \toprule
    &Final state&$\BR$& & Ref.\\ \midrule
    $\omega\to P_1P_2P_3$&$\pi^+\pi^-\pi^0$& $89.3(6)\%$&&\\
    $\phi\to P_1P_2P_3$&$\pi^+\pi^-\pi^0$& $15.24(33)\%$&&\\
    \midrule
    $J/\psi\to P_1P_2P_3$&$\pi^+\pi^-\pi^0$& $2.10(8)\%$&&\\
    &$K^+K^-\pi^0$& $2.88(12)\times10^{-3}$&&\\
    &$K_SK_L\pi^0$& $2.06(27)\times10^{-3}$&&\\ 
    &$K_SK^\pm\pi^\mp$& $5.6(5)\times10^{-3}$&&\\
     &$K_SK_L\eta$& $1.44(34)\times10^{-3}$&&\\        
    &$\pi^+\pi^-\eta$& $3.8(7)\times10^{-4}$&&\\  
    &$\pi^+\pi^-\eta'$& $8.1(8)\times10^{-4}$&&\cite{Ablikim:2017edw}\\ 
    \midrule 
    $\psi'\to P_1P_2P_3$&$\pi^+\pi^-\pi^0$& $2.01(17)\times10^{-4}$&&\\
    &$\pi^+\pi^-\eta'$& $1.9\big({}^{+1.7}_{-1.2}\big)\times10^{-5}$&&\cite{Ablikim:2017edw}\\ 
    &$K_SK_L\pi^0$& $<3.4\times10^{-4}$&&\\    
    &$K_SK_L\eta$& $1.3(5)\times10^{-3}$&&\\
    &$K^+K^-\pi^0$& $4.07(31)\times10^{-5}$&&\\  
    &$\pi^+\pi^-(\rho)\eta$& $2.2(6)\times10^{-4}$&&\\ \bottomrule
  \end{tabular}
  \renewcommand{\arraystretch}{1.0}
\end{center}
\end{table}
The decay distributions of three-pseudoscalar-meson decays of the narrow isoscalar vector mesons $\omega$, $\phi$, $J/\psi$, $\psi'$ are particularly apt to study the dynamics of the final-state interactions in detail and with high precision, most prominently three-pion final states.  An overview of the corresponding branching fractions is given in Table~\ref{tab:VtghPP}.

The matrix element of a vector meson decay into three pions $V(\varepsilon^\mu)\to\pi^+(q_+)\pi^-(q_-)\pi^0(q_0)$ can be written in exactly the same form as the three-pion hadronic current in Eq.~\eqref{eq:3picurr},
\begin{equation}
   {\cal M}(s,t,u)=\epsilon_{\mu\nu\alpha\beta} \, \varepsilon^\mu q_+^\nu q_-^\alpha q_0^\beta {\cal F}(s,t,u)\,.
\end{equation}
The matrix element squared of the $e^+e^-\to V \to \pi^+\pi^-\pi^0$ process can be represented as a product of the $P$-wave phase space  ${\cal P}(s,t,u)$ (see, e.g., Refs.~\cite{Leupold:2008bp,Niecknig:2012sj,Danilkin:2014cra}) and the squared modulus of the scalar function ${\cal F}(s,t,u)$, 
\begin{equation}
\sum\limits_{\lambda=\pm1} \vert {\cal M}_\lambda\vert^2 \propto {\cal P} \, \vert {\cal F}(s,t,u)\vert^2 \,,
\label{eq:distr-w-p}  
\end{equation}
where $\lambda$ is the helicity of the vector meson $V$, restricted in $e^+e^-$ annihilations to $\lambda=\pm 1$.
The ${\cal P}$ term vanishes at the Dalitz plot boundary~\cite{Zemach:1963bc}. It can also account for kinematic isospin violation due to the difference between the masses of the uncharged and charged pions.  The scalar function ${\cal F}(s,t,u)$ can be decomposed into series of odd partial waves only, and is hence entirely dominated by $P$-waves.  Neglecting $F$- and higher waves (or, more precisely, the discontinuities therein), it can be written as a sum of so-called single-variable functions, which themselves are free of cuts from crossed channels~\cite{Niecknig:2012sj}:
\begin{equation}
    {\cal F}(s,t,u)={\cal F}(s)+{\cal F}(t)+{\cal F}(u)\,.
    \label{eq:RecThm-V3pi}
\end{equation}
The $P$-wave projection $f_1(s)$ in, e.g., the $s$-channel is hence given by the sum of two terms,
\begin{equation}
    f_1(s) = \mathcal{F}(s)+ \hat{\mathcal{F}}(s) \,, \label{eq:F+Fhat}
\end{equation}
where the second term is due to the projection of ${\cal F}(t)+{\cal F}(u)$ onto the $s$-channel $P$-wave and contains the left-hand cuts in $f_1(s)$.  The separation of right- and left-hand cuts in Eq.~\eqref{eq:F+Fhat} allows us to translate a Watson-like unitarity relation for the partial wave into a discontinuity equation for the single-variable function, 
\begin{equation}
 \mathrm{disc}\,{\cal F}(s)=2i\left({\cal F}(s)+\hat{\mathcal{F}}(s)\right)\theta(s-4m_\pi^2)\sin\delta_1^1(s)\exp\left\{-i\delta_1^1(s)\right\}\,.
 \label{eq:discF}
\end{equation}
In the absence of the so-called inhomogeneity $\hat{\mathcal{F}}(s)$, 
Eq.~\eqref{eq:discF} is solved simply in terms of an Omn\`es function calculated from the $\pi\pi$ $P$-wave phase shift $\delta_1^1(s)$.  The inclusion of the crossed-channel effects is interpreted intuitively as the inclusion of all pairwise rescattering effects in the final state.  The solution has the form of so-called Khuri--Treiman equations~\cite{Khuri:1960zz}, and their concrete form for the system~\eqref{eq:RecThm-V3pi} has by now been discussed in the literature many times~\cite{Niecknig:2012sj,Hoferichter:2012pm,Danilkin:2014cra,Hoferichter:2017ftn,Dax:2018rvs,Albaladejo:2020smb}.

\paragraph{\boldmath $\omega\to \pi^+\pi^-\pi^0$}

Back in 1961, the $\omega$ meson's  
spin and parity were determined by observing that the Dalitz plot of the main decay mode $\omega \to \pi^+ \pi^- \pi^0$ is consistent with a $P$-wave phase space distribution~\cite{Maglic:1961nz}.
This is a very good approximation due to the limited phase space volume for the three pions. Only recently, precision Dalitz plot analyses have shown there is a deviation from the $P$-wave phase space distribution~\cite{Adlarson:2016wkw,Ablikim:2018yen}.
Given the small decay region and the lack of resonances within the kinematic limits, the deviation from $P$-wave phase space $|{\cal F}(s,t,u)|^2$ can be parameterized, in the isospin limit, using an expansion about the center of the Dalitz plot in the $z$ and $\phi$ variables~\cite{Niecknig:2012sj}:
\begin{equation}
\vert {\cal F}\vert^2 = {\cal N}\left(1+ 2\alpha z+ 2\beta z^{3/2}\sin3\phi+ 2\gamma z^2+\mathcal{O}\big(z^{5/2}\big)\right)  \, ,  \label{eqn:DPzphi}
\end{equation}
where  $\alpha$ (quadratic slope parameter), $\beta$, and $\gamma$ are Dalitz plot parameters.
The WASA-at-COSY experiment using $\omega$ mesons produced in hadronic processes~\cite{Adlarson:2016wkw} has found the Dalitz plot consistent with $\alpha=147(36)\times 10^{-3}$ and the remaining parameters equal to zero. The most precise study is from BESIII, using $2.6\times10^5$ $\omega\to \pi^+\pi^-\pi^0$ events tagged by the $J/\psi\to\omega\eta$ decay~\cite{Ablikim:2018yen}. In the best fit, $\alpha=120(8)\times 10^{-3}$ and there is evidence for the next term in the expansion Eq.~\eqref{eqn:DPzphi} with $\beta=29(10)\times 10^{-3}$. 

The process is important as a cross check of three-body dispersive calculations using pion--pion phase shifts. 
In the absence of available Dalitz plot data at the time, Refs.~\cite{Niecknig:2012sj,Danilkin:2014cra} employed a single subtraction constant in their amplitude representation, which amounts to an overall normalization and hence can be fixed from the partial width; the energy dependence of the Dalitz plot is then a parameter-free prediction.
The BESIII result is consistent with dispersive calculations without crossed-channel rescattering: 
$\alpha=(125\ldots135)\times 10^{-3}$, $\beta=(29\ldots33)\times 10^{-3}$~\cite{Niecknig:2012sj} and  
$\alpha=125\times 10^{-3}$, $\beta=30\times 10^{-3}$~\cite{Danilkin:2014cra}. However, there seems to be tension when the full calculations including the effect of crossed-channel rescattering are compared to: $\alpha=(74\ldots84)\times 10^{-3}$, $\beta=(24\ldots28)\times 10^{-3}$~\cite{Niecknig:2012sj} and $\alpha=84\times 10^{-3}$, $\beta=28\times 10^{-3}$~\cite{Danilkin:2014cra}. Thus, the investigation of this decay's dynamics with higher precision by analyzing the full $J/\psi$ data sample of 10 billion events at BESIII is expected to clarify this issue. 

\begin{figure}[t!]
\begin{center}
\includegraphics[width=0.38\textwidth]{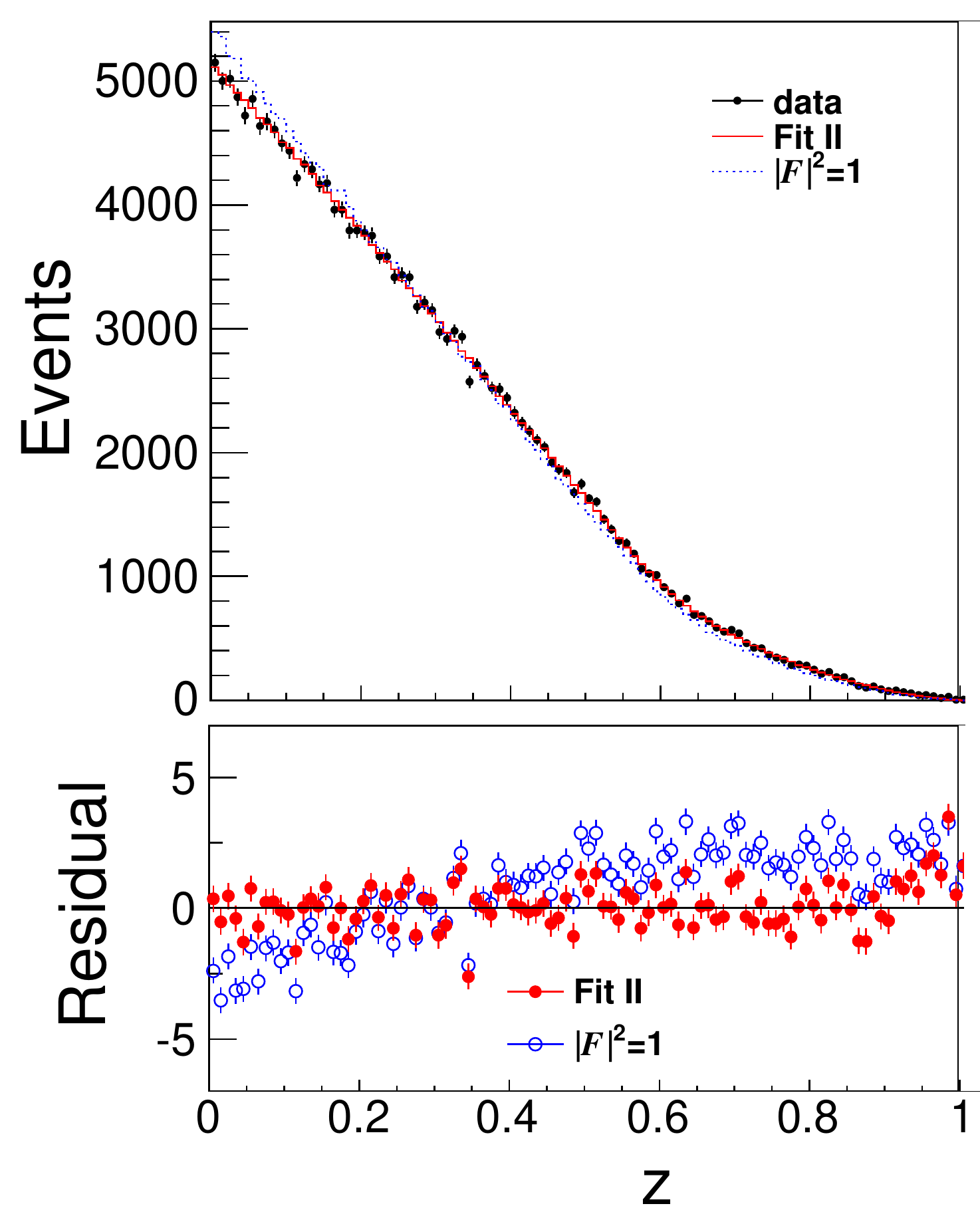}\put (-100,220){\large\bf(a)}\hspace{1cm}
\includegraphics[width=0.45\textwidth]{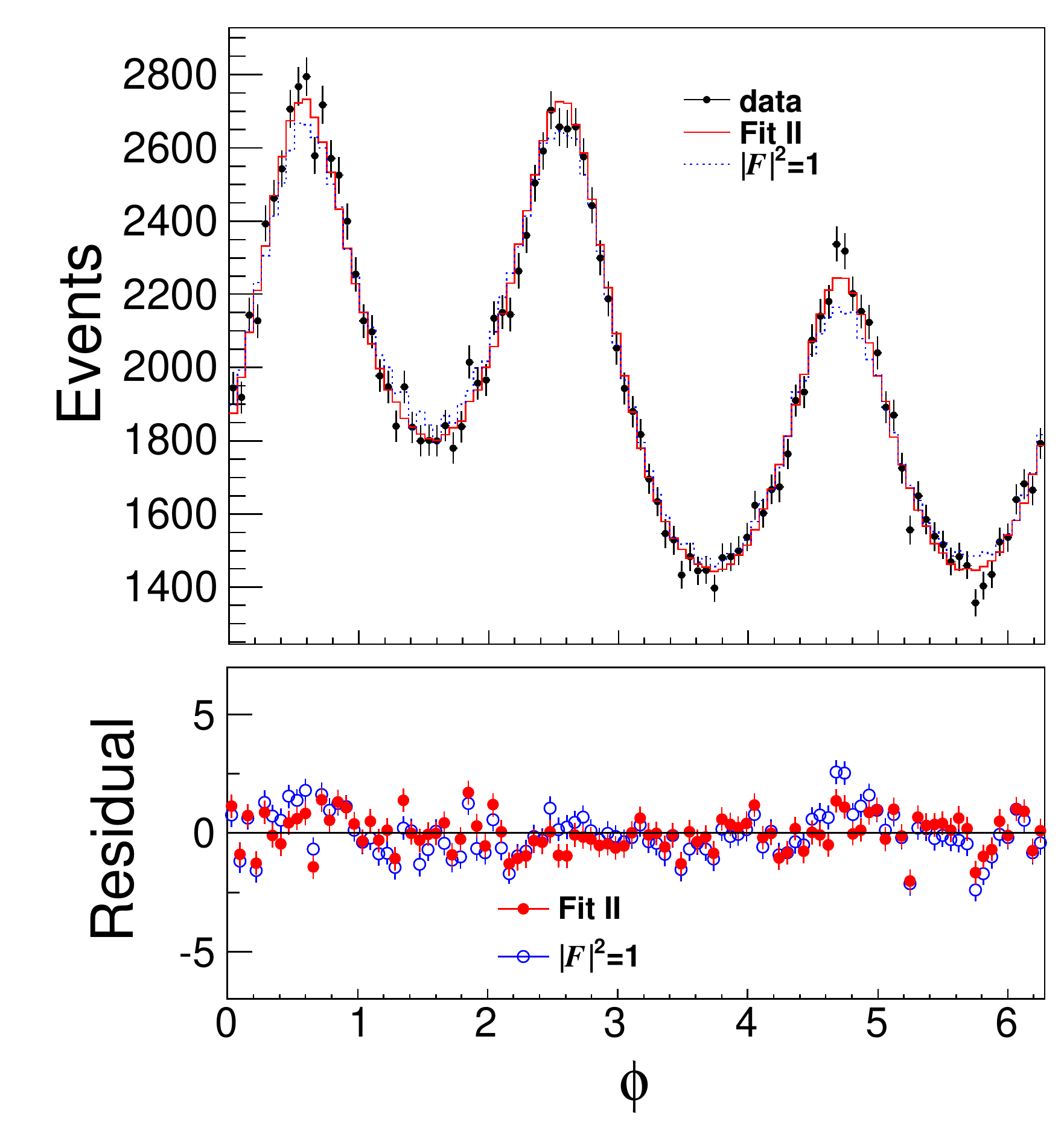}\put (-35,220){\large\bf(b)} \\
\caption[$\omega\to \pi^+\pi^-\pi^0$ data from BESIII]{BESIII data on $\omega\to \pi^+\pi^-\pi^0$ compared to the best fit and the $P$-wave phase space for: (a) $z$ and (b) $\phi$ distributions. The black points with error bars in the upper panels are data, the solid line is the fit, the dashed line the $P$-wave. The solid and hollow dots in the lower panels denote the residuals for the fit and  the $P$-wave, respectively. }
\label{fig:omega14}
\end{center}
\end{figure}

As a consequence, a recent update has demonstrated that the experimental Dalitz plot parameters can be reproduced with a second (complex) subtraction constant, at the expense of reducing the high predictive power of the original formulation~\cite{Albaladejo:2020smb}.
The projections on $z$ and $\phi$ axes of the Dalitz plot density are shown and compared to the fit results and to $P$-wave phase space in Fig.~\ref{fig:omega14}.
  
\paragraph{\boldmath $\phi\to \pi^+\pi^-\pi^0$}
\begin{figure}
      \centering
      \includegraphics[width=0.6\textwidth]{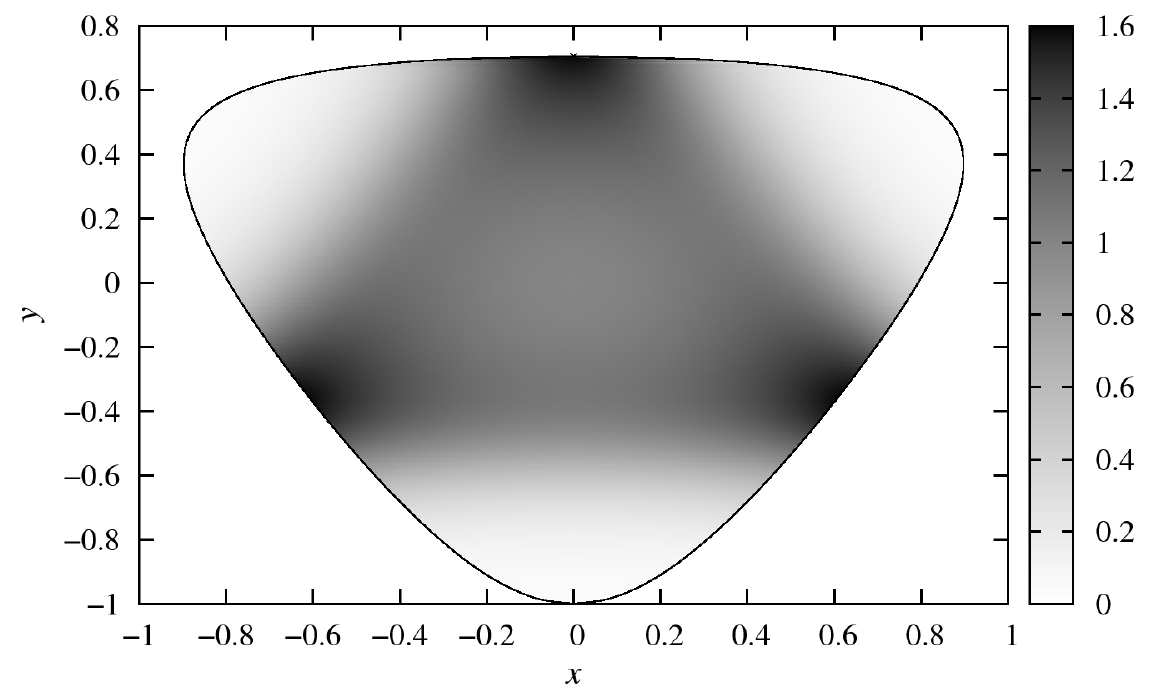}
      \caption[Dalitz plot for $\phi\to\pi^+\pi^-\pi^0$]{Dalitz plot distribution for $\phi\to\pi^+\pi^-\pi^0$ divided by $P$-wave phase space~\cite{Niecknig:2012sj}. 
      Three broad bands corresponding to the three $\rho$ states are seen. The kinematical boundary is also shown.}
      \label{fig:KPhiTo3P}
\end{figure}
The three-pion decay of the $\phi$ is one of its main decay modes despite OZI-rule suppression.
Two amplitude analyses were performed by the CMD2 experiment: the first using $10790(110)$ events~\cite{Akhmetshin:1998se} and the second $78780(280)$~\cite{Akhmetshin:2006sc}.
The most precise data on $\phi\to \pi^+\pi^-\pi^0$ is from KLOE~\cite{Aloisio:2003ur}, using $1.98\times10^6$ selected events. A theoretical fit of the Dalitz plot, divided by the phase space, is shown in Fig.~\ref{fig:KPhiTo3P}, where bands due to the $\rho\pi$ intermediate states are clearly visible. 
 
The description of the decay within
the isobar model uses  ${\cal F}(t)$ and ${\cal F}(u)$ as $\BW_\rho^{GS}$ or $\BW_\rho^{KS}$ form factors. For  ${\cal F}(s)$, where $\rho^0$--$\omega$ mixing is expected, a $\BW_{\rho+\omega}^{GS}$ form should be used to describe the precision data.\footnote{The $\omega$ features as a background effect via $e^+e^-\to\omega\pi^0$, with subsequent, isospin-violating, decay $\omega\to\pi^+\pi^-$, and does not necessarily appear as an actual decay product of the $\phi$, which would be doubly isospin-suppressed.} In addition, a contact term has to be introduced, representing a direct coupling or an effect of a higher-mass $\rho$ resonance like the $\rho(1450)$, which is outside the kinematic decay range.  The use of a contact term (or the simple addition of a heavier resonance) without subsequent rescattering violates the universality of final-state interactions and should hence be discarded from a theoretical point of view.   

The dispersive description of $\phi\to3\pi$ is formally identical to the one of $\omega\to3\pi$ discussed above, the only change being the mass of the decaying vector state.  Interestingly enough, the nontrivial rescattering effects beyond the simple $\rho\pi$ picture do depend on the decay mass, and vary with it. It is observed that here, rescattering effects significantly improve on the KLOE Dalitz plot description~\cite{Niecknig:2012sj}: the contact term employed in the model fit can be understood as an effective replacement of the final-state interactions that are incorporated in the Khuri--Treiman amplitudes, but omitted in the Breit--Wigner-type isobar model.  To achieve a perfect fit of the high-statistics data, a second subtraction constant was used, which however only induced a very minor change in the Dalitz plot, in contrast to what seems to be required for $\omega\to3\pi$~\cite{Ablikim:2018yen,Albaladejo:2020smb}.

  \begin{figure}[tb!]
        \centering
                \includegraphics[width=0.45\textwidth]{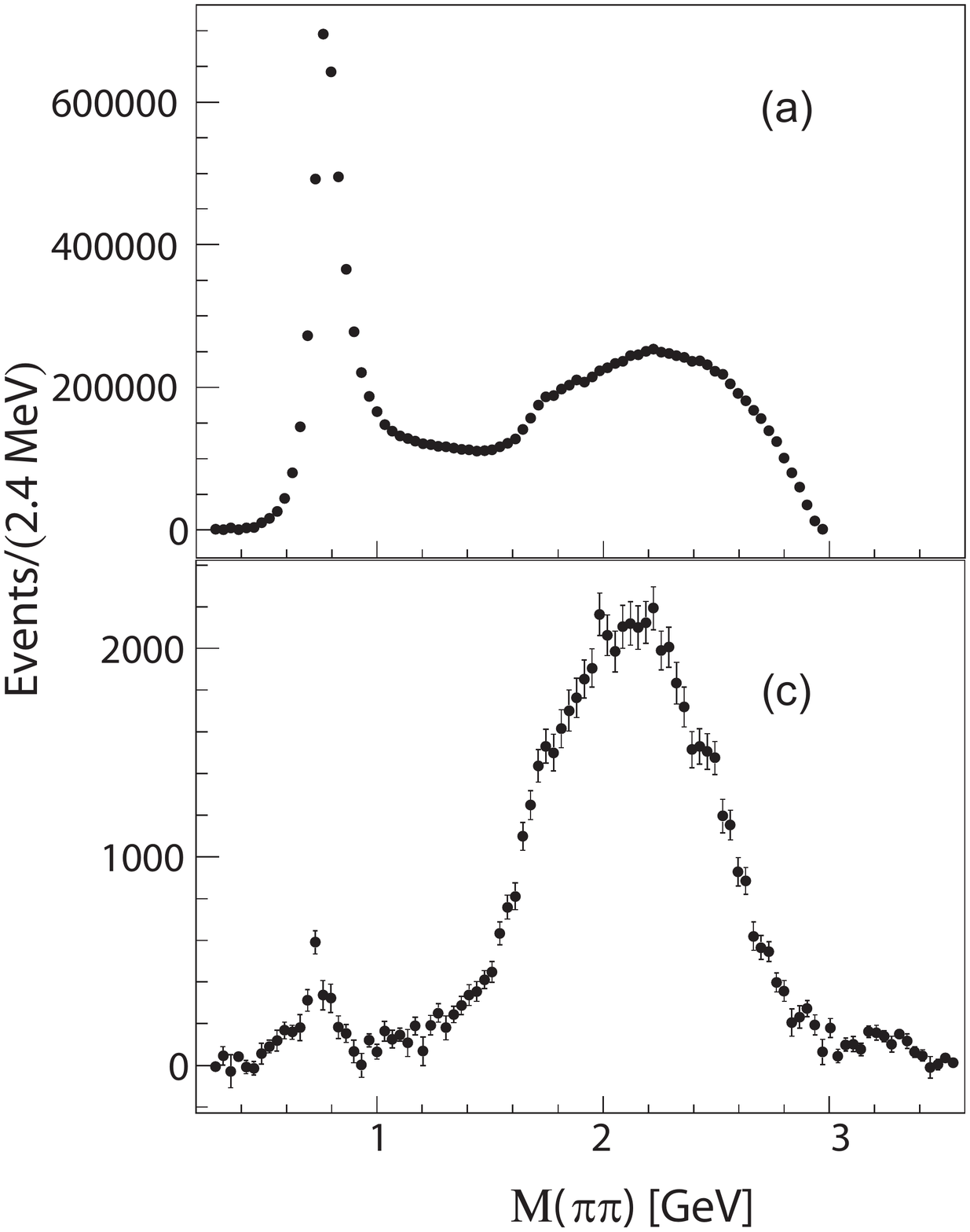}
                \includegraphics[width=0.45\textwidth]{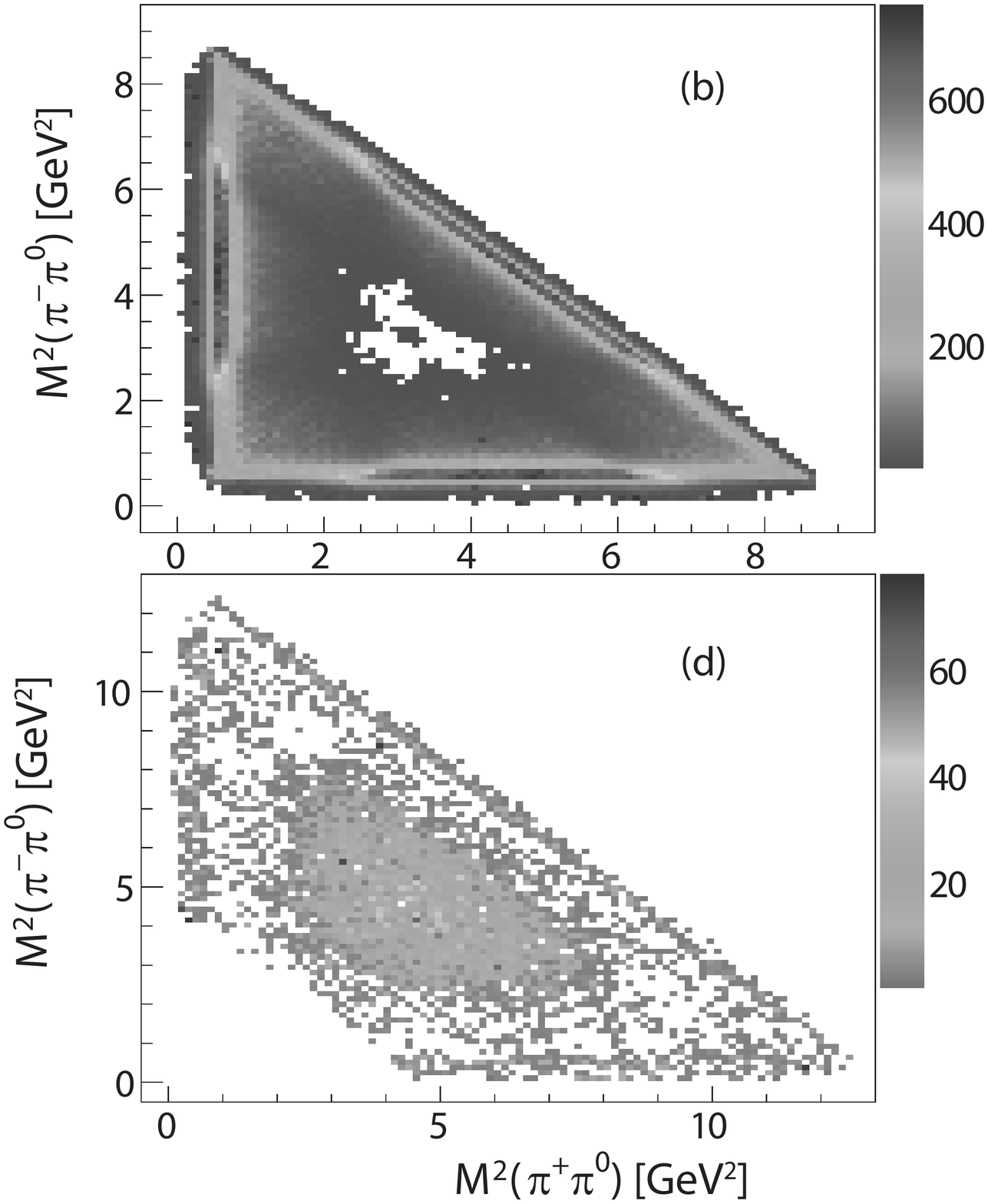}
        \caption[Three pion decays of $J/\psi$ and $\psi'$ ]{Results on three-pion decays of $J/\psi$ and $\psi'$ from BESIII~\cite{Ablikim:2012dx}. Invariant-mass distributions $M(\pi\pi)$ (a) and (c) and Dalitz plots (b) and (d)  with backgrounds subtracted and corrected
 for efficiency. The top and bottom panels are for the $J/\psi\to \pi^+\pi^-\pi^0$
 and $\psi'\to \pi^+\pi^-\pi^0$ decays, respectively.}
        \label{fig:resultpsi3p}
\end{figure}

\paragraph{\boldmath $J/\psi,\psi'\to \pi^+\pi^-\pi^0$}
 
It is instructive to also consider and compare the three-pion decays of the vector charmonia.  The decay $J/\psi \to \pi^+ \pi^- \pi^0$,  despite its huge phase space in principle allowing for the production of many $\rho$ excited states, is entirely dominated by the $\rho(770)\pi$ intermediate state, 
see Fig.~\ref{fig:resultpsi3p}(b), with almost no events in the center of the Dalitz plot. The situation changes dramatically for the $\psi'$ decay: the partial decay width of $\psi'\to \pi^+\pi^-\pi^0$ is only 3\% of $J/\psi \to \pi^+ \pi^- \pi^0$, and the Dalitz plot distribution is completely different; see Fig.~\ref{fig:resultpsi3p}(d). Here, the fraction of  $\rho(770)\pi$ events is suppressed and pions are often produced with equal energies, populating the center of the Dalitz plot.  This effect is known as the ``$\rho\pi$ puzzle,'' which is difficult to explain.

Partial-wave analyses of both decays might shed some light on the puzzle. For  $J/\psi \to \pi^+ \pi^- \pi^0$, in addition to the BESIII data shown in Fig.~\ref{fig:resultpsi3p} there is an amplitude analysis from BaBar using $19560(160)$  $J/\psi \to \pi^+ \pi^- \pi^0$ events~\cite{Lees:2017ndy}. 
The isobar model analysis for $J/\psi$ indicates a contribution of the $\rho(1450)\pi$ intermediate state (ca.\ 10\%), which interferes destructively with the main $\rho(770)\pi$ contribution. An alternative is that $\rho(1450)\pi$ represents an effective contribution of excited $\rho$ resonances as expected in the Veneziano model~\cite{Szczepaniak:2014bsa}.
The $\psi'\to \pi^+ \pi^- \pi^0$ Dalitz plot analysis was performed by BESII~\cite{Ablikim:2005jy}.  Here, the main contribution is from the $\rho(2150)$ resonance, where the mass and the width were fixed to $2.149\GeV$ and $363\MeV$, respectively. The $\rho(770)\pi$ contribution is only 28\%.

\paragraph{\boldmath $J/\psi\to K\bar K\pi$}
Amplitudes of the decays $J/\psi \to K^+K^- \pi^0$ and $J/\psi \to K_SK^\pm \pi^\mp$ have been studied at BES and BaBar. The processes are dominated by $K^*(892) \bar K$ and  $K_2^*(1430)\bar K$ intermediate states.
However, in 2005 BESII observed a broad structure in the $K^+K^-$ invariant mass in $J/\psi \to K^+K^- \pi^0$~\cite{Ablikim:2006hp}, dubbed $X(1575)$, using a data sample of $1000(100)$ events.  The $J/\psi$ decays to strangeonium states plus $\pi^0$ are  doubly OZI suppressed and the branching fractions should be very small, e.g., $\BR(J/\psi\to\phi\pi^0)<10^{-5}$. Therefore the broad $X(1570)$ state triggered speculations on its nature, including multi-quark states and the interference between $\rho(1450)$ and $\rho(1700)$.  Subsequently the Dalitz plot analysis using $2002(48)$ events performed by BaBar~\cite{Lees:2017ndy} attributed the broad structure to the presence of the $\rho(1450)$.  

\begin{figure}[t]
\centerline{\includegraphics[width=0.9\textwidth]{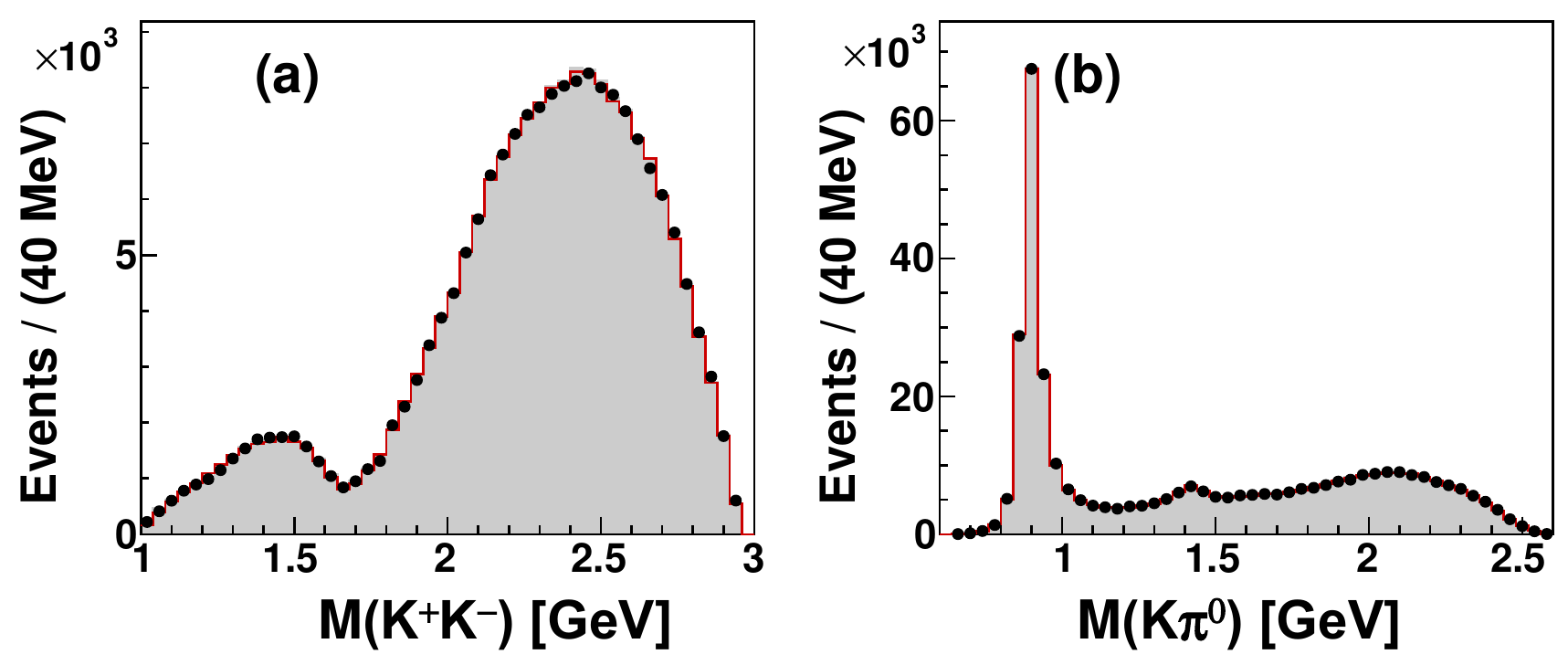}}
\caption[$J/\psi\to K^+K^-\pi^0$]{Invariant mass of the $K^+K^-$ (a) and $K^\pm\pi^0$ (b) systems for $J/\psi\to K^+K^-\pi^0$ from BESIII~\cite{Ablikim:2019tqd}  compared to two PWA solutions
PWA solution~I (shaded histograms) and the PWA solution~II (solid line).
}
\label{fig:kin_distributions_full}
\end{figure}

The most recent partial-wave analysis of the decay $J/\psi \to K^+K^-\pi^0$ has been performed at \mbox{BESIII}~\cite{Ablikim:2019tqd}, using $181550(430)$ signal events.
The invariant-mass distribution of both the $K^+K^-$ and the $K^\pm\pi^0$ systems are shown in Fig.~\ref{fig:kin_distributions_full}. 
In addition to the dominant $K^*(892)^\pm K^\mp$ and $K^*_2(1430)^\pm K^\mp$ channels, small contributions from $K^*_2(1980)^\pm K^\mp$ and $K^*_4(2045)^\pm K^\mp$ are also necessary to provide a good description of the data.  The broad structure clearly observed in the  $K^+K^-$ mass spectrum cannot be described with a single state. The PWA solution cannot be saturated with the well-known states included as Breit--Wigner resonances and constant contributions in the lowest partial waves. At the same time, the ``missing part'' of the PWA solution cannot be reliably attributed to a single resonance and mainly manifests itself as a slowly changing background in the $J^{PC}=3^{--}$  partial wave of the $K^\pm\pi^0$ pairs at high  $K^\pm\pi^0$ masses. 

Two solutions constructed with and without the smooth contribution in this partial wave were provided to demonstrate that the conclusions of this analysis are not strongly affected by the assumptions on the ``missing part'' of the PWA solution. For the first solution,  one contribution is from a vector structure  with  $M=1643(3)\MeV$ and  $\Gamma=167(12)\MeV$ as indicated in the Dalitz plot, which could be attributed to the $\rho(1700)$ or the $\omega(1650)$.  The second contribution that can be reliably identified is a vector resonance with a mass of $2078(6)\MeV$ and width of $149(21)\MeV$.  However, to describe the data additional contributions are necessary.
Tests including contributions from  $\rho(770)$, $\rho_3(1690)$, $\rho(1450)$, as well as a smooth background in the  $J^{PC}=1^{--}$ $K^+K^-$
partial wave, are all found to be significant. 

For the second solution, a striking feature is the presence of a nonresonance component in the $J^P=3^-$  $K\pi^0$ partial waves, which cannot be clearly interpreted as an interference between Breit--Wigner states. A possible interpretation is that this component is the manifestation of nonresolved contributions present in the $F$-wave $K\pi$ scattering amplitude~\cite{Aston:1987ir}. This may include the presence of several resonances, nonresonant production, and final-state particle rescattering effects. In this case, two $J^{PC}=1^{--}$ states around $1.65 \GeV$ and $2.04 \GeV$ were introduced to describe the enhancement at low  $M(K^+K^-)$ mass. 
The first state may be the $^3D_1$ isovector ground state, but at the same time its mass, width, and small relative contribution to the decay are reasonably consistent with the $\omega(1650)$ produced in $J/\psi$ decays through a virtual photon. The second state can be interpreted as the $\rho(2150)$ or as another isovector vector state.

\subsection[Other $J/\psi\to\gamma X$ and $J/\psi \to V X$ reactions]{\boldmath Other $J/\psi\to\gamma X$ and $J/\psi \to V X$ reactions}\label{sec:Xstates}
In this section, we discuss extensions of the methods for studies of the two-pseudoscalar-meson system $X$ in $J/\psi\to\gamma X$ and $J/\psi\to V X$ presented in Secs.~\ref{sec:VtoPPg} and \ref{sec:VtoPPV} to final states involving more particles. In the first place, the system $X$ can decay to three pseudoscalar mesons. 
In radiative decays of $J/\psi$ to three (and more) pseudoscalar mesons, hadronic systems $X$ with $C=+1$ and a variety of $J$ and $P$ quantum numbers can be studied. For example, for $\pi^+\pi^-\eta$ all $J^{P}$ values are possible: $0^{-}$, $1^{+}$, $1^{-}$, $2^{+}$, $2^{-}$, etc. 
An additional motivation for the studies of $J/\psi$ radiative decays is the large fraction of two-gluon processes $J/\psi\to\gamma gg$, estimated to be $8(1)\%$~\cite{PDG}. They are therefore considered  as promising  modes to search for glueballs~\cite{Brodsky:1977du}.

\subsubsection[Observation and properties of the $X(1835)$ state]{\boldmath Observation and properties of the $X(1835)$ state}
\paragraph{\boldmath $J/\psi\to\gamma\pi^+\pi^-\eta'$}

\begin{figure}
    \centering
 \includegraphics[height=0.17\textheight]{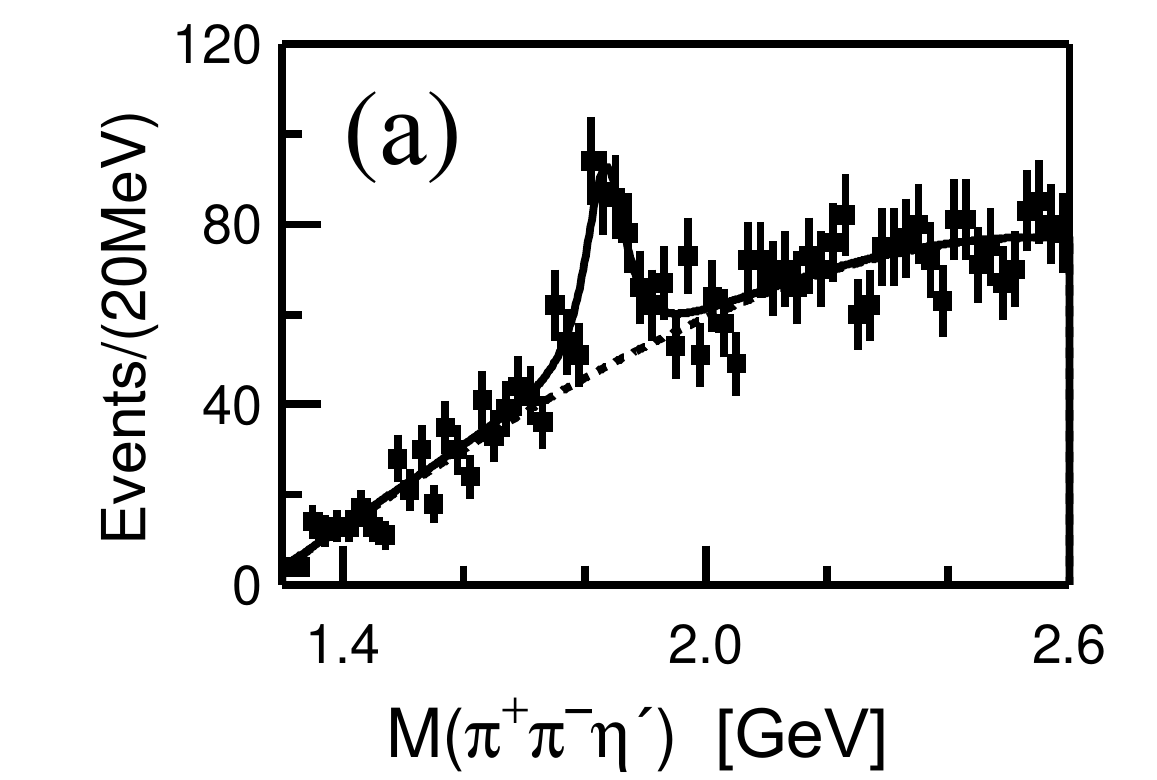}
 \includegraphics[height=0.18\textheight]{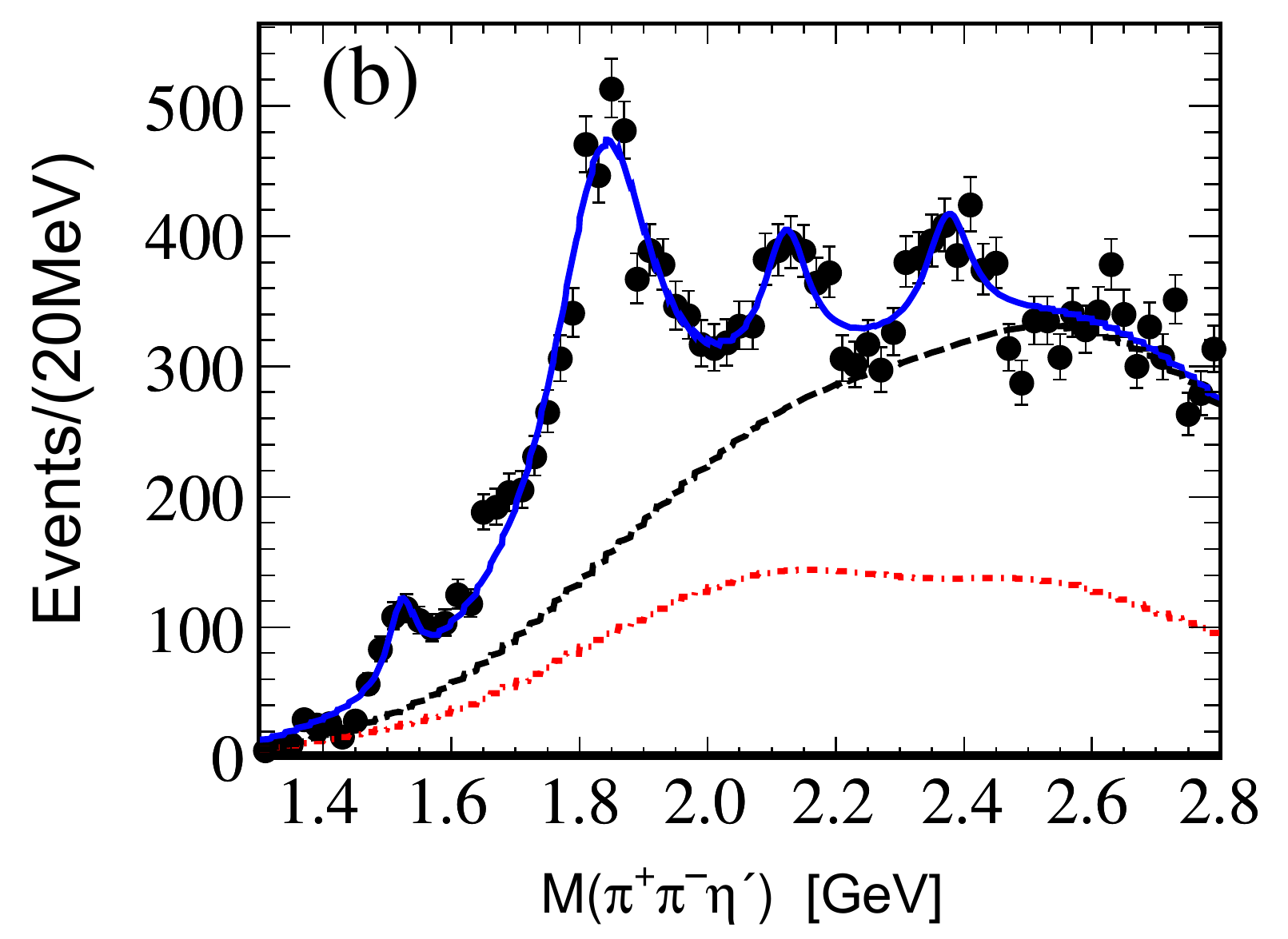}
 \includegraphics[height=0.18\textheight]{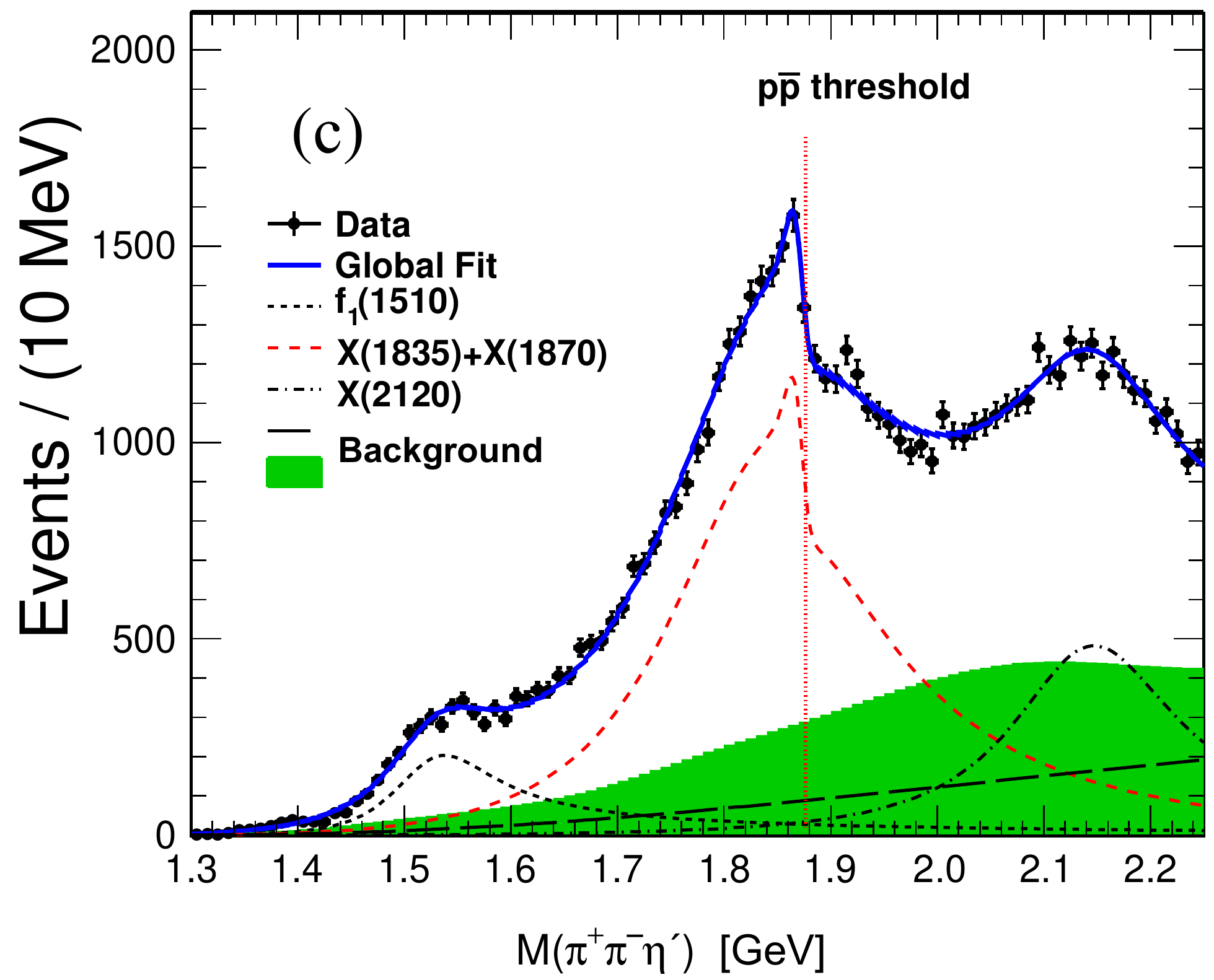}
    \caption[$X(1835)$ and other structures observed in the invariant mass $M(\pi^{+}\pi^{-}\eta^\prime)$]{$X(1835)$ and other structures observed in the invariant mass of the $\pi^{+}\pi^{-}\eta^\prime$ system $M(\pi^{+}\pi^{-}\eta^\prime)$ produced in $J/\psi$ radiative decays. (a) First observation at BESII~\cite{Ablikim:2005um}; (b) BESIII 2009 data~\cite{Ablikim:2010au}; (c) BESIII 2009 and 2012 data~\cite{Ablikim:2016itz}.}
    \label{fig:x1835}
\end{figure}

In 2005, the BESII experiment observed a structure in the invariant mass of the $\pi^{+}\pi^{-}\eta^\prime$ system $M(\pi^{+}\pi^{-}\eta^\prime)$ produced in $J/\psi$ radiative decay closely below nucleon--antinucleon threshold~\cite{Ablikim:2005um}, see Fig.~\ref{fig:x1835}(a). This allows one to speculate about a relation to the threshold enhancement in radiative $J/\psi$ decays to a $J^P=0^+$ $p\bar p$ pair, observed by BESII in 2003~\cite{Bai:2003sw} and subsequently confirmed by BESIII~\cite{BESIII:2011aa}. This interesting structure, called $X(1835)$, attracted many theoretical interpretations, including a $p\bar{p}$ bound state, a glueball, a radial excitation of the $\eta^\prime$ meson, etc.~\cite{Ding:2005ew,Zhu:2005ns,Huang:2005bc,Li:2005vd,Ding:2005gh}. 

Three structures, $X(1835)$, $X(2120)$, and $X(2370)$, were observed by the BESII and BESIII experiments in the invariant mass of the $\pi^{+}\pi^{-}\eta^\prime$ system produced in $J/\psi$ radiative decays~\cite{Ablikim:2005um,Ablikim:2010au},
see Fig.~\ref{fig:x1835}(b). 
The $X(1835)$ is probably the most interesting among those, due to its possible connection to the $p\bar p$ threshold, with many experimental findings pointing to this connection. The new BESIII analysis using 2012 data clearly shows that the 
shape of the $X(1835)$ is sharply affected by the $p\bar p$ threshold, see
Fig.~\ref{fig:x1835}(c)~\cite{Ablikim:2016itz}. The spin-parity of $X(1835)$ of $J^P=0^+$ was determined  by an amplitude analysis of the $J/\psi$ radiative decay to $\eta K_{S} K_{S}$~\cite{Ablikim:2015toc}. 
Moreover,  no evidence of the $X(1835)$ is found in $J/\psi\to\omega\eta^{\prime}\pi^{+}\pi^{-}$~\cite{BESIII:2019sfz}. 
The state is observed in $J/\psi\to\gamma\gamma\phi$ together with the $\eta(1475)$~\cite{Ablikim:2018hxj}, indicating a sizable $s\bar{s}$ component. 

The two higher-mass $X$ structures were studied in the decay $J/\psi\to\gamma K\bar K\eta'$. The higher-mass state $X(2370)$ was found with statistical significance of $8.3\sigma$~\cite{Ablikim:2019zyw}. In addition, the 90\% C.L.\ upper limit for the branching ratio of the decay chain  $J/\psi\to\gamma X(2120)\to\gamma  K\bar K\eta'$ was set at the $10^{-5}$ level.

The huge number of the  accumulated $J/\psi$ events allows BESIII to perform much more detailed studies of the $X(1835)$ and the other structures to understand their nature. 
Two decay modes of the $\eta^\prime$, $\eta^\prime\rightarrow \pi^+\pi^-\gamma$ and $\eta^\prime\rightarrow \pi^+ \pi^- \eta$, are used to reconstruct the $\eta^\prime$, which helps to cross-check the measurement as well as to improve on the statistical precision. Figure~\ref{fig:x1835} shows the $\pi^+\pi^-\eta^\prime$ invariant-mass spectrum for the two $\eta^\prime$ decay modes combined.  The $X(1835)$ structure with a width of $190(40)\MeV$ is seen~\cite{Ablikim:2010au}, cf.\ Fig.~\ref{fig:x1835}(b). In addition a little peak corresponding to the $f_1(1510)$ and two other unknown structures $X(2120)$ and $X(2370)$ are observed.   
  The masses and widths of these states are for the
   $X(1835)$ $M=1836(6)\MeV$, $\Gamma=190(49)\MeV$; for the  $X(2120)$ $M=2122(8)\MeV$, $\Gamma=83(35)\MeV$; and for the
 $X(2370)$  $M=2376(10)\MeV$, $\Gamma=83(47)\MeV$.  

However, an anomalous shape of the structure around $1.85\GeV$ was observed in the $\pi^+\pi^-\eta^\prime$ mass spectrum~\cite{Ablikim:2016itz} that cannot be accommodated  by an ordinary Breit--Wigner resonance
function. Two example models for such a line shape are used to describe the data. The first model assumes the state couples to $p \bar{p}$ and the distortion reflects the opening of the
$p \bar{p}$ decay channel. The fit result for this model yields a strong coupling between the broad structure and $p \bar{p}$. 
The pole nearest to the $p \bar{p}$ mass threshold of this state is located at $M_\text{pole}=1909(19)\MeV$ and $\Gamma_\text{pole}=273(40)\MeV$.
The second model assumes that the distortion reflects interference between the $X(1835)$ and another resonance with mass close to the $p \bar{p}$ mass threshold.  A fit with this model uses a coherent
sum of two interfering Breit--Wigner amplitudes and yields a narrow resonance below the $p \bar{p}$ mass threshold with $M=1870(3)\MeV$ and $\Gamma=13.0(67)\MeV$. With present accuracy, both models fit the data well and both suggest the existence of a new state: either a broad one with strong couplings to $p \bar{p}$, or a narrow one just below the $p \bar{p}$ mass threshold. For the broad state above the $p \bar{p}$ mass threshold, its strong couplings to $p\bar{p}$ suggests the existence of a $p \bar{p}$ molecule-like state. On the other hand the narrow width of the state just below $p \bar{p}$ mass threshold suggests an unconventional meson, most likely a $p \bar{p}$ bound state. 
In order to further elucidate the nature of the states around $1.85\GeV$,
more data is needed to further study the $J/\psi\to\gamma\eta^\prime\pi^+\pi^-$ process together 
with an analysis of line shapes for other $J/\psi$ decays near the $p \bar{p}$ mass threshold, including $J/\psi\to\gamma p \bar{p}$ and $J/\psi\to\gamma\eta K^{0}_{S}K^{0}_{S}$ reactions.

\begin{figure}[t]
    \centering
    \includegraphics[width=0.35\textwidth]{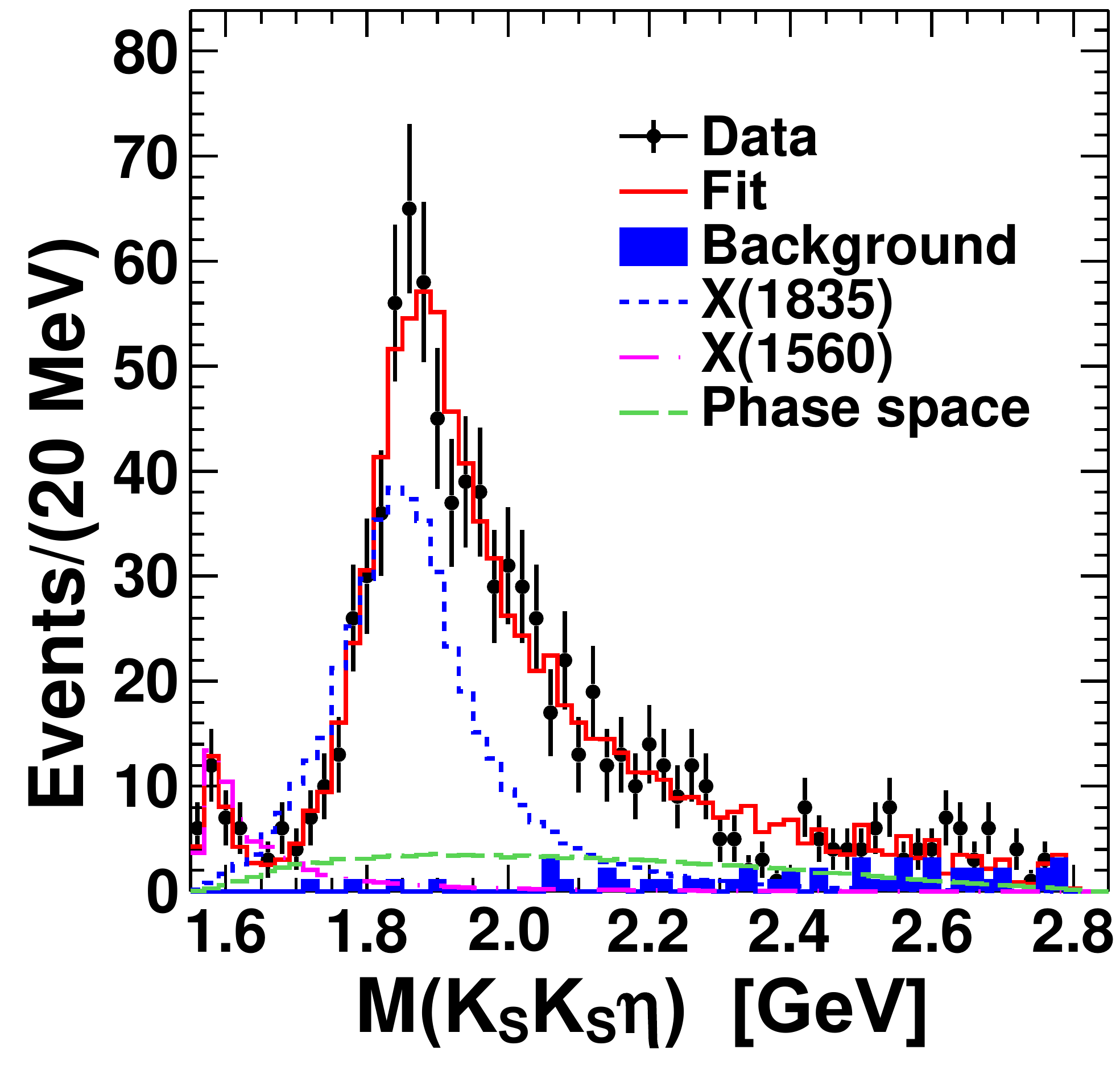}
    \put(-140,150){\large(a)}\hspace*{1cm}
    \includegraphics[width=0.35\textwidth]{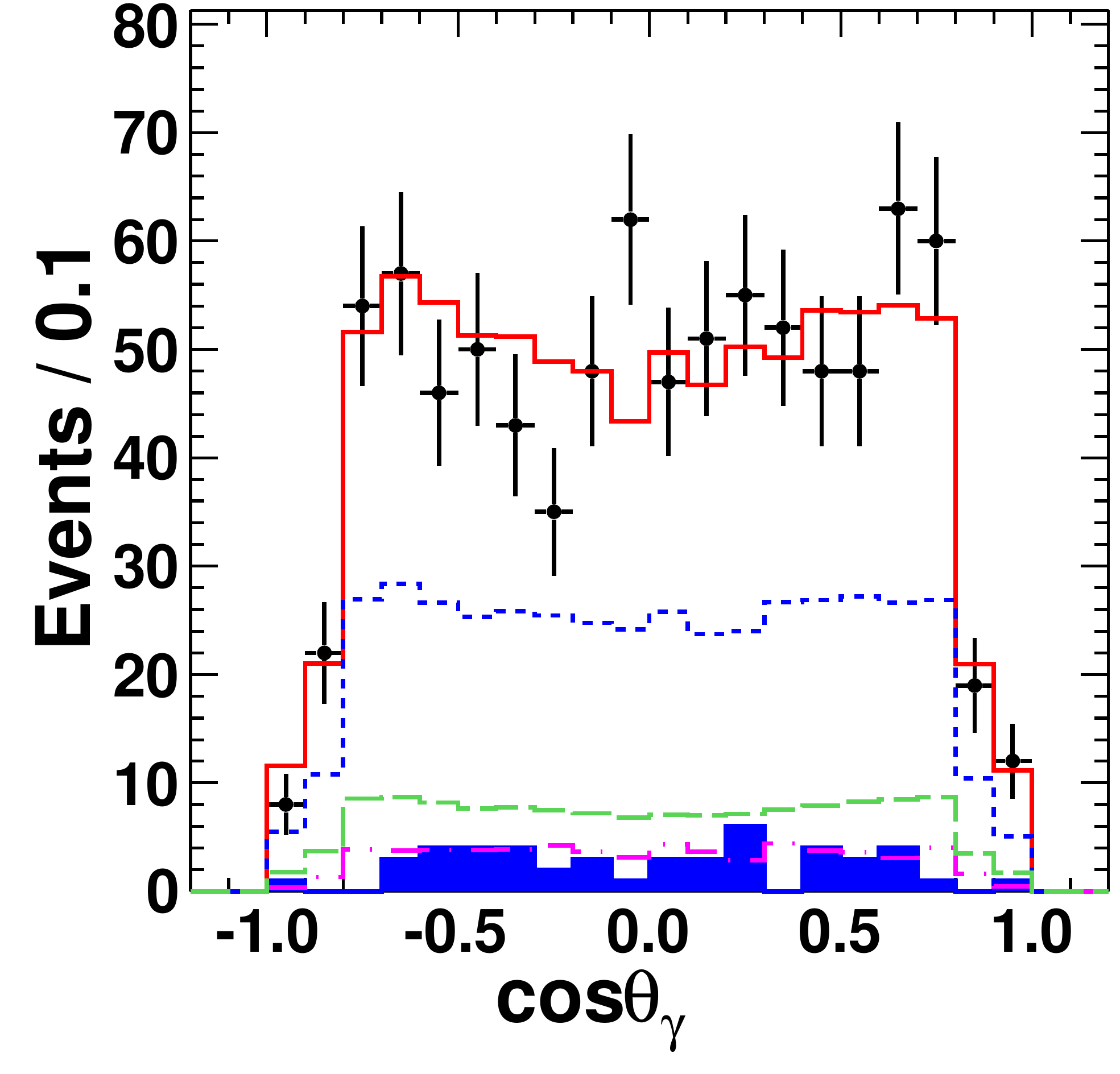}
    \put(-140,150){\large(b)}
   \caption[$J/\psi \to \gamma K_S^0K_S^0\eta$ from BESIII]{\label{fig:KKeta}
   (a)  $K_SK_S\eta$ invariant-mass spectrum for $J/\psi \to \gamma K_S^0K_S^0\eta$  events from BESIII with 
      the requirement $M{(K_SK_S)}<1.1\GeV$ and (b) the angular distribution of the photon.
      }
\end{figure}

\paragraph{\boldmath $J/\psi \to \gamma K_S^0K_S^0\eta$}
Further studies of  the structures observed in $\pi^+\pi^-\eta^\prime$ include searches for  other decay modes like $K_{S}K_{S}\eta$. A BESIII study of $J/\psi\rightarrow\gamma K_{S}K_{S}\eta$~\cite{Ablikim:2015toc} with invariant masses $M(K_S K_S)$ in the $f_0(980)$ mass region, $M(K_S K_S)<1.1\GeV$, a structure around $1.85\GeV$ is clearly seen, cf.\  Fig.~\ref{fig:KKeta}(a). The performed partial-wave analysis allows one to determine the spin and parity to be $J^{PC}=0^{-+}$ (pseudoscalar), the same as the $X(p\bar p)$ structure~\cite{BESIII:2011aa}. The mass and width are $1844(22)\MeV$ and  
$192(56)\MeV$, respectively, consistent with the $J/\psi\to\gamma\pip\pim\etap$ results.
\subsubsection[$J/\psi\to \gamma \pi^+\pi^-\pi^0 (3\pi^0)$]{\boldmath $J/\psi\to \gamma \pi^+\pi^-\pi^0 (3\pi^0)$}
\begin{figure}[t]
  \centering
 \includegraphics[width=0.4\textwidth]{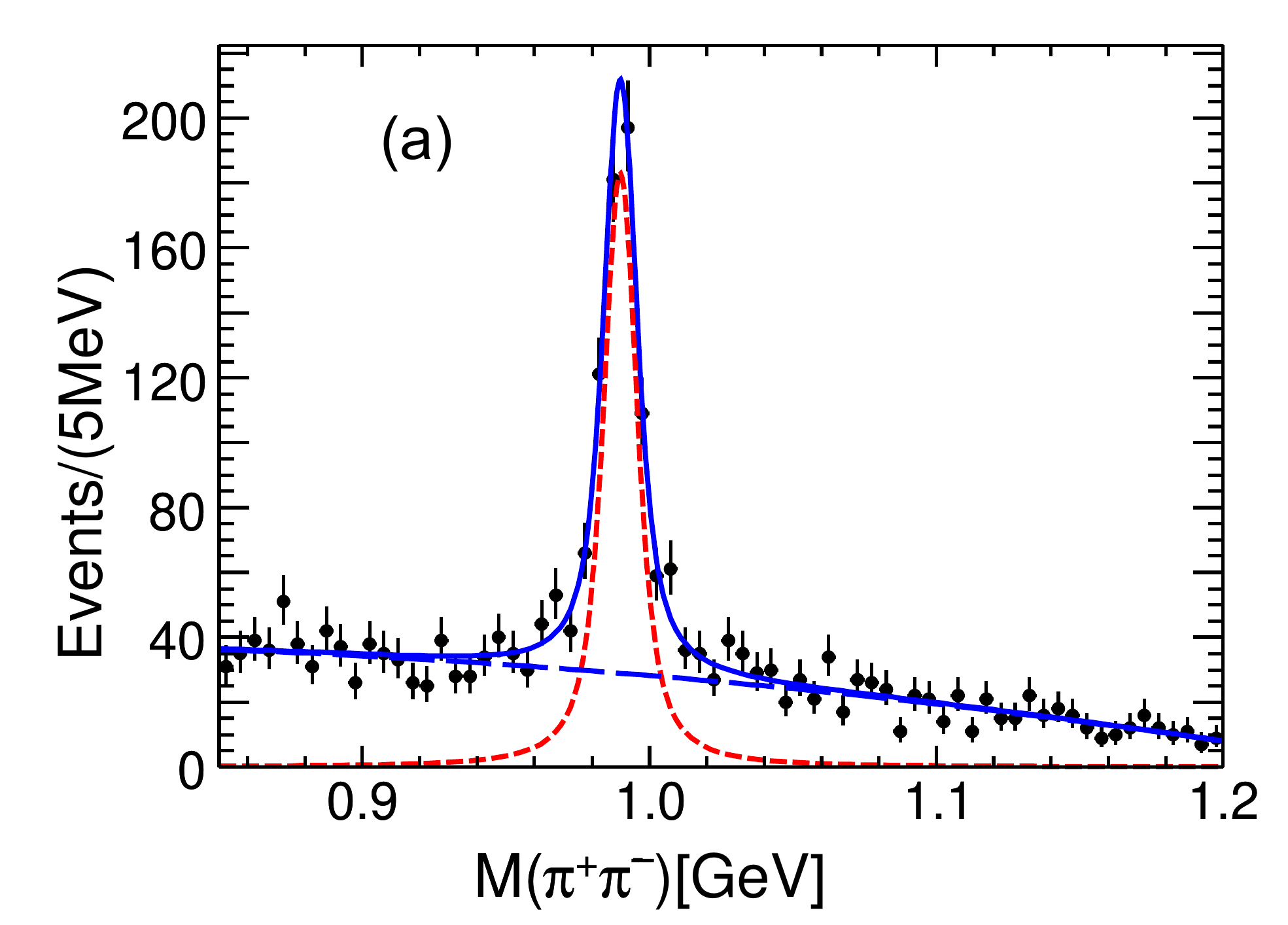}    \includegraphics[width=0.4\textwidth]{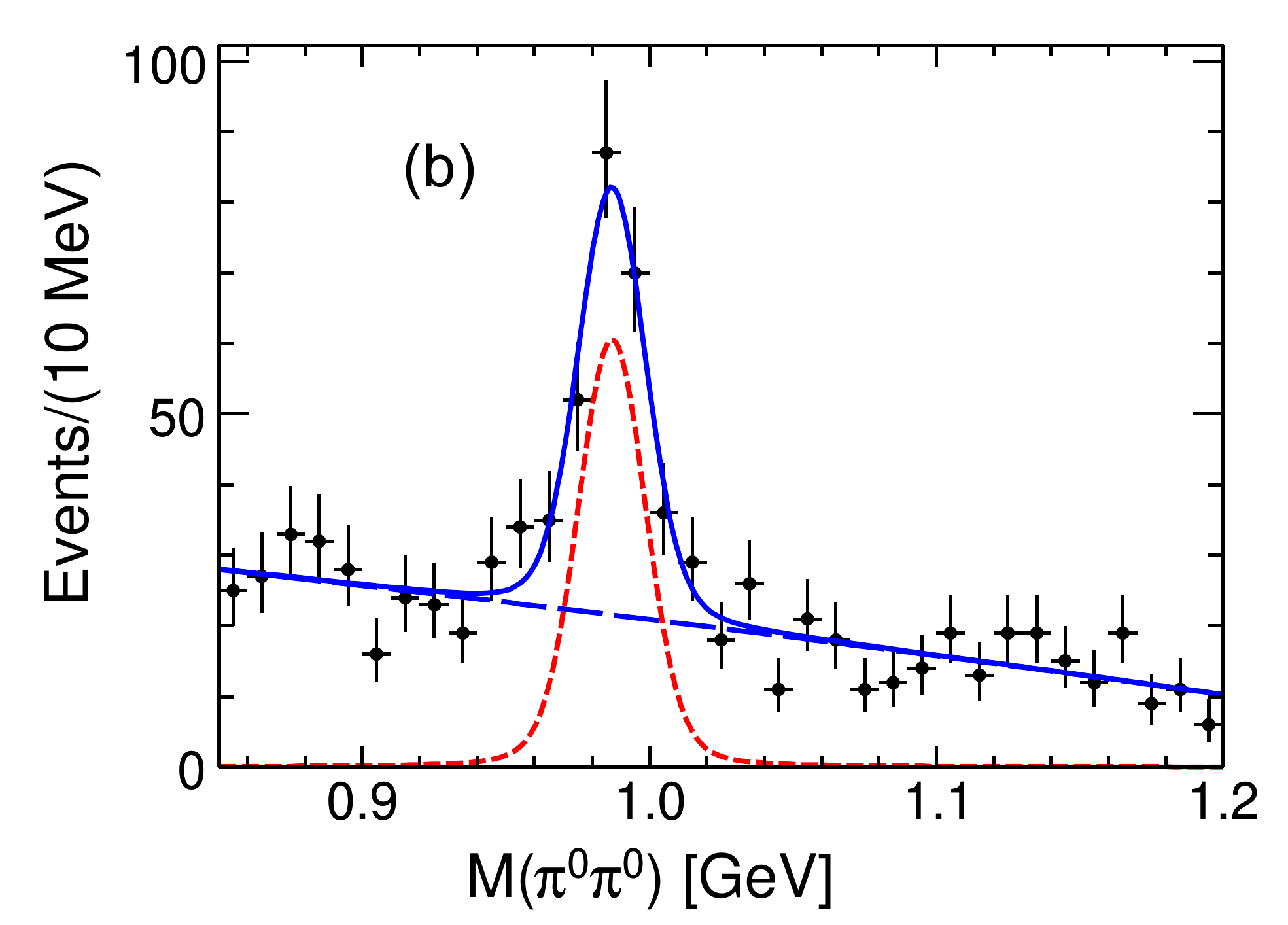}  
    \caption[$J/\psi\rightarrow \gamma \pi^+\pi^-\pi^0 (3\pi^0)$]{\label{fig:fit_result_2pi}
    The $\pi^+\pi^-$ (a) and $\pi^{0}\pi^{0}$ (b) invariant-mass
spectra with the $\pi^+\pi^-\pi^0$ and $3\pi^0$ systems in the $\eta(1405)$ mass
region, respectively. The dotted curves are the $f_{0}(980)$ signals, while the dashed curves denote the background. 
}
\end{figure}

Analyses of $J/\psi \rightarrow \gamma \pi^+\pi^-\pi^0$ and $J/\psi \rightarrow \gamma 3\pi^0$ at BESIII~\cite{BESIII:2012aa} can be used to study $\eta(1405)\to f_{0}(980)\pi^0$ with isospin violation. The apparent signatures of this  $\eta(1405)$ decay mode in the final states are shown in Fig.~\ref{fig:fit_result_2pi}. 
The ratio of $\BR(\eta(1405)\rightarrow
f_{0}(980)\pi^{0} \rightarrow\pi^{+}\pi^{-}\pi^{0})$ determined from this measurement to
$\BR(\eta(1405)\rightarrow a_{0}^{0}(980)\pi^{0} \rightarrow
\eta\pi^{0}\pi^{0})$~\cite{PDG} is $17.9(4.2)\%$, which
is one order of magnitude larger than the $a_0^{0}(980)-f_0(980)$
mixing intensity (less than 1\%) determined at BESIII previously
~\cite{Ablikim:2010aa}. In addition, the measured width of the $f_{0}(980)$
observed in the dipion mass spectra is anomalously narrow, $9 \MeV$, compared to the world average.
There is also evidence for an enhancement around
$1.3\GeV$ (likely from the $f_{1}(1285)$ or the $\eta(1295)$) in the charged mode.

\subsection{Strong three-body decays of nonvector states} \label{sec:strong3p}

Nonvector states cannot be directly produced in $e^+e^-$ collisions via single-photon processes,\footnote{Production via two-photon intermediate states is obviously strongly suppressed, but has been searched for, e.g., at the Novosibirsk experiments to put upper limits on $e^+e^-\to\eta'(958)$~\cite{Akhmetshin:2014hxv,Achasov:2015mek} or to establish the reaction $e^+e^-\to f_1(1285)$~\cite{Achasov:2019wtd}.} but as discussed in Sec.~\ref{sec:prodVmesons}, they are abundantly produced in $\phi$, $J/\psi$, and $\psi'$ decays. The decays of $\eta$, $\eta'$, $\eta_c$, $\chi_{cJ}$ ($J^{PC}=J^{++}$), and $h_c$ ($J^{PC}=1^{+-}$) into three pseudoscalars allow us to study the interactions in the configurations corresponding to the  quantum numbers of the initial states. The special cases are those of $\eta$ and $\eta'$ decays, which only involve the pseudoscalar nonet.
In the case of the $\eta_c$, the large decay width $32.0(7)\MeV$ presents 
certain challenges for amplitude analyses: depending on the decay mode there may be significant contributions from interference with the nonresonant  $J/\psi\to\gamma  X$ decays to the identical 
final states~\cite{Mitchell:2008aa,BESIII:2011ab,Anashin:2014wva}. Therefore a viable alternative, where the
 $\eta_c$ is produced in two-photon processes, is discussed in this section.

\paragraph{\boldmath $\eta\to3\pi$}
The $\eta$ meson decays into three pions ($\eta\to\pi^+\pi^-\pi^0$ and $\eta\to3\pi^0$)  are two of the main decay modes (the branching ratios are 32.7\% and 22.9\%, respectively). 
They are isospin-violating decays; since the electromagnetic contribution is suppressed~\cite{Sutherland:1966zz, Bell:1996mi, Baur:1995gc, Ditsche:2008cq}, the decay mechanism is instead driven by the light quark mass difference $m_u-m_d$.
\begin{figure}
\centering
\includegraphics[width=0.99\textwidth]{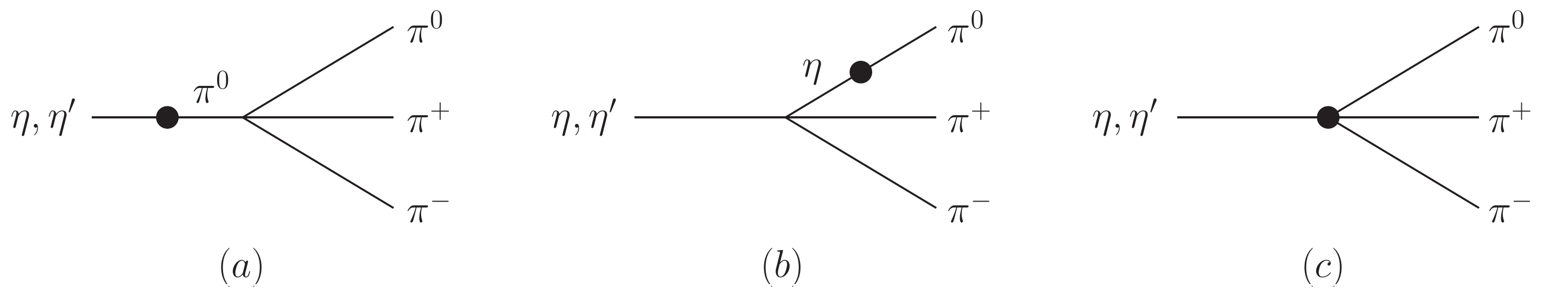}
\caption[ChPT LO diagrams for  $\eta^{(\prime)}\to  3\pi$]{\label{fig:grapheta3pi} 
The isospin-violating decay $\eta^{(\prime)} \to  \pi^+\pi^-\pi^0$ in leading-order ChPT (or current algebra)~\cite{Osborn:1970nn}: (a)+(b) $\eta^{(\prime)}$--$\pi^0$ mixing, (c) contact term.
}
\end{figure}
The lowest-order contributions~\cite{Osborn:1970nn} to the decay are shown in Fig.~\ref{fig:grapheta3pi}. 
The resulting decay amplitude for $\eta\to\pi^+\pi^-\pi^0$, ${\cal A}_c(s,t,u)$, is a linear function of the Mandelstam variable $s=(p_{\pi^+}+p_{\pi^-})^2$ in LO ChPT~\cite{Bardeen:1967rfy,Osborn:1970nn,Gasser:1984pr},
\begin{equation}
{\cal A}_c(s,t,u) 
={\cal A}_c(x,y) 
\propto \frac{3s-4m_\pi^2}
{m_\eta^2-m_\pi^2}=1-\frac{2m_\eta(m_\eta-3m_\pi)}{m_\eta^2-m_\pi^2}y \,,
\label{eq:pcac}
\end{equation}
where the Dalitz plot variables $x$ and $y$ are defined in Eq.~\eqref{eq:DaltzXY}. 
As ${\cal A}_c(s,t,u)\propto m_u-m_d$, 
the partial decay width can be formulated as a precise and sensitive probe of the light quark mass ratio $Q^2=(m_s^2-\hat m^2)/(m_d^2-m_u^2)$~\cite{Leutwyler:1996qg} and is used as a benchmark of ChPT and dispersive calculations. 
The need for the latter arises through the strong role played by final-state $\pi\pi$ rescattering~\cite{Roiesnel:1980gd}, in particular in the isospin $I=0$ $S$-wave: the perturbative chiral loop expansion converges only very slowly~\cite{Gasser:1984pr,Bijnens:2007pr}.
For this reason, dispersive representations akin to the one we have briefly described earlier for $\omega/\phi\to3\pi$, see Sec.~\ref{sec:3pidynamics}, have been studied extensively in the literature~\cite{Kambor:1995yc,Anisovich:1996tx,Kampf:2011wr,Guo:2015zqa,Guo:2016wsi,Colangelo:2016jmc,Colangelo:2018jxw,Kampf:2019bkf}:
rescattering is resummed in Khuri--Treiman-type representations, based on analyticity and unitarity and using precisely known $\pi \pi$  phase shifts as the essential input.  The remaining free parameters in the form of subtraction constants can be obtained from fits to the experimental Dalitz plot distributions and subsequently matched to ChPT at a kinematical point where the chiral expansion works best.  
As these theoretical studies have recently been reviewed and summarized elsewhere~\cite{Gan:2020aco}, we refrain from describing them here in any detail.

For $\eta\to\pi^+\pi^-\pi^0$, a polynomial parameterization is often used to represent the Dalitz plot distribution,
\begin{equation}
|\mathcal{A}_c(x,y)|^{2}    \propto 1   +ay    +
by^2+cx+dx^2+exy+fy^3+gx^2y+hx^3 +\ldots \,,\label{eq:DPpolform}
\end{equation}
where  the coefficients of the polynomial in $x$ and $y$ are called Dalitz  plot parameters.
The LO result Eq.~\eqref{eq:pcac} results in 
$a=-1.04$, $b=a^2/4=0.27$, and all other parameters vanish.  
The terms with odd powers of the $x$ variable, like
$c$,  $e$,  $h$, \ldots, are zero in the strong and electromagnetic interactions, as they imply charge  conjugation
violation.  The observation of C-violating asymmetries is of significant interest for searches of physics beyond the Standard Model~\cite{Gan:2020aco}; see Ref.~\cite{Gardner:2019nid} for an analysis of C-violating $\eta\to\pi^+\pi^-\pi^0$ amplitudes beyond polynomial parameters.

At $\Order(m_u-m_d)$, the amplitudes for $\mathcal{A}_c(\eta\to\pi^+\pi^-\pi^0)$ and $\mathcal{A}_n(\eta\to3\pi^0)$ are related by~\cite{Zemach:1963bc}
\begin{equation}
{\cal A}_n(s,t,u)=-{\cal A}_c(s,t,u)-{\cal A}_c(t,u,s)-{\cal A}_c(u,s,t) \,.
\label{eq:threepiamp}
\end{equation}
Equation~\eqref{eq:threepiamp} is valid at leading order in isospin breaking; corrections to this relation have been investigated and found to be tiny~\cite{Ditsche:2008cq,Schneider:2010hs}. 
Therefore, Eq.~\eqref{eq:pcac} implies ${\cal A}_n(s,t,u)$ to be a constant at leading order.
\begin{figure}
\centering
\vspace*{-0.8cm}
\includegraphics[width=0.45\textwidth]{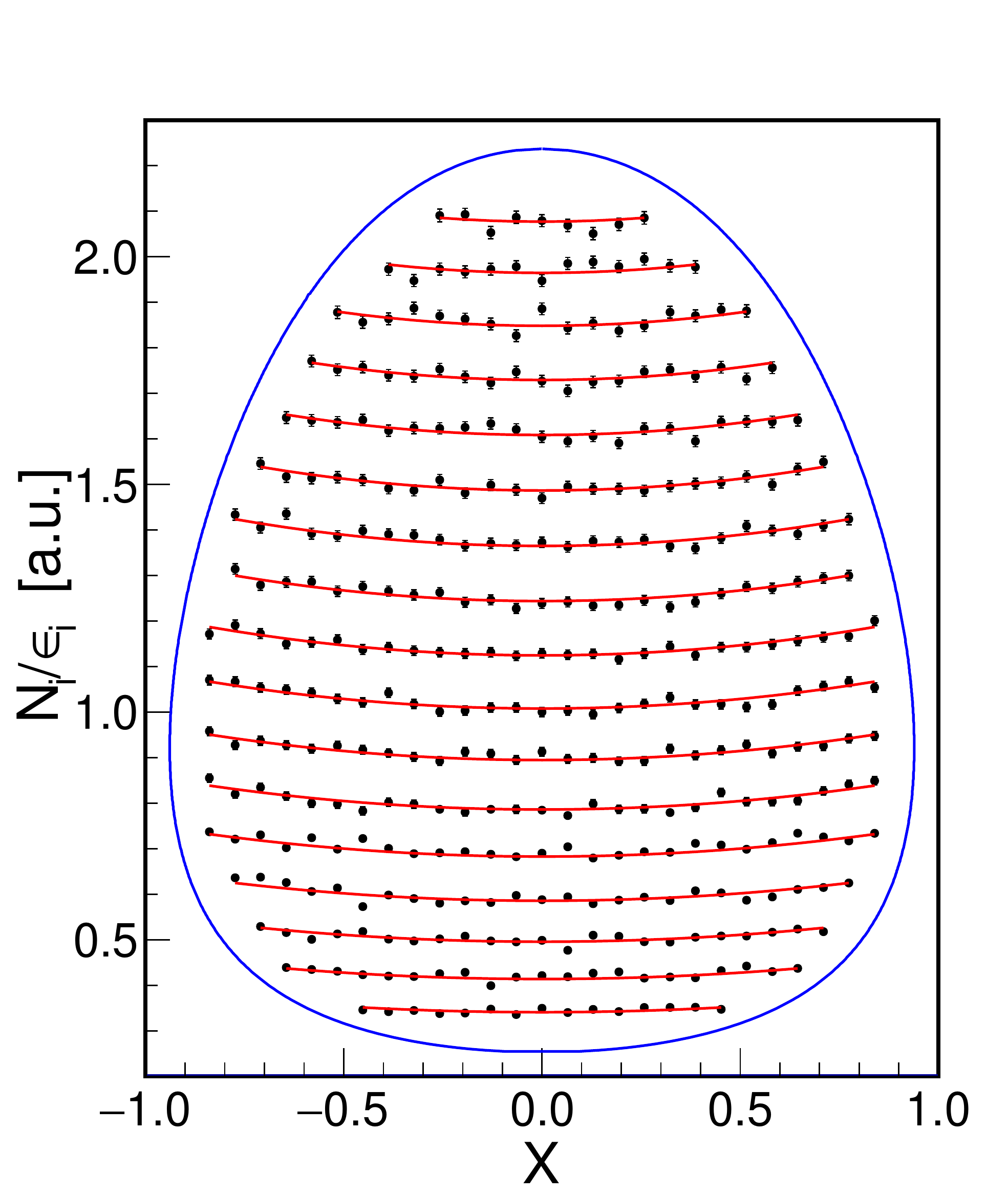}
\caption[KLOE $\eta\to\pi^+\pi^-\pi^0$ Dalitz plot]{Acceptance corrected $\eta \to \pi^+ \pi^- \pi^0$ Dalitz plot from KLOE~\cite{Anastasi:2016cdz} together with the result 
of the polynomial fit. Separate rows correspond to different $y$ values, where the number of events decreases with increasing $y$.}
\label{fig:DPeta3pi}
\end{figure}

The most precise result for the $\eta \to \pi^+ \pi^- \pi^0$ Dalitz plot is from KLOE-2~\cite{Anastasi:2016cdz}, using $4.7\times 10^6$ $\eta\to\pi^+ \pi^- \pi^0$ candidate events from $e^+e^-\to \phi\to\eta\gamma$ with background admixture of less than one percent.  The Dalitz plot distribution is constructed using 31 and 20 bins for the $x$ and $y$ ranges, respectively. The acceptance corrected data is presented in Fig.~\ref{fig:DPeta3pi} and compared to the fit using the polynomial parameterization of Eq.~\eqref{eq:DPpolform}. This presentation uses the fact that the $y$ dependence is monotonic, and therefore each row of data points (the fit line) corresponds to the subsequent value of the $y$ bin ranges. The lowest rows correspond to the largest $y$ bin values.
The results for the Dalitz plot parameters are $a = -1.095(4)$, $b = 0.145(6)$, $d = +0.081(7)$, $f = +0.141(11)$, and $g=-0.044(15)$. They are a factor of 2--3 more precise than previous measurements, and for the first time a statistically significant contribution of the $\propto g x^2y$ term is found. The data was used for the determination of the quark mass ratio by the Bern group~\cite{Colangelo:2016jmc,Colangelo:2018jxw}.

\paragraph{\boldmath $\etap\to\eta\pi\pi$}
The combined branching fraction of the two main hadronic decays of the $\eta'$, $\eta'\to\eta\pi^{+}\pi^{-}$ and $\eta'\to\eta\pi^{0}\pi^{0}$, is nearly $2/3$.  The ratio $\Gamma(\eta'\to\eta\pi^+\pi^-)/\Gamma(\eta'\to\eta\pi^0\pi^0)$ should be exactly two in the isospin limit.  The decays involve both $\eta$ and pions in the final state and therefore are sensitive to the largely unknown $\pi\eta$ interactions (dominantly in the $S$-wave). However, the excess energy of the processes is relatively small: $131\MeV$ and $140\MeV$ for $\eta\pi^+\pi^-$ and $\eta\pi^0\pi^0$, respectively. This means that precision high-statistics experimental data on the Dalitz plots together with the appropriate theory framework are needed to understand the dependence on $\pi\eta$ phase shifts.

\begin{figure*}[t]
\begin{center}
\includegraphics[width=0.98\textwidth]{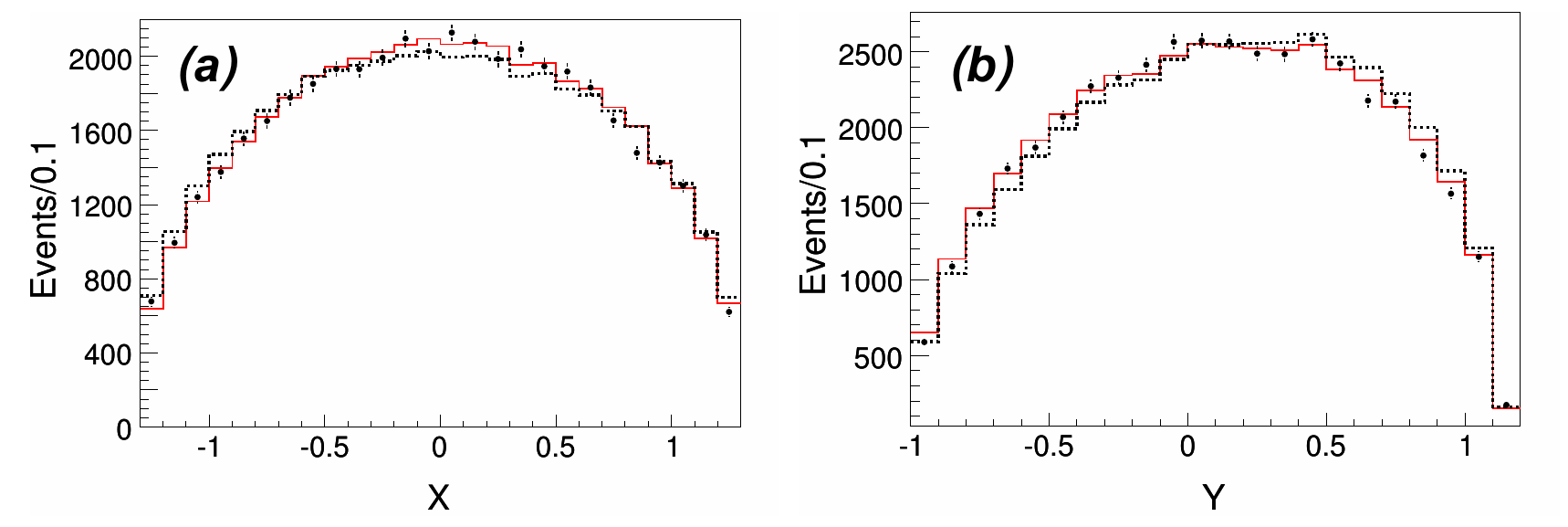}
\caption[Results for $\eta'\to\eta\pi^+\pi^-$ from BESIII] { Projections on the Dalitz plot variables $x$ and $y$ from the BESIII  measurement of $\eta'\to\eta\pi^+\pi^-$~\cite{Ablikim:2017irx}. The dashed
histograms are for the MC events generated with constant matrix element. The solid histograms are the result of the fit of the polynomial parameterization Eq.~\eqref{eq:DPpolform}.} \label{fig:etapxy}
\end{center}
\end{figure*}

There are two recent measurements of these decays from the BESIII~\cite{Ablikim:2017irx} and A2~\cite{Adlarson:2017wlz} experiments. The A2 measurement uses a photoproduction reaction on a proton target and considers only the  $\eta'\to\eta\pio\pio$ channel, with a signal of $1.2\times10^{5}$ events. The BESIII analysis includes both channels simultaneously, using $3.5\times 10^5$ $\eta^\prime\rightarrow\eta\pi^+\pi^-$ and $5.6\times 10^4$ $\eta^\prime\rightarrow\eta\pi^0\pi^0$ events. The general feature of the Dalitz plots is that the deviation from a uniform distribution is small, as illustrated for the $\eta^\prime\rightarrow\eta\pi^+\pi^-$ data in Fig.~\ref{fig:etapxy}.
Therefore, a polynomial parameterization analogous to the one for $\eta\to\pi^+\pi^-\pi^0$, see Eq.~\eqref{eq:DPpolform}, is frequently employed, with the similar restriction that terms odd in $x$ are forbidden either by $C$-conservation (for charged pions) or Bose symmetry (neutral).
The BESIII data is not consistent with the previous measurement from VES~\cite{Dorofeev:2006fb} and the predictions within the $U(3)$ chiral effective Lagrangian combined with a relativistic coupled-channel treatment of the final-state interactions~\cite{Borasoy:2005du}; this holds in particular for the Dalitz plot coefficient $a$, where the discrepancies are about four standard deviations. 
For  $\eta'\to\eta\pio\pio$, the results are in general consistent with previous measurements and the predictions within uncertainties. A discrepancy of 2.6 standard deviations for the parameter $a$ between the $\eta'\to\eta\pip\pi^-$ and $\eta'\to\eta\pio\pio$ modes is seen.
The significance is not sufficient to conclusively establish isospin violation. Effects like radiative corrections~\cite{Kubis:2009sb} and the $\pi^+/\pi^0$ mass difference should be considered consistently to study isospin conservation at the amplitude level; cf.\ Ref.~\cite{Isken:2017dkw}.
More recent theoretical analyses of these decays, based on unitarized versions of ChPT~\cite{Escribano:2010wt,Gonzalez-Solis:2018xnw} or dispersion theory with free subtraction constants~\cite{Isken:2017dkw}, all require experimental Dalitz plot information as input to fix various parameters. 

The most prominent isospin-violating feature occurs in the $\eta'\to\eta\pi^0\pi^0$ Dalitz plot and is due to $\pip\pim\to\pio\pio$ $S$-wave rescattering, which produces a cusp at the $\pi^+\pi^-$ threshold~\cite{Kubis:2009sb}.  This effect is analogous to the one observed with very high precision in $K^+\to\pi^+\pi^0\pi^0$ decays, where it has been used to extract a combination of pion--pion scattering lengths~\cite{Batley:2000zz}, based on a nonrelativistic EFT framework~\cite{Colangelo:2006va,Bissegger:2008ff,Gasser:2011ju} (cf.\ also Refs.~\cite{Cabibbo:2004gq,Cabibbo:2005ez}).
A search for the  cusp in  $\eta'\to\eta\pi^0\pi^0$ is performed by inspecting the $\pi^0\pi^0$ invariant-mass spectrum close to $2m_{\pi^+}$.  The A2 collaboration~\cite{Adlarson:2017wlz} reports evidence for the cusp with statistical significance of $\sim 2.5\sigma$, while BESIII~\cite{Ablikim:2017irx} does not observe a statistically significant effect.
A simultaneous direct fit of the dispersive amplitudes from Ref.~\cite{Isken:2017dkw} to the Dalitz plot data for the two decay modes would provide more precise information and in particular an improved prediction for the cusp effect.

\paragraph{\boldmath $\eta' \to 3\pi$} 
The Sutherland theorem~\cite{Sutherland:1966zz,Bell:1996mi} that ensures the suppression of electromagnetic effects as the source of isospin breaking in $\eta\to3\pi$ is equally applicable in the analogous decay $\eta'\to3\pi$,
which therefore depends similarly on the light quark mass difference $m_u-m_d$.
It was hypothesized early~\cite{Gross:1979ur} that the decay amplitude for
$\eta'\to \pi^+\pi^-\pi^0$ may be understood in terms of the dominating  $\eta^\prime\rightarrow\pi^+\pi^-\eta$ mode, followed by 
$\pi^0$--$\eta$ mixing. This would offer a straightforward possibility for an independent quark mass determination from the branching fraction ratio of the two processes. However, the theoretical assumptions underlying such a picture are far too simplistic~\cite{Borasoy:2006uv}, in particular the one of ``essentially constant'' decay amplitudes across the Dalitz plot; the decays $\eta'\to3\pi$ are strongly affected by resonances in the final state. 

\begin{figure}[t]
\centering
\includegraphics[width=0.98\textwidth]{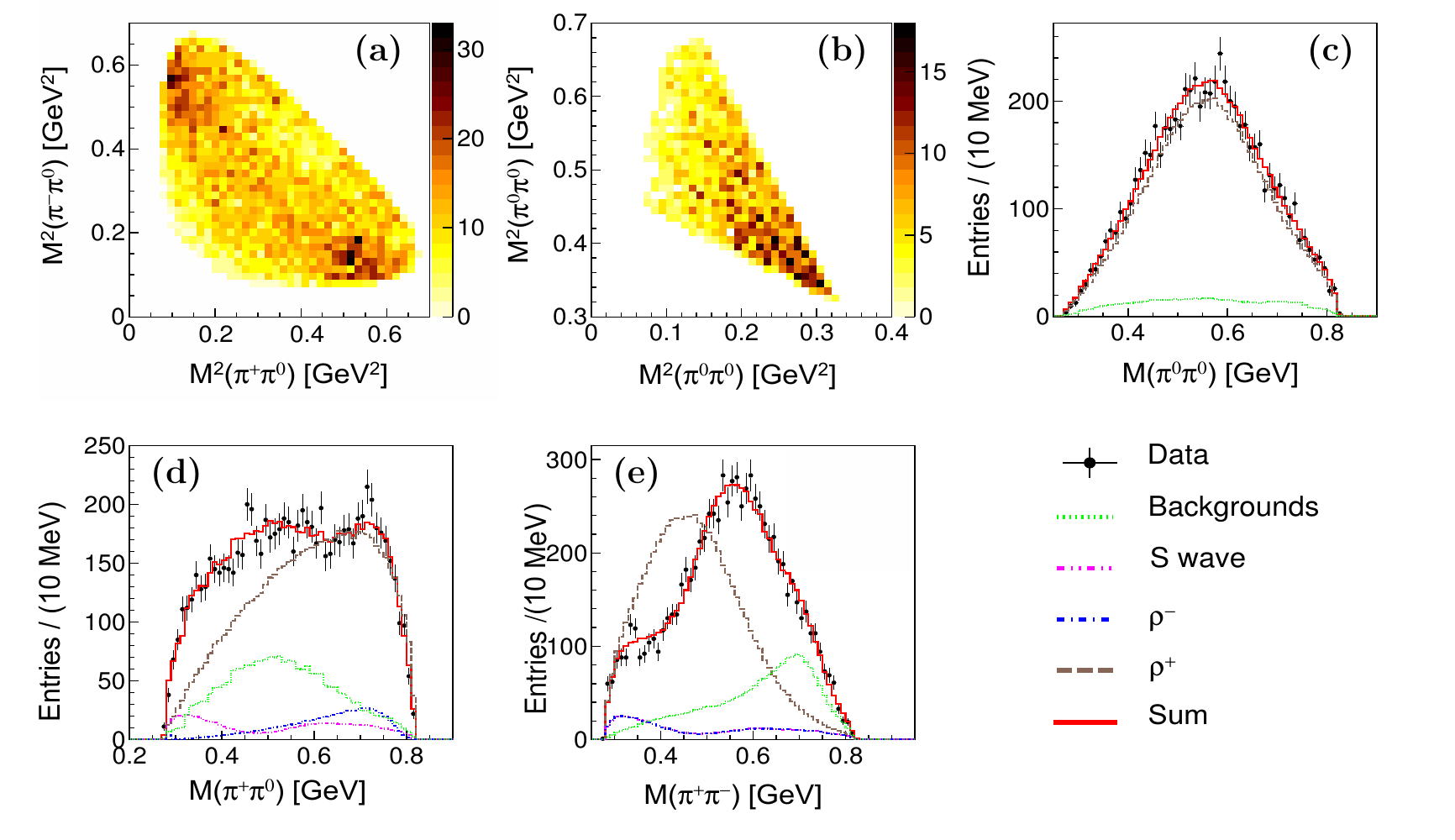}
\caption[Results for $\eta'\to 3\pi$ from BESIII]{Dalitz plots for (a)
$\eta'\to\pip\pim\pio$ and (b) $\eta'\to3\pio$
    from the BESIII analysis~\cite{Ablikim:2016frj}. Two-pion invariant-mass distributions for  (c)  $\eta'\to3\pio$ and (d)+(e) $\eta'\to\pip\pim\pio$ are also shown. The experimental distributions are compared to the projections of the amplitude fit. The  solid line is the sum of all contributions.
	 The dotted, dashed, dashed--dotted, and dashed--dot-dotted histograms show
	 the contributions from background, $S$-wave, $\rho^-$, and $\rho^+$, respectively.
\label{fig:etap3pi}}
\end{figure}
The Dalitz plot for $\eta'\to \pi^+\pi^-\pi^0$ constructed from BESIII data shows clear $P$-wave contributions, with $\eta'\to\rho^\pm\pi^\mp$ visible as the two clusters in Fig.~\ref{fig:etap3pi}(a).
The first amplitude analysis of the decays  $\eta'\to\pi^+\pi^-\pi^0$ and $\eta'\to3\pi^0$ was performed at BESIII using an isobar model~\cite{Ablikim:2016frj}. The fit results are illustrated by the invariant-mass spectra of $\pip\pi^-$, $\pip\pio$, and $\pi^-\pio$ in Fig.~\ref{fig:etap3pi}(d)--(e). A significant $P$-wave contribution from $\eta'\to\rho^{\pm}\pi^{\mp}$ in $\eta'\to\pip\pi^-\pio$ is needed.  In addition to a nonresonant $S$-wave, the resonant $\pi\pi$ $S$-wave with a pole at $512(15)-i\,188(12)\MeV$, interpreted as the broad $f_0(500)$ meson, plays an essential role
in the description of the $\eta'\to3\pi$ decays. 
The (unphysical) separation of the nonresonant and resonant $S$-wave contributions is hindered by the large interference between the two. 
A consistent dispersion-theoretical interpretation with universal $\pi\pi$ final-state interactions is still in progress~\cite{Isken:2019tdj}. 

The Dalitz plot for $\eta'\to3\pio$ is shown in Fig.~\ref{fig:etap3pi}(b) (due to the symmetry only one hexagonal section is shown) and a projection of the amplitude fit is displayed in Fig.~\ref{fig:etap3pi}(c). A significant resonant $S$-wave contribution is necessary to provide an explanation for the negative slope parameter $\alpha = -0.640(66)$ of the $\eta'\to3\pio$ Dalitz plot~\cite{Ablikim:2015cmz}. The value deviates significantly from zero, which implies that final-state interactions play an important role. The ratio between the $S$-wave components of the two decay modes, $\BR(\eta'\to3\pio)/\BR(\eta'\to\pip\pi^-\pio)_S$, is determined to be $0.94(32)$, where the common systematics cancel. 

\begin{figure}
    \centering
    \includegraphics[width=0.8\textwidth]{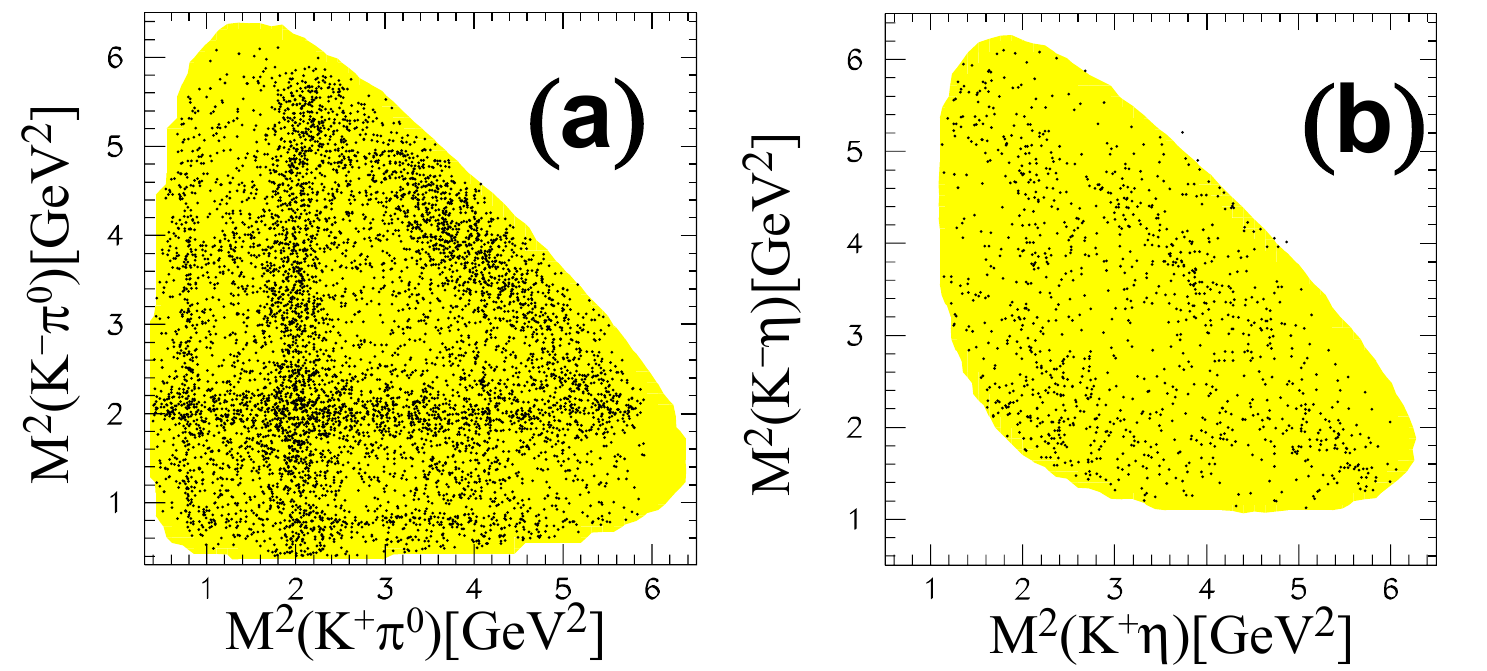}
    \caption[Dalitz plot for $\eta_c$ decays]{Dalitz plots for (a)  $\eta_c \to K^+ K^- \pi^0$ and (b)  $\eta_c \to K^+ K^- \eta$~\cite{Lees:2014iua}.}
    \label{fig:EtacKKP}
\end{figure}

\begin{figure}[t]
    \centering
    \includegraphics[width=0.9\textwidth]{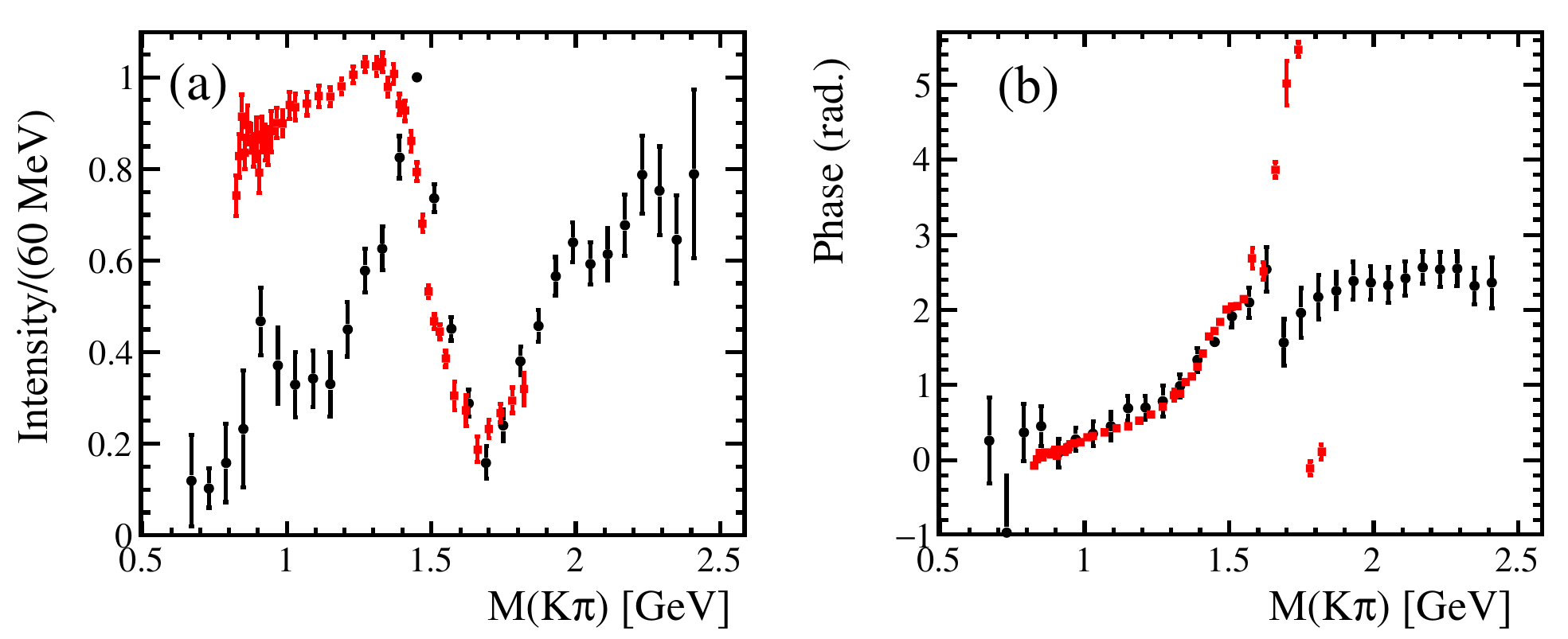}
    \caption[$I=1/2$ $K \pi$  $S$-amplitude]{The $I=1/2$ $K \pi$  $S$-wave amplitude extracted from the $\eta_c\to K_SK^\pm\pi^\mp$ BaBar Dalitz plot analysis~\cite{Lees:2015zzr} compared to the results from the LASS experiment~\cite{Aston:1987ir}.}
    \label{fig:BaBarEtac}
\end{figure}

\paragraph{\boldmath{$\eta_c \to K \bar K \pi$ and $\eta_c \to K^+ K^- \eta$}}
The BaBar analysis of the  $K^+ K^- \eta$ and $K \bar K \pi$ systems originating from $\eta_c$ decays~\cite{Lees:2014iua,Lees:2015zzr} is an interesting example of the use of the two-photon production mechanism to study meson dynamics at electron--positron colliders, with data samples of $4520(140)$ events for $\eta_c \to K^+ K^- \pi^0$,  $8260(110)$ for $\eta_c \to K_S K^\pm \pi^\mp$, and $1145(44)$ 
for $\eta_c \to K^+ K^- \eta$ (combined $\eta\to\gamma\gamma$ and $\eta\to\pi^+\pi^-\pi^0$ channels). The Dalitz plots are shown in Fig.~\ref{fig:EtacKKP}.
In particular, a contribution of the $K^*_0(1430) \to K \eta$ decay is observed. 
Dalitz plot analyses of the two final states were performed using a model-independent partial-wave approach for the $I=1/2$ $K \pi$  $S$-wave amplitude from Ref.~\cite{Aitala:2005yh}. In such analyses all other channels and partial waves are expressed as in the isobar-model approach. The model-independent component is extracted for each  $M(K\pi)$ invariant-mass bin as a complex number represented by the intensity and phase. The results are shown in Fig.~\ref{fig:BaBarEtac}. A comparison with the LASS experiment indicates similar behavior for the phase up to a mass of $1.5\GeV$. In contrast, the intensities show marked differences. The data requires the presence of a new $a_0(1950)$ resonance with parameters $m=1931(26) \MeV$ and $\Gamma=271(36) \MeV$. 

\begin{figure}[t]
    \centering
    \includegraphics[width=0.4\textwidth]{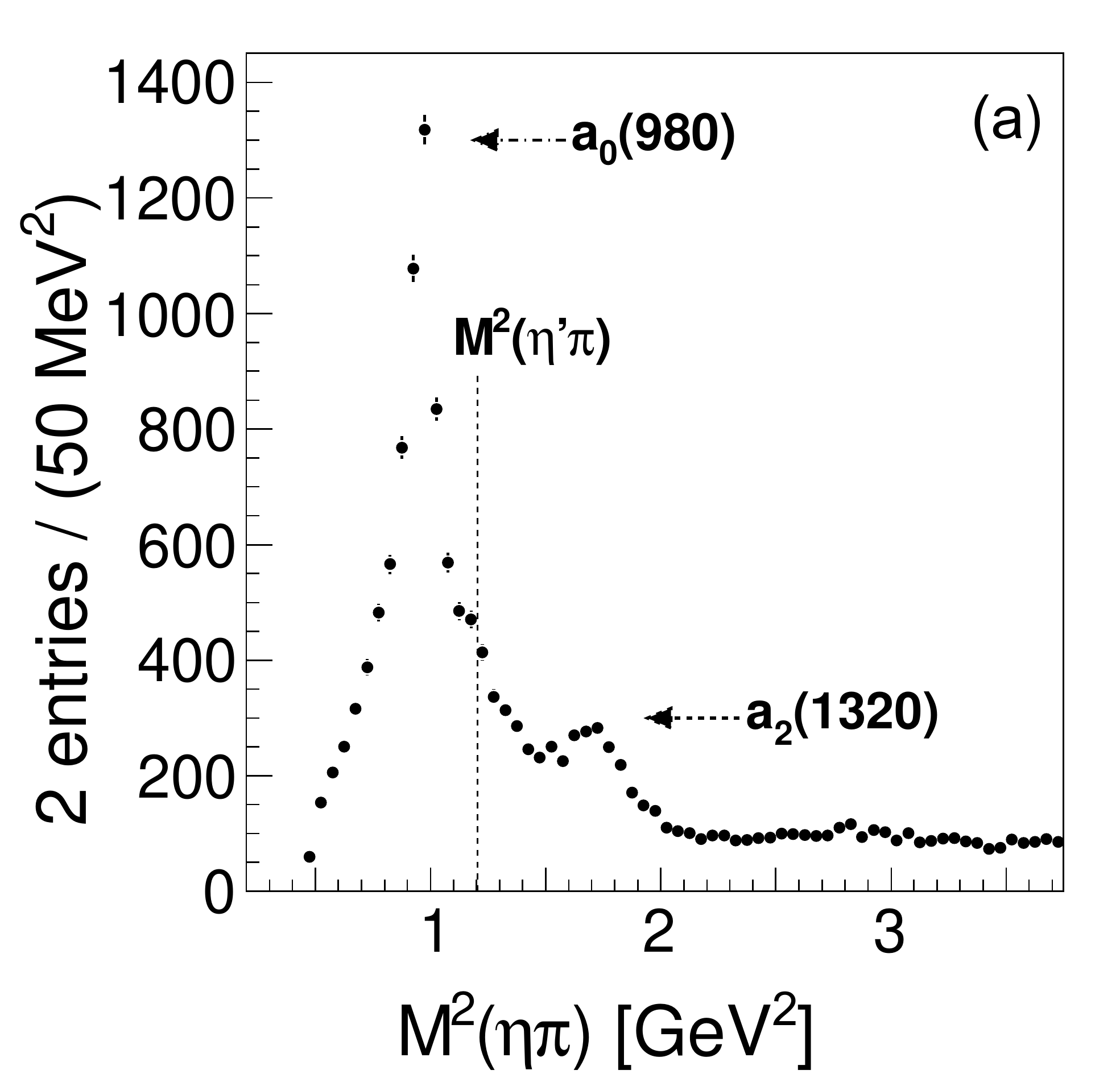}
    \includegraphics[width=0.4\textwidth]{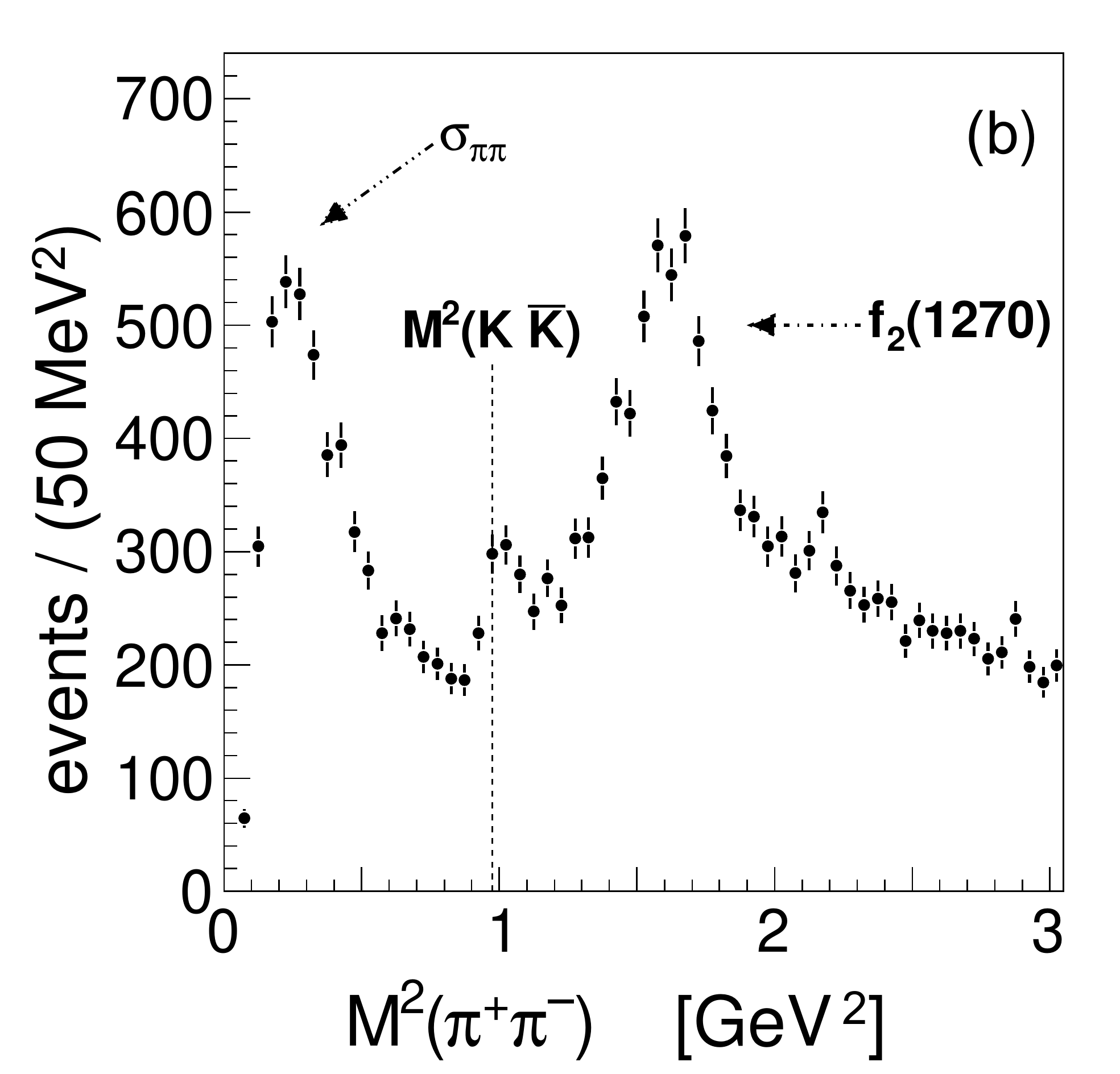}
    \caption[Dalitz plot projections for $\chi_{c1}\to\eta\pi^+\pi^-$]{Dalitz plot projections for $\chi_{c1}\to\eta\pi^+\pi^-$: (a) $M^2(\eta\pi)$; (b) $M^2(\pi\pi)$. The relevant thresholds are indicated and the main contributions to the spectrum listed.}
    \label{fig:chi1c}
\end{figure}

\paragraph{\boldmath{$\chi_{c1}$ decays }}
Other nonvector states that are useful for the study of meson dynamics in three-body decays are  $\chi_{cJ}$ states produced in radiative decays of the $\psi'$ at BESIII. In $\chi_{c1}\to\eta'K^+K^-$~\cite{Ablikim:2014tww}, 529(26) events are observed and used for an amplitude analysis. Intermediate processes $\chi_{c1}\to\eta' f_0(980)$, $\chi_{c1}\to\eta' f_0(1710)$, $\chi_{c1}\to\eta' f_2'(1525)$, and $\chi_{c1}\to K^*_0(1430)^{\pm}K^{\mp}$ ($K^*_0(1430)^{\pm}\to\eta' K^{\pm}$) are observed with statistical significances larger than $5\sigma$.
The decay $\chi_{c1}\to\eta\pi^+\pi^-$ was studied  using 32920(180) events~\cite{Kornicer:2016axs}. The dominant two-body structure is $a_0(980)^{\pm}\pi^{\mp}$ ($a_0(980)^{\pm}\to\eta\pi^{\pm}$). The $a_0(980)$ line shape is modeled using a dispersive approach, and a significant nonzero $a_0(980)$ coupling to the $\eta^{\prime}\pi$ channel is measured for the first time. The ratio of the couplings of $a_0(980)\to\eta'\pi$ and  $a_0(980)\to\eta\pi$ is expected to be large, since it is driven by $\eta$--$\eta'$ mixing.  In the Dalitz plot projection shown in Fig.~\ref{fig:chi1c}(a), the effect of the $\eta'\pi$ coupling is seen as a shoulder at $(m_{\eta'}+m_\pi)^2$. 
In addition, a $\chi_{c1}\to a_2(1700)\pi$ subprocess is identified.

\subsection[Weak decays $D\to P_1P_2P_3$ and $\tau\to  P_1P_2P_3\nu_\tau$]{\boldmath Weak decays $D\to P_1P_2P_3$ and $\tau\to  P_1P_2P_3\nu_\tau$}\label{sec:weak3P}
We have already argued that $\tau$ decays provide a unique place to study the dynamics of light mesons.  This is especially the case for decay modes into three pseudoscalars. 
Some of the branching fractions for these decays are given in Table~\ref{tab:taudecay3}. 
The hadronic systems not accessible in single-photon processes are produced by the elementary $V-A$ current, and the invariant-mass dependence can be studied just in one experiment due to the presence of the $\nu_\tau$, which carries away part of the energy. The three-pion decay modes probe the axial-vector-meson spectrum and are dominated by the $a_1(1260)$ meson, with the subsequent decay dominantly into $\rho\pi$~\cite{Tsai:1971vv}. The two final channels $\pi^+\pi^-\pi^-$ and  $\pi^0\pi^0\pi^-$ add up to 20\% of all $\tau$ decays.
\begin{table}[t]
    \caption{Example decay modes of the $\tau$ lepton into three light pseudoscalars (+ neutrino)~\cite{PDG}.}
    \label{tab:taudecay3}
    \renewcommand{\arraystretch}{1.3}
    \begin{center}
    \begin{tabular}{lr}
    \toprule
    Decay mode& $\BR$\\ \midrule
      $\tau^-\to \pi^-\pi^0\pi^0\nu_\tau$      & $9.26(10)\times10^{-2}$\\
      $\tau^-\to \pi^-\pi^+\pi^-\nu_\tau$      & $9.31(5)\times10^{-2}$\\
      $\tau^-\to \pi^-\bar K^0\pi^0\nu_\tau$      & $3.82(13)\times10^{-3}$\\
      $\tau^-\to K^-\pi^+\pi^-\nu_\tau$     & $3.45(7)\times10^{-3}$\\
      $\tau^-\to K^-K^0\pi^0\nu_\tau$      & $1.50(7)\times10^{-3}$\\
      $\tau^-\to\pi^- K^0\bar K^0\nu_\tau$      & $1.55(24)\times10^{-3}$\\
      $\tau^-\to\pi^- K^+K^-\nu_\tau$      & $1.496(33)\times10^{-3}$\\
      $\tau^-\to\eta\pi^-\pi^0\nu_\tau$      & $1.39(7)\times10^{-3}$\\
      \bottomrule
    \end{tabular}
    \end{center}
    \renewcommand{\arraystretch}{1.0}
\end{table}
\begin{figure}
    \centering
 \includegraphics[width=0.45\textwidth]{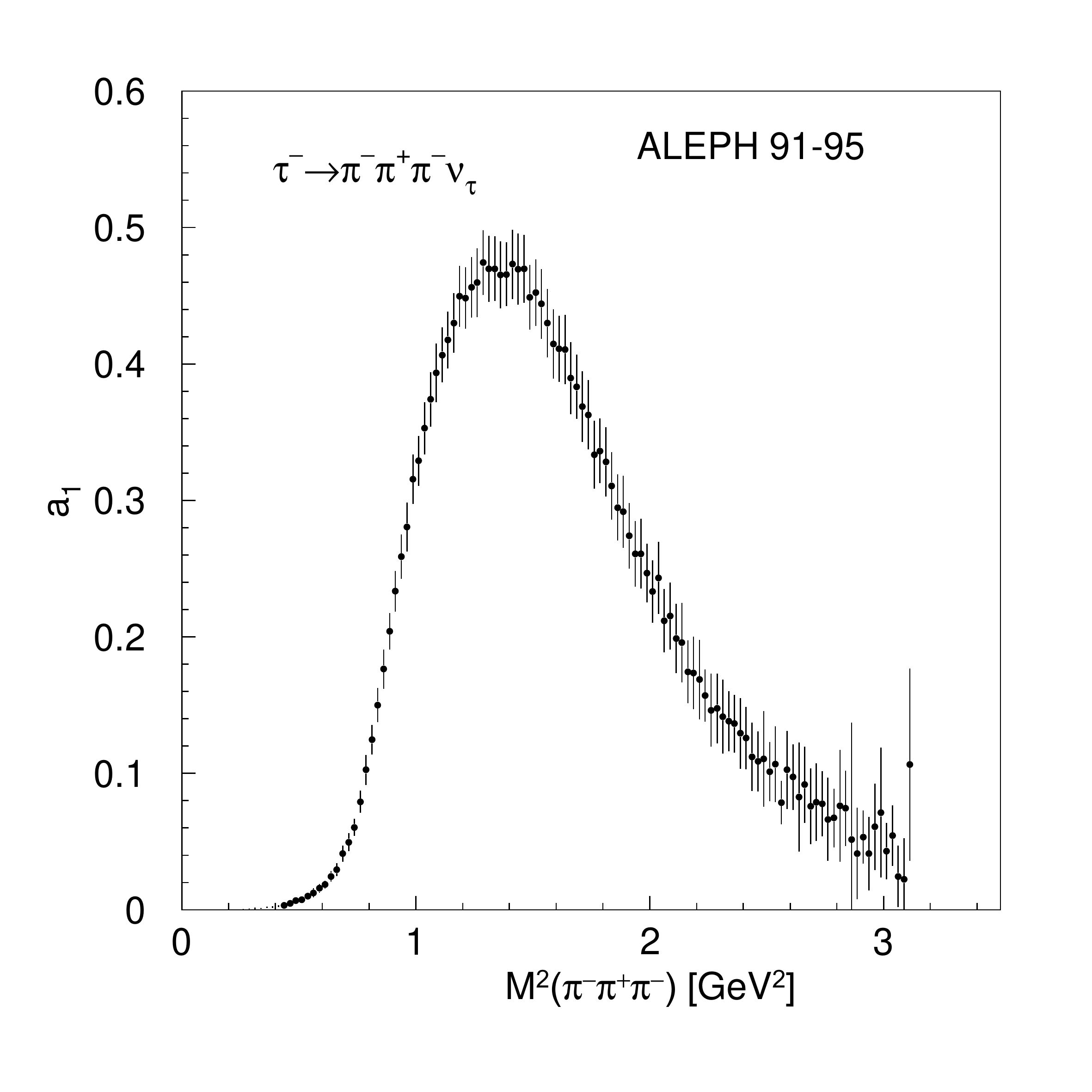}
    \caption[Spectral function for three pion]{Spectral function $a_1$ for $\tau^-\to\pi^-\pi^+\pi^-\nu_\tau$ from ALEPH~\cite{Schael:2005am}.}
    \label{fig:tauA1}
\end{figure}%
The decay rate 
${\diff\Gamma}/{\diff s}$ in $\tau^-\to\pi^-\pi^{+}\pi^{-}\nu_\tau$ is expressed using the axial-vector spectral  function $a_1(s)$, related by the same kinematic prefactor as in Eq.~\eqref{eq:dVtau}. In Fig.~\ref{fig:tauA1}, the $a_1(s)$ spectral function from the ALEPH experiment~\cite{Schael:2005am} is shown. The peak is almost exclusively due to the $a_1(1260)$ meson contribution. The most recent results on the dynamics in the three-meson decays of $\tau$ leptons are from the Belle~\cite{Lee:2010tc} and BaBar~\cite{Nugent:2013ij} experiments.
Theoretical analyses have been performed both using chiral Lagrangians with resonances~\cite{Nugent:2013hxa,Sanz-Cillero:2017fvr} and models based on unitarity~\cite{Lorenz:2017svg,Mikhasenko:2018bzm}. 

Weak three-body decays of $D$ mesons $D\to P_1P_2P_3$ are so far much better explored at \CT\ factories. 
In the isobar model the amplitude for the weak decay $D\to P_1P_2P_3$ can be written as 
\begin{align}
    \bra{P_1P_2P_3}H_{S+W}\ket{D}&=\sum_{R_1}\bra{P_2P_3}H_{S} \ket{R_1}\bra{R_1P_1}H_{W}\ket{D}+
    \sum_{R_2}\bra{P_1P_3}H_{S} \ket{R_2}\bra{R_2P_2}H_{W}\ket{D} \notag\\
    &+\sum_{R_3}\bra{P_1P_2}H_{S} \ket{R_3}\bra{R_3P_3}H_{W}\ket{D} \,,
\end{align}
where $H_{W}$ and $H_{S}$ refer to a weak and a strong transition, respectively. $R_1$ represents a meson resonance in the 
$P_2P_3$ system etc.  However, the  model does not include the most general form of final-state interactions between the three pseudoscalars. 

The two $D^+\to \bar K\pi\pi^+$ decays demonstrate why rescattering beyond the simplest isobar model leads to essential additional insights:
$D^+\to K^-\pi^+\pi^+$ and $D^+ \to K^0_S \pi^+ \pi^0$ are coupled to each other by charge exchange, and the various subamplitudes obey isospin symmetry; however, only the latter includes direct contributions of resonances in the $\pi\pi$ system (most prominently the $\rho(770)$), while $\pi^+\pi^+$ in the former necessarily has isospin 2 and is hence nonresonant.  

The extensive data base on $D^+\to K^-\pi^+\pi^+$ by E791~\cite{Aitala:2005yh}, FOCUS~\cite{Pennington:2007se,Link:2009ng}, and CLEO~\cite{Bonvicini:2008jw} has spurred several theoretical analyses, many of them focusing on the $K\pi$ $S$-wave amplitudes therein~\cite{Oller:2004xm,Boito:2009qd,Boito:2017jav}.  In Refs.~\cite{Magalhaes:2011sh,Nakamura:2015qga}, nontrivial rescattering effects beyond two-body interactions were considered theoretically, using a loop model and Faddeev equations, respectively.  A dispersion-theoretical approach based on Khuri--Treiman equations has been worked out in Ref.~\cite{Niecknig:2015ija}, and subsequently applied to a simultaneous analysis of $D^+\to K^-\pi^+\pi^+$ with the BESIII Dalitz plot data on $D^+ \to K^0_S \pi^+ \pi^0$~\cite{Ablikim:2014cea} in Ref.~\cite{Niecknig:2017ylb}.  The number of partial waves and subtraction constants, as well as the heightened importance of inelastic effects in the two-body subsystems in parts of the Dalitz plots (which are not included in the theoretical formalism), make the dispersive representation and the control of the various phase motions not quite as strict as in the low-energy decays such as $\eta,\omega,\phi\to3\pi$ or $\eta'\to\eta\pi\pi$ described earlier.  However, Refs.~\cite{Niecknig:2015ija,Niecknig:2017ylb} still demonstrate that three-body rescattering effects induce additional phase motion that, e.g., make the $K\pi$ $I=1/2$ $S$-wave rise more steeply than the corresponding elastic phase shift.  This has been observed in experimental analyses before~\cite{Aitala:2005yh,Link:2009ng}, but does not violate Watson's theorem: the presence of the third meson taking part in the final-state interactions allows us to explain these modifications.

In the long run, the efforts to develop physically sound descriptions of $D$-decay Dalitz plots are motivated by future studies of $CP$ violation in the charm sector, where the hope is that the resonance-rich environment of three-body decays with rapid, but ultimately controllable strong phase variation may (locally) enhance the supposedly very small weak $CP$ phases from the CKM matrix.

\section{Summary and perspectives}\label{sec:outlook}

To date, about seven decades since the discovery of the first light mesons (the pion and the kaon), studies of light-meson interactions continue to provide opportunities for a variety of physics at low energy scales, including precision tests of effective theoretical models, investigations of the quark structure of the light mesons, tests of fundamental symmetries, and searches for new physics beyond the Standard Model. With the advantages of high production rates and the excellent performance of the detectors, 
in this review we have tried to give a broad overview of the existing data from $e^+e^-$ colliders, which can be used to investigate systems of light mesons. 
The massive progress in precision studies in recent years is driven to a large extent by the need to understand hadronic backgrounds for low-energy tests of the Standard Model and for the interpretation of the many new exotic states.
Meanwhile, the theoretical advancement of rigorous tools, such as those based on dispersion-theoretical methods and effective field theories, for providing a good description of experimental data are highlighted. 

In order to better understand and appreciate the progress in the field, it is illuminating to compare the physics focus of the present review to the snapshot of the physics at electron--positron colliders in 1976, as described in Ref.~\cite{PerezYJorba:1977rh}. 
This was the time after the discoveries of the first charmonium states, with some observations hinting at the existence of charmed mesons and the $\tau$ lepton. 
In the beginning of the 1990s, the $J/\psi$ factories were motivated mainly by the search for QCD exotics, more specifically for gluonium states~\cite{Kopke:1988cs}. 
The physics of light mesons was described in terms of VMD and quark models.

Some of the strongest motivation for, and strongest impact of, the high-precision, high-statistics data taken at $e^+e^-$ data taken in the new millennium has been on precision observables: hadronic contributions to $(g-2)_\mu$ are the most obvious example, with the almost direct determination of hadronic vacuum polarization contributions, and the new strategies to constrain hadronic light-by-light scattering therefrom~\cite{Aoyama:2020ynm}.  
Maybe less obvious is the strong role of Dalitz plot studies of unprecedented accuracy, which in the case of $\eta\to3\pi$ has lead to the most precise phenomenological determination of the ratios of the light quark masses~\cite{Anastasi:2016cdz,Colangelo:2018jxw}.  Both of these examples have pushed the development of theoretical precision tools to describe form factors and scattering amplitudes, 
using combinations of dispersion-theoretical methods and QCD constraints as embodied in ChPT.
These developments allow us, in turn, to improve on the input for many of these dispersive studies, the elementary meson--meson phase shift information: the pion vector form factor measurements provide the sharpest experimental information available on the pion--pion $P$-wave phase shift, thanks to the universal relation between scattering and production~\cite{Colangelo:2018mtw}.  Precisely in $\pi\pi$ $P$-wave systems, the modification of the spectral form of the $\rho(770)$ is now well understood, e.g., in radiative $\eta'$ decays~\cite{Hanhart:2016pcd,Ablikim:2017fll}, using the same dispersive techniques, and thus paving the way towards a consistent high-precision extraction of $\rho$-pole properties from many different reactions~\cite{PDG}.
The two-meson radiative decays of $\phi$, $J/\psi$, and $\psi'$ may take a similar role for even partial waves of pairs of light pseudoscalars in the future; they have already provided crucial arguments for the modern interpretation of the scalar states.

Clearly, the theoretical progress varies a lot, depending on the specific processes: many-body interactions, scattering beyond the ground-state pseudoscalars, and inelastic reactions including many coupled channels still pose severe challenges to a truly rigorous theoretical treatment.  But also here, new efforts and developments are being triggered by new high-precision data.
The collider data allowed for the development and implementation of dispersive methods for three-pion final states discussed in Secs.~\ref{sec:ee-PPP} and \ref{sec:strong3p}, due to the relatively simple partial-wave structure.  These may be extended to more complicated three-pion systems in the future~\cite{Adolph:2015tqa,Ketzer:2019wmd}.
Also, form factors traditionally thought to be described sufficiently well using VMD approaches, such as vector--pseudoscalar transition form factors, now show deviations from such a simplistic picture and require more advanced theoretical tools; see Sec.~\ref{sec:eePV}. 
An increasing number of modern theory approaches to describe light-meson interactions 
includes frameworks to estimate uncertainties and to improve calculations in a systematic way.  And while several of the abovementioned problems are still out of reach for present-day lattice QCD calculations, we observe a rapid development of methods and algorithms for vastly improved ab initio studies of meson--meson interactions.

Despite this impressive progress, light-hadron physics is still a rich field to be explored and there remains considerable room for improvement of experimental precision.
With the abundant and clean event samples accumulated at the $e^+e^-$ colliders, light-meson interactions will continue to be studied at facilities such as CMD-3, KLOE-2, BESIII, and SND.
Taking BESIII for example,  a data sample of $10^{10}$ $J/\psi$ events is now available, eight times larger than the one used in the present publications, which offers great additional opportunities for research in light-meson decays, especially for pseudoscalar and vector mesons, with unprecedented precision. 
Moreover,  the presently operating facilities BEPCII and VEPP2000 have plans for new data to be collected. Particularly interesting from our perspective is BESIII: $20\fb^{-1}$ at $\psi(3770)$ for $D\bar D$ production and the full data set of $10^{9}$ $\psi'$ resonance events. The physics goals and plans for the further BESIII experiments are discussed in Refs.~\cite{Asner:2008nq,Ablikim:2019hff}. These experiments require at most minor improvements in the accelerator or in the detector and will be carried out in the coming few years. In addition, the huge data samples at the new $B$ factory, Belle-II, at SuperKEKB will also support investigations of light-meson interactions with the different opportunities of ISR and two-photon production techniques. In general,   
more surprises are expected to be produced at these precision experiments, which will further strengthen our present paradigms of light-hadron physics.
Complementary measurements are possible in experiments at hadron and photon beam machines: not only MAMI at Mainz university~\cite{Denig:2016dqo}, CLAS~\cite{Mecking:2003zu}, GlueX~\cite{Lawrence:2009zz}, and the planned KLF~\cite{Amaryan:2020xhw} experiments at JLab, LHCb~\cite{Alves:2008zz} and COMPASS~\cite{Ketzer:2019wmd} experiments at CERN, but also HESR at FAIR~\cite{Lutz:2009ff}, J-PARC at KEK~\cite{Sato:2009zze}, and others.
 
Experimentally, the future investigations of light-meson interactions, in particular the search for new physics, demand the creation of new-generation facilities that have excellent performance and can obtain experimental data of much higher statistics and precision. Indeed, in recent years there are discussions to construct super $\tau-$charm factories (SCTFs) in Novosibirsk~\cite{Bondar:2013cja} and in China~\cite{Peng:2020orp} to extensively explore the $\tau$-charm physics, such as charmonium decays, charm-meson physics, hadron spectroscopy, and the so-called $XYZ$ particles.  Naturally, these main goals of the experimental programs are connected with studies of light-hadron physics as we perform them at the data-taking experiments of BESIII and Belle-II.
With the advantages of high luminosity, about two orders of magnitude larger than BEPCII, and other new techniques such as polarization of beams and an energy spread compensation scheme,
the SCTFs will give access to previously unexplored regions and bring the study of light-meson interactions into a very-high-precision era. Many new precision tests of the Standard Model in the $s$- and $c$-quark systems will be possible. Examples for (open or hidden) strangeness systems are $\eta$, $\eta'$, and hyperon decays. For $c$-quarks, entangled systems of $D\bar D$ mesons and ground-state charmed baryons are of high interest. In addition, for the SCTF projects electron beam polarization of up to 80\% is considered. 
The polarization can be utilized in $\tau$-decay studies for the determination of structure functions~\cite{Kuhn:1992nz} and for $CP$ symmetry tests~\cite{Tsai:1994rc}. 
These new experimental activities in the low-energy region also strongly call for more detailed theoretical studies in this area, which will play an important role in the developments of chiral effective field theory and  lattice QCD, and make significant contributions to our understanding of hadron physics in the nonperturbative regime. 

\section*{Acknowledgements}
This publication is part of a project that has received funding from the European Union's Horizon 2020 research and innovation programme under grant agreement STRONG-2020 -- No 824093.
This project has received funding from the National Natural Science Foundation of China (NSFC) and the Deutsche Forschungsgemeinschaft (DFG) through the funds provided to the Sino--German Collaborative Research Center CRC 110 ``Symmetries and the Emergence of Structure in QCD'' (DFG -- Project-ID 196253076 -- TRR 110),
and by the Munich Institute for Astro- and Particle Physics (MIAPP), which is funded by the DFG under Germany's Excellence Strategy -- EXC-2094 -- 390783311. This work is also supported in part by the NSFC under Contracts No.~11735014 and No.~11675184. AK acknowledges a grant of the Chinese Academy of Science President's
International Fellowship Initiative (PIFI) for Visiting Scientists.
 We thank our colleagues for sharing the results that were included in this review, and apologize for all those that could not be discussed any more. 
\bibliographystyle{elsarticle-num}
\bibliography{base,kloe,bes,Novosibirsk,phasedb}

\end{document}